\documentclass[twoside]{report}

\usepackage{theorem}
\usepackage{amsmath}
\usepackage[psamsfonts,mathcal]{eucal}
\usepackage[psamsfonts]{amssymb}
\usepackage{latexsym}
\usepackage{cmmib57}
\usepackage[small,center,compact,outermarks]{titlesec}
\usepackage{a4}
\usepackage{amscd}
\usepackage{graphicx}
\usepackage{pstricks}
\usepackage{pst-coil}
\usepackage{pst-node}
\usepackage{float}
\usepackage{wrapfig}
\usepackage{epsf}

\newgray{meascolor}{.5}
\newgray{opercolor}{.5}
\newgray{entacolor}{.5}
\newgray{qbitcolor}{.1}
\newgray{mklight}{.7}
\newgray{mkfill}{.9}
\psset{unit=\unitlength}
\psset{arrowsize=2pt 5}
\psset{coilarm=.4}

\newcommand{\goth}[1]{\mathfrak{#1}}
\newcommand{\scr}[1]{\mathcal{#1}}
\def\idty{{\leavevmode{\rm 1\ifmmode\mkern -4.8mu\else\kern -.3em\fi
      I}}}
\renewcommand{\Bbb}[1]{\if1#1\idty\else\mathbb{#1}\fi}
\renewcommand{\tilde}{\widetilde}
\renewcommand{\hat}{\widehat}
\renewcommand{\Re}{\operatorname{Re}}
\renewcommand{\Im}{\operatorname{Im}}
\newcommand{\kb}[1]{|#1\rangle\langle#1|}
\newcommand{\KB}[2]{|#1\rangle\langle#2|}
\newcommand{\ket}[1]{|#1\rangle}
\newcommand{\tr}{\operatorname{tr}}
\newcommand{\supp}{\operatorname{supp}}

\newcommand{\Lz}{\operatorname{L}^2}

\newcommand{\SP}{\operatorname{span}}
\newcommand{\CCP}{\mathbf{P}}
\newcommand{\CCNP}{\mathbf{NP}}
\newcommand{\CCBPP}{\mathbf{BPP}}
\newcommand{\CCBQP}{\mathbf{BQP}}
\newcommand{\diag}{\operatorname{diag}}

\newcommand{\U}{\operatorname{U}}
\newcommand{\Or}{\operatorname{O}}
\newcommand{\SU}{\operatorname{SU}}
\newcommand{\Sym}{\operatorname{S}}

\newcommand{\fac}{\operatorname{fac}}
\newcommand{\Id}{\operatorname{Id}}
\newcommand{\Evn}{E_{\rm vN}}
\newcommand{\ED}{E_{\rm D}}
\newcommand{\EC}{E_{\rm C}}
\newcommand{\EOF}{E_{\rm F}}
\newcommand{\ER}{E_{\rm R}}

\newcommand{\pure}{{\rm pure}}
\newcommand{\all}{{\rm all}}
\newcommand{\tUUbar}{{\rm U\bar{U}}}
\newcommand{\tUU}{{\rm UU}}
\newcommand{\tOO}{{\rm OO}}
\newcommand{\cb}{{\rm cb}}

\newcommand{\mmod}{{\rm mod}}

\newcommand{\Cite}[1]{\cite{#1}}

\titleformat{name=\chapter}[display]{\normalfont\Large\filcenter}
{\titlerule[1pt]\vspace{1pt}\titlerule\vspace{1ex}\Large\chaptername\ \thechapter}{1ex}
{\LARGE\bfseries}[\vspace{1ex}\titlerule]
\titleformat{name=\chapter,numberless}[display]{\normalfont\Large\filcenter}
{\titlerule[1pt]\vspace{1pt}\titlerule}{-2ex}{\LARGE\bfseries}[\vspace{1ex}\titlerule]
\titlespacing{\chapter}{0pt}{-2em}{2em}
\titleformat{\section}[block]{\bfseries\filcenter\large}{\thesection}{1em}{}
\titleformat{\subsection}[runin]{\bfseries\filright}{\thesubsection.}{.5em}{}[.\ ---]

\renewpagestyle{plain}{}
\newpagestyle{main}[\small\sffamily]{
  \headrule
  \sethead[\textsl{\thechapter.\ \chaptertitle}][][\usepage]%
  {\usepage}{}{\textsl{\thesection.\ \sectiontitle}}
  \setfoot[][][]{}{}{}}
\newpagestyle{special}[\small\sffamily]{
  \headrule
  \sethead[\textsl{\chaptertitle}][][\usepage]{\usepage}{}{\textsl{\chaptertitle}}
  \setfoot[][][]{}{}{}}

\theoremheaderfont{\scshape}
\newtheorem{thm}{Theorem}[section]
\newtheorem{defi}[thm]{Definition}
\newtheorem{prop}[thm]{Proposition}

\newtheorem{kor}[thm]{Corollary}
\newenvironment{proof}{\par\noindent\textit{Proof.\ }}{\hfill $\Box$ \vspace{1em}}

\newtheorem{aX}{Axiom} 
\newenvironment{ax}[2]{\renewcommand{\theaX}{#1}\if1#2\begin{aX}\else\begin{aX}[#2]\fi}{\end{aX}}

\begin{document}

\title{Fundamentals of Quantum Information Theory}
\author{Michael Keyl\thanks{electronic mail: \texttt{M.Keyl@TU-BS.DE}}\\
\small
TU-Braunschweig, Inst. Math. Phys, Mendelssohnstra{\ss}e 3, D-38106 Braunschweig}
\date{\today}
\maketitle

\begin{abstract}
  In this paper we give a self contained introduction to the conceptional and mathematical foundations of
  quantum information theory. In the first part we introduce the basic notions like entanglement,
  channels, teleportation etc. and their mathematical description. The second part is focused on a
  presentation of the quantitative aspects of the theory. Topics discussed in this context include:
  entanglement measures, channel capacities, relations between both, additivity and continuity properties
  and asymptotic rates of quantum operations. Finally we give an overview on some recent developments and
  open questions.
\end{abstract}
\clearpage
\tableofcontents
\clearpage

\pagestyle{main}

\chapter{Introduction}

Quantum information and quantum computation have recently attracted a lot of interest. The promise of
new technologies like safe cryptography and new ``super computers'', capable of handling otherwise
untractable problems, has excited not only researchers from many different fields like physicists,
mathematicians and computer scientists, but also a large public audience. On a practical level all these
new visions are based on the ability to control the quantum states of (a small  number of) micro systems
individually and to use them for information transmission and processing. From a more fundamental point
of view the crucial point is a reconsideration of the foundations of quantum mechanics  in an information
theoretical context. The purpose of this work is to follow the second path and to guide physicists into
the theoretical foundations of quantum information and some of the most relevant topics of current
research.  

To this end the outline of this paper is as follows: The rest of this introduction is devoted to a rough
and informal overview of the field, discussing some of its tasks and experimental
realizations. Afterwards, in Chapter \ref{cha:basic-concepts}, we will consider the basic formalism which
is necessary to present more detailed results. Typical keywords in this context are: systems, states,
observables, correlations,  entanglement and quantum channels. We then clarify these concepts (in
particular entanglement and channels) with several examples in Chapter \ref{cha:basic-examples}, and in
Chapter \ref{cha:basic-tasks} we discuss the most important tasks of quantum information in greater
detail. The last three Chapters are devoted to a more quantitative analysis, where we make closer
contact to current research: In Chapter \ref{cha:quant-theory-i} we will discuss how entanglement can be
measured. The topic of Chapter \ref{cha:quant-theory-ii} are channel capacities, i.e. we are looking at
the amount of information which can maximally be transmitted over a noisy channel and in Chapter
\ref{cha:quant-theory-iii} we consider state estimation, optimal cloning and related tasks.

Quantum information is a rapidly developing field and the present work can of course reflect only a small
part of it. An incomplete list of other general sources the reader should consult is: the books of Lo
\Cite{LoBook}, Gruska \Cite{Gruska}, Nielsen and Chuang \Cite{NC}, Bouwmeester et. al. \Cite{BEZBook} and
Alber et. al. \Cite{DPGBook}, the lecture notes of Preskill \Cite{P219} and the collection of references
by Cabello \Cite{QInfLit} which particularly contains many references to other reviews.

\section{What is quantum information?}
\label{sec:impossible-machines}

Classical information is, roughly speaking, everything which can be transmitted from a sender to a receiver
with ``letters'' from a ``classical alphabet'' e.g. the two digits ``0'' and ``1'' or any other finite
set of symbols. In the context of classical information theory, it is completely irrelevant which type of
physical system is used to perform the transmission. This abstract approach is successful because it
is easy to transform information between different types of carriers like electric currents in a wire,
laser pulses in an optical fiber, or symbols on a piece of paper without loss of data; and even if 
there are losses they are well understood and it is known how to deal with them. However, quantum
information theory breaks with this point of view. It studies, loosely speaking, that kind of information
(``quantum information'') which is transmitted by micro particles from a preparation device (sender) to a
measuring apparatus (receiver) in a quantum mechanical experiment -- in other words the distinction
between carriers of classical and quantum information becomes essential. This approach is justified by the
observation that a lossless conversion of quantum information into classical information is in the above
sense not possible. Therefore, quantum information is a \emph{new kind of information}. 

In order to explain why there is no way from quantum to classical information and back, let us discuss
how such a conversion would look like. To convert quantum to classical information we need a device which
takes quantum systems as input and produces classical information as output -- this is nothing else than
a measuring apparatus. The converse translation from classical to quantum information can be rephrased
similarly as ``parameter dependent preparation'', i.e. the classical input to such a device is used to
control the state (and possibly the type of system) in which the micro particles should be
prepared. A combination of these two elements can be done in two ways. Let us first consider a device
which goes from classical to quantum to classical information.  This is a possible task and in fact
technically  realized already. A typical example is the transmission of classical information via an optical
fiber. The information transmitted through the fiber is carried by micro particles (photons) and is
therefore quantum information (in the sense of our preliminary definition). To send classical information
we have to prepare first photons in a certain state send them through the channel and measure an
appropriate observable at the output side. This is exactly the combination of a classical $\to$ quantum
with a quantum $\to$ classical device just described. 

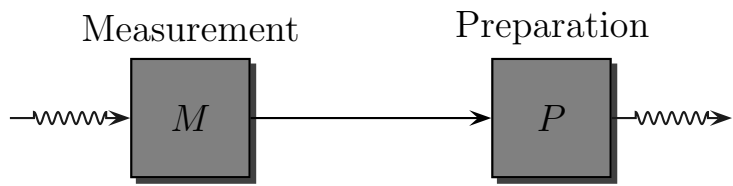
\begin{figure}[h]
  \begin{center}
    \begin{picture}(12,3)
      \Large
      \rput(3,2.5){Measurement}
      \rput(9,2.5){Preparation}
      \psframe[fillcolor=meascolor,fillstyle=solid,shadow=true](2,0)(4,2)
      \psframe[fillcolor=opercolor,fillstyle=solid,shadow=true](8,0)(10,2)
      \pccoil[linecolor=qbitcolor,coilaspect=0,coilheight=1,coilwidth=.2]{->}(0,1)(2,1)
      \psline[linecolor=black]{->}(4,1)(8,1)
      \pccoil[linecolor=qbitcolor,coilaspect=0,coilheight=1,coilwidth=.2]{->}(10,1)(12,1)
      \rput(3,1){$M$}
      \rput(9,1){$P$}
    \end{picture}
    \caption{Schematic representation of classical teleportation. Here and in the following diagrams a
      curly arrow stands for quantum systems and a straight one for the flow of classical information.}
    \label{fig:cl-tele-1}
  \end{center}
\end{figure}

The crucial point is now that the converse composition -- performing the measurement $M$ first and the
preparation $P$ afterwards (cf. Figure \ref{fig:cl-tele-1}) -- is more problematic. Such a process is
called \emph{classical teleportation}, if the particles produced by $P$ are ``indistinguishable'' from
the input systems. We will show the impossibility of such a device via a hierarchy of other ``impossible
machines'' which traces the problem back to the fundamental structure of quantum mechanics. This finally
will prove our statement that quantum information is a new kind of information\footnote{The following
  chain of arguments is taken from   \Cite{Werner01}, where it is   presented in greater detail. This
  concerns in particular the construction of Bell's telephone from a joint measurement, which we have
  omitted here.}.  

\begin{figure}[t]
  \begin{center}
    \begin{picture}(14,6)
      \Large
      \psframe[fillcolor=mklight,fillstyle=solid,shadow=true](3,3)(11,6)
      \psframe[fillcolor=meascolor,fillstyle=solid,shadow=true](4,3.5)(6,5.5)
      \psframe[fillcolor=meascolor,fillstyle=solid,shadow=true](8,3.5)(10,5.5)
      \psframe[fillcolor=meascolor,fillstyle=solid,shadow=true](0,3.5)(2,5.5)
      \psframe[fillcolor=meascolor,fillstyle=solid,shadow=true](12,3.5)(14,5.5)
      \psframe[fillcolor=meascolor,fillstyle=solid,shadow=true](0,0)(2,2)
      \psframe[fillcolor=meascolor,fillstyle=solid,shadow=true](12,0)(14,2)
      \pccoil[linecolor=qbitcolor,coilaspect=0,coilheight=1,coilwidth=.2]{->}(2,1)(12,1)
      \pccoil[linecolor=qbitcolor,coilaspect=0,coilheight=1,coilwidth=.2]{->}(2,4.5)(4,4.5)
      \pccoil[linecolor=qbitcolor,coilaspect=0,coilheight=1,coilwidth=.2]{->}(10,4.5)(12,4.5)
      \psline[linecolor=black]{->}(6,4.5)(8,4.5)
      \rput(5,4.5){$M$}
      \rput(9,4.5){$P$}
      \rput(1,4.5){$P'$}
      \rput(13,4.5){$M'$}
      \rput(1,1){$P'$}
      \rput(13,1){$M'$}
      \rput(7,2){\LARGE $\cong$}
    \end{picture}
    \caption{A teleportation process should not affect the results of a statistical experiment with
      quantum systems. A more precise explanation of the diagram is given in the text.\label{fig:cl-teleB}}
  \end{center}
\end{figure}
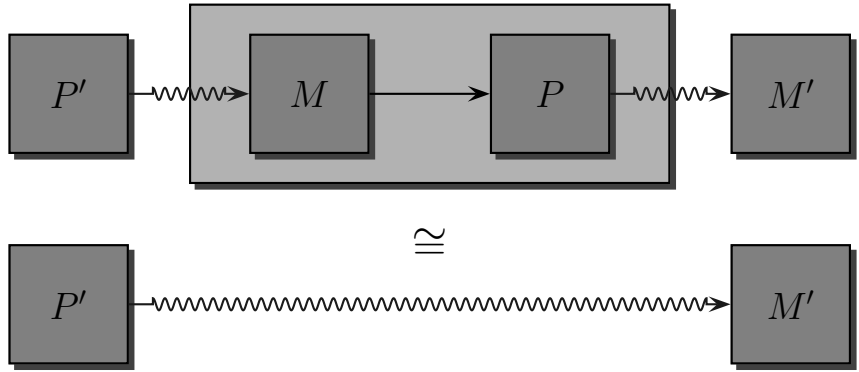

To start with, we have to clarify the precise meaning of ``indistinguishable'' in this context. This has
to be done in a statistical way, because the only possibility to compare quantum mechanical systems is
in terms of \emph{statistical experiments}. Hence we need an additional preparation device $P'$ and an
additional measuring apparatus $M'$. Indistinguishable now means that it does not matter whether we
perform $M'$ measurements directly on $P'$ outputs or whether we switch a teleportation device in
between; cf. Figure \ref{fig:cl-teleB}. In both cases we should get the same \emph{distribution of
  measuring   results} for a large number of repetitions of the corresponding experiment. This
requirement should hold for any preparation $P'$ and any measurement $M'$, but for fixed $M$ and $P$. The
latter means that we are not allowed to use a priori knowledge about $P'$ or $M'$ to adopt the
teleportation process (otherwise we can choose in the most extreme case always $P'$ for $P$ and the
whole discussion becomes meaningless).   

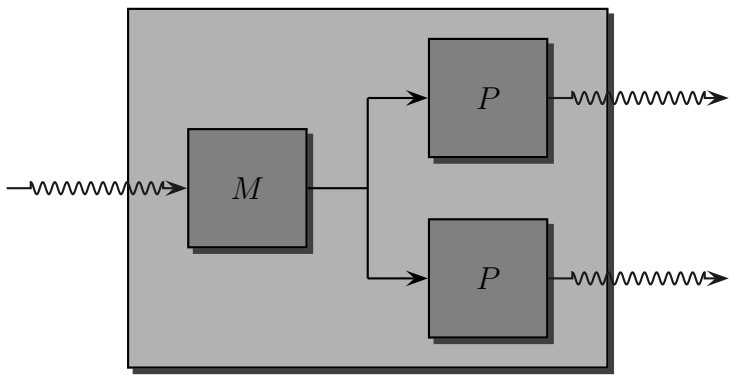
\begin{figure}[b]
  \begin{center}
    \begin{picture}(12,6)
      \large
      \psframe[fillcolor=mklight,fillstyle=solid,shadow=true](2,0)(10,6)
      \psframe[fillcolor=meascolor,fillstyle=solid,shadow=true](3,2)(5,4)
      \psframe[fillcolor=meascolor,fillstyle=solid,shadow=true](7,3.5)(9,5.5)
      \psframe[fillcolor=meascolor,fillstyle=solid,shadow=true](7,0.5)(9,2.5)
      \pccoil[linecolor=qbitcolor,coilaspect=0,coilheight=1,coilwidth=.2]{->}(0,3)(3,3)
      \pccoil[linecolor=qbitcolor,coilaspect=0,coilheight=1,coilwidth=.2]{->}(9,4.5)(12,4.5)
      \pccoil[linecolor=qbitcolor,coilaspect=0,coilheight=1,coilwidth=.2]{->}(9,1.5)(12,1.5)
      \psline[linecolor=black]{-}(5,3)(6,3)
      \psline[linecolor=black]{-}(6,1.5)(6,4.5)
      \psline[linecolor=black]{->}(6,4.5)(7,4.5)
      \psline[linecolor=black]{->}(6,1.5)(7,1.5)
      \rput(4,3){$M$}
      \rput(8,4.5){$P$}
      \rput(8,1.5){$P$}
    \end{picture}
    \caption{Constructing a quantum copying machine from a teleportation device.}
    \label{fig:clon}
  \end{center}
\end{figure}

The second impossible machine we have to consider is a \emph{quantum copying machine}. This is a device $C$
which takes one quantum system $p$ as input and produces two systems $p_1, p_2$ of the same type as
output. The limiting condition on $C$ is that $p_1$ and $p_2$ are indistinguishable from the input, where
``indistinguishable'' has to be understood in the same way as above: Any statistical experiment performed
with one of the output particles (i.e. always with $p_1$ or always with $p_2$) yields the same result as
applied directly to the input $p$. To get such a device from teleportation is easy: We just have to
perform an $M$ measurement on $p$, make two copies of the classical data obtained, and run the preparation
$P$  on each of them; cf. Figure \ref{fig:clon}. Hence if teleportation is possible copying is possible
as well. 

 \begin{figure}[t]
  \begin{center}
    \begin{picture}(12,6)
      \large
      \psframe[fillcolor=mklight,fillstyle=solid,shadow=true](2,0)(10,6)
      \psframe[fillcolor=meascolor,fillstyle=solid,shadow=true](3,2)(5,4)
      \psframe[fillcolor=meascolor,fillstyle=solid,shadow=true](7,3.5)(9,5.5)
      \psframe[fillcolor=meascolor,fillstyle=solid,shadow=true](7,0.5)(9,2.5)
      \pccoil[linecolor=qbitcolor,coilaspect=0,coilheight=1,coilwidth=.2]{->}(0,3)(3,3)
      \psline[linecolor=black]{->}(9,4.5)(12,4.5)
      \psline[linecolor=black]{->}(9,1.5)(12,1.5) 
      \pccoil[linecolor=qbitcolor,coilaspect=0,coilheight=1,coilwidth=.2]{->}(5,3)(7,4.5)
      \pccoil[linecolor=qbitcolor,coilaspect=0,coilheight=1,coilwidth=.2]{->}(5,3)(7,1.5)
      \rput(4,3){$C$}
      \rput(8,4.5){$A$}
      \rput(8,1.5){$B$}
    \end{picture}
    \caption{Constructing a joint measurement for the observables $A$ and $B$ from a quantum copying
      machine.} 
    \label{fig:jointm}
  \end{center}
\end{figure}
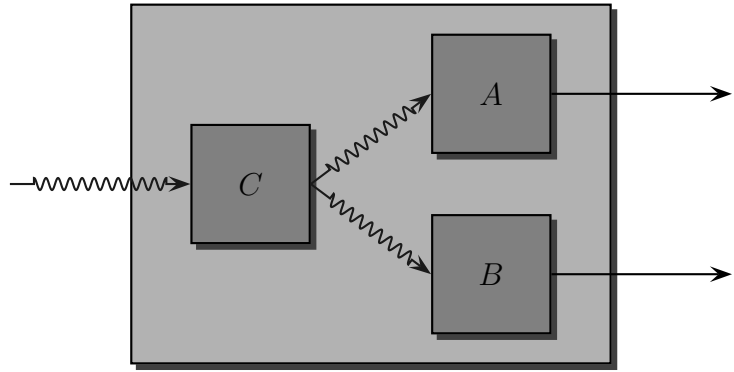

According to the ``no-cloning theorem'' of Wootters and Zurek \Cite{WoZu}, however, a quantum
copy machine does not exist and this basically concludes our proof. However we will give an easy
argument for this theorem in terms of a third impossible machine -- a \emph{joint measuring device}
$M_{AB}$ for two arbitrary observables $A$ and $B$. This is a measuring apparatus which produces each
time it is invoked a pair $(a,b)$ of classical outputs, where $a$ is a possible output of $A$ and $b$ a
possible output of $B$. The crucial requirement for $M_{AB}$ again is of statistical nature: The
statistics of the $a$ outcomes is the same as for device $A$, and similarly for $B$. It is known from
elementary quantum mechanics that many quantum observables are not jointly measurable in this way. The
most famous examples are position and momentum or different components of angular momentum. Nevertheless
a device $M_{AB}$ could be constructed for arbitrary $A$ and $B$ from a quantum copy machine $C$. We
simply have to operate with $C$ on the input system $p$ producing two outputs $p_1$ and $p_2$ and to
perform an $A$ measurement on $p_1$ and a $B$ measurement on $p_2$; cf. Figure \ref{fig:jointm}. Since
the outputs $p_1$, $p_2$ are, by assumption indistinguishable from the input $p$ the overall device
constructed this way would give a joint measurement for $A$ and $B$. Hence a quantum copying machine
cannot exist, as stated by the no-cloning theorem. This in turn implies that classical teleportation is
impossible, and therefore we can not transform quantum information lossless into classical information
and back. This concludes our chain of arguments.  

\section{Tasks of quantum information}
\label{sec:tasks-quant-inform}

So we have seen that quantum information is something new, but what can we do with it? There are three
answers to this question which we want to present here. First of all let us remark that in fact all
information in a modern data processing environment is carried by micro particles (e.g. electrons or
photons). Hence quantum information comes automatically into play. Currently it is safe to ignore this
and to use classical information theory to describe all relevant processes. If the size of the structures
on a typical circuit decreases below a certain limit, however, this is no longer true and quantum information
will become relevant.

This leads us to the second answer. Although it is far too early to say which concrete technologies will
emerge from quantum information in the future, several interesting proposals show that devices based on
quantum information can solve certain practical tasks much better than classical ones. The most well
known and exciting one is, without a doubt, quantum computing. The basic idea is, roughly speaking, that
a quantum computer can operate not only on one number per register but on \emph{superpositions of
  numbers}. This possibility leads to an ``exponential speedup'' for some computations which makes
problems feasible which are considered intractable by any classical algorithm. This is most impressively
demonstrated by Shor's factoring algorithm \Cite{Shor94,Shor97}. A second example which is quite close to
a concrete practical realization (i.e. outside the laboratory; see next Section) is quantum
cryptography. The fact that it is impossible to perform a quantum mechanical measurement without
disturbing the state of the measured system is used here for the secure transmission of a cryptographic
key (i.e. each eavesdropping attempt can be detected with certainty). Together with a subsequent
application of a classical encryption method known as the ``one-time'' pad this leads to a cryptographic
scheme with provable security -- in contrast to currently used public key systems whose security relies
on possibly doubtful assumptions about (pseudo) random number generators and prime numbers. We will come
back to both subjects -- quantum computing and quantum  cryptography in Sections
\ref{sec:quantum-computing} and \ref{sec:quantum-cryptography}. 

The third answer to the above question is of more fundamental nature. The discussion of questions
from information theory in the context of quantum mechanics leads to a deeper and in many cases more
quantitative understanding of quantum theory. Maybe the most relevant example for this statement is the
study of entanglement,  i.e. non-classical correlations between quantum systems, which lead to
violations of Bell inequalities\footnote{This is only a very rough characterization. A more precise one
  will be given in Section \ref{sec:tens-prod-entangl}.}. Entanglement is a fundamental aspect of quantum
mechanics and demonstrates the differences between quantum and classical physics in the most drastical
way -- this can be seen from Bell-type experiments, like the one of Aspect et. al. \Cite{ADR82},
and the discussion about. Nevertheless, for a long time it was only considered as an exotic feature of
the foundations of quantum mechanics which is not so relevant from a practical point of view. Since
quantum information attained broader interest, however, this has changed completely. It has turned out
that entanglement is an essential resource whenever classical information processing is outperformed by
quantum devices. One of the most remarkable examples is the experimental realization of ``entanglement
enhanced'' teleportation \Cite{TeleExp1,TeleExp2}. We have argued in Section
\ref{sec:impossible-machines} that \emph{classical} teleportation, i.e. transmission of quantum 
information through a classical information channel, is impossible. If sender and receiver share,
however, an entangled pair of particles (which can be used as an additional resource) the impossible task
becomes, most surprisingly, possible \Cite{Bennett93}! (We will discuss this fact in detail in Section 
\ref{sec:telep-dense-coding}.) The study of entanglement and in particular the question \emph{how it can
  be quantified} is therefore a central topic within quantum information theory (cf. Chapter
\ref{cha:quant-theory-i}). Further examples for fields where quantum information has led to a deeper and
in particular more quantitative insight include  ``capacities'' of quantum information channels and
``quantum cloning''. A detailed discussion of these topics will be given in Chapter
\ref{cha:quant-theory-ii} and \ref{cha:quant-theory-iii}. Finally let us remark that classical
information theory benefits in a similar way from the synthesis with quantum mechanics. Beside the just
mentioned channel capacities this concerns for example the theory of computational complexity which
analyzes the scaling behavior of time and space consumed by an algorithm in dependence of the size of the input
data. Quantum information challenges here in particular the fundamental Church-Turing hypotheses
\Cite{Church36,Turing36} which claims that each computation can be simulated ``efficiently'' on a Turing
machine; we come back to this topic in Section \ref{sec:quantum-computing}.  

\section{Experimental realizations}
\label{sec:exper-real}

Although this is a theoretical paper, it is of course necessary to say something about experimental
realizations of the ideas of quantum information. Let us consider quantum computing first. Whatever way we
go here, we need systems which can be prepared very precisely in few distinct states (i.e. we need
``qubits''), which can be manipulated afterwards individually (we have to realize ``quantum gates'') and
which can finally be measured with an appropriate observable (we have to ``read out'' the result). 

One of the most far developed approaches to quantum computing is the ion trap technique (see
Section 4.3 and 5.3 in \Cite{BEZBook} and Section 7.6 of \Cite{NC} for an overview and further
references). A ``quantum register'' is realized here by a string of ions kept by electromagnetic fields
in high vacuum inside a Paul trap, and two long-living states of each ion are chosen to represent ``0''
and ``1''.  A single ion can be manipulated by  laser beams and this allows the implementation of all
``one-qubit gates''. To get  two-qubit gates as well (for a quantum computer we need at least one two
qubit gate together with all one-qubit operations; cf. Section \ref{sec:quantum-computing}) the
collective motional state of the ions has to be used. A ``program'' on an ion trap quantum computer
starts now with a preparation of the register in an initial state -- usually the ground state of the
ions. This is done by optical pumping and laser cooling (which is in fact one of the most difficult parts
of the whole procedure, in particular if many ions are involved). Then the ``network'' of quantum gates
is applied, in terms of a (complicated) sequence of laser pulses. The readout finally is done by laser
beams which illuminate the ions subsequently. The beams are tuned to a fast transition which affects only
one of the qubit states and the fluorescent light is detected. Concrete implementations (see
e.g. \Cite{IOT1,IOT2}) are currently restricted to two qubits, however there is some hope that we will be
able to control up to 10 or 12 qubits in the not too distant future.     
 
A second quite successful technique is NMR quantum computing (see Section 5.4 of \Cite{BEZBook} and
Section 7.7 of \Cite{NC} together with the references therein for details). NMR stands for ``nuclear
magnetic resonance'' and it is the study of transitions between Zeeman levels of an atomic nucleus in a
magnetic field. The qubits are in this case different spin states of the nuclei in an appropriate
molecule and quantum gates are realized by high frequency oscillating magnetic fields in pulses of
controlled duration. In contrast to ion traps however we do not use \emph{one} molecule but a whole cup
of liquid containing some $10^{20}$ of them. This causes a number of problems, concerning in particular
the preparation of an initial state, fluctuations in the free time evolution of the molecules and the
readout. There are several ways to overcome these difficulties and we refer the reader again to
\Cite{BEZBook} and \Cite{NC} for details. 
 Concrete implementations of NMR quantum computers are capable to use up to five qubits
 \Cite{NMR4}. Other realizations include the implementation of several known quantum algorithms on two and
 three qubits; see e.g. \Cite{NMR1,NMR2,NMR3}.  

The fundamental problem of the two methods for quantum computation discussed so far, is their lack of
scalability. It is realistic to assume that NMR and ion-trap quantum computer with up to tens of qubits
will exist somewhen in the future but not with thousands of qubits which are necessary for ``real world''
applications. There are, however, many other alternative proposals available and some of them might be
capable to avoid this problem. 
The following is a small (not at all exhaustive) list: 
atoms in optical lattices \Cite{QCOptLat}, semiconductor nanostructures such as quantum dots (there are
many works in this area, some recent are \Cite{QDot1,QDot2,QDot3,QDot4}) and arrays of Josephson junctions
\Cite{QCJosJun}.

A second circle of experiments we want to mention here is grouped around quantum communication and
quantum cryptography (for a more detailed overview let us refer to \Cite{WeZe} and \Cite{QCRev}). 
Realizations of quantum cryptography are fairly far developed and it is currently possible to span up to
50km with optical fibers (e.g. \Cite{QCExp2}). Potentially greater distances can be bridged by ``free
space cryptography'' where the quantum information is transmitted through the air (e.g
\Cite{QCExp3}). With this technology satellites can be used as some sort of ``relays'', thus enabling
quantum key distribution over arbitrary distances. In the meantime there are quite a lot of successful
implementations. For a detailed discussion we will refer the reader to the review of Gisin
et. al. \Cite{QCRev} and the references therein. Other experiments concern the usage of entanglement in
quantum communication.   
The creation and detection of entangled photons is here a fundamental building block. Nowadays this is no
problem  and the most famous experiment in this context is the one of Aspect et. al. \Cite{ADR82}, where
the maximal violation of Bell inequalities was demonstrated with polarization correlated photons. Another
spectacular experiment is the creation of entangled photons over a distance of 10 km using standard
telecommunication optical fibers by the Geneva group \Cite{EntGen}. Among the most exciting applications
of entanglement is the realization of entanglement based quantum key distribution \Cite{QCExp1}, the
first successful ``teleportation'' of a photon \Cite{TeleExp1,TeleExp2}  and the implementation of
``dense coding'' \Cite{DCExp}; cf. Section \ref{sec:telep-dense-coding}.  

\chapter{Basic concepts}
\label{cha:basic-concepts}

After we have got a first, rough impression of the basic ideas and most relevant subjects of quantum
information theory, let us start with a more detailed presentation. First we have to introduce the fundamental
notions of the theory and their mathematical description. Fortunately, much of the material we should have
to present here, like Hilbert spaces, tensor products and density matrices, is known already from quantum
mechanics
and we can focus our discussion to those concepts which are less familiar
like POV measures, completely positive maps and entangled states.

\section{Systems, States and Effects}
\label{sec:syst-stat-observ}

As classical probability theory quantum mechanics is a \emph{statistical theory}. Hence its predictions
are of probabilistic nature and can only be tested if the same experiment is repeated very often and the
relative frequencies of the outcomes are calculated. In more operational terms this means: the experiment
has to be repeated according to the same \emph{procedure} as it can be set out in a detailed laboratory
manual. If we consider a somewhat idealized model of such a \emph{statistical experiment} we get in fact
two different types of procedures: first \emph{preparation procedures} which prepare a certain kind of
physical system in a distinguished \emph{state} and second \emph{registration procedures} measuring a
particular \emph{observable}.

A mathematical description of such a setup basically consists of two sets $\scr{S}$ and $\scr{E}$ and a
map $\scr{S} \times \scr{E} \ni (\rho,A) \to \rho(A) \in [0,1]$. The elements of $\scr{S}$ describe the states,
i.e. preparations, while the $A \in \scr{E}$ represent all yes/no measurements (\emph{effects}) which can
be performed on the system. The probability (i.e. the relative frequency for a large number of
repetitions) to get the result ``yes'', if we are measuring the effect $A$ on a system prepared in the
state $\rho$, is given by $\rho(A)$. This is a very general scheme applicable not only to quantum mechanics but
also to a very broad class of statistical models, containing in particular classical probability. In
order to make use of it we have to specify of course the precise structure of the sets $\scr{S}$ and
$\scr{E}$ and the map $\rho(A)$ for the types of systems we want to discuss. 

\subsection{Operator algebras}
\label{sec:operator-algebras}

Throughout this paper we will encounter three different kinds of systems: quantum and classical systems
and hybrid systems which are half classical, half quantum (cf. Subsection \ref{sec:hybrid-systems}). In
this subsection we will describe a general way to define states and effects which is applicable to all
three cases and which therefore provides a handy way to discuss all three cases simultaneously (this will
become most useful in Section \ref{sec:tens-prod-entangl} and \ref{sec:channels}).

The scheme we are going to discuss is based on an algebra $\scr{A}$ of bounded operators acting on a
Hilbert space $\scr{H}$. More precisely $\scr{A}$ is a (closed) linear subspace of $\scr{B}(\scr{H})$,
the algebra of bounded operates on $\scr{H}$, which contains the identity ($\Bbb{1} \in \scr{A}$) and is
closed under products ($A,B \in \scr{A}$ $\Rightarrow AB \in \scr{A}$) and adjoints ($A \in \scr{A}$ $\Rightarrow A^* \in
\scr{A}$). For simplicity we will refer to each such $\scr{A}$ as an \emph{observable algebra}. The key
observation is now that each type of system we will study in the following can be \emph{completely
  characterized} by its observable algebra $\scr{A}$, i.e. once $\scr{A}$ is known there 
is a systematic way to derive the sets $\scr{S}$ and $\scr{E}$ and the map $(\rho,A) \mapsto \rho(A)$ from it. We
frequently make use of this fact by referring to systems in terms of their observable algebra $\scr{A}$,
or even by identifying them with their algebra and saying that $\scr{A}$ \emph{is the system}.  

 Although $\scr{A}$ and $\scr{H}$ can be infinite dimensional in general, we will consider
only finite dimensional Hilbert spaces, as long as nothing else is explicitly stated. Since most research
in quantum information is done up to now for finite dimensional systems (the only exception in this work 
is the discussion of Gaussian systems in Section \ref{sec:quant-mech-phase}) this is not a too severe loss
of generality. Hence we can choose $\scr{H} = \Bbb{C}^d$ and $\scr{B}(\scr{H})$ is just the algebra of
complex $d \times d$ matrices.
 Since $\scr{A}$ is a subalgebra of $\scr{B}(\scr{H})$ it operates naturally on
$\scr{H}$ and it inherits from $\scr{B}(\scr{H})$ the \emph{operator norm} $\|A\| = \sup_{\|\psi\|=1} \|A\psi\|$ and
the \emph{operator ordering} $A \geq B \Leftrightarrow \langle \psi, A \psi\rangle \geq \langle \psi, B \psi\rangle$ $\forall \psi \in \scr{H}$. Now we can define: 
\begin{equation}\label{eq:26}
  \scr{S}(\scr{A}) = \{ \rho \in \scr{A}^* \, | \, \rho \geq 0, \rho(\Bbb{1}) = 1 \} 
\end{equation}
where $\scr{A}^*$ denotes the \emph{dual space} of $\scr{A}$, i.e. the set of all linear functionals on
$\scr{A}$, and $\rho \geq 0$ means $\rho(A) \geq 0$ $\forall A \geq 0$. Elements of $\scr{S}(\scr{A})$ describe the
states of the system in question while effects are given by
\begin{equation}\label{eq:24}
  \scr{E}(\scr{A}) = \{ A \in \scr{A} \, | \, A \geq 0,\ A \leq \Bbb{1}\}.
\end{equation}
The probability to measure the effect $A$ in the state $\rho$ is $\rho(A)$. More generally we can look at
$\rho(A)$ for an arbitrary $A$ as the \emph{expectation value} of $A$ in the state $\rho$. Hence the idea
behind Equation (\ref{eq:26}) is to define states in terms of their expectation value functionals.

Both spaces are \emph{convex}, i.e. $\rho, \sigma \in \scr{S}(\scr{A})$ and $0 \leq \lambda \leq 1$ implies $\lambda\rho + (1-\lambda)\sigma \in
\scr{S}(\scr{A})$ and similarly for $\scr{E}(\scr{A})$. The \emph{extremal points} of $\scr{S}(\scr{A})$
respectively $\scr{E}(\scr{A})$, i.e. those elements which do not admit a proper convex decomposition ($x
= \lambda y + (1-\lambda)z$ $\Rightarrow$ $\lambda=1$ or $\lambda=0$ or $y = z = x$), play a distinguished role: the extremal points of
$\scr{S}(\scr{A})$ are \emph{pure states} and those of $\scr{E}(\scr{A})$ are the \emph{propositions} of
the system in question.  The latter represent those effects which register a property with certainty in 
contrast to non-extremal effects which admit some ``fuzziness''. As a simple example for the latter
consider a detector which registers particles not with certainty but only with a probability which is
smaller than one.  

Finally let us note that the complete discussion of this section can be generalized easily to infinite
dimensional systems, if we replace $\scr{H} = \Bbb{C}^d$ by an infinite dimensional Hilbert space
(e.g. $\scr{H} = \Lz(\Bbb{R})$). This would require however more material about C* algebras and measure
theory than we want to use in this paper. 

\subsection{Quantum mechanics}
\label{sec:quantum-mechanics}

For quantum mechanics we have
\begin{equation}
  \scr{A} = \scr{B}(\scr{H}),  
\end{equation}
where we have chosen again $\scr{H} = \Bbb{C}^d$.  The corresponding systems are called \emph{$d$-level
  systems} or \emph{qubits} if $d=2$ holds. To avoid clumsy notations we frequently write
$\scr{S}(\scr{H})$ and $\scr{E}(\scr{H})$ instead of $\scr{S}\bigl[\scr{B}(\scr{H})\bigr]$ and
$\scr{E}\bigl[\scr{B}(\scr{H})\bigr]$. From Equation (\ref{eq:24}) we immediately see that an operator $A
\in  \scr{B}(\scr{H})$ is an effect  iff it is positive and bounded from above by $\Bbb{1}$. An element $P
\in \scr{E}(\scr{H})$ is a propositions iff $P$ is a projection operator ($P^2=P$).   

States are described in quantum mechanics usually by density matrices, i.e. positive
and normalized trace class\footnote{On a finite dimensional Hilbert space this attribute is of course
  redundant, since each operator is of trace class in this case. Nevertheless we will frequently use this
  terminology, due to greater consistency with the infinite dimensional case.} operators. To 
make contact to the general definition in Equation (\ref{eq:26}) note first that $\scr{B}(\scr{H})$ is a
Hilbert space with the Hilbert-Schmidt scalar product $\langle A,B\rangle = \tr(A^*B)$. Hence each linear functional
$\rho \in \scr{B}(\scr{H})^*$ can be expressed in terms of a (trace class) operator $\tilde{\rho}$
by\footnote{\label{fn:1} If we consider infinite dimensional systems this is not true. In this case the
  dual space of the observable algebra is much larger and Equation (\ref{eq:26}) leads to states which
  are not necessarily given by trace class operators. Such ``singular states'' play an important role in
  theories which admit an infinite number of degrees of freedom like quantum statistics and quantum field
  theory; cf. \Cite{BraRob1,BraRob2}. For applications of singular states within quantum information see
  \Cite{InfEnt}.} $A \mapsto \rho(A) = \tr(\tilde{\rho}A)$. It is obvious that each $\tilde{\rho}$ defines a unique
functional $\rho$. If we start on the other hand with $\rho$ we can recover the matrix elements of $\tilde{\rho}$
from $\rho$ by $\tilde{\rho}_{kj} = \tr(\tilde{\rho}\KB{j}{k}) = \rho(\KB{j}{k})$, where $\KB{j}{k}$ denotes the
canonical basis of $\scr{B}(\scr{H})$ (i.e. $\KB{j}{k}_{ab} = \delta_{ja}\delta_{kb}$). More generally we get for
$\psi, \phi \in \scr{H}$ the relation $\langle\phi,\tilde{\rho}\psi\rangle = \rho(\KB{\psi}{\phi})$, where $\KB{\psi}{\phi}$ now denotes the rank
one operator which maps $\eta \in \scr{H}$ to $\langle\phi,\eta\rangle\psi$. In the following we drop the $\sim$ and use the same
symbol for the operator and the functional whenever confusion can be avoided. Due to the same abuse of
language we will interpret elements of $\scr{B}(\scr{H})^*$ frequently as (trace class) operators instead
of linear functionals (and write $\tr(\rho A)$ instead of $\rho(A)$).  However we do not identify
$\scr{B}(\scr{H})^*$ with $\scr{B}(\scr{H})$ in general, because the two different notations help to keep
track of the distinction between spaces of states and spaces of observables. In addition we equip
$\scr{B}^*(\scr{H})$ with the trace-norm $\|\rho\|_1 = \tr|\rho|$ instead of the operator norm. 

Positivity of the \emph{functional} $\rho$ implies positivity of the \emph{operator} $\rho$ due to $0 \leq
\rho(\kb{\psi}) = \langle\psi,\rho\psi\rangle$ and the same holds for normalization: $1 = \rho(\Bbb{1}) = \tr(\rho)$. Hence we can
identify the state space from Equation (\ref{eq:26}) with the set of density matrices, as expected for
quantum mechanics. Pure states of a quantum system are the one dimensional projectors. As usual we will
frequently identify the density matrix $\kb{\psi}$ with the wave function $\psi$  and call the latter in abuse
of language a state.  

To get a useful parameterization of the state space consider again the Hilbert-Schmidt scalar product 
$\langle\rho,\sigma\rangle = \tr(\rho^*\sigma)$, but now on $\scr{B}^*(\scr{H})$. The space of trace free matrices in
$\scr{B}^*(\scr{H})$ (alternatively the functionals with $\rho(\Bbb{1}) = 0$) is the corresponding
orthocomplement $\Bbb{1}^\bot$ of the unit operator. If we choose a basis $\sigma_1,\ldots,\sigma_{d^2-1}$ with $\langle\sigma_j,\sigma_k\rangle
= 2 \delta_{jk}$ in $\Bbb{1}^\bot$ we can write each selfajoint (trace class) operator $\rho$ with $\tr(\rho) = 1$ as 
\begin{equation}
  \rho = \frac{\Bbb{1}}{d} + \frac{1}{2} \sum_{j=1}^{d^2-1} x_j \sigma_j =: \frac{\Bbb{1}}{d} + \frac{1}{2} \vec{x}
  \cdot \vec{\sigma} ,\ \text{with}\ \vec{x} \in \Bbb{R}^{d^2-1}.  
\end{equation}
If $d=2$ or $d= 3$ holds, it is most natural to choose the Pauli matrices respectively the Gell-Mann
matrices (cf. e.g. Sect. 13.4 of \Cite{Cornwall2}) for the $\sigma_j$. In the qubit case it is easy to see
that $\rho \geq 0$ holds iff $|\vec{x}|\leq1$. Hence the state space $\scr{S}(\Bbb{C}^2)$ coincides with the
\emph{Bloch ball} $\{ \vec{x} \in \Bbb{R}^3 \, | \, |\vec{x}|\leq 1\}$, and the set of pure states with its
boundary, the \emph{Bloch sphere} $\{ \vec{x} \in \Bbb{R}^3 \, | \, |\vec{x}| = 1\}$. This shows in a very
geometric way that the pure states are the extremal points of the convex set $\scr{S}(\scr{H})$. If $\rho$
is more generally a pure state of a $d$-level system we get  
\begin{equation}
  1 = \tr(\rho^2) = \frac{1}{d} + \frac{1}{2} |\vec{x}|^2 \Rightarrow |\vec{x}| = \sqrt{2 \left(1 - 1/d\right)}.
\end{equation}
This implies that all states are contained in the ball with radius $2^{1/2} (1 - 1/d)^{1/2}$, however not
all operators in this set are positive. A simple example is $d^{-1}\Bbb{1} \pm 2^{1/2} (1 - 1/d)^{1/2}
\sigma_j$, which is positive only if $d=2$ holds.

\subsection{Classical probability}
\label{sec:class-prob}

Since the difference between classical and quantum systems is an important issue in this work let us
reformulate classical probability theory according to the general scheme from Subsection
\ref{sec:operator-algebras}. The restriction to finite dimensional observable algebras leads now to the
assumption that all systems we are considering admit a finite set $X$ of \emph{elementary
  events}. Typical examples are: throwing a dice $X=\{1,\ldots,6\}$, tossing a coin $X=\{\text{``head''},
\text{``number''}\}$ or \emph{classical bits} $X = \{0,1\}$. To simplify the notations we write (as in
quantum mechanics) $\scr{S}(X)$ and $\scr{E}(X)$ for the spaces of states and effects.

The observable algebra $\scr{A}$ of such a system is the space 
\begin{equation}
  \scr{A} = \scr{C}(X) = \{ f: X \to \Bbb{C} \} 
\end{equation}
of complex valued functions on $X$. To interpret this as an operator algebra acting on a Hilbert space
$\scr{H}$ (as indicated in Subsection \ref{sec:operator-algebras}) choose an arbitrary but fixed
orthonormal basis $\ket{x}, x \in X$ in $\scr{H}$ and identify \emph{the function} $f \in \scr{C}(X)$ with
\emph{the operator} $f =\sum_x f_x \kb{x} \in \scr{B}(\scr{H})$ (we use the same symbol for the function and
the operator, provided confusion can be avoided). Most frequently we have $X = \{1,\ldots,d\}$ and we can choose
$\scr{H} = \Bbb{C}^d$  and the canonical basis for $\ket{x}$. Hence $\scr{C}(X)$ becomes the algebra of
\emph{diagonal} $d \times d$ matrices. 
Using Equation (\ref{eq:24}) we immediately see that $f \in \scr{C}(X)$ is an
effect iff $0 \leq f_x \leq 1$, $\forall x  \in X$. Physically we can interpret $f_x$ as the probability that the effect
$f$ registers the elementary event $x$. This makes the distinction between propositions and ``fuzzy''
effects very transparent: $P \in \scr{E}(X)$ is a proposition iff we have either $P_x =
1$ or $P_x=0$ for all $x \in X$. Hence the propositions $P \in \scr{C}(X)$ are in one to one correspondence
with the subsets $\omega_P = \{ x \in X \, | \, P_x = 1\} \subset X$ which in turn describe the \emph{events} of the
system. Hence $P$ registers the event $\omega_P$ with certainty, while a fuzzy effect $f < P$ does this only
with a probability less then one. 

Since $\scr{C}(X)$ is finite dimensional and admits the distinguished basis $\kb{x}, x \in X$ it is
naturally isomorphic to its dual $\scr{C}^*(X)$. More precisely: each linear functional $\rho \in
\scr{C}^*(X)$ defines and is uniquely defined by the function $x \mapsto \rho_x = \rho(\kb{x})$ and we have $\rho(f) =
\sum_x f_x \rho_x$. As in the quantum case we will identify the function $\rho$ with the linear functional and
use the same symbol for both, although we keep the notation $\scr{C}^*(X)$ to indicate that we are
talking about states rather than observables.

Positivity of $\rho \in \scr{C}^*(X)$ is given by $\rho_x \geq 0$ for all $x$ and normalization leads to $1 =
\rho(\Bbb{1}) = \rho\left(\sum_x \kb{x}\right) = \sum_x \rho_x$. Hence to be a state $\rho \in \scr{C}^*(X)$ must be a
\emph{probability distribution} on $X$ and $\rho_j$ is the probability that the elementary event $x$ occurs
during statistical experiments with systems in the state $\rho$. More generally $\rho(f) = \sum_j \rho_j f_j$ is the
probability to measure the effect $f$ on systems in the state $\rho$. If $P$ is in particular a proposition, 
$\rho(P)$ gives the probability for the event $\omega_P$. The pure states of the system
are the \emph{Dirac measures} $\delta_x$, $x \in X$; with $\delta_x(\kb{y}) = \delta_{xy}$. Hence each $\rho \in
\scr{S}(X)$ can be decomposed \emph{in a unique way} into a convex linear combination
of pure states. 

\subsection{Observables}
\label{sec:observables}

Up to now we have discussed only effects, i.e. yes/no experiments. In this subsection we will have a
first short look at more general observables. We will come back to this topic in Section
\ref{sec:observables-2} after we have introduced channels. 
We can think of an observable $E$ taking its values in a finite set $X$ as a map which
associates to each possible outcome $x \in X$ the effect $E_x \in \scr{E}(\scr{A})$ (if $\scr{A}$ is the
observable algebra of the system in question) which is true if $x$ is measured and false otherwise. If
the measurement is performed on systems in the state $\rho$ we get for each $x \in X$ the probability $p_x =
\rho(E_x)$ to measure $x$. Hence the family of the $p_x$ should be a probability distribution on $X$, and
this implies that $E$ should be a \emph{POV measure} on $X$.

\begin{defi} \label{def:1}
  Consider an observable algebra $\scr{A} \subset \scr{B}(\scr{H})$ and a finite\footnote{This is if course an
    artifical restriction and in many situations not justified (cf. in particular the discussion of
    quantum state estimation in Section \ref{sec:estimating-copying} and Chapter
    \ref{cha:quant-theory-iii}). However, it helps us to avoid measure theoretical subtleties;
    cf. Holevo's book \Cite{HolBook} for a more general discussion.} set $X$. A family $E=(E_x)_{x \in X}$
  of effects in $\scr{A}$ (i.e. $0 \leq E_x \leq \Bbb{1}$) is called a \emph{positive operator valued measure}
  (POV measure) on $X$ if $\sum_{x \in X} E_x = \Bbb{1}$ holds. If all $E_x$ are projections, $E$ is called
  \emph{projection valued measure} (PV measure).  
\end{defi}

From basic quantum mechanics we know that observables are described by self adjoint operators on a
Hilbert space $\scr{H}$. But, how does this point of view fit into the previous definition? The answer is
given by the spectral theorem (Thm. VIII.6 \Cite{RESI1}): Each selfadjoint operator $A$ on a finite
dimensional Hilbert space $\scr{H}$ has the form $A = \sum_{\lambda \in \sigma(A)} \lambda P_\lambda$ where $\sigma(A)$ denotes the
\emph{spectrum} of $A$, i.e. the set of eigenvalues and $P_\lambda$ denotes the projection onto the
corresponding eigenspace. Hence there is a unique PV measure $P = (P_\lambda)_{\lambda \in \sigma(A)}$ associated to
$A$ which is called the \emph{spectral measure} of $A$. It is uniquely characterized by the property that
the \emph{expectation value} $\sum_\lambda \lambda \rho(P_\lambda)$ of $P$ in the state $\rho$ is given for any state $\rho$ by $\rho(A) =
\tr(\rho A)$; as it is well known from quantum mechanics. Hence the traditional way to define observables
within quantum mechanics perfectly fits into the scheme  just outlined, however it only covers the
projection valued case and therefore admits no fuzziness. For this reason POV measures are sometimes
called \emph{generalized observables}. 

Finally note that the eigenprojections $P_\lambda$ of $A$ are elements of an observable algebra $\scr{A}$ iff
$A \in \scr{A}$. This shows two things: First of all we can consider selfadjoint elements of \emph{any}
*-subalgebra $\scr{A}$ of $\scr{B}(\scr{H})$ as observables of $\scr{A}$-systems, and this is precisely
the reason why we have called $\scr{A}$ \emph{observable} algebra. Secondly we see why it is essential
that $\scr{A}$ is really a subalgebra of $\scr{B}(\scr{H})$: if it is only a linear subspace of
$\scr{B}(\scr{H})$ the relation $A \in \scr{A}$ does not imply $P_\lambda \in \scr{A}$.

\section{Composite systems and entangled states}
\label{sec:tens-prod-entangl}

composite systems occur in many places in quantum information theory. A typical example is a register of
a quantum computer, which can be regarded as a system consisting of $N$ qubits (if $N$ is the length of
the register). The crucial point is that this opens the possibility for correlations and entanglement
between subsystems. In particular entanglement is of great importance, because it is a central resource
in many applications of quantum information theory like entanglement enhanced teleportation or quantum
computing -- we already discussed this in Section \ref{sec:tasks-quant-inform} of the introduction. To
explain entanglement in greater detail and to introduce some necessary formalism we have to complement
the scheme developed in the last section by a procedure which allows us to construct states and
observables of the composite system from its subsystems.  In quantum mechanics this is done of course in
terms of tensor products, and we will review in the following some of the  most relevant material.

\subsection{Tensor products}

Consider two (finite dimensional) Hilbert spaces $\scr{H}$ and $\scr{K}$. To each pair of vectors $\psi_1 \in
\scr{H}$, $\psi_2 \in \scr{K}$ we can associate a bilinear form $\psi_1 \otimes \psi_2$ called the \emph{tensor product} of
$\psi_1$ and $\psi_2$ by $\psi_1 \otimes \psi_2(\phi_1,\phi_2) = \langle\psi_1,\phi_1\rangle\langle\psi_2,\phi_2\rangle$. For two product vectors $\psi_1 \otimes \psi_2$ and
$\eta_1 \otimes \eta_2$ their scalar product is defined by $\langle\psi_1 \otimes \psi_2, \eta_1 \otimes \eta_2\rangle = \langle\psi_1,\eta_1\rangle\langle\psi_2,\eta_2\rangle$ and it can
be shown that this definition extends in a unique way to the span of all $\psi_1 \otimes \psi_2$ which therefore
defines the tensor product $\scr{H} \otimes \scr{K}$. If we have more than two Hilbert spaces $\scr{H}_j$,
$j=1,\ldots,N$  their tensor product $\scr{H}_1 \otimes \cdots \otimes \scr{H}_N$ can be defined similarly. 

The tensor product $A_1 \otimes A_2$ of two bounded operators $A_1 \in \scr{B}(\scr{H})$, $A_2 \in
\scr{B}(\scr{K})$ is defined first for  product vectors $\psi_1 \otimes \psi_2 \in \scr{H} \otimes \scr{K}$ by
$A_1 \otimes A_2 (\psi_1 \otimes \psi_2) = (A_1 \psi_1) \otimes (A_2 \psi_2)$ and then extended by linearity. The space
$\scr{B}(\scr{H} \otimes \scr{K})$ coincides with the span of all $A_1 \otimes A_2$. If $\rho \in \scr{B}(\scr{H} \otimes
\scr{K})$ is not of product form (and of trace class for infinite dimensional $\scr{H}$ and $\scr{K}$)
there is nevertheless a way to define ``restrictions'' to $\scr{H}$ respectively $\scr{K}$ called the 
\emph{partial trace} of $\rho$. It is defined by the equation
\begin{equation} \label{eq:2}
  \tr[\tr_\scr{K}(\rho) A] = \tr(\rho A \otimes \Bbb{1})\quad \forall A \in \scr{B}(\scr{H})
\end{equation}
where the trace on the left hand side is over $\scr{H}$ and on the right hand side over $\scr{H} \otimes
\scr{K}$. 

If two orthonormal bases $\phi_1,\ldots,\phi_n$ and $\psi_1,\ldots,\psi_m$ are given in $\scr{H}$ respectively 
$\scr{K}$  we can consider the product basis $\phi_1 \otimes \psi_1,\ldots,\phi_n \otimes \psi_m$ in $\scr{H} \otimes \scr{K}$, and we can
expand each $\Psi \in \scr{H} \otimes \scr{K}$ as $\Psi = \sum_{jk} \Psi_{jk} \phi_j \otimes \psi_k$ with $\Psi_{jk} = \langle\phi_j \otimes \psi_k, \Psi\rangle$. This
procedure works for an arbitrary number of tensor factors. However, if we have exactly a twofold tensor
product, there is a more economic way to expand $\Psi$, called \emph{Schmidt decomposition} in which only
diagonal terms of the form $\phi_j \otimes \psi_j$ appear. 

\begin{prop} \label{prop:2}
  For each element $\Psi$ of the \emph{twofold} tensor product $\scr{H} \otimes \scr{K}$ there are orthonormal
  systems $\phi_j$, $j=1,\ldots,n$ and $\psi_k$, $k=1,\ldots,n$ (not necessarily bases, i.e. $n$ can be smaller than
  $\dim\scr{H}$ and $\dim\scr{K}$) of $\scr{H}$ and $\scr{K}$ respectively such that $\Psi = \sum_j \sqrt{\lambda}_j
  \phi_j \otimes \psi_j$ holds. The $\phi_j$ and $\psi_j$ are uniquely determined by $\Psi$. The expansion is called
  \emph{Schmidt decomposition} and the numbers $\sqrt{\lambda}_j$ are the \emph{Schmidt coefficients}. 
\end{prop}

\begin{proof}
  Consider the partial trace $\rho_1 = \tr_\scr{K}(\kb{\Psi})$ of the one dimensional projector $\kb{\Psi}$
  associated to $\Psi$. It can be decomposed in terms of its eigenvectors $\phi_n$ and we get
  $\tr_\scr{K}(\kb{\Psi}) = \rho_1 = \sum_n \lambda_n \kb{\phi_n}$. Now we can choose an orthonormal basis $\psi'_k$,
  $k=1,\ldots,m$ in $\scr{K}$ and expand $\Psi$ with respect to $\phi_j \otimes \psi'_k$. Carrying out the $k$ summation we 
  get a family of vectors $\psi''_j = {\sum_k} \langle \Psi, \phi_j \otimes \psi'_k\rangle \psi'_k$ with the property $\Psi = \sum_j \phi_j \otimes
  \psi''_j$. Now we can calculate the partial trace and get for any $A \in 
  \scr{B}(\scr{H}_1)$:
  \begin{equation}
    \sum_j \lambda_j \langle\phi_j, A \phi_j\rangle = \tr(\rho_1 A) = \langle\Psi, (A \otimes \Bbb{1}) \Psi\rangle = {\sum}_{j,k} \langle\phi_j, A \phi_k\rangle\langle\psi''_j, \psi''_k\rangle.
  \end{equation}
  Since $A$ is arbitrary we can compare the left and right hand side of this equation term by term and we
  get $\langle\psi''_j,\psi''_k\rangle = \delta_{jk} \lambda_j$. Hence $\psi_j = \lambda_j^{-1/2} \psi''_j$ is the desired orthonormal system.  
\end{proof} 

As an immediate application of this result we can show that each mixed state $\rho \in \scr{B}^*(\scr{H})$ (of
the quantum system $\scr{B}(\scr{H})$) can be regarded as a pure state on a larger Hilbert space $\scr{H}
\otimes \scr{H}'$. We just have to consider the eigenvalue expansion $\rho = \sum_j \lambda_j \kb{\phi_j}$ of $\rho$ and to choose
an arbitrary orthonormal system $\psi_j$, $j=1,\ldots n$ in $\scr{H}'$. Using Proposition \ref{prop:2} we get  

\begin{kor} \label{kor:2} 
  Each state $\rho \in \scr{B}^*(\scr{H})$ can be extended to a pure state $\Psi$ on a larger system with Hilbert
  space $\scr{H} \otimes \scr{H}'$ such that $\tr_{\scr{H}'}\kb{\Psi} = \rho$ holds.  
\end{kor}

\subsection{Compound and hybrid systems}
\label{sec:hybrid-systems}

To discuss the composition of two arbitrary (i.e. classical or quantum) systems it is very convenient to
use the scheme developed in Subsection \ref{sec:operator-algebras} and to talk about the two subsystems
in terms of their observable algebras $\scr{A} \subset \scr{B}(\scr{H})$ and $\scr{B} \subset \scr{B}(\scr{K})$. The
observable algebra of the composite system is then simply given by the tensor product of $\scr{A}$ and
$\scr{B}$, i.e. 
\begin{equation}
  \scr{A} \otimes \scr{B} := \SP \{ A \otimes B \, | \, A \in \scr{A},\ B \in \scr{B}\} \subset \scr{B}(\scr{K} \otimes \scr{H}).
\end{equation}
The dual of $\scr{A} \otimes \scr{B}$ is generated by product states, $(\rho \otimes \sigma)(A \otimes B) = \rho(A)\sigma(B)$ and we
therefore write $\scr{A}^* \otimes \scr{B}^*$ for $(\scr{A} \otimes \scr{B})^*$.

The interpretation of the composed system $\scr{A} \otimes \scr{B}$ in terms of states and effects is
straightforward and therefore postponed to the next Subsection. We will consider
first the special cases arising from different choices for $\scr{A}$ and $\scr{B}$. If both systems are
quantum 
($\scr{A} = \scr{B}(\scr{H})$ and $\scr{B} = \scr{B}(\scr{K})$)  we get  
\begin{equation}
  \scr{B}(\scr{H}) \otimes \scr{B}(\scr{K}) = \scr{B}(\scr{H} \otimes \scr{K})
\end{equation}
as expected. For two classical systems $\scr{A}= \scr{C}(X)$ and $\scr{B} = \scr{C}(Y)$ recall that
elements of $\scr{C}(X)$ (respectively $\scr{C}(Y)$) are complex valued functions on $X$ (on $Y$). Hence
the tensor product $\scr{C}(X) \otimes \scr{C}(Y)$ consists of complex valued functions on $X \times
Y$, i.e. $\scr{C}(X) \otimes \scr{C}(Y) = \scr{C}(X \times Y)$.
In other words states and observables of the composite system $\scr{C}(X) \otimes \scr{C}(Y)$ are, in
accordance with classical probability theory, given by probability distributions and random variables on
the Cartesian product $X \times Y$.

If only one subsystem is classical and the other is quantum; e.g. a micro particle interacting with a
classical measuring device we have a hybrid system. The elements of its observable algebra $\scr{C}(X) \otimes
\scr{B}(\scr{H})$ can be regarded as operator valued functions on $X$, i.e. $X \ni x \mapsto A_x \in
\scr{B}(\scr{H})$  and $A$ is an effect iff $0 \leq A_x \leq \Bbb{1}$ holds for all $x \in X$. The elements of
the dual $\scr{C}^*(X) \otimes \scr{B}^*(\scr{H})$ are in a similar way $\scr{B}^*(X)$ valued functions $X \ni x
\mapsto \rho_x \in \scr{B}^*(\scr{H})$ and $\rho$ is a state iff each $\rho_x$ is a positive trace class operator on
$\scr{H}$ and $\sum_x \rho_x = \Bbb{1}$. The probability to measure the effect $A$ in the state $\rho$ is $\sum_x
\rho_x(A_x)$. 

\subsection{Correlations and entanglement}
\label{sec:corr-entangl}

Let us now consider two effects $A \in \scr{A}$  and $B \in \scr{B}$ then $A \otimes B$ is an effect of the
composite system $\scr{A} \otimes \scr{B}$. It is interpreted as the joint measurement of $A$ on the first
and $B$ on the second subsystem, where the ``yes'' outcome means ``both effects give yes''. In particular
$A \otimes \Bbb{1}$ means to measure $A$ on the first subsystem and to ignore the second one completely. If $\rho$
is a state of $\scr{A} \otimes \scr{B}$ we can define its \emph{restrictions} by $\rho^\scr{A}(A) = \rho(A \otimes \Bbb{1})$
and $\rho^\scr{B}(A) = \rho(\Bbb{1} \otimes A)$. If both systems are quantum the restrictions of $\rho$ are the
partial traces, while in the classical case we have to sum over the $\scr{B}$, respectively $\scr{A}$
variables. For two states
$\rho_1 \in \scr{S}(\scr{A})$ and $\rho_2 \in \scr{S}(\scr{B})$ there is always a state $\rho$ of $\scr{A} \otimes \scr{B}$
such that $\rho_1 = \rho^\scr{A}$ and $\rho_2 = \rho^\scr{B}$ holds: We just have to choose the product state $\rho_1 \otimes
\rho_2$. However in general we have $\rho \not= \rho^\scr{A} \otimes \rho^\scr{B}$ which means nothing else then $\rho$ also
contains \emph{correlations} between the two subsystems systems. 

\begin{defi}
   A state $\rho$ of a bipartite system $\scr{A} \otimes \scr{B}$ is called \emph{correlated} if there are some $A
  \in \scr{A}$, $B \in \scr{B}$ such that $\rho(A \otimes B) \not= \rho^\scr{A}(A)\rho^\scr{B}(B)$ holds.
\end{defi}

We immediately see that $\rho = \rho_1 \otimes \rho_2$ implies $\rho(A \otimes B) = \rho_1(A) \rho_2(B) = \rho^\scr{A}(A)\rho^\scr{B}(B)$
hence $\rho$ is not correlated. If on the other hand $\rho(A \otimes B) = \rho^\scr{A}(A)\rho^\scr{B}(B)$ holds we get $\rho =
\rho^\scr{A} \otimes \rho^\scr{B}$. Hence, the definition of correlations just given perfectly fits into our intuitive
considerations.  

An important issue in quantum information theory is the comparison of correlations between quantum
systems on the one hand and classical systems on the other. Hence let us have a closer look on the state
space of a system consisting of at least one classical subsystem.

\begin{prop} \label{prop:1}
  Each state $\rho$ of a composite system $\scr{A} \otimes \scr{B}$ consisting of a classical ($\scr{A} = 
  \scr{C}(X)$) and an arbitrary system ($\scr{B}$) has the form
  \begin{equation} \label{eq:18}
    \rho =\sum_{j \in X} \lambda_j \rho^\scr{A}_j \otimes \rho^\scr{B}_j
  \end{equation}
  with positive weights $\lambda_j > 0$ and $\rho_j^\scr{A} \in \scr{S}(\scr{A})$,
  $\rho_j^\scr{B} \in \scr{S}(\scr{B})$.
\end{prop}

\begin{proof}
  Since $\scr{A} = \scr{C}(X)$ is classical, there is a basis $\kb{j} \in \scr{A}$, $j \in X$ of mutually
  orthogonal one-dimensional projectors and we can write each $A \in \scr{A}$ as $\sum_j a_j \kb{j}$
  (cf. Subsection \ref{sec:class-prob}). For each state $\rho \in \scr{S}(\scr{A} \otimes \scr{B})$ we can now
  define $\rho^\scr{A}_j \in \scr{S}(\scr{A})$ with $\rho_j^\scr{A}(A) = \tr(A\kb{j}) = a_j$ and $\rho^\scr{B}_j \in
  \scr{S}(\scr{B})$ with $\rho^\scr{B}_j(B) = \lambda_j^{-1} \rho(\kb{j} \otimes B)$ and $\lambda_j = \rho(\kb{j} \otimes \Bbb{1})$. Hence
  we get $\rho = \sum_{j \in X} \lambda_j \rho^\scr{A}_j \otimes \rho^\scr{B}_j$ with positive $\lambda_j$ as stated.    
 \end{proof}

If $\scr{A}$ and $\scr{B}$ are two quantum systems it is still possible for them to be correlated in the
way just described. We can simply prepare them with a classical random generator which triggers two
preparation devices to produce systems in the states $\rho_j^A, \rho_j^B$ with probability $\lambda_j$. The overall
state produced by this setup is obviously the $\rho$ from Equation (\ref{eq:18}). However, the crucial point
is that \emph{not all} correlations of quantum systems are of this type! This is an immediate consequence
of the definition of pure states $\rho = \kb{\Psi} \in \scr{S}(\scr{H})$: Since there is no proper convex
decomposition of $\rho$, it can be written as in Proposition \ref{prop:1} iff $\Psi$ is a product vector,
i.e. $\Psi = \phi \otimes \psi$. This observation motivates the following definition.

\begin{defi} \label{def:3}
  A state $\rho$ of the composite system $\scr{B}(\scr{H}_1) \otimes \scr{B}(\scr{H}_2)$ is called
  \emph{separable} or \emph{classically correlated} if it can be written as
  \begin{equation}
    \rho ={\sum}_j \lambda_j \rho^{(1)}_j \otimes \rho^{(2)}_j
  \end{equation}
  with states $\rho^{(k)}_j$ of $\scr{B}(\scr{H}_k)$ and weights $\lambda_j > 0$. Otherwise $\rho$ is called
  \emph{entangled}. The set of all separable states is denoted by $\scr{D}(\scr{H}_1 \otimes \scr{H}_2)$ or
  just $\scr{D}$ if $\scr{H}_1$ and $\scr{H}_2$ are understood.
\end{defi}

\subsection{Bell inequalities}
\label{sec:bell-inequalities}

We have just seen that it is quite easy for pure states to check whether they are entangled or not. In
the mixed case however this is a much bigger, and in general unsolved, problem. In this subsection we
will have a short look at Bell inequalities, which are maybe the oldest criterion for
entanglement (for a more detailed review see \Cite{BellRev}). Today more powerful methods, most of them
based on positivity properties, are available. We will postpone the corresponding discussion to the end
of the following section, after we have studied (completely) positive maps (cf. Section
\ref{sec:sepr-crit-posit}). 


Bell inequalities are traditionally discussed in the framework of ``local hidden variable
theories''. More precisely we will say that a state $\rho$ of a bipartite system $\scr{B}(\scr{H} \otimes
\scr{K})$ admits a hidden variable model, if there is a probability space $(X,\mu)$ and (measurable)
response functions $X \ni x \mapsto F_A(x,k), F_B(x,l) \in \Bbb{R}$ for all discrete PV measures $A=A_1,\ldots,A_N \in
\scr{B}(\scr{H})$ respectively $B = B_1,\ldots, B_M \in \scr{B}(\scr{K})$ such that 
\begin{equation}
  \int_X F_A(x,k)F_B(x,l) \mu(dx) = \tr(\rho A_k \otimes B_l)
\end{equation}
holds for all, $k,l$ and $A,B$. The value of the functions $F_A(x,k)$ is interpreted as the probability
to get the value $k$ during an $A$ measurement with known ``hidden parameter'' $x$. The set of states
admitting a hidden variable model is a convex set and as such it can be described by an (infinite)
hierarchy of correlation inequalities. Any one of these inequalities is usually called (generalized) Bell
inequality. The most well known one is those given by Clauser, Horne, Shimony and Holt \Cite{CHSH}: The
state $\rho$ satisfies the CHSH-inequality if
\begin{equation} \label{eq:34}
  \rho\bigl(A \otimes (B + B') + A' \otimes (B - B')\bigr) \leq 2
\end{equation}
holds for all $A,A' \in \scr{B}(\scr{H})$ respectively $B,B' \in \scr{B}(\scr{K})$, with $-\Bbb{1} \leq A,A' \leq
\Bbb{1}$ and $-\Bbb{1} \leq B,B' \leq \Bbb{1}$. For the special case of two dichotomic observables the CHSH
inequalities are sufficient to characterize the states with a hidden variable model. In the general case
the CHSH-inequalities are a necessary but not a sufficient condition and a complete characterization is
not known. 

It is now easy to see that each separable state $\rho =\sum_{j=1}^n \lambda_j \rho^{(1)}_j \otimes \rho^{(2)}_j$ admits a hidden
variable model: we have to choose $X={1,\ldots,n}$, $\mu(\{j\})=\lambda_j$, $F_A(x,k)=\rho^{(1)}_x(A_k)$ and $F_B$
analogously. Hence we immediately see that each state of a composite system with at least one classical
subsystem satisfies the Bell inequalities (in particular the CHSH version) while this is not the case for
pure quantum systems. The most prominent examples are ``maximally entangled states'' (cf. Subsection
\ref{sec:pure-states}) which violate the CHSH inequality (for appropriately chosen $A,A',B,B'$) with a
maximal value of $2\sqrt{2}$. This observation is the starting point for many discussions concerning the
interpretation of quantum mechanics, in particular because the maximal violation of $2\sqrt{2}$ was
observed in 1982 experimentally by Aspect and coworkers \Cite{ADR82}. We do not want to follow this path
(see \Cite{BellRev} and the the references therein instead). Interesting for us is the fact that Bell
inequalities, in particular the CHSH case in Equation (\ref{eq:34}), provide a \emph{necessary condition}
for a state $\rho$ to be separable. However there exist entangled states admitting a hidden variable
model \Cite{Werner89}. Hence, Bell inequalities are not sufficient for separability.

\section{Channels}
\label{sec:channels}

Assume now that we have a number of quantum systems, e.g. a string of ions in a trap. To ``process'' the 
quantum information they carry we have to perform in general many steps of a quite different
nature. Typical examples are: free time evolution, controlled time evolution (e.g. the application of a
``quantum gate'' in a quantum computer), preparations and measurements. The purpose of this section is to
provide a unified framework for the description of all these different operations. The basic idea is to
represent each processing step by a ``channel'', which converts input systems, described by an observable
algebra $\scr{A}$ into output systems described by a possibly different algebra $\scr{B}$. Henceforth we
will call $\scr{A}$ the \emph{input} and $\scr{B}$ the \emph{output algebra}. If we consider e.g. the
free time evolution, we need quantum systems of the same type on the input and the output side, hence in
this case we have $\scr{A} = \scr{B} = \scr{B}(\scr{H})$ with an appropriately chosen Hilbert space
$\scr{H}$. If on the other hand we want to describe a measurement we have to map quantum systems (the
measured system) to classical information (the measuring result). Therefore we need in this example
$\scr{A} = \scr{B}(\scr{H})$ for the input and $\scr{B} = \scr{C}(X)$ for the output algebra, where $X$
is the set of possible outcomes of the measurement (cf. Subsection \ref{sec:observables}).

Our aim is now to get a mathematical object which can be used to describe a channel. To this end consider
an effect $A \in \scr{B}$ of the output system. If we invoke first a channel which transforms $\scr{A}$
systems into $\scr{B}$ systems, and measure $A$ afterwards on the output systems, we end up with a 
measurement of an effect $T(A)$ on the input systems. Hence we get a map $ T: \scr{E}(\scr{B}) \to
\scr{E}(\scr{A})$ which \emph{completely describes the channel}\footnote{Note that the direction of the
  mapping arrow is reversed compared to the natural ordering of processing.}. Alternatively we can look
at the states and interpret a channel as a map $T^*: \scr{S}(\scr{A}) \to \scr{S}(\scr{B})$ which
transforms $\scr{A}$ systems in the state $\rho \in \scr{S}(\scr{A})$ into $\scr{B}$ systems in the state
$T^*(\rho)$. To distinguish between both maps we can say that $T$ describes the channel in the
\emph{Heisenberg picture} and $T^*$ in the \emph{Schr{\"o}dinger picture}. On the level of the statistical
interpretation both points of view should coincide of course, i.e. the probabilities\footnote{To keep
  notations more readable we will follow   frequently the usual convention to drop the parenthesis around
  arguments of linear operators. Hence we   will write $TA$ and $T^*\rho$ instead of $T(A)$ and
  $T^*(\rho)$. Similarly we will simply write $TS$ instead   of $T \circ S$ for compositions.} $(T^*\rho)(A)$ and
$\rho(TA)$ to get the result ``yes'' during an $A$ measurement on $\scr{B}$ systems in the state $T^*\rho$,
respectively a $TA$ measurement on $\scr{A}$ systems in the state $\rho$, should be the same. Since
$(T^*\rho)(A)$ is linear in $A$ we see immediately that $T$ must be an \emph{affine map}, i.e. $T(\lambda_1 A_1 +
\lambda_2 A_2) = \lambda_1T(A_1) + \lambda_2 T(A_2)$ for each convex linear combination $\lambda_1 A_1 + \lambda_2 A_2$ of effects in
$\scr{B}$, and this in turn implies that $T$ can be extended naturally to a \emph{linear map}, which we
will identify in the following with the channel itself, i.e. we say that $T$ \emph{is} the
channel.

\subsection{Completely positive maps}
\label{sec:compl-posit-maps}

Let us change now slightly our point of view and start with a linear operator $T: \scr{A} \to \scr{B}$.
To be a channel, $T$ must map effects to effects, i.e. $T$ has to be positive: $T(A) \geq 0$ $\forall A\geq0$ and
bounded from above by $\Bbb{1}$, i.e. $T(\Bbb{1}) \leq \Bbb{1}$. In addition it is natural to require that
two channels in parallel are again a channel. More precisely, if two channels $T: \scr{A}_1 \to \scr{B}_1$
and $S: \scr{A}_2 \to  \scr{B}_2$ are given we can consider the map $T \otimes S$ which associates to each $A \otimes B
\in \scr{A}_1 \otimes \scr{A}_2$  the tensor product $T(A) \otimes S(B) \in \scr{B}_1 \otimes \scr{B}_2$. It is natural to
assume that $T \otimes S$ is a channel which converts composite systems of type $\scr{A}_1 \otimes \scr{A}_2$  into
$\scr{B}_1 \otimes \scr{B}_2$ systems. Hence $S \otimes T$ should be positive as well \Cite{Paulsen}.

\begin{defi} \label{def:6}
  Consider two observable algebras $\scr{A}$, $\scr{B}$ and a linear map $T: \scr{A} \to
  \scr{B} \subset \scr{B}(\scr{H})$.   
  \begin{enumerate}
  \item 
    $T$ is called \emph{positive} if $T(A) \geq 0$ holds for all positive $A \in \scr{A}$.
  \item \label{item:6}
    $T$ is called \emph{completely positive} (cp) if $T \otimes \Id: \scr{A} \otimes \scr{B}(\Bbb{C}^n) \to  
    \scr{B}(\scr{H}) \otimes \scr{B}(\Bbb{C}^n)$ is positive for all $n \in \Bbb{N}$. Here $\Id$ denotes the
    identity map on $\scr{B}(\Bbb{C}^n)$. 
  \item 
    $T$ is called \emph{unital} if $T(\Bbb{1}) = \Bbb{1}$ holds.
  \end{enumerate}
\end{defi}

Consider now the map $T^*: \scr{B}^* \to \scr{A}^*$ which is \emph{dual} to $T$, i.e. $T^*\rho(A) = \rho(TA)$ for
all $\rho \in \scr{B}^*$ and $A \in \scr{A}$. It is called the Schr{\"o}dinger picture representation of the
channel $T$, since it maps states to states provided $T$ is unital. (Complete) positivity can be defined
in the Schr{\"o}dinger picture as in the Heisenberg picture and we immediately see that $T$ is (completely)
positive iff $T^*$ is. 

It is natural to ask whether the distinction between positivity and complete positivity is really
necessary, i.e. whether there are positive maps which are not completely positive. If at least one of the
algebras $\scr{A}$ or $\scr{B}$ is classical the answer is no: each positive map is completely positive
in this case. If both algebras are quantum however complete positivity is \emph{not implied} by
positivity alone. We will discuss explicit examples in Subsection \ref{sec:transposition}.

If item \ref{item:6} holds only for a fixed $n \in \Bbb{N}$ the map $T$ is called
\emph{$n$-positive}. This is obviously a weaker condition then complete positivity. However,
$n$-positivity implies $m$-positivity for all $m \leq n$, and for $\scr{A} = \scr{B}(\Bbb{C}^d)$ complete
positivity is implied by $n$-positivity, provided $n \geq d$ holds. 

Let us consider now the question whether a channel should be unital or not. We have already mentioned 
that $T(\Bbb{1}) \leq \Bbb{1}$ must hold since effects should be mapped to effects. If $T(\Bbb{1})$ is not
equal to $\Bbb{1}$ we get $\rho(T\Bbb{1}) = T^*\rho(\Bbb{1}) < 1$ for the probability to measure the effect
$\Bbb{1}$ on systems in the state $T^*\rho$, but this is impossible for channels \emph{which produce an
  output with certainty}, because $\Bbb{1}$ is the effect which is always true. In other words: If a cp
map is not unital it describes a channel which sometimes produces no output at all and $T(\Bbb{1})$ is
the effect which measures whether we have got an output. We will assume in the future that channels are
unital if nothing else is explicitly stated.

\subsection{The Stinespring theorem}
\label{sec:stinespring-theorem}

Consider now channels between quantum systems, i.e. $\scr{A}=\scr{B}(\scr{H}_1)$ and
$\scr{B}=\scr{B}(\scr{H}_2)$.  
A fairly simple example (not necessarily unital) is given in terms of an operator $V: \scr{H}_1 \to
\scr{H}_2$ by $\scr{B}(\scr{H}_1) \ni A \mapsto VAV^* \in \scr{B}(\scr{H}_2)$. A second example is the restriction
to a subsystem, which is given in the Heisenberg picture by $\scr{B}(\scr{H}) \ni A \mapsto A \otimes \Bbb{1}_\scr{K} \in
\scr{B}(\scr{H} \otimes \scr{K})$. Finally the composition $S \circ T =  ST$ of two channels is again a
channel. The following theorem, which is the most fundamental structural result about cp
maps\footnote{Basically there is a more general version of this   theorem which works with arbitrary
  output algebras. It needs however some material from representation   theory of C*-algebras which we
  want to avoid here. See e.g. \Cite{Paulsen,HolBook2}.}, says that each channel can be represented as a
composition of these two examples \Cite{StSpr}.   

\begin{thm}[Stinespring dilation theorem] \label{thm:2}
  Every completely positive map $T: \scr{B}(\scr{H}_1) \to \scr{B}(\scr{H}_2)$ has the form 
  \begin{equation} \label{eq:8}
    T(A) = V^* (A \otimes \Bbb{1}_\scr{K}) V,
  \end{equation}
  with an additional Hilbert space $\scr{K}$ and an operator $V: \scr{H}_2 \to \scr{H}_1 \otimes \scr{K}$. 
  Both (i.e. $\scr{K}$ and $V$) can be chosen such that the span of all $(A \otimes \Bbb{1})V\phi$ with $A \in  
  \scr{B}(\scr{H}_1)$ and $\phi \in \scr{H}_2$ is dense in $\scr{H}_1 \otimes \scr{K}$. This particular
  decomposition is unique (up to unitary equivalence) and called the \emph{minimal} decomposition. If
  $\dim \scr{H}_1 = d_1$ and $\dim \scr{H}_2 = d_2$ the minimal $\scr{K}$ satisfies $\dim \scr{K} \leq d^2_1d_2$. 
 
\end{thm}

By introducing a family $\kb{\chi_j}$ of one dimensional projectors with $\sum_j \kb{\chi_j} = \Bbb{1}$ we can
define the ``Kraus operators'' $\langle\psi,V_j\phi\rangle = \langle\psi \otimes \chi_j,V\phi\rangle$. In terms of them we can rewrite Equation
(\ref{eq:8}) in the following form \Cite{Kraus}: 

\begin{kor}[Kraus form] \label{kor:1}
  Every completely positive map $T: \scr{B}(\scr{H}_1) \to \scr{B}(\scr{H}_2)$ can be written in the
  form 
  \begin{equation} \label{eq:40}
    T(A) = \sum_{j=1}^N V_j^* A V_j
  \end{equation}
  with operators $V_j: \scr{H}_2 \to \scr{H}_1$ and $N \leq \dim(\scr{H}_1) \dim(\scr{H}_2)$.
\end{kor}

\subsection{The duality lemma}
\label{sec:duality-lemma}

We will consider a fundamental relation between positive maps and bipartite systems, which will allow us
later on 
to translate properties of entangled states to properties of channels and vice versa. The basic idea
originates from elementary linear algebra: A bilinear form $\phi$ on a $d$-dimensional vector space $V$ can
be represented by a $d \times d$-matrix, just as an operator on $V$. Hence, we can transform $\phi$ into an
operator simply by reinterpreting the matrix elements. In our situation things are more difficult,
because the positivity constraints for states and channels should match up in the right way. Nevertheless
we have the following theorem.  

\begin{thm} \label{thm:6}
  Let $\rho$ be a density operator on $\scr{H} \otimes \scr{H}_1$. Then there is a Hilbert space $\scr{K}$ a
  pure state $\sigma$ on $\scr{H} \otimes \scr{K}$ and a channel $T: \scr{B}(\scr{H}_1) \to \scr{B}(\scr{K})$ with  
  \begin{equation} \label{eq:11}
    \rho = \left(\Id \otimes T^*\right) \sigma,
  \end{equation}
  where $\Id$ denotes the identity map on $\scr{B}^*(\scr{H})$. The pure state $\sigma$ can be chosen such
  that $\tr_\scr{H}(\sigma)$ has no zero eigenvalue. In this case $T$ and $\sigma$ are uniquely determined (up to
  unitary equivalence) by Equation (\ref{eq:11}); i.e. if $\tilde{\sigma}$, $\tilde{T}$ with $\rho =
  \left(\Id \otimes \tilde{T}^*\right) \tilde{\sigma}$ are given, we have $\tilde{\sigma} = (\Bbb{1} \otimes U)^* \sigma
  (\Bbb{1} \otimes U)$ and  $\tilde{T}(\,\cdot\,) = U^* T(\,\cdot\,) U$ with an appropriate unitary operator $U$.  
\end{thm}

\begin{proof}
  The state $\sigma$ is obviously the purification of $\tr_{\scr{H}_1} (\rho)$. Hence if $\lambda_j$ and $\psi_j$ are
  eigenvalues and eigenvectors of $\tr_{\scr{H}_1}(\rho)$ we can set $\sigma = \kb{\Psi}$ with $\Psi = \sum_j \sqrt{\lambda_j}
  \psi_j \otimes \phi_j$ where $\phi_j$ is an (arbitrary) orthonormal basis in $\scr{K}$. It is clear that $\sigma$ is
  uniquely determined up to a unitary. Hence we only have to show that a unique $T$ exists if $\Psi$ is
  given. To satisfy Equation (\ref{eq:11}) we must have
  \begin{align}
    \rho\bigl(\KB{\psi_j \otimes \eta_k}{\psi_l \otimes \eta_l}\bigr) &= \left\langle\Psi, (\Id \otimes T)\bigl(\KB{\psi_j \otimes \eta_k}{\psi_l \otimes
          \eta_l}\bigr)\Psi\right\rangle  \\
    &= \left\langle\Psi, \KB{\psi_j}{\psi_l} \otimes T\bigl(\KB{\eta_k}{\eta_p}\bigr) \Psi\right\rangle \\
      &= \sqrt{\lambda_j\lambda_l} \left\langle\phi_j, T\bigl(\KB{\eta_k}{\eta_p}\bigr)\phi_l\right\rangle, \label{eq:63}
  \end{align}
  where $\eta_k$ is an (arbitrary) orthonormal basis in $\scr{H}_1$. Hence $T$ is uniquely determined by $\rho$ in
  terms of its matrix elements and we only have to check complete positivity. To this end it is useful to
  note that the map $\rho \mapsto T$ is linear if the $\lambda_j$ are fixed. Hence it is sufficient to consider the case
  $\rho = \kb{\chi}$. Inserting this in Equation (\ref{eq:63}) we immediately see that $T(A) = V^*AV$ with $\langle V
  \phi_j, \eta_k\rangle = \lambda_j^{-1/2} \langle\psi_j \otimes \eta_k, \chi\rangle$ holds. Hence $T$ is completely positive. Since normalization
  $T(\Bbb{1}) = \Bbb{1}$ follows from the choice of the $\lambda_j$ the theorem is proved. 
\end{proof}

\section{Separability criteria and positive maps}
\label{sec:sepr-crit-posit}

We have already stated in Subsection \ref{sec:compl-posit-maps} that positive but not completely positive
maps exist, whenever input and output algebra are quantum. No such map represents a valid quantum
operation, nevertheless they are of great importance in quantum information theory, due to their deep
relations to entanglement properties. Hence, this Section is a continuation of the study of
separability criteria which we have started in \ref{sec:bell-inequalities}. In contrast to the rest of
this section, all maps are considered in the Schr{\"o}dinger rather than in the Heisenberg picture.

\subsection{Positivity}
\label{sec:positivity}

Let us consider now an arbitrary positive, but not necessarily completely positive map $T^*:
\scr{B}^*(\scr{H}) \to \scr{B}^*(\scr{K})$. If $\Id$ again denotes the identity map, it  is easy to see
that $(\Id \otimes T^*)(\sigma_2 \otimes \sigma_2) = \sigma_1 \otimes T^*(\sigma_2) \geq  0$ holds for each product state $\sigma_1 \otimes \sigma_2 \in
\scr{S}(\scr{H} \otimes \scr{K})$. Hence $(Id \otimes T^*) \rho \geq 0$ for each positive $T^*$ is a necessary
condition for $\rho$ to be separable. The following theorem proved  in \Cite{H3PosSep} shows that
sufficiency holds as well. 

\begin{thm} \label{thm:4}
 A state $\rho \in \scr{B}^*(\scr{H} \otimes \scr{K})$ is separable iff for any  positive map $T^*: \scr{B}^*(\scr{K}) \to
 \scr{B}^*(\scr{H})$ the operator $(\Id \otimes T^*)\rho$ is positive. 
\end{thm}

\begin{proof}  
  We will only give a sketch of the proof see \Cite{H3PosSep} for details. The condition is obviously
  necessary since $(\Id \otimes T^*) \rho_1 \otimes \rho_2 \geq 0$ holds for any product state provided $T^*$ is
  positive. The proof of sufficiency relies on the fact that it is always possible to separate a point
  $\rho$ (an entangled state) from a convex set $\scr{D}$ (the set of separable states) by a hyperplane. A
  precise formulation of this idea leads to the following proposition. 

  \begin{prop} \label{prop:3}
    For any entangled state $\rho \in \scr{S}(\scr{H} \otimes \scr{K})$ there is an operator $A$ on $\scr{H} \otimes
    \scr{K}$ called \emph{entanglement witness} for $\rho$, with the property $\rho(A) < 0$ and $\sigma(A) \geq 0$ for
    all separable $\sigma \in \scr{S}(\scr{H} \otimes \scr{K})$. 
  \end{prop}

  \begin{proof}
    Since $\scr{D} \subset \scr{B}^*(\scr{H} \otimes \scr{K})$ is a closed convex set, for each $\rho \in \scr{S} \subset
    \scr{B}^*(\scr{H} \otimes \scr{K})$ with $\rho \not\in \scr{D}$ there exists a linear functional $\alpha$ on 
    $\scr{B}^*(\scr{H} \otimes \scr{K})$, such that $\alpha(\rho) < \gamma \leq \alpha(\sigma)$ for each $\sigma \in \scr{D}$ with a constant
    $\gamma$. This holds as well in infinite dimensional Banach spaces and is a consequence of the Hahn-Banach
    theorem (cf. \Cite{Rudin} Theorem 3.4). Without loss of generality we can assume that $\gamma = 0$
    holds. Otherwise we just have to replace $\alpha$ by $\alpha - \gamma \tr$. Hence the result follows from the fact
    that each linear functional on $\scr{B}^*(\scr{H} \otimes \scr{K})$ has the form $\alpha(\sigma) = \tr(A\sigma)$ with $A \in
    \scr{B}(\scr{H} \otimes \scr{K})$.  
  \end{proof}

  To continue the proof of Theorem \ref{thm:4} associate now to any operator $A \in \scr{B}(\scr{H} \otimes
  \scr{K})$ the map $T_A^*: \scr{B}^*(\scr{K}) \to \scr{B}^*(\scr{H})$ with
  \begin{equation}
     \tr(A \rho_1 \otimes \rho_2) = \tr(\rho_1^T T_A^*(\rho_2)),
  \end{equation}
  where $(\,\cdot\,)^T$ denotes the transposition in an arbitrary but fixed orthonormal basis $\ket{j}$,
  $j=1,\ldots,d$. It is easy to see that $T_A^*$ is positive if $\tr(A \rho_1 \otimes \rho_2) \geq 0$ for all product states 
  $\rho_1 \otimes \rho_2 \in \scr{S}(\scr{H} \otimes \scr{K})$ \Cite{Jamiolkowski}. A straightforward calculation
  \Cite{H3PosSep} shows in addition that  
  \begin{equation} \label{eq:1}
    \tr(A \rho) = \tr\bigl(\kb{\Psi} (\Id \otimes T_A^*)(\rho)\bigr)
  \end{equation}
  holds, where $\Psi = d^{-1/2} \sum_j \ket{j}\otimes\ket{j}$.
  Assume now that $(\Id \otimes T^*)\rho \geq 0$ for all positive $T^*$. Since $T_A^*$ is positive this implies
  that the left hand site of (\ref{eq:1}) is positive, hence $\tr(A\rho) \geq 0$ provided $\tr(A\sigma) \geq 0$ 
  holds for all separable $\sigma$, and the statement follows from Proposition \ref{prop:3}.
\end{proof}

\subsection{The partial transpose}
\label{sec:transposition}

The most typical example for a positive non-cp map is the transposition $\Theta A = A^T$ of $d \times d$ matrices,
which we have just used in the proof of Theorem \ref{thm:4}. $\Theta$ is obviously a positive map, but the
\emph{partial transpose} 
\begin{equation}
  \scr{B}^*(\scr{H} \otimes \scr{K}) \ni \rho \mapsto (\Id \otimes \Theta)(\rho) \in \scr{B}^*(\scr{H} \otimes \scr{K})
\end{equation}
is not. The latter can be easily checked with the maximally entangled state (cf. Subsection
\ref{sec:pure-states}). 
\begin{equation} \label{eq:55}
  \Psi = \frac{1}{\sqrt{d}} \sum_j \ket{j} \otimes \ket{j}
\end{equation}
where $\ket{j} \in \Bbb{C}^d$, $j=1,\ldots,d$ denote the canonical basis vectors. In low dimensions
the transposition is basically the only positive map which is not cp. Due to results of St{\o}rmer 
\Cite{Stormer} and Woronowicz \Cite{Woronowicz} we have: $\dim \scr{H} = 2$ and $\dim \scr{K} = 2,3$
imply that each positive map $T^* : \scr{B}^*(\scr{H}) \to \scr{B}^*(\scr{K})$ has the form $T^* = T_1^* + T_2^* \Theta$
with two cp maps $T_1^*, T_2^*$ and the transposition on $\scr{B}(\scr{H})$. This immediately implies that
positivity of the partial transpose is necessary \emph{and sufficient} for separability of a state $\rho \in
\scr{S}(\scr{H} \otimes \scr{K})$ (cf. \Cite{H3PosSep}): 

\begin{thm} \label{thm:5}
  Consider a bipartite system $\scr{B}(\scr{H} \otimes \scr{K})$ with $\dim\scr{H} = 2$ and $\dim
  \scr{K} = 2,3$. A state $\rho \in \scr{S}(\scr{H} \otimes \scr{K})$ is separable iff its partial transpose
  is positive. 
\end{thm}

To use positivity of the partial transpose as a separability criterion was proposed for the first time by
Peres \Cite{Peres}, and he conjectured that it is a necessary and sufficient condition in arbitrary
finite dimension. Although it has turned out in the meantime that this conjecture is wrong in general
(cf. Subsection \ref{sec:bound-entangl-stat}), partial transposition has become a crucial tool within
entanglement theory and we define: 

\begin{defi} \label{def:4}
  A state $\rho \in \scr{B}^*(\scr{H} \otimes \scr{K})$ of a bipartite quantum system is called \emph{ppt-state}
  if $(\Id \otimes \Theta)\rho \geq 0$ holds and \emph{npt-state} otherwise (ppt=``positive partial transpose'' and
  npt=``negative partial transpose'').
\end{defi}

\subsection{The reduction criterion}
\label{sec:reduction-criterion}

Another frequently used example of a non-cp but positive map is $\scr{B}^*(\scr{H}) \ni \rho \mapsto T^*(\rho) = (\tr
\rho)\Bbb{1} - \rho \in \scr{B}^*(\scr{H})$. The eigenvalues of $T^*(\rho)$ are given by $\tr \rho - \lambda_i$, where $\lambda_i$
are the eigenvalues of $\rho$. If $\rho \geq 0$ we have $\lambda_i \geq 0$ and therefore $\sum_j \lambda_j - \lambda_k \geq 0$. Hence $T^*$ is 
positive. That $T^*$ is not completely positive follows if we consider again the example $\kb{\psi}$ from
Equation (\ref{eq:55}), hence we get
\begin{equation} \label{eq:56}
  \Bbb{1} \otimes \tr_2(\rho) - \rho \geq 0, \quad \tr_1(\rho) \otimes \Bbb{1} - \rho \geq 0
\end{equation}
for any separable state $\rho \in \scr{B}^*(\scr{H} \otimes \scr{K})$, These equations are another non-trivial
separability criterion, which is called the \emph{reduction criterion} \Cite{H2RedKrit,CAGRedCrit}. It is
closely related to the ppt criterion, due to the following proposition (see \Cite{H2RedKrit}) for a
proof). 

\begin{prop}
  Each ppt-state $\rho \in \scr{S}(\scr{H} \otimes \scr{K})$ satisfies the reduction criterion. If $\dim \scr{H} =
  2$ and $\dim \scr{K} = 2,3$ both criteria are equivalent.
\end{prop}

Hence we see with Theorem  \ref{thm:5} that a state $\rho$ in $2 \times 2$ or $2 \times 3$ dimensions is separable iff
it satisfies the reduction criterion.

\chapter{Basic examples} 
\label{cha:basic-examples}

After the somewhat abstract discussion in the last chapter we will become more concrete now. In the
following we will present a number of examples which help on the one hand to understand the structures
just introduced, and which are of fundamental importance within quantum information on the other. 

\section{Entanglement}
\label{sec:corr-entangl-finite}

Although our definition of entanglement (Definition \ref{def:3}) is applicable in arbitrary dimensions,
detailed knowledge about entangled states is available only for low dimensional systems or for states
with very special properties. In this section we will discuss some of the most basic examples.

\subsection{Maximally entangled states}
\label{sec:pure-states}

Let us start with a look on pure states of a composite systems $\scr{A} \otimes \scr{B}$ and their possible
correlations. If one subsystem is classical, i.e. $\scr{A} = \scr{C}\bigl(\{1,\ldots,d\}\bigr)$, the state space
is given according to Subsection \ref{sec:hybrid-systems} by $\scr{S}(\scr{B})^d$ and $\rho \in
\scr{S}(\scr{B})^d$ is pure iff $\rho = (\delta_{j1}\tau, \ldots, \delta_{jd}\tau)$ with $j=1,\ldots,d$ and a pure state $\tau$ of the
$\scr{B}$ system. Hence the restrictions of $\rho$ to $\scr{A}$ respectively $\scr{B}$ are the
Dirac measure $\delta_j \in \scr{S}(X)$ or $\tau \in \scr{S}(\scr{B})$, in other words both restrictions are
pure. This is completely different if $\scr{A}$ and $\scr{B}$ are quantum, i.e. $\scr{A} \otimes \scr{B} =
\scr{B}(\scr{H} \otimes \scr{K})$: Consider $\rho = \kb{\Psi}$ with $\Psi \in \scr{H} \otimes \scr{K}$ and Schmidt 
decomposition (Proposition \ref{prop:2}) $\Psi = \sum_j \lambda_j^{1/2} \phi_j \otimes \psi_j$. Calculating the $\scr{A}$
restriction, i.e. the partial trace over $\scr{K}$ we get
\begin{equation}
  \tr[\tr_\scr{K}(\rho) A] = \tr[ \kb{\Psi} A \otimes \Bbb{1}] = \sum_{jk} \lambda_j^{1/2}\lambda_k^{1/2} \langle\phi_j,A\phi_k\rangle \delta_{jk},
\end{equation}
hence $\tr_\scr{K}(\rho) = \sum_j \lambda_j \kb{\phi_j}$ is mixed iff $\Psi$ is entangled. The most extreme case arises if
$\scr{H} = \scr{K} = \Bbb{C}^d$ and $\tr_\scr{K}(\rho)$ is maximally mixed, i.e. $\tr_\scr{K}(\rho) =
\frac{\Bbb{1}}{d}$. We get for $\Psi$
\begin{equation}
  \Psi = \frac{1}{\sqrt{d}} \sum_{j=1}^d \phi_j \otimes \psi_j
\end{equation}
with two orthonormal bases $\phi_1,\ldots,\phi_d$ and $\psi_1,\ldots,\psi_d$. In $2n \times 2n$ dimensions these states violate
maximally the CHSH inequalities, with appropriately chosen operators $A,A',B,B'$. Such states are
therefore called \emph{maximally entangled}. The most prominent examples of maximally entangled states
are the four ``Bell states'' for two qubit systems, i.e. $\scr{H} =\scr{K} = \Bbb{C}^2$, $\ket{1},
\ket{0}$ denotes the canonical basis and  
\begin{equation} \label{eq:45}
  \Phi_0 = \frac{1}{2} \left( \ket{11} + \ket{00} \right), \quad \Phi_j = i(\Bbb{1} \otimes \sigma_j) \Phi_0, \quad j=1,2,3 
\end{equation}
where we have used the shorthand notation $\ket{jk}$ for $\ket{j} \otimes \ket{k}$ and the $\sigma_j$ denote the
Pauli matrices.

The Bell states, which form an orthonormal basis of $\Bbb{C}^2 \otimes \Bbb{C}^2$, are the best studied and
most relevant examples of entangled states within quantum information. A mixture of them, i.e. a density
matrix $\rho \in \scr{S}(\Bbb{C}^2 \otimes \Bbb{C}^2)$ with eigenvectors $\Phi_j$ and eigenvalues $0 \leq \lambda_j \leq 1$, $\sum_j
\lambda_j = 1$ is called a \emph{Bell diagonal state}. It can be shown \Cite{BDiVSW} that $\rho$ is entangled iff
$\max_j \lambda_j > 1/2$ holds. We omit the proof of this statement here, but we will come back to this point
in Chapter \ref{cha:quant-theory-i} within the discussion of entanglement measures.

Let us come back to the general case now and consider an arbitrary $\rho \in \scr{S}(\scr{H} \otimes
\scr{H})$. Using maximally entangled states, we can introduce another separability criterion in terms of
the \emph{maximally entangled fraction} (cf. \Cite{BDiVSW})  
\begin{equation} \label{eq:57}
  \scr{F}(\rho) = \sup_{\Psi\ \text{max. ent.}} \langle \Psi, \rho \Psi\rangle.
\end{equation}
If $\rho$ is separable the reduction criterion (\ref{eq:56}) implies $\langle\Psi, [\tr_1(\rho) \otimes \Bbb{1} - \rho] \Psi\rangle \geq 0$
for any maximally entangled state. Since the partial trace of $\kb{\Psi}$ is $d^{-1} \Bbb{1}$ we get
\begin{equation}
   d^{-1} = \langle\Psi, \tr_1(\rho) \otimes \Bbb{1} \Psi\rangle \leq \langle\Psi, \rho \Psi\rangle,
\end{equation}
hence $\scr{F}(\rho) \leq 1/d$. This condition is not very sharp however. Using the ppt criterion it can be
shown that $\rho = \lambda \kb{\Phi_1} + (1-\lambda) \kb{00}$ (with the Bell state $\Phi_1$) is entangled for all $0 < \lambda \leq 1$
but a straightforward calculation shows that $\scr{F}(\rho) \leq 1/2$ holds for $\lambda \leq 1/2$.

Finally, we have to mention here a very useful parameterization of the set of pure states on $\scr{H} \otimes
\scr{H}$ in terms of maximally entangled states: If $\Psi$ is an arbitrary but fixed maximally entangled
state, each $\phi \in \scr{H} \otimes \scr{H}$ admits (uniquely determined) operators $X_1, X_2$ such that
\begin{equation}
  \phi = (X_1 \otimes \Bbb{1}) \Psi = (\Bbb{1} \otimes X_2) \Psi
\end{equation}
holds. This can be easily checked in a product basis.

\subsection{Werner states}
\label{sec:stat-under-symm}

If we consider entanglement of mixed states rather than pure ones, the analysis becomes quite difficult,
even if the dimensions of the underlying Hilbert spaces are low. The reason is that the state space
$\scr{S}(\scr{H}_1 \otimes \scr{H}_2)$ of a two-partite system with $\dim \scr{H}_i = d_i$ is a geometric
object in a $d_1^2d_2^2-1$ dimensional space. Hence even in the simplest non-trivial case (two qubits)
the dimension of the state space becomes very high (15 dimensions) and naive geometric intuition can be
misleading. Therefore it is often useful to look at special classes of model states, which can be
characterized by only few parameters. A quite powerful tool is the study of symmetry properties; i.e. to
investigate the set of states which is invariant under a group of local unitaries. A general discussion
of this scheme can be found in \Cite{VW1}. In this paper we will present only three of the most prominent
examples.

Consider first a state $\rho \in \scr{S}(\scr{H} \otimes \scr{H})$ (with $\scr{H} = \Bbb{C}^d$) which is invariant
under the group of all $U \otimes U$ with a unitary $U$ on $\scr{H}$; i.e. $[U \otimes U, \rho]=0$ for all $U$. Such a
$\rho$ is usually called a \emph{Werner state} \Cite{Werner89,Popescu94} and its structure can be analyzed
quite easily using a well known result of group theory which goes back to Weyl \Cite{WeylCG} (see also
Theorem IX.11.5 of \Cite{Simon96}), and which we will state in detail for later reference: 

\begin{thm} \label{thm:1}
  Each operator $A$ on the $N$-fold tensor product $\scr{H}^{\otimes N}$ of the (finite dimensional) Hilbert space
  $\scr{H}$ which commutes with all unitaries of the form $U^{\otimes N}$ is a linear combination of
  permutation operators, i.e. $A = \sum_\pi \lambda_\pi V_\pi$, where the sum is taken over all permutations $\pi$ of $N$ 
  elements, $\lambda_\pi \in \Bbb{C}$ and $V_\pi$ is defined by
  \begin{equation} \label{eq:105}
    V_\pi \phi_1 \otimes \cdots  \otimes \phi_N = \phi_{\pi^{-1}(1)} \otimes \cdots \otimes \phi_{\pi^{-1}(N)}.
  \end{equation}
\end{thm}

In our case ($N=2$) there are only two permutations: the identity $\Bbb{1}$ and the flip $F(\psi \otimes
\phi) = \phi \otimes \psi$. Hence $\rho = a \Bbb{1} + b F$ with appropriate coefficients $a,b$. Since $\rho$ is a density
matrix, $a$ and $b$ are not independent. To get a transparent way to express these constraints, it is
reasonable to consider the eigenprojections $P_\pm$ of $F$ rather then $\Bbb{1}$ and $F$; i.e. $FP_\pm\psi =
\pm P_\pm\psi$ and $P_\pm=(\Bbb{1} \pm F)/2$. The $P_\pm$ are the projections on the subspaces $\scr{H}^{\otimes2}_\pm \subset
\scr{H} \otimes \scr{H}$ of symmetric respectively antisymmetric tensor products (Bose- respectively
Fermi-subspace). If we write $d_\pm = d(d\pm1)/2$ for the dimensions of $\scr{H}^{\otimes2}_\pm$ we get for each
Werner state $\rho$ 
\begin{equation} \label{eq:38}
  \rho = \frac{\lambda}{d_+} P_+ + \frac{(1-\lambda)}{d_-} P_-,\quad \lambda \in [0,1].
\end{equation}
On the other hand it is obvious that each state of this form is $U \otimes U$ invariant, hence a Werner
state.

If $\rho$ is given, it is very easy to calculate the parameter $\lambda$ from the expectation value of $\rho$ and the
flip $\tr(\rho F) = 2\lambda - 1 \in [-1,1]$. Therefore we can write for an \emph{arbitrary} state $\sigma \in
\scr{S}(\scr{H} \otimes \scr{H})$
\begin{equation} \label{eq:68}
  P_\tUU(\sigma) = \frac{\tr(\sigma F)+1}{2d_+} P_+ + \frac{(1-\tr{\sigma F})}{2d_-} P_-,
\end{equation}
and this defines a projection from the full state space to the set of Werner states which is called the
\emph{twirl operation}. In many cases it is quite useful that it can be written alternatively as a group
average of the form
\begin{equation} \label{eq:37}
  P_\tUU(\sigma) = \int_{\U(d)} (U \otimes U) \sigma (U^* \otimes U^*) dU, 
\end{equation}
where $dU$ denotes the normalized, left invariant Haar measure on $\U(d)$. To check this identity note
first that its right hand side is indeed $U \otimes U$ invariant, due to the invariance of the volume element
$dU$. Hence we have to check only that the trace of $F$ times the integral coincides with $\tr(F\sigma)$:
\begin{align}
  \tr\left[F \int_{\U(d)} (U \otimes U) \sigma (U^* \otimes U^*) dU \right] &= \int_{\U(d)} \tr \left[F (U \otimes U) \sigma (U^* \otimes U^*)
  \right] dU \\
  &= \tr(F\sigma) \int_{\U(d)} dU= \tr(F\sigma), \label{eq:14}
\end{align}
where we have used the fact that $F$ commutes with $U \otimes U$ and the normalization of $dU$. We can apply
$P_\tUU$ obviously to arbitrary operators $A \in \scr{B}(\scr{H} \otimes \scr{H})$ and, as an integral over
unitarily implemented operations, we get a channel. Substituting $U \to U^*$ in (\ref{eq:37}) and cycling
the trace $\tr(AP_\tUU(\sigma))$ we find $\tr(P_\tUU(A)\rho) = \tr(AP_\tUU(\rho))$, hence $P_\tUU$ has the same form
in the Heisenberg and the Schr{\"o}dinger picture (i.e. $P_\tUU^* = P_\tUU$). 

If $\sigma \in \scr{S}(\scr{H} \otimes \scr{H})$ is a separable state the integrand of $P_\tUU(\sigma)$ in Equation
(\ref{eq:37}) consists entirely of separable states, hence $P_\tUU(\sigma)$ is separable. Since each Werner
state $\rho$ is the twirl of itself, we see that $\rho$ is separable iff it is the twirl $P_\tUU(\sigma)$ of a
separable state $\sigma \in \scr{S}(\scr{H} \otimes \scr{H})$. To determine the set of separable Werner states we
therefore have to calculate only the set of all $\tr(F\sigma) \in [-1,1]$ with separable $\sigma$. Since each such $\sigma$
admits a convex decomposition into pure product states it is sufficient to look at
\begin{equation} \label{eq:39}
  \langle\psi \otimes \phi, F \psi \otimes \phi\rangle = |\langle\psi,\phi\rangle|^2
\end{equation}
which ranges from $0$ to $1$. Hence $\rho$ from Equation (\ref{eq:38}) is separable iff $1/2 \leq \lambda \leq 1$ and
entangled otherwise (due to $\lambda = (\tr(F\rho) + 1)/2$). If $\scr{H} = \Bbb{C}^2$ holds, each Werner state is
Bell diagonal and we recover the result from Subsection \ref{sec:pure-states} (separable if highest
eigenvalue less or equal than $1/2$). 

\subsection{Isotropic states}
\label{sec:isotropic-states}

To derive a second class of states consider the partial transpose $(\Id  \otimes \Theta)\rho$ (with respect to a
distinguished base $\ket{j} \in \scr{H}$, $j=1,\ldots,d$) of a Werner state $\rho$. Since $\rho$ is, by definition, $U
\otimes U$ invariant, it is easy to see that $(\Id  \otimes \Theta)\rho$ is $U \otimes \bar{U}$ invariant, where $\bar{U}$
denotes component wise complex conjugation in the base $\ket{j}$ (we just have to use that $U^* =
\bar{U}^T$ holds). Each state $\tau$ with this kind of symmetry is called an \emph{isotropic state}
\Cite{Rains99}, and our previous discussion shows that $\tau$ is a linear combination of $\Bbb{1}$ and the
partial transpose of the flip, which is the rank one operator
\begin{equation} \label{eq:73}
  \tilde{F} = (\Id \otimes \Theta) F = \kb{\Psi} = \sum_{jk=1}^d \KB{jj}{kk},
\end{equation}
where $\Psi=\sum_j \ket{jj}$ is, up to normalization a maximally entangled state. Hence each isotropic $\tau$ can
be written as
\begin{equation} \label{eq:51}
  \tau = \frac{1}{d}\left(\lambda \frac{\Bbb{1}}{d} + (1-\lambda) \tilde{F}\right),\quad \lambda \in
  \left[0,\frac{d^2}{d^2-1}\right], 
\end{equation}
where the bounds on $\lambda$ follow from normalization and positivity. As above we can determine the parameter
$\lambda$ from the expectation value 
\begin{equation}
  \tr(\tilde{F} \tau) = \frac{1-d^2}{d}  \lambda + d
\end{equation}
which ranges from $0$ to $d$ and this again leads to a twirl operation: For an arbitrary state $\sigma \in
\scr{S}(\scr{H} \otimes \scr{H})$ we can define 
\begin{equation} \label{eq:69}
  P_\tUUbar(\sigma) = \frac{1}{d(1-d^2)} \biggl( \bigl[\tr(\tilde{F} \sigma) - d\bigr] \Bbb{1} + \bigl[1 - d
  \tr(\tilde{F} \sigma)\bigr]  \tilde{F} \biggr),  
\end{equation}
and as for Werner states $P_\tUUbar$ can be rewritten in terms of a group average
\begin{equation} \label{eq:53}
  P_\tUUbar(\sigma) = \int_{\U(d)} (U \otimes \bar{U}) \sigma (U^* \otimes \bar{U}^*) dU.
\end{equation}
Now we can proceed in the same way as above: $P_\tUUbar$ is a channel with $P_\tUUbar^* = P_\tUUbar$,
its fixed points $P_\tUUbar(\tau) = \tau$ are exactly the isotropic states, and the image of the set of separable
states under $P_\tUUbar$ coincides with the set of separable isotropic states. To determine the latter we
have to consider the expectation values (cf. Equation (\ref{eq:39}))
\begin{equation} 
  \langle\psi \otimes \phi, \tilde{F} \psi \otimes \phi\rangle = \left| \sum_{j=1}^d \psi_j\phi_j\right| = |\langle\psi,\bar{\phi}\rangle|^2 \in [0,1].
\end{equation}
This implies that $\tau$ is separable iff 
\begin{equation}
  \frac{d(d-1)}{d^2-1} \leq \lambda \leq \frac{d^2}{d^2-1}
\end{equation}
holds and entangled otherwise. For $\lambda = 0$ we recover the maximally entangled state. For $d=2$, again we
recover again the special case of Bell diagonal states encountered already in the last subsection. 

\subsection{OO-invariant states}
\label{sec:oo-invariant-states}

\begin{figure}[t]
  \begin{center}
    \begin{pspicture}(12,12)
      \rput(3,3){
        \psset{unit=2cm}
        \rput[b](-1,3.2){\Large $\tr(\tilde{F}\rho)$}
        \rput[l](3.2,-1){\Large $\tr(F\rho)$}
        \psgrid[unit=2cm,subgriddiv=1,griddots=20](-1,-1)(-1,-1)(3,3)        
        \pspolygon[unit=2cm,linewidth=2pt,linearc=.5pt,fillstyle=none](-1,0)(1,3)(1,0)
        \pspolygon[unit=2cm,linewidth=2pt,linestyle=dotted,dotsep=1.5pt,linearc=.5pt,fillstyle=none](0,-1)(3,1)(0,1)
        \pspolygon[unit=2cm,linewidth=2pt,fillstyle=solid,fillcolor=gray](0,0)(1,0)(1,1)(0,1)
        \psline(-1,0)(1,0.5)
        \psline[linestyle=dotted](1,0.5)(3,1)
        \psline[linestyle=dotted](0,-1)(0.25,0)
        \psline(0.25,0)(1,3)
      }
    \end{pspicture}
  \end{center}
  \caption{State space of OO-invariant states (upper triangle) and its partial transpose (lower
    triangle) for $d=3$. The special cases of isotropic and Werner states are drawn as thin lines.}
  \label{fig:oo-inv-states}
\end{figure}
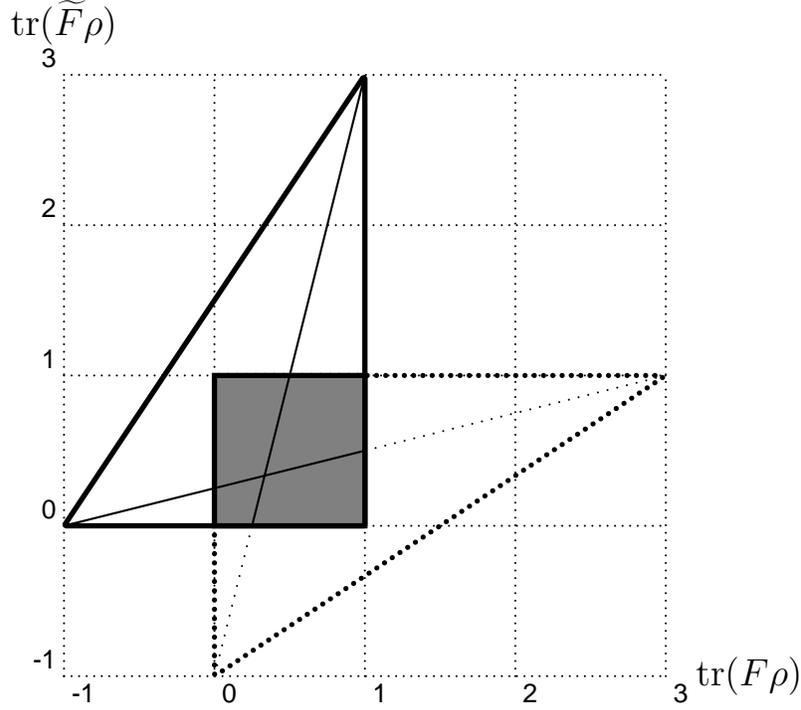

Let us combine now Werner states with isotropic states, i.e. we look for density matrices $\rho$ which can be
written as $ \rho = a \Bbb{1} + b F + c \tilde{F}$, or, if we introduce the three mutually orthogonal
projection operators 
\begin{equation} \label{eq:15}
  p_0 = \frac{1}{d} \tilde{F},\quad p_1 = \frac{1}{2}(\Bbb{1} - F),\quad \frac{1}{2}(\Bbb{1} + F) -
  \frac{1}{d} \tilde{F}
\end{equation}
as a convex linear combination of $\tr(p_j)^{-1} p_j$, $j=0,1,2$:
\begin{equation} \label{eq:13}
  \rho = (1 - \lambda_1 - \lambda_2) p_0 + \lambda_1 \frac{p_1}{\tr(p_1)} + \lambda_2 \frac{p_2}{\tr(p_2)}, \quad \lambda_1,\lambda_2 \geq 0,\ \lambda_1 + \lambda_2 \leq 1
\end{equation}
Each such operator is invariant under all transformations of the form $U \otimes U$ if $U$ is a unitary with $U
= \bar{U}$, in other words: $U$ should be a real orthogonal matrix. A little bit representation theory of
the orthogonal group shows that in fact all operators with this invariance property have the form given
in (\ref{eq:13}); cf. \Cite{VW1}. The corresponding states are therefore called \emph{OO-invariant}, and
we can apply basically the same machinery as in Subsection \ref{sec:stat-under-symm} if we replace the
unitary group $\U(d)$ by the orthogonal group $\Or(d)$. This includes in particular the definition of a
twirl operation as an average over $\Or(d)$ (for an arbitrary $\rho \in \scr{S}(\scr{H} \otimes \scr{H})$):
\begin{equation}
  P_{\tOO}(\rho) = \int_{\Or(d)} U \otimes U \rho U \otimes U^* dU
\end{equation}
which we can express alternatively in terms of the expectation values $\tr(F\rho)$, $\tr(\tilde{F}\rho)$ by
\begin{equation} \label{eq:70}
  P_{\tOO}(\rho) = \frac{\tr(\tilde{F}\rho)}{d}p_0 + \frac{1-\tr(F\rho)}{2\tr(p_1)} p_1 + 
  \left(\frac{1+\tr(F\rho)}{2} - \frac{\tr(\tilde{F}\rho)}{d}\right) \frac{p_2}{\tr(p_2)}.
\end{equation}
The range of allowed values for $\tr(F\rho)$, $\tr(\tilde{F}\rho)$ is given by
\begin{equation}
  -1 \leq \tr(F\rho) \leq 1,\quad 0 \leq \tr(\tilde{F}\rho) \leq d,\quad \tr(F\rho) \geq \frac{2 \tr(\tilde{F}\rho)}{d} - 1.
\end{equation}
For $d=3$ this is the upper triangle in Figure \ref{fig:oo-inv-states}.
 
The values in the lower (dotted) triangle belong to partial transpositions of OO-invariant states. The
intersection of both, i.e. the gray shaded square $Q = [0,1] \times [0,1]$, represents therefore the set of 
OO-invariant ppt states, and at the same time the set of separable states, since each OO-invariant ppt
state is separable. To see the latter note that separable OO-invariant states form a convex subset of
$Q$. Hence, we only have to show that the corners of $Q$ are separable. To do this note that 1. $P_{\tOO}(\rho)$
is separable whenever $\rho$ is and 2. that $\tr\bigl(FP_{\tOO} (\rho)\bigr) = \tr(F\rho)$ and 
$\tr\bigl(\tilde{F} P_{\tOO} (\rho)\bigr) = \tr(F\rho)$ holds (cf. Equation (\ref{eq:14})). We can
consider pure product states $\kb{\phi \otimes \psi}$ for $\rho$ and get $\bigl( |\langle\phi,\psi\rangle|^2, \langle\phi,\bar{\psi}\rangle|^2 \bigr)$ for
the tuple $\bigl(\tr(F\rho), \tr(\tilde{F}\rho)\bigr)$. Now the point $1,1)$ in $Q$ is obtained if $\psi = \phi$ is
real, the point $(0,0)$ is obtained for real and orthogonal $\phi,\psi$ and the point $(1,0)$ belongs to the
case $\psi = \phi$ and $\langle\phi,\bar{\phi}\rangle = 0$. Symmetrically we get $(0,1)$ with the same $\phi$ and $\psi = \bar{\phi}$.

\subsection{PPT states}
\label{sec:bound-entangl-stat}

We have seen in Theorem \ref{thm:5} that separable states and ppt states coincide in $2 \times 2$ and $2 \times 3$
dimensions. Another class of examples with this property are OO-invariant states just studied.
Nevertheless, separability and a positive partial transpose are \emph{not} equivalent. An easy way to
produce such examples of states which are entangled and ppt is given in terms of \emph{unextendible
  product bases} \Cite{BDiVMSST}. An orthonormal family $\phi_j \in \scr{H}_1 \otimes \scr{H}_2$, $j=1,\ldots,N < d_1d_2$
(with $d_k = \dim \scr{H}_k$) is called an unextendible product basis\footnote{This name is somewhat
  misleading because the $\phi_j$ are \emph{not} a base of $\scr{H}_1 \otimes \scr{H}_2$.} (UPB) iff 1. all $\phi_j$
are product vectors and 2. there is no product vector orthogonal to all $\phi_j$. Let us denote the
projector to the span of all $\phi_j$ by $E$, its orthocomplement by $E^\bot$, i.e. $E^\bot = \Bbb{1} - E$, and
define the state $\rho = (d_1d_2 - N)^{-1} E^\bot$. It is entangled because there is by construction no product
vector in the support of $\rho$, and it is ppt. The latter can be seen as follows: The projector $E$ is a
sum of the one dimensional projectors $\kb{\phi_j}$, $j=1,\ldots,N$. Since all $\phi_j$ are product vectors the
partial transposes of the $\kb{\phi_j}$ are of the form $\kb{\tilde{\phi}_j}$, with another UPB $\tilde{\phi}_j$, 
$j=1,\ldots,N$ and the partial transpose $(\Bbb{1} \otimes \Theta)E$ of $E$ is the sum of the $\kb{\tilde{\phi}_j}$. Hence
$(\Bbb{1} \otimes \Theta)E^\bot = \Bbb{1} - (\Bbb{1} \otimes \Theta)E$ is a projector and therefore positive. 

To construct entangled ppt states we have to find UPBs. The following two examples are taken from
\Cite{BDiVMSST}. Consider first the five vectors
\begin{equation}
  \phi_j = N(\cos(2\pi j/5), \sin(2\pi j/5), h),\quad j =0,\ldots,4,
\end{equation}
with $N = 2/ \sqrt{5 + \sqrt{5}}$ and $h = \frac{1}{2}   \sqrt{1 + \sqrt{5}}$. They form the apex of a
regular pentagonal pyramid with height $h$. The latter is chosen such that nonadjacent
vectors are orthogonal. It is now easy to show that the five vectors
\begin{equation}
 \Psi_j = \phi_j \otimes  \phi_{2j \mmod 5},\quad j=0,\ldots,4  
\end{equation}
form a UPB in the Hilbert space $\scr{H} \otimes \scr{H}$, $\dim \scr{H} = 3$ (cf. \Cite{BDiVMSST}). A second
example, again in $3 \times 3$ dimensional Hilbert space are the following five vectors (called ``Tiles'' in
\Cite{BDiVMSST}): 
\begin{gather}
  \frac{1}{\sqrt{2}} \ket{0} \otimes \bigl(\ket{0} - \ket{1}\bigr),\quad \frac{1}{\sqrt{2}} \ket{2} \otimes
  \bigl(\ket{1} - \ket{2}\bigr),\quad \frac{1}{\sqrt{2}} \bigl(\ket{0} - \ket{1}\bigr) \otimes \ket{2},
  \nonumber \\ 
  \frac{1}{\sqrt{2}} \bigl(\ket{1} - \ket{2}\bigr) \otimes \ket{0},\quad \frac{1}{3} \bigl(\ket{0} + \ket{1} +
  \ket{2}\bigr) \otimes \bigl(\ket{0} + \ket{1} + \ket{2}\bigr), 
\end{gather}
where $\ket{k}$, $k =0,1,2$ denotes the standard basis in $\scr{H} = \Bbb{C}^3$.

\subsection{Multipartite states}
\label{sec:tri-partitestates}

In many applications of quantum information rather big systems, consisting of a large number of
subsystems, occur (e.g. a quantum register of a quantum computer) and it is necessary to study the
corresponding correlation and entanglement properties. Since this is a fairly difficult task, there is
not much known about -- much less as in the two-partite case, which we mainly consider in this
paper. Nevertheless, in this subsection we will give a rough outline of some of the most relevant
aspects.

At the level of pure states the most significant difficulty is the lack of an analog of the Schmidt
decomposition \Cite{Peres95}. More precisely there are elements in an $N$-fold tensor product
$\scr{H}^{(1)} \otimes \cdots \otimes \scr{H}^{(N)}$ (with $N > 2$) which can not be written as\footnote{There is however
  the possibility to choose the bases $\phi^{(k)}_1, \ldots, \phi^{(k)}_d$ such that the number of summands becomes
  minimal. For tri-partite systems this ``minimal canonical form'' is study in \Cite{AACJLT}.}
\begin{equation} \label{eq:4}
  \Psi = \sum_{j=1}^d \lambda_j \phi_j^{(1)} \otimes \cdots \otimes \phi_j^{(N)}
\end{equation}
with $N$ orthonormal bases $\phi^{(k)}_1, \ldots, \phi^{(k)}_d$ of $\scr{H}^{(k)}$, $k=1,\ldots,N$. To get examples for such
states in the tri-partite case, note first that any partial trace of $\kb{\Psi}$ with $\Psi$ from Equation
(\ref{eq:4}) has separable eigenvectors. Hence, each purification (Corollary \ref{kor:2}) of an entangled,
two-partite, mixed state with inseparable eigenvectors (e.g. a Bell diagonal state) does not admit a
Schmidt decomposition. This implies on the one hand that there are interesting new properties to be
discovered, but on the other we see that many techniques developed for bipartite pure states can
be generalized in a straightforward way only for states which are \emph{Schmidt decomposable} in the
sense of Equation (\ref{eq:4}). The most well known representative of this class for a tripartite qubit
system is the GHZ state \Cite{GHZ}
\begin{equation}
  \Psi = \frac{1}{\sqrt{2}} \bigl(\ket{000}+ \ket{111}\bigr),
\end{equation}
which has the special property that contradictions between local hidden variable theories and quantum
mechanics occur even for non-statistical predictions (as opposed to maximally entangled states of
bipartite systems; \Cite{GHZ,Mermin90a,Mermin90b}). 

A second new aspect arising in the discussion of multiparty entanglement is the fact that several
different notions of separability occur. A state $\rho$ of an $N$-partite system $\scr{B}(\scr{H}_1) \otimes \cdots \otimes
\scr{B}(\scr{H}_N)$ is called \emph{$N$-separable} if
\begin{equation}
  \rho = \sum_J \lambda_J \rho_{j_1} \otimes \cdots \otimes \rho_{j_N},
\end{equation}
with states $\rho_{j_k} \in \scr{B}^*(\scr{H}_k)$ and multi indices $J=(j_1,\ldots,j_k)$. Alternatively, however, we
can decompose $\scr{B}(\scr{H}_1) \otimes \cdots \otimes \scr{B}(\scr{H}_N)$ in two subsystems (or even into $M$
subsystems if $M < N$) and call $\rho$ \emph{biseparable} if it is separable with respect to this
decomposition. It is obvious that $N$-separability implies biseparability with respect to all possible
decompositions. The converse is -- not very surprisingly -- not true. One way to construct a
corresponding counterexample is to use an unextendable product base (cf. Subsection
\ref{sec:bound-entangl-stat}). In \Cite{BDiVMSST} it is shown that the tripartite qubit state
complementary to the UPB  
\begin{equation}
  \ket{0,1,+}, \ket{1,+,0}, \ket{+,0,1}, \ket{-,-,-}\ \text{with}\ \ket{\pm} = \frac{1}{\sqrt{2}}
  \left(\ket{0} \pm \ket{1}\right)
\end{equation}
is entangled (i.e. tri-inseparable) but biseparable with respect to any decomposition into two
subsystems (cf. \Cite{BDiVMSST} for details). 

Another, maybe more systematic, way to find examples for multipartite states with interesting properties
is the generalization of the methods used for Werner states (Subsection \ref{sec:stat-under-symm}),
i.e. to look for density matrices $\rho \in \scr{B}^*(\scr{H}^{\otimes N})$ which commute with all unitaries of the
form $U^{\otimes N}$. Applying again theorem \ref{thm:1} we see that each such $\rho$ is a linear combination of
permutation unitaries. Hence the structure of the set of all $U^{\otimes N}$ invariant states can be derived
from representation theory of the symmetric group (which can be tedious for large $N$!). For $N=3$ this
program is carried out in \Cite{Tilo} and it turns out that the corresponding set of invariant states is
a five dimensional (real) manifold. We skip the details here and refer to \Cite{Tilo} instead. 

\section{Channels}
\label{sec:examples-channels}

In Section \ref{sec:channels} we have introduced channels as very general objects transforming arbitrary
types of information (i.e. classical, quantum and mixtures of them) into one another. In the following we
will consider some of the most important special cases.

\subsection{Quantum channnels}
\label{sec:ideal-channels-noisy}

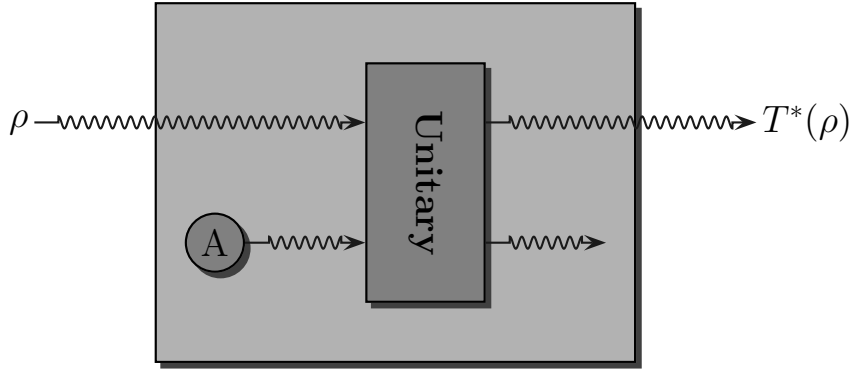
\begin{figure}[t]
  \begin{center}
    \begin{picture}(16,6)
      \rput[bl](-0.5,0){
      \psframe[fillcolor=mklight,fillstyle=solid,shadow=true](4,0)(12,6)
      \psframe[fillcolor=entacolor,fillstyle=solid,shadow=true](7.5,1)(9.5,5)
      \pccoil[linecolor=qbitcolor,coilaspect=0,coilheight=1,coilwidth=.2]{->}(2,4)(7.5,4)
      \pccoil[linecolor=qbitcolor,coilaspect=0,coilheight=1,coilwidth=.2]{->}(9.5,4)(14,4)
      \pscircle[fillcolor=meascolor,fillstyle=solid,shadow=true](5,2){0.5}
      \rput(5,2){\Large A}
      \pccoil[linecolor=qbitcolor,coilaspect=0,coilheight=1,coilwidth=.2]{->}(5.5,2)(7.5,2)
      \pccoil[linecolor=qbitcolor,coilaspect=0,coilheight=1,coilwidth=.2]{->}(9.5,2)(11.5,2)
      \rput(8.5,3){\rotateright{\Large \bf Unitary}}
      \rput[r](1.9,4){\Large $\rho$}
      \rput[l](14.1,4){\Large $T^*(\rho)$}}
    \end{picture}
    \caption{Noisy channel}
    \label{fig:noisy-channel}
  \end{center}
\end{figure}

Many tasks of quantum information theory require the transmission of quantum information over long
distances, using devices like optical fibers or storing quantum information in some sort of memory. Both
situations can be described by a channel or quantum \emph{operation} $T: \scr{B}(\scr{H}) \to
\scr{B}(\scr{H})$, where $T^*(\rho)$ is the quantum information which will be received when $\rho$ was sent, or
alternatively: which will be read off the quantum memory when $\rho$ was written. Ideally we would prefer those
channels which do not affect the information at all, i.e. $T = \Bbb{1}$, or, as the next best choice, a
$T$ whose action can be undone by a physical device, i.e. $T$ should be invertible and $T^{-1}$ is again
a channel. The Stinespring Theorem (Theorem \ref{thm:2}) immediately shows that this implies $T^*\rho = U\rho
U^*$ with a unitary $U$; in other words the systems carrying the information do not interact with the
environment. We will call such a kind of channel an \emph{ideal channel}. In real situations however
interaction with the environment, i.e. additional, unobservable degrees of freedom, can not be
avoided. The general structure of such a \emph{noisy channel} is given by  
\begin{equation} \label{eq:32}
  T^*(\rho) = \tr_\scr{K} \bigl(U (\rho \otimes \rho_0) U^*\bigr)
\end{equation}
where $U: \scr{H} \otimes \scr{K} \to \scr{H} \otimes \scr{K}$ is a unitary operator describing the common evolution of
the system (Hilbert space $\scr{H}$) and the environment (Hilbert space $\scr{K}$) and $\rho_0 \in
\scr{S}(\scr{K})$ is the initial state of the environment (cf. Figure \ref{fig:noisy-channel}).
It is obvious that the quantum information originally stored in $\rho \in \scr{S}(\scr{H})$ can not be
completely recovered from $T^*(\rho)$ if \emph{only   one system is available}. It is an easy consequence of
the Stinepspring theorem that each channel can be expressed in this form 

\begin{kor}[Ancilla form] \label{kor:3}
  Assume that $T: \scr{B}(\scr{H}) \to \scr{B}(\scr{H})$ is a channel. Then there is a   Hilbert space 
  $\scr{K}$, a pure state $\rho_0$ and a unitary map $U: \scr{H} \otimes \scr{K} \to \scr{H} \otimes \scr{K}$ such  
  that Equation (\ref{eq:32}) holds. It is allways possible, to choose $\scr{K}$ such that $\dim(\scr{K})
  = \dim(\scr{H})^3$ holds.
\end{kor}

\begin{proof}
Consider the Stinepspring form $T(A) = V^* (A \otimes \Bbb{1}) V$ with $V : \scr{H} \to \scr{H} \otimes \scr{K}$ of $T$
and choose a vector $\psi \in \scr{K}$ such that $U (\phi \otimes \psi) = V(\phi)$ can be extended to a unitary map $U:
\scr{H} \otimes \scr{K} \to \scr{H} \otimes \scr{K}$ (this is always possible since $T$ is unital and $V$ therefore
isometric). If $e_j \in \scr{H}$, $j=1,\ldots,d_1$ and $f_k \in \scr{K}$, $k=1,\ldots,d_2$ are orthonormal bases with
$f_1  = \psi$ we get
\begin{align}
  \tr\bigl[ T(A) \rho \bigr] &= \tr\bigl[ \rho V^* (A \otimes \Bbb{1}) V \bigr] = \sum_j \langle V \rho e_j, (A \otimes \Bbb{1}) V e_j\rangle \\
  &= \sum_{jk} \Bigl\langle U (\rho \otimes \kb{\psi}) (e_j \otimes f_k), (A \otimes \Bbb{1}) U (e_j \otimes f_k)\Bigr\rangle \\
  &= \tr\Bigl[\tr_\scr{K}\bigl[U (\rho \otimes \kb{\psi}) U^*\bigr] A \Bigr],
\end{align}
which proves the statement.
\end{proof}

Note that there are in general many ways to express a channel this way, e.g. if $T$ is an ideal channel
$\rho \mapsto T^*\rho = U\rho U^*$ we can rewrite it with an arbitrary unitary $U_0 :\scr{K} \to \scr{K}$ by $T^* \rho =
\tr_2( U \otimes U_0 \rho \otimes \rho_0 U^* \otimes U^*_0)$. This is the weakness of the ancilla form compared to the Stinespring
representation of Theorem \ref{thm:2}. Nevertheless Corollary \ref{kor:3} shows that each channel which
is not an ideal channel is noisy in the described way.

The most prominent example for a noisy channel is the \emph{depolarizing channel} for $d$-level systems
(i.e. $\scr{H} = \Bbb{C}^d$)
\begin{equation} 
  \scr{S}(\scr{H}) \ni \rho \mapsto \vartheta \rho + (1-\vartheta) \frac{\Bbb{1}}{d} \in \scr{S}(\scr{H}),\quad 0 \leq \vartheta \leq 1
\end{equation}
or in the Heisenberg picture
\begin{equation} \label{eq:94}
  \scr{B}(\scr{H}) \ni A \mapsto\vartheta A + (1-\vartheta) \frac{\tr(A)}{d} \Bbb{1} \in \scr{B}(\scr{H}).
\end{equation}
A Stinespring dilation of $T$ (not the minimal one -- this can be checked by counting dimensions) is
given by $\scr{K} = \scr{H} \otimes \scr{H} \oplus \Bbb{C}$ and $V: \scr{H} \to \scr{H} \otimes \scr{K} = \scr{H}^{\otimes 3} \oplus
\scr{H}$ with  
\begin{equation} \label{eq:16}
  \ket{j} \mapsto V\ket{j} = \left[ \sqrt{\frac{1-\vartheta}{d}} \sum_{k=1}^d \ket{k}\otimes\ket{k}\otimes\ket{j} \right] \oplus \left[
    \sqrt{\vartheta} \ket{j}\right],
\end{equation}
where $\ket{k}$, $k=1,\ldots,d$ denotes again the canonical basis in $\scr{H}$.  An ancilla form of $T$ with
the same $\scr{K}$ is given by the (pure) environment state  
\begin{equation}
  \psi = \left[ \sqrt{\frac{1-\vartheta}{d}} \sum_{k=1}^d \ket{k}\otimes\ket{k} \right] \oplus \left[\sqrt{\vartheta} \ket{0}\right] \in
  \scr{K} 
\end{equation}
and the unitary operator $U : \scr{H} \otimes \scr{K} \to \scr{H} \otimes \scr{K}$ with
\begin{equation}
   U(\phi_1 \otimes \phi_2 \otimes \phi_3 \oplus \chi) = \phi_2 \otimes \phi_3 \otimes \phi_1 \oplus \chi,
\end{equation}
i.e. $U$ is the direct sum of a permutation unitary and the identity. 

\subsection{Channels under symmetry}
\label{sec:chann-under-symm}

Similarly to the discussion in Section \ref{sec:corr-entangl-finite} it is often useful to consider
channels with special symmetry properties. To be more precise, consider a group $G$ and two unitary
representations $\pi_1, \pi_2$ on the Hilbert spaces $\scr{H}_1$ and $\scr{H}_2$ respectively. A channel $T:
\scr{B}(\scr{H}_1) \to \scr{B}(\scr{H}_2)$ is called \emph{covariant} (with respect to $\pi_1$ and $\pi_2$) if 
\begin{equation}
  T[\pi_1(U)A\pi_1(U)^*] = \pi_2(U)T[A]\pi_2(U)^*\quad \forall A \in \scr{B}(\scr{H}_1)\ \forall U \in G
\end{equation}
holds. The general structure of covariant channels is governed by a fairly powerful variant of
Stinesprings theorem which we will state below (and which will be very useful for the study of the
cloning problem in Chapter \ref{cha:quant-theory-iii}). Before we do this let us have a short look on a
particular class of examples which is closely related to OO-invariant states. 

Hence consider a channel $T: \scr{B}(\scr{H}) \to \scr{B}(\scr{H})$ which is covariant with respect to the
orthogonal group, i.e. $T(UAU^*) = UT(A)U^*$ for all unitaries $U$ on $\scr{H}$ with $\bar{U} = U$ in a
distinguished basis $\ket{j}$, $j=1,\ldots,d$. The maximally entangled state $\psi = d^{-1/2} \sum_j \ket{jj}$ is
OO-invariant, i.e. $U \otimes U \psi = \psi$ for all these $U$. Therefore each state $\rho = (\Id \otimes T^*) \kb{\psi}$ is
OO-invariant as well and by the duality lemma (Theorem \ref{thm:6}) $T$ and $\psi$ are uniquely determined
(up to unitary equivalence) by $\rho$. This means we can use the structure of OO-invariant states derived in
Subsection \ref{sec:oo-invariant-states} to characterize all orthogonal covariant channels. As a first
step consider the linear maps $X_1(A) = d \tr(A) \Bbb{1}$, $X_2(A) = d A^T$ and $X_3(A) = dA$. They are
not channels (they are not unital and $X_2$ is not cp) but they have the correct covariance property and
it is easy to see that they correspond to the operators $\Bbb{1}, F, \tilde{F} \in \scr{B}(\scr{H} \otimes
\scr{H})$, i.e. 
\begin{equation}
  (\Id \otimes X_1) \kb{\psi} = \Bbb{1},\ (\Id \otimes X_2) \kb{\psi} = F,\ (\Id \otimes X_3) \kb{\psi} = \tilde{F}.
\end{equation}
Using Equation (\ref{eq:15}) we can determine therefore the channels which belong to the three extremal
OO-invariant states (the corners of the upper triangle in Figure \ref{fig:oo-inv-states}):
\begin{gather}
  T_0(A) = A,\ T_1(A) = \frac{\tr(A) \Bbb{1} - A^T}{d-1}\\ 
  T_2(A) = \frac{2}{d(d+1)-2} \left[\frac{d}{2}\bigl(\tr(A) \Bbb{1} + A^T\bigr) - A\right] 
\end{gather}
Each OO-invariant channel is a convex linear combination of these three. Special cases are the channels
corresponding to Werner and isotropic states. The latter leads to depolarizing channels 
$T(A) = \vartheta A + (1-\vartheta) d^{-1} \tr(A) \Bbb{1}$ with $\vartheta \in [0,d^2/(d^2-1)]$; cf. Equation (\ref{eq:51}), while
Werner states  correspond to  
\begin{equation}  
  T(A) = \frac{\vartheta}{d+1}\bigl[\tr(A)\Bbb{1} + A^T\bigr] + \frac{1-\vartheta}{d-1} \bigr[\tr(A) \Bbb{1} -
  A^T\bigr],\ \vartheta \in[0,1];
\end{equation}
cf. Equation (\ref{eq:38}).

Let us come back now to the general case. We will state here the covariant version of the Stinespring
theorem (see \Cite{Klo2} for a proof). The basic idea is that all covariant channels  are parameterized by
representations on the dilation space. 

\begin{thm} \label{thm:3}
  Let $G$ be a group with finite dimensional unitary representations $\pi_j: G \to \U(\scr{H}_j)$ and $T:
  \scr{B}(\scr{H}_1) \to \scr{B}(\scr{H}_2)$ a $\pi_1, \pi_2$ - covariant channel. Then there is a finite
  dimensional unitary representation $\tilde{\pi}: G \to \U(\scr{K})$ and an operator $V: \scr{H}_2 \to
  \scr{H}_1 \otimes \scr{K}$ with $V \pi_2(U) = \pi_1(U) \otimes \tilde{\pi}(U)$ and $T(A) = V^* A \otimes \Bbb{1} V$.
\end{thm}

To get an explicit example consider the dilation of a depolarizing channel given in Equation
(\ref{eq:16}). In this case we have $\pi_1(U) = \pi_2(U) = U$ and $\tilde{\pi}(U) = (U \otimes \bar{U}) \oplus
\Bbb{1}$. The check that the map $V$ has indeed the intertwining property $V \pi_2(U) = \pi_1(U) \otimes
\tilde{\pi}(U)$ stated in the theorem is left as an exercise to the reader.

\subsection{Classical channels}
\label{sec:classical-channels}

The classical analog to a quantum operation is a channel $T: \scr{C}(X) \to \scr{C}(Y)$ which describes the 
transmission or manipulation of classical information. As we have mentioned already in Subsection
\ref{sec:compl-posit-maps} positivity and complete positivity are equivalent in this case. Hence we have
to assume only that $T$ is positive and unital. Obviously $T$ is characterized by its matrix elements
$T_{xy} = \delta_y(T \kb{x})$, where $\delta_y \in \scr{C}^*(X)$ denotes the Dirac measure at $y \in Y$ and $\kb{x} \in
\scr{C}(X)$ is the canonical basis in $\scr{C}(X)$ (cf. Subsection \ref{sec:class-prob}). Positivity and
normalization of $T$ imply that $0 \leq T_{xy} \leq 1$ and  
\begin{equation}
  1 = \delta_y(\Bbb{1}) =  \delta_y\bigl(T(\Bbb{1})\bigr) = \delta_y\left[T\left({\sum}_x \kb{x}\right)\right] = {\sum}_x
  T_{xy} 
\end{equation}
holds. Hence the family $(T_{xy})_{x \in X}$ is a probability distribution on $X$ and $T_{xy}$ is therefore
the probability to get the information $x \in X$ at the output side of the channel if $y \in Y$ was
send. Each classical channel is uniquely determined by its matrix of \emph{transition probabilities}. For
$X = Y$ we see that the information is transmitted without error iff $T_{xy}=\delta_{xy}$, i.e. $T$ is
an ideal channel if $T=\Id$ holds and noisy otherwise. 

\subsection{Observables and Preparations}
\label{sec:observables-2}

Let us consider now a channel which transforms quantum information $\scr{B}(\scr{H})$ into classical
information $\scr{C}(X)$. Since positivity and complete positivity are again equivalent,
we just have to look at a positive and unital map $E: \scr{C}(X) \to \scr{B}(\scr{H})$. With the canonical
basis $\kb{x}$, $x \in X$ of $\scr{C}(X)$ we get a family $E_x = E(\kb{x})$, $x \in X$ of positive operators
$E_x \in \scr{B}(\scr{H})$ with $\sum_{x \in X} E_x = \Bbb{1}$. Hence the $E_x$ form a POV measure, i.e. an
observable. If on the other hand a POV measure $E_x \in \scr{B}(\scr{H})$, $x \in X$ is given we can 
define a quantum to classical channel $E: \scr{C}(X) \to \scr{B}(\scr{H})$ by $E(f) = \sum_x f(x)
E_x$. This shows that the observable $E_x, x \in X$ and the channel $E$ can be identified and we say $E$
\emph{is the observable}. 

With this interpretation in mind it is possible to have a short look at continuous observables without the
need of abstract measure theory: We only have to say how the classical algebra $\scr{C}(X)$ is defined
for a set $X$ which is not finite or discrete. For simplicity we assume that $X = \Bbb{R}$ holds, however
the generalization to other locally compact spaces is straightforward. We choose for $\scr{C}(\Bbb{R})$ the
space of continuous, complex valued functions vanishing at infinity, i.e. $|f(x)| < \epsilon$ for each $\epsilon > 0$
provided $|x|$ is large enough. $\scr{C}(\Bbb{R})$ can be equipped with the sup-norm and becomes an
Abelian C*-algebra (cf. \Cite{BraRob1}). To interpret it as an operator algebra as assumed in Subsection 
\ref{sec:operator-algebras} we have to identify $f \in \scr{C}(\Bbb{R})$ with the corresponding
multiplication operator on $\Lz(\Bbb{R})$. An observable taking arbitrary real values can be defined now
as a positive map $E: \scr{C}(\Bbb{R}) \to \scr{B}(\scr{H})$. The probability to get a result in the
interval $[a,b] \subset \Bbb{R}$ during an $E$ measurement on systems in the state $\rho$ is\footnote{Due to the
  Riesz-Markov theorem (cf. Theorem IV.18 of \Cite{RESI1}) the set function $\mu$ extends in unique way to
  a probability measure on the real line.} 
\begin{equation}
  \mu([a,b]) = \sup\, \{ \tr (E(f) \rho) \, | \, f \in \scr{C}(\Bbb{R}),\ 0 \leq f \leq \Bbb{1},\ \supp f \subset [a,b] \} 
\end{equation}
where $\supp$ denotes the \emph{support} of $f$. The most well known example for $\Bbb{R}$ valued
observables are of course position $Q$ and momentum $P$ of a free particle in one dimension. In this case
we have $\scr{H} = \Lz(\Bbb{R})$ and the channels corresponding to $Q$ and $P$ are (in position
representation) given by $\scr{C}(\Bbb{R}) \ni f \mapsto E_Q(f) \in \scr{B}(\scr{H})$ with $E_Q(f)\psi = f\psi$
 respectively $\scr{C}(\Bbb{R}) \ni f \mapsto E_P(f) \in \scr{B}(\scr{H})$ with $E_P(f)\psi = (f\hat{\psi})^{\lor}$ where
$\land$ and $\lor$ denote the Fourier transform and its inverse.
 
Let us return now to a finite set $X$ and exchange the role of $\scr{C}(X)$ and $\scr{B}(\scr{H})$; in
other words let us consider a channel $R: \scr{B}(\scr{H}) \to \scr{C}(X)$ with a classical input and a
quantum output algebra. In the Schr{\"o}dinger picture we get a family of density matrices $\rho_x := R^*(\delta_x) \in
\scr{B}^*(\scr{H})$, $x \in X$, where $\delta_x \in \scr{C}^*(X)$ denote again the Dirac measures (cf. Subsection
\ref{sec:class-prob}). Hence we get a \emph{parameter dependent preparation} which can be used to encode
the classical information $x \in X$ into the quantum information $\rho_x \in \scr{B}^*(\scr{H})$.

\subsection{Instruments and Parameter Dependent Operations}
\label{sec:prep-instr}

\begin{figure}[b]
  \begin{center}
    \begin{picture}(12,3.5)
      \psframe[fillcolor=meascolor,fillstyle=solid,shadow=true](5,0)(7,3)
      \pscoil[linecolor=qbitcolor,coilaspect=0,coilheight=1,coilwidth=.2]{->}(1,1.5)(5,1.5)
      \pscoil[linecolor=qbitcolor,coilaspect=0,coilheight=1,coilwidth=.2]{->}(7,2.5)(11,2.5)
      \psline[linecolor=black]{->}(7,0.5)(11,0.5)
      \rput(6,1.5){\Large $T$}
      \rput[b](3,1.8){$\rho \in \scr{B}^*(\scr{K})$}
      \rput[b](9,2.8){$T^*_x(\rho) \in \scr{B}^*(\scr{H})$}
      \rput[b](9,0.8){$x \in X$}
    \end{picture}
    \caption{Instrument}
    \label{fig:instrument}
  \end{center}
\end{figure}
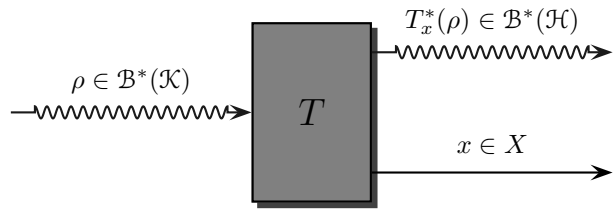

An observable describes only the statistics of measuring results, but contains no information about the
state of the system after the measurement. To get a description which fills this gap we have to consider
channels which operates on quantum systems and produces hybrid systems as output, i.e. $T:
\scr{B}(\scr{H}) \otimes \scr{M}(X) \to \scr{B}(\scr{K})$. Following Davies \Cite{Davies} we will call such an
object an \emph{instrument}. From $T$ we can derive the subchannel
\begin{equation}
  \scr{C}(X) \ni f \mapsto T(\Bbb{1} \otimes f) \in \scr{B}(\scr{K})
\end{equation}
which is the observable measured by $T$, i.e. $\tr\bigl[T\bigl(\Bbb{1} \otimes \kb{x}\bigr)\rho\bigr]$ is the
probability to measure $x \in X$ on systems in the state $\rho$. On the other hand we get for each $x \in X$ a
quantum channel (which is \emph{not} unital) 
\begin{equation} \label{eq:5}
  \scr{B}(\scr{H}) \ni A \mapsto T_x(A) = T(A \otimes \kb{x}) \in \scr{B}(\scr{K}).
\end{equation}
It describes the operation performed by the instrument $T$ if $x \in X$ was measured. More precisely if a
measurement on systems in the state $\rho$ gives the result $x \in X$ we get (up to normalization) the state
$T^*_x(\rho)$ \emph{after the measurement} (cf. Figure \ref{fig:instrument}), while
\begin{equation}
  \tr\left(T^*_x(\rho)\right) = \tr\left(T^*_x(\rho) \Bbb{1}\right) = \tr\bigl(\rho T(\Bbb{1} \otimes \kb{x})\bigr)
\end{equation}
is (again) the probability to measure $x \in X$ on $\rho$. The instrument $T$ can be expressed in terms of the
operations $T_x$ by 
\begin{equation} \label{eq:17}
  T(A \otimes f) = \sum_x f(x) T_x(A);
\end{equation}
hence we can identify $T$ with the family $T_x$, $x \in X$. Finally we can consider the second marginal of
$T$  
\begin{equation}
  \scr{B}(\scr{H}) \ni A \mapsto T(A \otimes \Bbb{1}) = \sum_{x \in X} T_x(A) \in \scr{B}(\scr{K}).
\end{equation}
It describes the operation we get if the outcome of the measurement is ignored.

The most well known example of an instrument is a \emph{von Neumann-L{\"u}ders measurement} associated
to a PV measure given by family of projections $E_x$, $x=1,\ldots d$; e.g. the eigenprojections of a
selfadjoint operator $A \in \scr{B}(\scr{H})$. It is defined as the channel 
\begin{equation}
  T : \scr{B}(\scr{H}) \otimes \scr{C}(X) \to \scr{B}(\scr{H})\ \text{with}\ X = \{1,\ldots,d\}\  \text{and}\ T_x(A) =
  E_xAE_x,
\end{equation}
Hence we get the final state $\tr(E_x\rho)^{-1} E_x \rho E_x$ if we measure the value $x \in X$ on systems
initially in the state $\rho$ -- this is well known from quantum mechanics. 

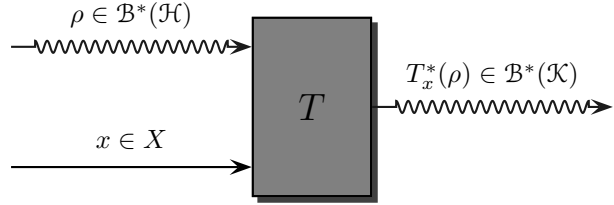
\begin{figure}[h]
  \begin{center}
    \begin{picture}(12,3.5)
      \psframe[fillcolor=meascolor,fillstyle=solid,shadow=true](5,0)(7,3)
      \pscoil[linecolor=qbitcolor,coilaspect=0,coilheight=1,coilwidth=.2]{->}(7,1.5)(11,1.5)
      \pscoil[linecolor=qbitcolor,coilaspect=0,coilheight=1,coilwidth=.2]{->}(1,2.5)(5,2.5)
      \psline[linecolor=black]{->}(1,0.5)(5,0.5)
      \rput(6,1.5){\Large $T$}
      \rput[b](9,1.8){$T^*_x(\rho) \in \scr{B}^*(\scr{K})$}
      \rput[b](3,2.8){$\rho \in \scr{B}^*(\scr{H})$}
      \rput[b](3,0.8){$x \in X$}
    \end{picture}
    \caption{Parameter dependent operation}
    \label{fig:par-dep-op}
  \end{center}
\end{figure}

Let us change now the role of $\scr{B}(\scr{H}) \otimes \scr{C}(X)$ and $\scr{B}(\scr{K})$; in other words
consider a channel $T: \scr{B}(\scr{K}) \to \scr{B}(\scr{H}) \otimes \scr{C}(X)$ with hybrid input and quantum
output. It describes a device which changes the state of a system depending on additional classical
information. As for an instrument, $T$ decomposes into a family of (unital!) channels $T_x:
\scr{B}(\scr{K}) \to \scr{B}(\scr{H})$ such that we get $T^* (\rho \otimes p) = \sum_x p_x T^*_x(\rho)$ in the Schr{\"o}dinger
picture. Physically $T$ describes a \emph{parameter dependent operation}: depending on the classical
information $x \in X$ the quantum information $\rho \in \scr{B}(\scr{K})$ is transformed by the operation $T_x$
(cf. figure \ref{fig:par-dep-op})

\begin{figure}[h]
  \begin{center}
    \begin{picture}(12,3.5)
      \psframe[fillcolor=meascolor,fillstyle=solid,shadow=true](5,0)(7,3)
      \pscoil[linecolor=qbitcolor,coilaspect=0,coilheight=1,coilwidth=.2]{->}(7,2.5)(11,2.5)
      \pscoil[linecolor=qbitcolor,coilaspect=0,coilheight=1,coilwidth=.2]{->}(1,2.5)(5,2.5)
      \psline[linecolor=black]{->}(7,0.5)(11,0.5)
      \psline[linecolor=black]{->}(1,0.5)(5,0.5)
      \rput(6,1.5){\Large $T$}
      \rput[b](9,2.8){$T^*_{y,x}(\rho) \in \scr{B}^*(\scr{K})$}
      \rput[b](9,0.8){$x \in X$}
      \rput[b](3,2.8){$\rho \in \scr{B}^*(\scr{H})$}
      \rput[b](3,0.8){$y \in Y$}
    \end{picture}
    \caption{Parameter dependent instrument}
    \label{fig:par-dep-instr}
  \end{center}
\end{figure}
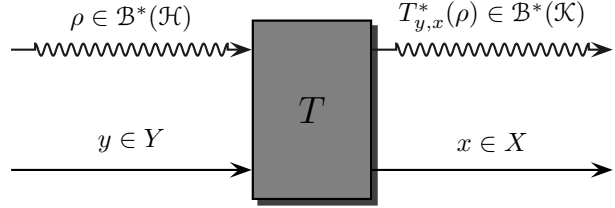

Finally we can consider a channel $T: \scr{B}(\scr{H}) \otimes \scr{C}(X) \to \scr{B}(\scr{K}) \otimes \scr{C}(Y) $
with hybrid input and output to get a \emph{parameter dependent instrument} (cf. figure
\ref{fig:par-dep-instr}): Similarly to the discussion in the last paragraph we can define a family of
instruments $T_y : \scr{B}(\scr{H}) \otimes \scr{C}(X) \to \scr{B}(\scr{K})$, $y \in Y$ by the equation $T^*(\rho \otimes
p) = \sum_y p_y T^*_y(\rho)$. Physically $T$ describes the following device: It receives the classical information
$y \in Y$ and a quantum system in the state $\rho \in \scr{B}^*(\scr{K})$ as input. Depending on $y$ a
measurement with the instrument $T_y$ is performed, which in turn produces the measuring value $x \in X$
and leaves the quantum system in the state (up to normalization) $T_{y,x}^*(\rho)$; with $T_{y,x}$ given as
in Equation (\ref{eq:5}) by $T_{y,x}(A) = T_y(A \otimes \kb{x})$. 

\subsection{LOCC and separable channels}
\label{sec:locc-separ-chann}

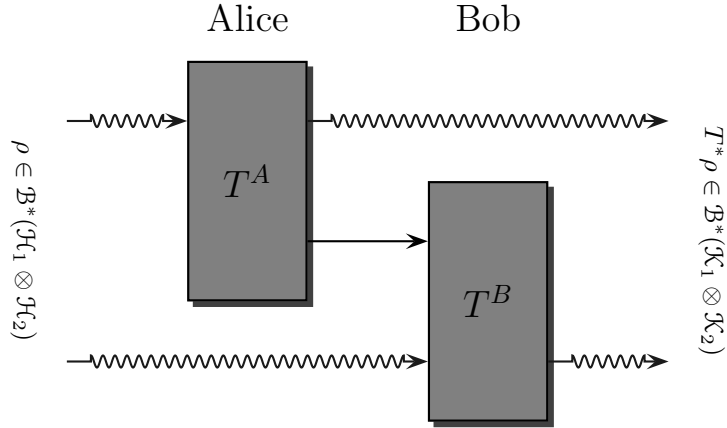
\begin{figure}[h]
  \begin{center}
    \begin{picture}(14,6.5)
      \psframe[fillcolor=meascolor,fillstyle=solid,shadow=true](4,2)(6,6)
      \pscoil[linecolor=qbitcolor,coilaspect=0,coilheight=1,coilwidth=.2]{->}(2,5)(4,5)
      \pscoil[linecolor=qbitcolor,coilaspect=0,coilheight=1,coilwidth=.2]{->}(6,5)(12,5)
      \psline[linecolor=black]{->}(6,3)(8,3)
      \psframe[fillcolor=meascolor,fillstyle=solid,shadow=true](8,0)(10,4)
      \pscoil[linecolor=qbitcolor,coilaspect=0,coilheight=1,coilwidth=.2]{->}(2,1)(8,1)
      \pscoil[linecolor=qbitcolor,coilaspect=0,coilheight=1,coilwidth=.2]{->}(10,1)(12,1)
      \rput(5,4){\Large $T^A$}
      \rput(9,2){\Large $T^B$}
      \rput[r](1.5,3){\rotateright{$\rho \in \scr{B}^*(\scr{H}_1 \otimes \scr{H}_2)$}}
      \rput[l](12.5,3){\rotateright{$T^*\rho \in \scr{B}^*(\scr{K}_1 \otimes \scr{K}_2)$}}
      \rput[b](5,6.5){\Large Alice}
      \rput[b](9,6.5){\Large Bob}
    \end{picture}
    \caption{One way LOCC operation; cf Figure \ref{fig:locc} for an explanation.}
    \label{fig:one-locc}
  \end{center}
\end{figure}

Let us consider now channels acting on finite dimensional bipartite systems: $T: \scr{B}(\scr{H}_1 \otimes
\scr{K}_2) \to \scr{B}(\scr{K}_1 \otimes \scr{K}_2)$. In this case we can ask the question whether a channel
preserves separability. Simple examples are \emph{local operations} (LO), i.e. $T=T^{A} \otimes T^{B}$ with two
channels $T^{A,B}: \scr{B}(\scr{H}_j) \to \scr{B}(\scr{K}_j)$. Physically we think of such a $T$ in terms
of two physicists Alice and Bob both performing operations on their own particle but without information
transmission neither classical nor quantum. The next difficult step are local operations with \emph{one
  way classical communications} (one way LOCC). This means Alice operates on her system with an
instrument, communicates the classical measuring result $j\in X = \{1,\ldots,N\}$ to Bob and he selects an
operation depending on these data. We can write such a channel as a composition
$T = (T^A \otimes \Id)(\Id \otimes T^B)$ of the instrument $T^A: \scr{B}(\scr{H}_1) \otimes
\scr{C}(X_1) \to \scr{B}(\scr{K}_1)$ and the parameter dependent operation $T^B: \scr{B}(\scr{H}_2) \to
\scr{C}(X_1) \otimes \scr{B}(\scr{K}_2)$ (cf. Figure \ref{fig:one-locc})
\begin{equation} \label{eq:41}
  \begin{CD}
    \scr{B}(\scr{H}_1 \otimes \scr{H}_2) @>\Id \otimes T^B>> \scr{B}(\scr{H}_1) \otimes \scr{C}(X) \otimes
    \scr{B}(\scr{K}_2) @>T^A \otimes \Id>> \scr{B}(\scr{K}_1 \otimes \scr{K}_2).
  \end{CD}
\end{equation}

\begin{figure}[t]
  \begin{center}
    \begin{picture}(12,7)
      \psset{coilarm=0}
      \psframe[fillcolor=meascolor,fillstyle=solid,shadow=true](1,2)(2,6)
      \pscoil[linecolor=qbitcolor,coilaspect=0,coilheight=1,coilwidth=.2]{-}(-1,5)(0.5,5)
      \psline{->}(0.5,5)(1,5)
      \pscoil[linecolor=qbitcolor,coilaspect=0,coilheight=1,coilwidth=.2]{-}(2,5)(5,5)
      \psline[linecolor=black]{->}(2,3)(3,3)
      \psframe[fillcolor=meascolor,fillstyle=solid,shadow=true](3,0)(4,4)
      \pscoil[linecolor=qbitcolor,coilaspect=0,coilheight=1,coilwidth=.2]{-}(-1,1)(2.5,1)
      \psline{->}(2.5,1)(3,1)
      \psline[linecolor=black]{-}(4,3)(5,3)
      \pscoil[linecolor=qbitcolor,coilaspect=0,coilheight=1,coilwidth=.2]{-}(4,1)(5,1)
      \pscoil[linecolor=qbitcolor,coilaspect=0,coilheight=1,coilwidth=.2,linestyle=dotted]{-}(5,1)(7,1)
      \pscoil[linecolor=qbitcolor,coilaspect=0,coilheight=1,coilwidth=.2,linestyle=dotted]{-}(5,5)(7,5)
      \psline[linestyle=dotted]{-}(5,3)(7,3)
      \pscoil[linecolor=qbitcolor,coilaspect=0,coilheight=1,coilwidth=.2]{-}(7,1)(9.5,1)
      \pscoil[linecolor=qbitcolor,coilaspect=0,coilheight=1,coilwidth=.2]{-}(7,5)(7.5,5)
      \psline{->}(9.5,1)(10,1)
      \psline{->}(7.5,5)(8,5)
      \psline{->}(7,3)(8,3)
      \psframe[fillcolor=meascolor,fillstyle=solid,shadow=true](8,2)(9,6)
      \psline[linecolor=black]{->}(9,3)(10,3)
      \pscoil[linecolor=qbitcolor,coilaspect=0,coilheight=1,coilwidth=.2]{-}(9,5)(12.5,5)
      \psline{->}(12.5,5)(13,5)
      \psframe[fillcolor=meascolor,fillstyle=solid,shadow=true](10,0)(11,4)
      \pscoil[linecolor=qbitcolor,coilaspect=0,coilheight=1,coilwidth=.2]{-}(11,1)(12.5,1)
      \psline{->}(12.5,1)(13,1)
      \rput[b](1.5,6.5){\Large Alice}
      \rput[b](8.5,6.5){\Large Alice}
      \rput[b](3.5,6.5){\Large Bob}
      \rput[b](10.5,6.5){\Large Bob}
      \psset{coilarm=.4}
    \end{picture}
    \caption{LOCC operation. The upper and lower curly arrows represent Alice's respectively
      Bob's quantum system, while the straight arrows in the middle stand for the classical information 
      Alice and Bob exchange. The boxes symbolize the channels applied by Alice and Bob.}
    \label{fig:locc}
  \end{center}
\end{figure}
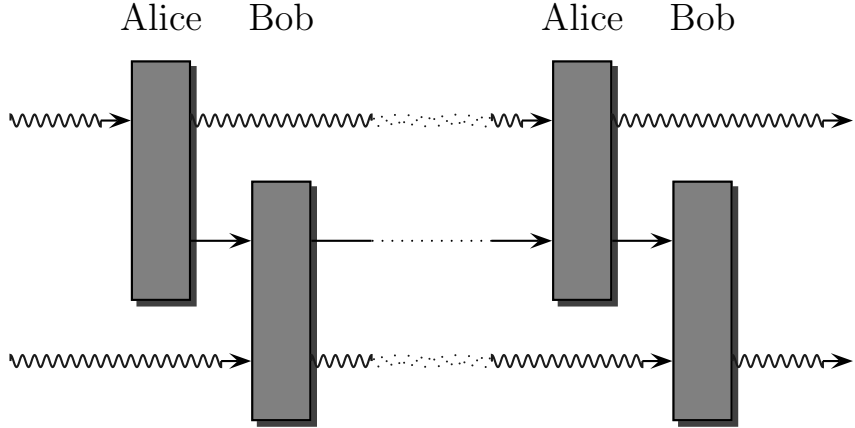

It is of course possible to 
continue the chain in Equation (\ref{eq:41}), i.e. instead of just operating on his system, Bob can
invoke a parameter dependent instrument depending on Alice's data $j_1 \in X_1$, send the corresponding
measuring results $j_2 \in X_2$  to Alice and so on. To write down the corresponding chain of maps (as in
Equation (\ref{eq:41})) is simple but not very illuminating and therefore omitted; cf. Figure
\ref{fig:locc} instead. If we allow Alice and Bob to drop some of their particles, i.e. the operations
they perform need not to be unital, we get a \emph{LOCC channel} (``local operations and classical
communications''). It represents the most general physical process which can be performed on a two
partite system if only classical communication (in both directions) is available.

LOCC channels play a significant role in entanglement theory (we will see this in Section
\ref{sec:dist-entangl}), but they are difficult to handle. Fortunately it is often possible to replace
them by closely related operations with a more simple structure: A \emph{not necessarily unital} channel
$T: \scr{B}(\scr{H}_1 \otimes \scr{K}_2) \to \scr{B}(\scr{K}_1 \otimes \scr{K}_2)$ is called \emph{separable}, if it is
a sum of (in general non-unital) local operations, i.e.
\begin{equation}
  T = \sum_{j=1}^N  T^{A}_j \otimes T^{B}_j. 
\end{equation}
It is easy to see that a separable $T$ maps separable states to separable states (up to normalization)
and that each LOCC channel is separable (cf. \Cite{BDiVFMRSSW}). The converse however is (somewhat
surprisingly) not true: there are separable channels which are not LOCC, see \Cite{BDiVFMRSSW} for a
concrete example. 

\section{Quantum mechanics in phase space}
\label{sec:quant-mech-phase}

Up to now we have considered only finite dimensional systems and even in this extremely
idealized situation it is not easy to get nontrivial results. At a first look the discussion of
continuous quantum systems seems therefore to be hopeless. If we restrict our attention however to
small classes of states and channels, with sufficiently simple structure, many problems become
tractable. Phase space quantum mechanics, which will be reviewed in this Section (see Chapter 5 of
\Cite{HolBook} for details), provides a very powerful tool in this context. 

Before we start let us add some remarks to the discussion of Chapter \ref{cha:basic-concepts} which we
have restricted to finite dimensional Hilbert spaces. Basically most of the material considered there can
be generalized in a straightforward way, as long as topological issues like continuity and convergence
arguments are treated carefully enough. There are of course some caveats (cf. in particular Footnote
\ref{fn:1} of Chapter \ref{cha:basic-concepts}), however they do not lead to problems in the framework
we are going to discuss and can therefore be ignored.

\subsection{Weyl operators and the CCR}
\label{sec:weyl-operators-ccr}

The kinematical structure of a quantum system with $d$ degrees of freedom is usually described by a
separable Hilbert space $\scr{H}$ and $2d$ selfadjoint operators $Q_1,\ldots,Q_d,P_1,\ldots,P_d$ satisfying the
canonical commutation relations $[Q_j,Q_k]=0$, $[P_j,P_k]=0$, $[Q_j,P_k]=i\delta_{jk}\Bbb{1}$. The latter can
be rewritten in a more compact form as
\begin{equation} \label{eq:6}
  R_{2j-1} = Q_j, R_{2j} = P_j,\ j=1,\ldots,d,\ [R_j,R_k] = -i \sigma_{jk}.
\end{equation}
Here $\sigma$ denotes the \emph{symplectic matrix}
\begin{equation}
  \sigma = \diag(J,\ldots,J),\quad J = \left[ \begin{array}{cc} 0 & 1 \\ -1 & 0  \end{array}\right],
\end{equation}
which plays a crucial role for the geometry of classical mechanics. We will call the pair $(V,\sigma)$
consisting of $\sigma$ and the $2d$-dimensional real vector space $V=\Bbb{R}^{2d}$ henceforth the
\emph{classical phase space}.
  
The relations in Equation (\ref{eq:6}) are, however, not sufficient to fix the operators $R_j$ up to
unitary  equivalence. The best way to remove the remaining physical ambiguities is the study of the
unitaries 
\begin{equation} \label{eq:7}
  W(x) = \exp(i  x \cdot \sigma \cdot R),\ x \in V,\ x \cdot \sigma \cdot R = \sum_{jk=1}^{2d} x_j \sigma_{jk} R_k 
\end{equation}
instead of the $R_j$ directly. If the family $W(x)$, $x \in V$ is \emph{irreducible}
(i.e. $[W(x),A]=0$, $\forall x \in V$ implies $A=\lambda\Bbb{1}$ with $\lambda \in \Bbb{C}$) and
satisfies\footnote{Note that the CCR (\ref{eq:6}) are implied by the Weyl relations (\ref{eq:42}) but the
  converse is, in contrast to popular believe, not true: There are representations of the CCR which are
  unitarily inequivalent to the Schr{\"o}dinger representation; cf. \Cite{RESI1} Section VIII.5 for
  particular examples. Hence uniqueness can only be achieved on the level of Weyl operators -- which is
  one major reason to study them.} 
\begin{equation} \label{eq:42}
  W(x)W(x') = \exp\left(-\frac{i}{2}\, x \cdot \sigma \cdot x'\right) W(x+x'),
\end{equation}
it is called an (irreducible) representation of the \emph{Weyl relations} (on $(V,\sigma)$) and the operators
$W(x)$ are called \emph{Weyl operators}. By the well known Stone - von Neumann uniqueness theorem all
these representations are mutually unitarily equivalent, i.e. if we have two of them $W_1(x), W_2(x)$,
there is a unitary operator $U$ with $UW_1(x)U^* = W_2(x)$ $\forall x \in V$. This implies that it does not
matter from a physical point of view which representation we use. The most well known one is of course
the \emph{Schr{\"o}dinger representation} where $\scr{H} = \Lz(\Bbb{R}^d)$ and $Q_j$, $P_k$ are the usual
position and momentum operators.

\subsection{Gaussian states}
\label{sec:gaussian-states}

A density operator $\rho \in \scr{S}(\scr{H})$ has \emph{finite second moments} if the expectation values
$\tr(\rho Q_j^2)$ and $\tr(\rho P_j^2)$ are finite for all $j=1,\ldots,d$. In this case we can define the \emph{mean}
$m \in \Bbb{R}^{2d}$ and the \emph{correlation matrix} $\alpha$ by
\begin{equation} \label{eq:9}
  m_j = \tr(\rho R), \quad \alpha_{jk} + i \sigma_{jk}= 2 \tr\bigl[(R_j-m_j)\rho(R_k-m_k)].
\end{equation}
The mean $m$ can be arbitrary, but the correlation matrix $\alpha$ must be real and symmetric and the 
positivity condition 
\begin{equation} \label{eq:10}
  \alpha + i \sigma \geq 0  
\end{equation}
 must hold (this is an easy consequence of the canonical commutation relations (\ref{eq:6})). 

Our aim is now to distinguish exactly one state among all others with the same mean and correlation
matrix. This is the point where the Weyl operators come into play. Each state $\rho \in \scr{S}(\scr{H})$ can
be characterized uniquely by its \emph{quantum characteristic function} $X \ni x \mapsto \tr\bigl[W(x)\rho\bigr] \in
\Bbb{C}$ which should be regarded as the quantum Fourier transform of $\rho$ and is in fact the Fourier
transform of the Wigner function of $\rho$ \Cite{WeQMPS}. We call $\rho$ \emph{Gaussian} if 
\begin{equation}
  \tr\bigl[W(x)\rho\bigr] = \exp\left(im\cdot x - \frac{1}{4} x \cdot \alpha \cdot x\right)
\end{equation}
holds. By differentiation it is easy to check that $\rho$ has indeed mean $m$ and covariance matrix $\alpha$. 

The most prominent examples for Gaussian states are the ground state $\rho_0$ of a system of $d$ harmonic
oscillators (where the mean is $0$ and $\alpha$ is given by the corresponding classical Hamiltonian) and its 
phase space translates $\rho_m = W(m)\rho W(-m)$ (with mean $m$ and the same $\alpha$ as $\rho_0$), which are known
from quantum optics as \emph{coherent states}. $\rho_0$ and $\rho_m$ are pure states and it can be shown that a
Gaussian state is pure iff $\sigma^{-1}\alpha = -\Bbb{1}$ holds (see \Cite{HolBook}, Ch. 5). Examples for mixed
Gaussians are temperature states of harmonic oscillators. In one degree of freedom this is
\begin{equation} \label{eq:88}
  \rho_N = \frac{1}{N + 1} \sum_{n=0}^\infty \left(\frac{N}{N+1}\right)^n \kb{n}
\end{equation}
where $\kb{n}$ denotes the number basis and $N$ is the mean photon number. The characteristic function of
$\rho_N$ is
\begin{equation}
  \tr\bigl[W(x)\rho_N\bigr] = \exp \left[ -\frac{1}{2} \left( N + \frac{1}{2} \right) |x|^2 \right],
\end{equation}
and its correlation matrix is simply $\alpha = 2 (N + 1/2) \Bbb{1}$

\subsection{Entangled Gaussians}
\label{sec:entangled-gaussians}

Let us consider now bipartite systems. Hence the phase space $(V,\sigma)$ decomposes into a direct sum $V =
V_A \oplus V_B$ (where $A$ stands for ``Alice'' and $B$ for ``Bob'') and the symplectic matrix $\sigma = \sigma_A \oplus \sigma_B$
is block diagonal with respect to this decomposition. If $W_A(x)$ respectively $W_B(y)$ denote Weyl
operators, acting on the Hilbert spaces $\scr{H}_A$, $\scr{H}_B$, and corresponding to the phase spaces
$V_A$ and $V_B$, it is easy to see that the tensor product $W_A(x) \otimes W_B(y)$ satisfies the Weyl relations
with respect to $(V,\sigma)$. Hence by the Stone - von Neumann uniqueness theorem we can identify $W(x \oplus y)$,
$x \oplus y \in V_a \oplus V_B = V$ with $W_A(x) \otimes W_A(y)$. This shows immediately that a state $\rho$ on $\scr{H} =
\scr{H}_A \otimes \scr{H}_B$ is a product state iff its characteristic function factorizes.
Separability\footnote{In infinite dimensions we have to define separable states (in slight generalization
  to Definition \ref{def:3}) as a trace-norm convergent convex sum of product states.} is characterized
as follows (we omit the proof, see \Cite{BEG} instead).

\begin{thm}
  A Gaussian state with covariance matrix $\alpha$ is separable iff there are covariance matrices $\alpha_A, \alpha_B$
  such that
  \begin{equation}
    \alpha \geq \left[ \begin{array}{cc} \alpha_A & 0\\ 0 & \alpha_B \end{array}\right]
  \end{equation}
  holds.
\end{thm}

This theorem is somewhat similar to Theorem \ref{thm:4}: It provides a useful criterion as long as
abstract considerations are concerned, but not for explicit calculations. In contrast to finite
dimensional systems, however, separability of Gaussian states can be decided by an operational criterion
in terms of nonlinear maps between matrices \Cite{GaussSep}. To state it we have to introduce some
terminology first. The key tool is a sequence of $2n+2m \times 2n+2m$ matrices $\alpha_N$, $N \in 
\Bbb{N}$, written in block matrix notation as
\begin{equation}
  \alpha_N = \left[\begin{array}{ll} A_N & C_N \\ C_N^T & B_N \end{array}\right].
\end{equation}
Given $\alpha_0$ the other $\alpha_N$ are recursively defined by: 
\begin{equation}
  A_{N + 1} = B_{N + 1} = A_N - \Re(X_N)\ \text{and}\ C_{N+1} = - \Im(X_N)
\end{equation}
if $\alpha_N - i \sigma \geq 0$ and $\alpha_{N+1} = 0$ otherwise. Here we have set $X_N = C_N (B_N - i\sigma_B)^{-1} C_N^T$ and
the inverse denotes the \emph{pseudo inverse}\footnote{$A^{-1}$ is the pseudo inverse of a matrix $A$ if
  $AA^{-1} = A^{-1}A$ is the projector onto the range of $A$. If $A$ is invertible $A^{-1}$ is the usual
  inverse.}  if $B_N - i \sigma_B$ is not invertible. Now we can state the following theorem (see
\Cite{GaussSep} for a proof):

\begin{thm}
  Consider a Gaussian state $\rho$ of a bipartite system with correlation matrix $\alpha_0$ and the sequence
  $\alpha_N$, $N \in \Bbb{N}$ just defined.
  \begin{enumerate}
  \item \label{item:3} 
    If for some $N \in \Bbb{N}$ we have $A_N -i \sigma_A \not \geq 0$ then $\rho$ is not separable.
  \item \label{item:4}
    If there is on the other hand an $N \in \Bbb{N}$ such that $A_N - \|C_N\| \Bbb{1} - i \sigma_A \geq 0$, then the
    state $\rho$ is separable ($\|C_N\|$ denotes the operator norm of $C_N$).
  \end{enumerate}
\end{thm}

To check whether a Gaussian state $\rho$ is separable or not we have to iterate through the sequence $\alpha_N$
until either condition \ref{item:3} or \ref{item:4} holds. In the first case we know that $\rho$ is
entangled and separable in the second. Hence only the question remains whether the whole procedure
terminates after a finite number of iterations. This problem is treated in \Cite{GaussSep} and it
turns out that the set of $\rho$ for which separability is decidable after a finite number of steps is the
complement of a measure zero set (in the set of all separable states). Numerical calculations indicate in
addition that the method converges usually very fast (typically less than five iterations).

To consider ppt states we first have to characterize the transpose for infinite dimensional
systems. There are different ways to do that. We will use the fact that the adjoint of a matrix can be
regarded as transposition followed by componentwise complex conjugation. Hence we define for any
(possibly unbounded) operator $A^T = CA^*C$, where $C: \scr{H} \to \scr{H}$ denotes complex
conjugation of the wave function in position representation. This implies $Q^T_j = Q_j$ for position and
$P_j^T = -P_j$ for momentum operators. If we insert the \emph{partial} transpose of a bipartite state
$\rho$ into Equation (\ref{eq:9}) we see that 
the correlation matrix
$\tilde{\alpha}_{jk}$ of $\rho^T$ picks up a minus sign whenever one of the indices belongs to one of Alice's
momentum operators. To be a state $\tilde{\alpha}$ should satisfy $\tilde{\alpha} + i \sigma \geq 0$, but this is
equivalent to $\alpha + i \tilde{\sigma} \geq 0$, where in $\tilde{\sigma}$ the corresponding components are reversed
i.e. $\tilde{\sigma} = (-\sigma_A) \oplus \sigma_B$. Hence we have shown

\begin{prop} \label{prop:4}
  A Gaussian state is ppt iff its correlation matrix $\alpha$ satisfies 
  \begin{equation} \label{eq:12}
    \alpha + i \tilde{\sigma} \geq 0\ \text{with}\ \tilde{\sigma} = \left[ \begin{array}{cc} -\sigma_A & 0\\ 0 & \sigma_B
      \end{array}\right]. 
  \end{equation}
\end{prop}

The interesting question is now whether the ppt criterion is for a given number of degrees of freedom
equivalent to separability or not. The following theorem which was proved in \Cite{Simon00} for $1 \times 1$
systems and in \Cite{BEG} in $1 \times d$ case gives a complete answer. 

\begin{thm}
  A Gaussian state of a quantum system with $1 \times d$ degrees of freedom (i.e. $\dim X_A = 2$ and $\dim X_B
  = 2d$) is separable iff it is ppt; in other words iff the condition of Proposition \ref{prop:4} holds.
\end{thm}

For other kinds of systems the ppt criterion may fail which means that there are entangled Gaussian
states which are ppt. A systematic way to construct such states can be found in \Cite{BEG}. Roughly
speaking, it is based on the idea to go to the boundary of the set of ppt covariance matrices, i.e. $\alpha$
has to satisfy Equation (\ref{eq:10}) and (\ref{eq:12}) and it has to be a minimal matrix with this
property. Using this method explicit examples for ppt and entagled Gaussians are constructed for $2 \times 2$
degrees of freedom (cf. \Cite{BEG} for details). 



\subsection{Gaussian channels}
\label{sec:gaussian-channels}

Finally we want to give a short review on a special class of channels for infinite dimensional quantum 
systems (cf. \Cite{GaussCap} for details). To explain the basic idea note first that each finite set of
Weyl  operators ($W(x_j)$, $j=1,\ldots,N$, $x_j \not= x_k$ for $j \not= k$) is linear independent. This can be
checked easily using expectation values of $\sum_j \lambda_j W(x_j)$ in Gaussian states. Hence linear maps on
the space of finite linear combinations of Weyl operators can be defined by $T[W(x)] = f(x) W(Ax)$ where
$f$ is a complex valued function on $V$ and $A$ is a $2d \times 2d$ matrix. If we choose $A$ and $f$ carefully
enough, such that some continuity properties match $T$ can be extended in a unique way to a linear map on
$\scr{B}(\scr{H})$ -- which is, however, in general not completely positive.

This means we have to consider special choices for $A$ and $f$. The most easy case arises if $f \equiv 1$ and
$A$ is a symplectic isomorphism, i.e. $A^T\sigma A = \sigma$. If this holds the map $V \ni x \mapsto W(Ax)$ is a
representation of the Weyl relations and therefore unitarily equivalent to the representation we have
started with. In other words there is a unitary operator $U$ with $T[W(x)] = W(Ax) = UW(x)U^*$, i.e. $T$
is unitarily implemented, hence completely positive and, in fact, well known as \emph{Bogolubov
  transformation}. 

If $A$ does not preserve the symplectic matrix, $f \equiv 1$ is no option. Instead we have to choose $f$ such
that the matrices
\begin{equation}
  M_{jk} = f(x_j - x_k) \exp\left(-\frac{i}{2} x_j \cdot \sigma x_k + \frac{i}{2} Ax_j \cdot \sigma A x_k \right)
\end{equation}
are positive. Complete positivity of the corresponding $T$ is then a standard result of abstract
C*-algebra theory (cf. \Cite{Demoen77}). If the factor $f$ is in addition a Gaussian, i.e. $f(x) =
\exp\left(-\frac{1}{2} x \cdot \beta x\right)$ for a positive definite matrix $\beta$ the cp-map $T$ is called a
\emph{Gaussian channel}. 

A simple way to construct a Gaussian channel is in terms of an ancilla representation. More precisely, if
$A : V \to V$ is an arbitrary linear map we can extend it to a symplectic map $V \ni x \mapsto Ax \oplus A'x \in V \oplus
V'$, where the symplectic vector space $(V',\sigma')$ now refers to the environment. Consider now the Weyl
operator $W(x) \otimes W'(x') = W(x,x')$ on the Hilbert space $\scr{H} \otimes \scr{H}'$ associated to the phase
space element $x \oplus x' \in V \oplus V'$. Since $A \oplus A'$ is symplectic it admits a unitary Bogolubov
transformation $U: \scr{H} \otimes \scr{H}' \to \scr{H} \otimes \scr{H}'$ with $U^*W(x,x')U = W(Ax,A'x)$. If $\rho'$
denotes now a Gaussian density matrix on $\scr{H}'$ describing the initial state of the environment we
get a Gaussian channel by
\begin{equation}
  \tr\bigl[T^*(\rho) W(x)\bigr] = \tr \bigl[ \rho \otimes \rho' U^* W(x,x') U \bigr] = \tr\bigl[\rho W(Ax)\bigr]
  \tr\bigl[\rho' W(A'x)\bigr].
\end{equation}
Hence $T\bigl[W(x)\bigr] = f(x) W(Ax)$ with $f(x) = \tr\bigl[\rho'W(A'x)]$.

Particular examples for Gaussian channels in the case of one degree of freedom are attenuation and
amplification channels \Cite{HolGauss,GaussCap}. They are given in terms of a real parameter $k \not= 1$
by $\Bbb{R}^2 \ni x \mapsto Ax = kx \in \Bbb{R}^2$ 
\begin{equation}
  \Bbb{R}^2 \ni x \mapsto A'x = \sqrt{1 - k^2} x \in \Bbb{R}^2 < 1,
\end{equation}
for $k < 1$ and
\begin{equation}
   \Bbb{R}^2 \ni (q,p) \mapsto A'(q,p) = (\kappa q, -\kappa p) \in \Bbb{R}^2\ \text{with}\ \kappa = \sqrt{k^2 -1}
\end{equation}
for $k > 1$. If the environment is initially in a thermal state $\rho_{\tilde{N}}$ (cf. Equation
(\ref{eq:88})) this leads to  
\begin{equation} \label{eq:89}
  T\bigl[W(x)\bigr] = \exp\left[\frac{1}{2}\left(\frac{|k^2-1|}{2} + N_c\right)x^2\right] W(kx),
\end{equation}
where we have set $N_c = |k^2-1|\tilde{N}$. If we start initially with a thermal state $\rho_N$ it is
mapped by $T$ again to a thermal state $\rho_{N'}$ with mean photon number $N'$ given by
\begin{equation} \label{eq:91}
  N' = k^2 N + \max \{ 0, k^2 -1 \} + N_c.
\end{equation}
If $N_c = 0$ this means that $T$ amplifies ($k > 1$) or damps ($k < 1$) the mean photon number, while
$N_c > 0$ leads to additional classical, Gaussian noise. We will reconsider this channel in greater
detail in Chapter \ref{cha:quant-theory-ii}.

\chapter{Basic tasks}
\label{cha:basic-tasks}

After we have discussed the conceptual foundations of quantum information we will consider now some of
its basic tasks. The spectrum ranges here from elementary processes, like teleportation
\ref{sec:telep-dense-coding} or error correction \ref{sec:quant-error-corr}, which are building blocks
for more complex applications, up to possible future technologies like quantum cryptography
\ref{sec:quantum-cryptography} and quantum computing \ref{sec:quantum-computing}. 

\section{Teleportation and dense coding}
\label{sec:telep-dense-coding}

Maybe the most striking feature of entanglement is the fact that otherwise
impossible machines become possible if entangled states are used as an additional resource. The most
prominent examples are teleportation and dense coding which we want to discuss in this section.

\subsection{Impossible machines revisited: Classical teleportation}
\label{sec:imposs-mach-revis}

We have already pointed out in the introduction that \emph{classical} teleportation, i.e. transmission of
quantum information over a classical information channel is impossible. With the material introduced in
the last two chapters it is now possible to reconsider this subject in a slightly more mathematical way,
which makes the following treatment of \emph{entanglement enhanced} teleportation more transparent. To
``teleport'' the state $\rho \in  \scr{B}^*(\scr{H})$ Alice performs a measurement (described by a POV measure
$E_1,\ldots,E_N \in \scr{B}(\scr{H})$) on her system and gets a value $x \in X = \{1,\ldots,N\}$ with probability $p_x =
\tr(E_x \rho)$. These data she communicates to Bob and he prepares a $\scr{B}(\scr{H})$  system in the state
$\rho_x$. Hence the overall state Bob gets if the experiment is  repeated many times is: $\tilde{\rho} = \sum_{x \in
  X} \tr(E_x \rho) \rho_x$  (cf. Figure \ref{fig:cl-tele-1}).  
The latter can be rewritten as the \emph{composition} 
\begin{equation} \label{eq:47}
  \scr{B}^*(\scr{H}) \xrightarrow{E^*} \scr{C}(X)^* \xrightarrow{D^*} \scr{B}^*(\scr{H})^* 
\end{equation}
of the channels
\begin{equation}
  \scr{C}(X) \ni f \mapsto E(f) = \sum_{x\in X} f(x) E_x \in \scr{B}(\scr{H}) 
\end{equation}
and 
\begin{equation}
  \scr{C}^*(X) \ni p \mapsto D^*(p) = \sum_{x\in X} p_x \rho_x \in \scr{B}^*(\scr{H}),
\end{equation}
i.e. $\tilde{\rho} = D^*E^*(\rho)$ and this Equation makes sense even if $X$ is not finite. The teleportation
is successful if the output state $\tilde{\rho}$ can not be distinguished from the input state $\rho$ by any
statistical experiment, i.e. if $D^*E^*(\rho) = \rho$. Hence the impossibility of classical teleportation can be
rephrased simply as $ED \not= \Id$ for all observables $E$ and all preparations $D$.

\subsection{Entanglement enhanced teleportation}
\label{sec:teleportation}

\begin{figure}[b]
  \begin{center}
    \begin{picture}(12,6)
      \large
      \psframe[fillcolor=meascolor,fillstyle=solid,shadow=true](2,3)(4,5)
      \rput[b](3,5.5){Alice}
      \pccoil[linecolor=qbitcolor,coilaspect=0,coilheight=1,coilwidth=.2]{->}(0,4)(2,4)
      \rput[b](1,4.5){$\rho$}
      \psframe[fillcolor=opercolor,fillstyle=solid,shadow=true](8,3)(10,5)
      \rput[b](9,5.5){Bob}
      \psframe[fillcolor=entacolor,fillstyle=solid,shadow=true](5,0)(7,2)
      \rput(6,1){$\sigma$}
      \pccoil[linecolor=qbitcolor,coilaspect=0,coilheight=1,coilwidth=.2]{->}(5,1)(3,3) 
      \pccoil[linecolor=qbitcolor,coilaspect=0,coilheight=1,coilwidth=.2]{->}(7,1)(9,3)
      \rput(3,4){$E$}
      \pccoil[linecolor=qbitcolor,coilaspect=0,coilheight=1,coilwidth=.2]{->}(0,4)(2,4)
      \psline[linecolor=black]{->}(4,4)(8,4)
      \rput(6,4.5){$x \in X$}
      \rput(9,4){$D_x$}
      \pccoil[linecolor=qbitcolor,coilaspect=0,coilheight=1,coilwidth=.2]{->}(10,4)(12,4)
      \rput[b](11,4.5){$\rho$}
    \end{picture}
    \caption{Entanglement enhanced teleportation}
    \label{fig:ent-tele}
  \end{center}
\end{figure}
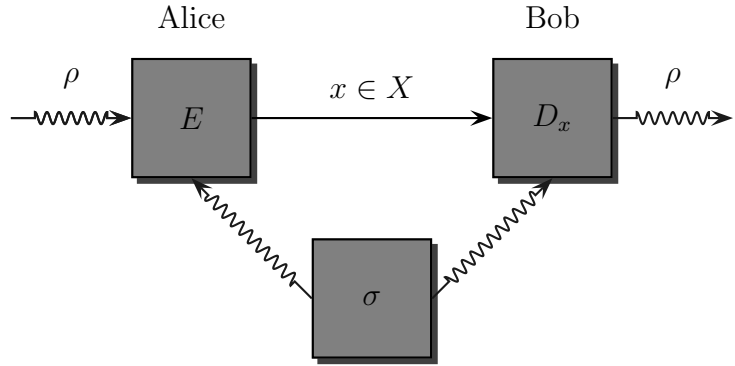

Let us change our setup now slightly. Assume that Alice wants to send a quantum state $\rho \in
\scr{B}^*(\scr{H})$ to Bob and that she shares an entangled state $\sigma \in \scr{B}^*(\scr{K} \otimes \scr{K})$ and
an ideal classical communication channel $\scr{C}(X) \to \scr{C}(X)$ with him. Alice can perform a
measurement $E: \scr{C}(X) \to \scr{B}(\scr{H} \otimes \scr{K})$ on the composite system $\scr{B}(\scr{H} \otimes
\scr{K})$ consisting of the particle to teleport ($\scr{B}(\scr{H})$) and her part of the entangled
system ($\scr{B}(\scr{K})$). Then she communicates the classical data $x \in X$ to Bob and he operates with
the parameter dependent operation $D: \scr{B}(\scr{H}) \to \scr{B}(\scr{K}) \otimes \scr{C}(X)$ appropriately on
his particle (cf. Figure \ref{fig:ent-tele}).  
Hence the overall procedure can be described by the channel $T = (E \otimes \Id)D$,
or in analogy to (\ref{eq:47})
\begin{equation} \label{eq:44}
  \scr{B}^*(\scr{H} \otimes \scr{K}^{\otimes2}) \xrightarrow{E^* \otimes \Id} \scr{C}^*(X) \otimes \scr{B}^*(\scr{K})
  \xrightarrow{D^*} \scr{B}^*(\scr{H}).
\end{equation}
The teleportation of $\rho$ is successful if 
\begin{equation} \label{eq:46}
  T^*(\rho \otimes \sigma) := D^*\bigl((E^* \otimes \Id)(\rho \otimes \sigma)\bigr)= \rho
\end{equation}
holds, in other words if there is no statistical measurement which can distinguish the final state
$T^*(\rho \otimes \sigma)$ of Bob's particle from the initial state $\rho$ of Alice's input system. The two
channels $E$ and $D$ and the entangled state $\sigma$ form a \emph{teleportation scheme} if Equation
(\ref{eq:46}) holds for all states $\rho$ of the $\scr{B}(\scr{H})$ system, i.e. if each state of a
$\scr{B}(\scr{H})$ system can be teleported without loss of quantum information.

Assume now that $\scr{H} = \scr{K} = \Bbb{C}^d$ and $X = \{0,\ldots,d^2-1\}$ holds. In this case we can define a
teleportation scheme as follows: The entangled state shared by Alice and Bob is a maximally entangled
state $\sigma = \kb{\Omega}$ and Alice performs a measurement which is given by the one dimensional projections
$E_j = \kb{\Phi_j}$, where $\Phi_j \in \scr{H} \otimes \scr{H}$, $j = 0,\ldots,d^2-1$ is a basis of maximally entangled
vectors. If her result is $j=0,\ldots,d^2-1$ Bob has to apply the operation $\tau \mapsto U_j^* \tau U_j$ on his
partner of the entangled pair, where the $U_j \in \scr{B}(\scr{H})$, $j = 0,\ldots,d^2-1$ are an
orthonormal family of unitary operators, i.e. $\tr(U_j^*U_k) = d \delta_{jk}$. Hence the parameter dependent
operation $D$ has the form (in the Schr{\"o}dinger picture): 
\begin{equation}
  \scr{C}^*(X) \otimes \scr{B}^*(\scr{H}) \ni (p,\tau) \mapsto D^*(p,\tau) = \sum_{j=0}^{d^2-1} p_j U_j^* \tau U_j \in
  \scr{B}^*(\scr{H}).
\end{equation}
Therefore we get for $T^*(\rho \otimes \sigma)$ from Equation (\ref{eq:46})
\begin{align}
  \tr\bigl[ T^*(\rho \otimes \sigma) A \bigr] &= \tr\bigl[ (E \otimes \Id)^*(\rho \otimes \sigma) D(A) \bigr] \\ &=
  \tr\left[\sum_{j=0}^{d^2-1} \tr_{12}\bigl[\kb{\Phi_j} (\rho \otimes \sigma) \Bigr] U_j^* A U_j\right] \\ &=
  \sum_{j=0}^{d^2-1} \tr\bigl[(\rho \otimes \sigma) \kb{\Phi_j} \otimes (U_j^* A U_j)\bigr] 
\end{align}
here $\tr_{12}$ denotes the partial trace over the first two tensor factors (= Alice's qubits). If $\Omega$,
the $\Phi_j$ and the $U_j$ are related by the equation
\begin{equation} \label{eq:49}
  \Phi_j = (U_j \otimes \Bbb{1}) \Omega
\end{equation}
it is a straightforward calculation to show that $T^*(\rho \otimes \sigma) = \rho$ holds as expected \Cite{WTele}. If
$d=2$ there is basically a unique choice: the $\Phi_j$, $j=0,\ldots,3$ are the four Bell states (cf. Equation
(\ref{eq:45}), $\Omega = \Phi_0$ and the $U_j$ are the identity and the three Pauli matrices. In this way we
recover the standard example for teleportation, published for the first time in \Cite{Bennett93}. The
first experimental realizations are \Cite{TeleExp1,TeleExp2}.

\subsection{Dense coding}
\label{sec:dense-coding}

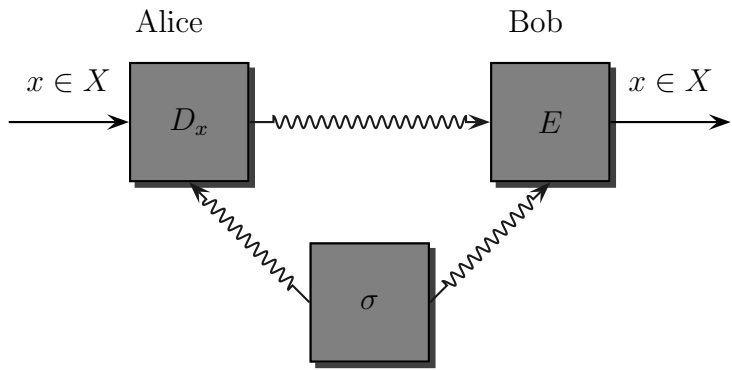
\begin{figure}[b]
  \begin{center}
    \begin{picture}(12,6)
      \large
      \psframe[fillcolor=entacolor,fillstyle=solid,shadow=true](5,0)(7,2)
      \rput(6,1){$\sigma$}
      \psframe[fillcolor=opercolor,fillstyle=solid,shadow=true](2,3)(4,5)
      \rput[bl](2.1,5.5){Alice}
      \rput(3,4){$D_x$}
      \psframe[fillcolor=meascolor,fillstyle=solid,shadow=true](8,3)(10,5)
      \rput[bl](8.3,5.5){Bob}
      \rput(9,4){$E$}
      \pccoil[linecolor=qbitcolor,coilaspect=0,coilheight=1,coilwidth=.2]{->}(5,1)(3,3) 
      \pccoil[linecolor=qbitcolor,coilaspect=0,coilheight=1,coilwidth=.2]{->}(7,1)(9,3)
      \psline[linecolor=black]{->}(0,4)(2,4)
      \large
      \rput[b](1,4.5){$x \in X$}
      \rput[b](11,4.5){$x \in X$}
      \psline[linecolor=black]{->}(10,4)(12,4)
      \pscoil[linecolor=qbitcolor,coilaspect=0,coilheight=1,coilwidth=.2]{->}(4,4)(8,4)
    \end{picture}
    \caption{Dense coding}
    \label{fig:dense-coding}
  \end{center}
\end{figure}

We have just shown how quantum information can be transmitted via a classical channel, if entanglement is
available as an additional resource. Now we are looking at the dual procedure: transmission of classical
information over a quantum channel. To send the classical information $x \in X = \{1,\ldots,n\}$ to Bob, Alice
can prepare a $d$-level quantum system in the state $\rho_x \in \scr{B}^*(\scr{H})$, sends it to Bob and he
measures an observable given by positive operators $E_1,\ldots,E_m$. The probability for Bob to receive
the signal $y \in X$ if Alice has sent $x \in X$ is $\tr(\rho_xE_y)$ and this defines a classical information
channel by  (cf. Subsection \ref{sec:classical-channels})
\begin{equation} 
  \scr{C}^*(X) \ni p \mapsto \left(\mbox{$\sum_{x \in X}$} p(x) \tr(\rho_xE_1), \ldots, \mbox{$\sum_{x \in X}$} p(x)
    \tr(\rho_xE_m)\right) \in \scr{C}^*(X).
\end{equation} 
To get an ideal channel we just have to choose mutually orthogonal pure states $\rho_x = \kb{\psi_x}$, $x
=1,\ldots,d$ on Alice's side and the corresponding one-dimensional projections $E_y = \kb{\psi_y}$, $y =1,\ldots,d$ on
Bob's. If $d=2$ and $\scr{H} = \Bbb{C}^2$ it is possible to send one bit classical information via one
qubit quantum information. The crucial point is now that the amount of classical information \emph{can be
  increased} (doubled in the qubit case) if Alice shares an entangled state $\sigma \in \scr{S}(\scr{H} \otimes
\scr{H})$ with Bob. To send the classical information  $x \in X = \{1,\ldots,n\}$ to Bob, Alice operates on her
particle with an operation  $D_x: \scr{B}(\scr{H}) \to \scr{B}(\scr{H})$, sends it through an (ideal)
quantum channel to Bob and he performs a measurement $E_1, \ldots, E_n \in \scr{B}(\scr{H} \otimes \scr{H})$ on
\emph{both} particles. The probability for Bob to measure $y \in X$ if Alice has send $x \in X$ is given by
\begin{equation} \label{eq:48}
  \tr\bigl[ (D_x \otimes \Id)^*(\sigma) E_y\bigr],
\end{equation}
and this defines the transition matrix of a classical communication channel $T$. If $T$ is an ideal
channel, i.e. if the transition matrix (\ref{eq:48}) is the identity, we will call $E$, $D$ and $\sigma$ a
\emph{dense coding scheme} (cf. Figure \ref{fig:dense-coding}).

In analogy to Equation (\ref{eq:44}) we can rewrite the channel $T$ defined by (\ref{eq:48}) in terms of
the composition 
\begin{equation}
  \scr{C}^*(X) \otimes \scr{B}^*(\scr{H}) \otimes \scr{B}^*(\scr{H}) \xrightarrow{D^* \otimes \Id} \scr{B}^*(\scr{H}) \otimes
  \scr{B}^*(\scr{H}) \xrightarrow{E^*} \scr{C}^*(X)
\end{equation}
of the parameter dependent operation 
\begin{equation}
  D: \scr{C}^*(X) \otimes \scr{B}^*(\scr{H})\to \scr{B}^*(\scr{H}), \quad p \otimes \tau \mapsto \sum_{j=1}^n p_j D_j(\tau)
\end{equation}
and the observable
\begin{equation}
  E: \scr{C}(X) \to \scr{B}(\scr{H} \otimes \scr{H}), \quad p \mapsto \sum_{j=1}^n p_j E_j,
\end{equation}
i.e. $T^*(p) = E^* \circ (D^* \otimes \Id) (p \otimes \sigma)$. The advantage of this point of view is that it works as
well for infinite dimensional Hilbert spaces and continuous observables. 

Finally let us consider again the case where $\scr{H} = \Bbb{C}^d$ and $X = \{1,\ldots,d^2\}$. If we choose as
in the last paragraph a maximally entangled vector $\Omega \in \scr{H} \otimes \scr{H}$, an orthonormal base $\Phi_x \in
\scr{H} \otimes \scr{H}$, $j=x,\ldots,d^2$ of maximally entangled vectors and an orthonormal family $U_x \in
\scr{B}(\scr{H} \otimes \scr{H})$, $x=1,\ldots,d^2$ of unitary operators, we can construct a dense coding scheme as 
follows: $E_x = \kb{\Phi_x}$, $D_x(A) = U_x^*AU_x$ and $\sigma = \kb{\Omega}$. If $\Omega$, the $\Phi_x$ and the $U_x$ are
related by Equation (\ref{eq:49}) it is easy to see that we really get a dense coding scheme
\Cite{WTele}. If $d=2$ holds, we have to set again the Bell basis for the $\Phi_x$, $\Omega = \Phi_0$ and
the identity and the Pauli matrices for the $U_x$. We recover in this case the standard example of dense
coding proposed in \Cite{Bennett92} and we see that we can transfer two bits via one qubit, as stated above.

\section{Estimating and copying}
\label{sec:estimating-copying}

The impossibility of classical teleportation can be rephrased as follows: It is impossible to get
complete information about the state $\rho$ of a quantum system by \emph{one} measurement on \emph{one}
system. However, if we have \emph{many systems}, say $N$, all prepared in the same state $\rho$ it should be
possible to get (with a clever measuring strategy) as much information on $\rho$ as possible, provided $N$
is large enough. In this way we can circumvent the impossibility of devices like classical teleportation
or quantum copying at least in an approximate way.

\subsection{Quantum state estimation}
\label{sec:quant-state-estim}

To discuss this idea in a more detailed way consider a number $N$ of $d$-level quantum systems, all of
them prepared in the same (unknown) state $\rho \in \scr{B}^*(\scr{H})$. Our aim is to \emph{estimate} the
state $\rho$ by measurements on the compound system $\rho^{\otimes N}$. This is described in terms of an observable
$E^N: \scr{C}(X_N) \to \scr{B}(\scr{H}^{\otimes N})$ 
with values in a finite subset\footnote{This is a severe restriction at this point and physically not
  very well motivated. There might be more general (i.e. continuous) observables taking their values in
  the whole state space $\scr{S}(\scr{H})$ which lead to much better estimates. However we do not
  discuss this possibility in order to keep mathematics more elementary.} $X_N \subset \scr{S}(\scr{H})$ of the
quantum state space $\scr{S}(\scr{H})$.  
According to Subsection \ref{sec:observables-2} each such $E^N$ is given in terms of a tuple $E^N_\sigma$, $\sigma
\in X_N$, by $E(f) = \sum_\sigma f(\sigma) E^N_\sigma$ hence we get for the expectation value of an $E_N$ measurement on
systems in the state $\rho^{\otimes N}$ the density matrix $\hat{\rho}_N \in \scr{S}(\scr{H})$ with matrix elements
\begin{equation}
  \langle\phi,\hat{\rho}_N\psi\rangle = \sum_{x \in X_N} \langle\phi,\sigma \psi\rangle E^N_\sigma.
\end{equation}
We will call the channel $E^N$ an \emph{estimator} and the criterion for a good estimator $E^N$ is that
for any one-particle density operator $\rho$, the value measured on a state $\rho^{\otimes N}$ is likely to be close
to $\rho$, i.e. that the probability  
\begin{equation} \label{eq:50}
  K^N(\omega) := \tr\bigl(E^N(\omega)\rho^{\otimes N}\bigr)\ \text{with}\ E^N(\omega) = \sum_{\sigma \in X_N \cap \omega} E^N_\sigma
\end{equation}
is small if $\omega \subset \scr{S}(\scr{H})$ is the complement of a small ball around $\rho$. Of course, we will
look at this problem for large $N$. So the task is to find a whole sequence of observables $E^N$,
$N=1,2,\ldots$, making error probabilities like  (\ref{eq:50}) go to zero as $N\to\infty$. 

The most direct way to get a family $E^N$, $N \in \Bbb{N}$ of estimators with this property is to perform a
sequence of measurements on each of the $N$ input systems separately. A finite set of observables which
leads to a successful estimation strategy is usually called a ``quorum''
(cf. e.g. \Cite{Leonhardt97,Weigert00}). E.g. for $d=2$ we can perform alternating measurements of the
three spin components. If $\rho = \frac{1}{2} (\Bbb{1} + \vec{x}\cdot\vec{\sigma})$ is the Bloch representation of
$\rho$ (cf. Subsection  \ref{sec:quantum-mechanics}) we see that the expectation values of these
measurements are given by  $\frac{1}{2}(1+x_j)$. Hence we get an arbitrarily good estimate if $N$ is
large enough. A similar procedure is possible for arbitrary $d$ if we consider the generalized Bloch
representation for $\rho$ (see again Subsection \ref{sec:quantum-mechanics}). There are however more
efficient strategies based on  ``entangled'' measurements (i.e. the $E_N(\sigma)$ can not be decomposed into
pure tensor products) on the whole input system $\rho^{\otimes N}$ (e.g. \Cite{Vidal99,KWEst}). Somewhat in
between are ``adaptive schemes'' \Cite{Fischer00} consisting of separate measurements but the $j^{\rm th}$
measurement depend on the results of $(j-1)^{\rm   th}$. We will reconsider this circle of questions in a
more quantitative way in Chapter \ref{cha:quant-theory-iii}.   

\subsection{Approximate cloning}
\label{sec:approximate-cloning}

By virtue of the no-cloning theorem \Cite{WoZu}, it is impossible to produce $M$ perfect copies of a
$d$-level quantum system if $N < M$ input systems in the common (unknown) state $\rho^{\otimes N}$ are given. More
precisely there is no channel $T_{MN}: \scr{B}(\scr{H}^{\otimes M}) \to \scr{B}(\scr{H}^{\otimes N})$ such that
$T_{MN}^*(\rho^{\otimes   N}) = \rho^{\otimes M}$ holds for all $\rho \in \scr{S}(\scr{H})$. Using state estimation, however, it
is easy to find a device $T_{MN}$ which produces at least approximate copies which become exact in the
limit $N,M \to \infty$: If $\rho^{\otimes N}$ is given, we measure the observable $E^N$ and get \emph{the classical data}
$\sigma \in X_N \subset \scr{S}(\scr{H})$, which we use subsequently to prepare $M$ systems in the state $\sigma^{\otimes M}$. In
other words, $T_{MN}$ has the form  
\begin{equation}
  \scr{B}^*(\scr{H}^{\otimes N}) \ni \tau \mapsto \sum_{\sigma \in X_N} \tr(E^N_\sigma \tau) \sigma^{\otimes M} \in \scr{B}^*(\scr{H}^{\otimes M}).
\end{equation}
We see immediately that the probability to get wrong copies coincides exactly with the error
probability of the estimator given in Equation (\ref{eq:50}). This shows first that we get exact copies
in the limit $N \to \infty$ and second that the quality of the copies does not depend on the number $M$ of
output systems, i.e. the asymptotic rate $\lim_{N,M \to \infty} M/N$ of output systems per input system can be
arbitrary large.  

\begin{figure}[t]
  \begin{center}
    \begin{picture}(16,8)
      \rput[bl](0,2){
        \psframe[fillcolor=meascolor,fillstyle=solid,shadow=true](4,0)(6,6)
        \psframe[fillcolor=meascolor,fillstyle=solid,shadow=true](7,2)(9,4)
        \psframe[fillcolor=meascolor,fillstyle=solid,shadow=true](10,0)(12,6)
        \pscoil[linecolor=qbitcolor,coilaspect=0,coilheight=1,coilwidth=.2]{->}(2,1)(4,1)
        \pscoil[linecolor=qbitcolor,coilaspect=0,coilheight=1,coilwidth=.2]{->}(2,2)(4,2)
        \pscoil[linecolor=qbitcolor,coilaspect=0,coilheight=1,coilwidth=.2]{->}(2,3)(4,3)
        \pscoil[linecolor=qbitcolor,coilaspect=0,coilheight=1,coilwidth=.2]{->}(2,4)(4,4)
        \pscoil[linecolor=qbitcolor,coilaspect=0,coilheight=1,coilwidth=.2]{->}(2,5)(4,5)
        \pscoil[linecolor=qbitcolor,coilaspect=0,coilheight=1,coilwidth=.2]{->}(12,0.5)(14,0.5)
        \pscoil[linecolor=qbitcolor,coilaspect=0,coilheight=1,coilwidth=.2]{->}(12,1.5)(14,1.5)
        \pscoil[linecolor=qbitcolor,coilaspect=0,coilheight=1,coilwidth=.2]{->}(12,2.5)(14,2.5)
        \pscoil[linecolor=qbitcolor,coilaspect=0,coilheight=1,coilwidth=.2]{->}(12,3.5)(14,3.5)
        \pscoil[linecolor=qbitcolor,coilaspect=0,coilheight=1,coilwidth=.2]{->}(12,4.5)(14,4.5)
        \pscoil[linecolor=qbitcolor,coilaspect=0,coilheight=1,coilwidth=.2]{->}(12,5.5)(14,5.5)
        \psline[linecolor=black]{->}(6,3)(7,3)
        \psline[linecolor=black]{->}(9,3)(10,3)
        \rput[r](1.5,3){\Large \rotateright{$\rho^{\otimes N} \in \scr{B}^*(\scr{H}^{\otimes N})$}}
        \rput[l](14.5,3){\Large \rotateright{$F(\sigma)^{\otimes M} \in \scr{B}^*(\scr{H}^{\otimes M})$}}
        \rput(5,3){\LARGE \rotateright{Estimation}}
        \rput(11,3){\LARGE \rotateright{Preparation}}
        \rput(8,3){\LARGE $F$}}
      \psline{->}(6.5,1.8)(6.5,4.8)
      \psline{->}(9.5,1.8)(9.5,4.8)
      \rput[t](8,1.6){\Large classical data}
      \rput(6.5,0.5){\large $\sigma \in X \subset \scr{S}$}
      \rput(9.5,0.5){\large $F(\sigma) \in \scr{S}$}
    \end{picture}
    \caption{Approximating the impossible machine $F$ by state estimation.}
    \label{fig:est}
  \end{center}
\end{figure}
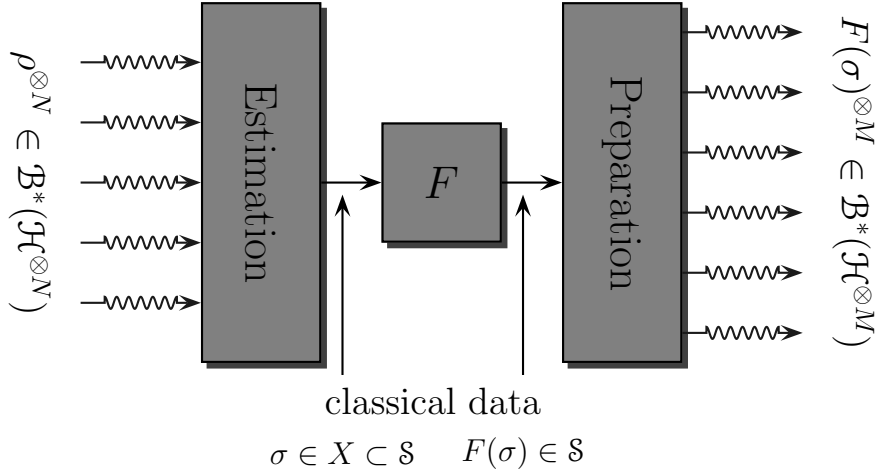

The fact that we get classical data at an intermediate step allows a further generalization of this
scheme. Instead of just preparing $M$ systems in the state $\sigma$ detected by the estimator, we can apply
first an \emph{arbitrary transformation} $F: \scr{S}(\scr{H}) \to \scr{S}(\scr{H})$ on the density
matrix $\sigma$ and prepare $F(\sigma)^{\otimes M}$ instead of $\sigma^{\otimes M}$. In this way we get the channel (cf. Figure
\ref{fig:est}) 
\begin{equation}
  \scr{B}^*(\scr{H}^{\otimes N}) \ni \tau \mapsto \sum_{\sigma \in X_N} \tr(E^N_\sigma \tau) F(\sigma)^{\otimes M} \in \scr{B}^*(\scr{H}^{\otimes M}),
\end{equation}
i.e. a \emph{physically realizable} device which approximates the \emph{impossible machine} $F$. The
probability to get a bad approximation of the state $F(\rho)^{\otimes M}$ (if the input state was $\rho^{\otimes N}$) is
again given by the error probability of the estimator and we get a perfect realization of $F$ at
arbitrary rate as $M,N \to \infty$. 

There are in particular two interesting tasks which become possible this way: The first is the
``universal not gate'' which associates to each pure state of a qubit the unique pure state orthogonal to
it \Cite{UNOT}. This is a special example of a antiunitarily implemented symmetry operation and therefore
not completely positive. The second example is the purification of states \Cite{CEM99,pur}. Here it
is assumed that the input states were once pure but have passed later on a depolarizing channel $\kb{\phi}
\mapsto \vartheta \kb{\phi} + (1-\vartheta) \Bbb{1}/d$. If $\vartheta > 0$ this map is invertible but its inverse does not describe an
allowed quantum operation because it maps some density operators to operators with negative
eigenvalues. Hence the reversal of noise is not possible with a one shot operation but can be done with
high accuracy if enough input systems are available. We rediscuss this topic in Chapter
\ref{cha:quant-theory-iii}.
 
\section{Distillation of entanglement}
\label{sec:dist-entangl}

Let us return now to entanglement. We have seen in Section \ref{sec:telep-dense-coding} that maximally
entangled states play a crucial role for processes like teleportation and dense coding. In practice
however entanglement is a rather fragile property: If Alice produces a pair of particles in a maximally
entangled state $\kb{\Omega} \in \scr{S}(\scr{H}_A \otimes \scr{H}_B)$  and distributes one of them
over a great distance to Bob, both end up with a mixed state $\rho$ which contains much less entanglement
then the original and which can not be used any longer for teleportation. The latter can be seen quite
easily if we try to apply the qubit teleportation scheme (Subsection \ref{sec:teleportation}) with a
non-maximally entangled isotropic state (Equation (\ref{eq:51}) with $\lambda > 0$) instead of $\Omega$.

Hence the question arises, whether it is possible to recover $\kb{\Omega}$ from $\rho$, or, following the
reasoning from the last section, at least a small number of (almost) maximally entangled states from a
large number $N$ of copies of $\rho$. However since the distance between Alice and Bob is big (and quantum
communication therefore impossible) only LOCC operations (Section \ref{sec:locc-separ-chann}) are 
available for this task (Alice and Bob can only operate on their respective particles, drop some of them
and communicate classically with one another). This excludes procedures like the purification scheme just
sketched, because we would need ``entangled'' measurements to get an asymptotically exact estimate for
the state $\rho$. Hence we need a sequence of LOCC channels  
\begin{equation}
  T_N : \scr{B}(\Bbb{C}^{d_N} \otimes \Bbb{C}^{d_N}) \to \scr{B}(\scr{H}^{\otimes N}_A \otimes \scr{H}^{\otimes N}_B)
\end{equation}
such that 
\begin{equation} \label{eq:52}
  \| T_N^*(\rho^{\otimes N}) - \kb{\Omega_N}\|_1 \to 0,\ \text{for}\ N \to \infty
\end{equation}
holds, with a sequence of maximally entangled vectors $\Omega_N \in \Bbb{C}^{d_N} \otimes \Bbb{C}^{d_N}$. Note that we
have to use here the natural isomorphism $\scr{H}^{\otimes N}_A \otimes \scr{H}^{\otimes N}_B \cong (\scr{H}_A \otimes \scr{H}_B)^{\otimes
  N}$, i.e. we have to reshuffle $\rho^{\otimes N}$ such that the first $N$ tensor factors belong to Alice
($\scr{H}_A$) and the last $N$ to Bob ($\scr{H}_B$). If confusion can be avoided we will use this
isomorphism in the following without a further note. We will call a sequence of LOCC channels,
$T_N$ satisfying (\ref{eq:52}) with a state $\rho \in \scr{S}(\scr{H}_A \otimes \scr{H}_B)$ a \emph{distillation
  scheme} for $\rho$ and $\rho$ is called \emph{distillable} if it admits a distillation scheme. The
\emph{asymptotic rate} with which maximally entangled states can be distilled with a given protocol is 
\begin{equation} \label{eq:31}
  \liminf_{n \to \infty} \log_2(d_N)/N.
\end{equation}
This quantity will become relevant in the framework of entanglement measures (Chapter
\ref{cha:quant-theory-i}). 

\subsection{Distillation of pairs of qubits}
\label{sec:dist-pairs-qubits}

Concrete distillation protocols are in general rather complicated procedures. We will sketch in the
following how any pair of entangled qubits can be distilled. The first step is a scheme  proposed for the
first time by Bennett et. al. \Cite{BBPSSW}. It can be applied if the maximally entangled fraction
$\scr{F}$ (Equation (\ref{eq:57})) is greater than $1/2$.  
As indicated above, we assume that Alice and Bob share a large amount of pairs in the
state $\rho$, so that the total state is $\rho^{\otimes N}$. To obtain a smaller number of pairs with a higher
$\scr{F}$ they proceed as follows: 
\begin{enumerate}
\item 
  First they take two pairs (let us call them pair 1 and pair 2), i.e. $\rho \otimes \rho$ and apply to each of them
  the twirl operation $P_\tUUbar$ associated to isotropic states (cf. Equation (\ref{eq:53})). This can be
  done by LOCC operations in the following way: Alice selects at random (respecting the Haar measure on
  $\U(2)$) a unitary operator $U$ applies it to her qubits and sends to Bob  which transformation she has 
  chosen; then he applies $\bar{U}$ to his particles. They end up with two isotropic states $\tilde{\rho} \otimes
  \tilde{\rho}$ with the same maximally entangled fraction as $\rho$. 
\item 
  Each party performs the unitary transformation
  \begin{equation}
     U_{\rm XOR} : \ket{a} \otimes \ket{b} \mapsto \ket{a} \otimes \ket{a+b \mod 2}
  \end{equation}
 on his/her members of the pairs.
\item 
  Finally Alice and Bob perform locally a measurement in the basis $\ket{0}, \ket{1}$ on pair 1 and
  discards it afterwards. If the measurements agree, pair 2 is kept and has a higher $\scr{F}$. Otherwise
  pair 2 is discarded as well. 
\end{enumerate}

If this procedure is repeated over and over again, it is possible to get states with an arbitrarily high
$\scr{F}$, but we have to sacrifice more and more pairs and the asymptotic rate is zero. To overcome
this problem we can apply the scheme above until $\scr{F}(\rho)$ is high enough such that $1 + \tr(\rho \ln\rho) \geq
0$ holds and then we continue with another scheme called hashing \Cite{BDiVSW} which leads to a
nonvanishing rate.

If finally $\scr{F}(\rho) \leq 1/2$ but $\rho$ is entangled, Alice and Bob can increase $\scr{F}$ for some of
their particles by \emph{filtering operations} \Cite{BBPS,Gisin96}. The basic idea is that Alice applies
an instrument $T: \scr{C}(X) \otimes \scr{B}(\scr{H}) \to \scr{B}(\scr{H})$ with two possible outcomes ($X =
\{1,2\}$) to her particles. Hence the state becomes $\rho \mapsto p_x^{-1} (T_x \otimes \Id)^*(\rho)$, $x=1,2$ with
probability $p_x = \tr\bigl[T_x^*(\rho)\bigr]$ (cf. Subsection \ref{sec:prep-instr} in particular Equation
(\ref{eq:5}) for the definition of $T_x$). Alice communicates her measuring result $x$ to Bob and if
$x=1$ they keep the particle otherwise ($x=2$) they discard it. If the instrument $T$ was correctly
chosen Alice and Bob end up with a state $\tilde{\rho}$ with higher maximally entangled fraction. To find an
appropriate $T$ note first that there are $\psi \in \scr{H} \otimes \scr{H}$ with $\langle\psi, (\Id \otimes \Theta) \rho \psi\rangle \leq 0$ (this
follows from Theorem \ref{thm:5} since $\rho$ is by assumption entangled) and second that we can write each
vector $\psi \in \scr{H} \otimes \scr{H}$ as $(X_\psi \otimes \Bbb{1})\Phi_0$ with the Bell state $\Phi_0$ and an appropriately chosen
operator $X_\psi$ (see Subsection \ref{sec:pure-states}). Now we can define $T$ in terms of the two
operations $T_1, T_2$ (cf. Equation (\ref{eq:17})) with 
\begin{equation} \label{eq:19}
  T_1(A) = X_\psi^* A X_\psi^{-1},\quad  \Id - T_1 = T_2  
\end{equation}
It is straightforward to check that we end up with 
\begin{equation}
  \tilde{\rho} = \frac{(T_x \otimes \Id)^*(\rho)}{\tr\bigl[ (T_x \otimes \Id)^*(\rho) \bigr]}
\end{equation}
 such that $\scr{F}(\tilde{\rho}) > 1/2$ holds and we can continue with the scheme described in the previous
 paragraph.

\subsection{Distillation of isotropic states}
\label{sec:dist-isotr-stat}

Consider now an entangled isotropic state $\rho$ in $d$ dimensions, i.e. we have $\scr{H} = \Bbb{C}^d$ and
$0 \leq \tr(\tilde{F}\rho) \leq 1$ (with the operator $\tilde{F}$ of Subsection \ref{sec:isotropic-states}). Each
such state is distillable via the following scheme \Cite{BCJLPS,H2RedKrit}: First Alice and Bob apply a
filter operation $T: \scr{C}(X) \otimes \scr{B}(\scr{H}) \to\scr{B}(\scr{H})$ on their respective particle given by
$T_1(A) = PAP$, $T_2 = 1 - T_1$ where $P$ is the projection onto a two dimensional subspace. If both
measure the value $1$ they get a qubit pair in the state $\tilde{\rho} = (T_1 \otimes T_1)(\rho)$. Otherwise they
discard their particles (this requires classical communication). Obviously the state $\tilde{\rho}$ is
entangled (this is easily checked) hence they can proceed as in the previous Subsection. 

The scheme just proposed can be used to show that each state $\rho$ which violates the reduction criterion 
(cf. Subsection \ref{sec:reduction-criterion}) can be distilled \Cite{H2RedKrit}. The basic idea is to
project $\rho$ with the twirl $P_{\tUUbar}$ (which is LOCC as we have seen above; cf. Subsection  
\ref{sec:dist-pairs-qubits}) to an isotropic state $P_{\tUUbar}(\rho)$ and to apply afterwards the
procedure from the last paragraph. We only have to guarantee that $P_{\tUUbar}(\rho)$ is entangled. To this
end use a vector $\psi \in \scr{H} \otimes \scr{H}$ with $\langle\psi, (\Bbb{1} \otimes \tr_1(\rho) - \rho) \psi\rangle < 0$ (which exists by 
assumption since $\rho$ violates the reduction criterion) and to apply the filter operation given by $\psi$ via
Equation (\ref{eq:19}).

\subsection{Bound entangled states}
\label{sec:bound-entangl-stat-1}

It is obvious that separable states are not distillable, because a LOCC operation map separable states to
separable states. However is each entangled state distillable? The answer, maybe somewhat surprising,
is no and an entangled state which is not distillable is called \emph{bound entangled}
\Cite{H3BE} (distillable states are sometimes called \emph{free entangled}, in analogy to
thermodynamics). Examples of bound entangled states are all ppt entangled states \Cite{H3BE}: This is an 
easy consequence of the fact that each separable channel (and therefore each LOCC channel as well) maps
ppt states to ppt states (this is easy to check), but a maximally entangled state is never ppt. It is not
yet known, whether bound entagled npt states exists, however, there are at least some partial results:
1. It is sufficient to solve this question for Werner states, i.e. if we can show that each npt Werner
state is distillable it follows that all npt  states are distillable \Cite{H2RedKrit}. 2. Each npt
Gaussian state is distllable \Cite{GaussDistill}. 3. For each $N \in \Bbb{N}$ there is an npt Werner state
$\rho$ which is not ``$N$-copy distillable'', i.e. $\langle\psi,\rho^{\otimes N}\psi\rangle \geq 0$ holds for each pure state $\psi$ with
exactly two Schmidt summands \Cite{DiVSSTT,DCLB}. This gives some evidence for the existence of bound
entangled npt states because $\rho$ is distillabile iff it is $N$-copy distillability for some $N$
\Cite{H3BE,DiVSSTT,DCLB}.

Since bound entangled states can not be distilled, they can not be used for teleportation. Nevertheless
bound entanglement can produce a non-classical effect, called ``activation of bound entanglement''
\Cite{H3ActBE}. To explain the basic idea, assume that Alice and Bob share \emph{one} pair of particles
in a distillable state $\rho_f$ and many particles in a bound entangled state $\rho_b$. Assume in addition that
$\rho_f$ can not be used for teleportation, or, in other words if $\rho_f$ is used for teleportation the
particle Bob receives is in a state $\sigma'$ which differs from the state $\sigma$ Alice has send. This problem can
not be solved by distillation, since Alice and Bob share only one pair of particles in the state
$\rho_f$. Nevertheless they can try to apply an appropriate filter operation on $\rho$ to get with a certain
probability a new state which leads to a better quality of the teleportation (or, if the filtering fails,
to get nothing at all). It can be shown however \Cite{H3Tele} that there are states $\rho_f$ such that the
error occuring in this process (e.g. measured by the trace norm distance of $\sigma$ and $\sigma'$) is always above
a certain threshold. This is the point where the bound entangled states $\rho_b$ come into play: If Alice
and Bob operate with an appropriate protocol on $\rho_f$ and many copies of $\rho_b$ the distance between $\sigma$
and $\sigma'$ can be made arbitrarily small (although the probability to be successful goes to zero). Another
example for an activation of bound entanglement is related to distillability of npt states: If Alice and
Bob share a certain ppt-entangled state as additional recource  each npt state $\rho$ becomes distillable 
(evem if $\rho$ is bound entangled) \Cite{EVWW,KLC01}. For a more detailed survey of the role of bound
entanglement and further references see \Cite{H3DPG}. 

\section{Quantum error correction}
\label{sec:quant-error-corr}

If we try to distribute quantum information over large distances or store it for a long time in some sort
of ``quantum memory'' we always have to deal with ``decoherence effects'', i.e. unavoidable interactions
with the environment. This results in a significant information loss, which is particularly bad for the
functioning of a quantum computer. Similar problems arise as well in a classical computer, but the
methods used there to circumvent the problems can not be transferred to the quantum regime. E.g. the most
simple strategy to protect classical information against noise is redundancy: instead of storing the
information once we make three copies and decide during readout by a majority vote which bit to take. It
is easy to see that this reduces the probability of an error from order $\epsilon$ to $\epsilon^2$. Quantum
mechanically however such a procedure is forbidden by the no cloning theorem. 

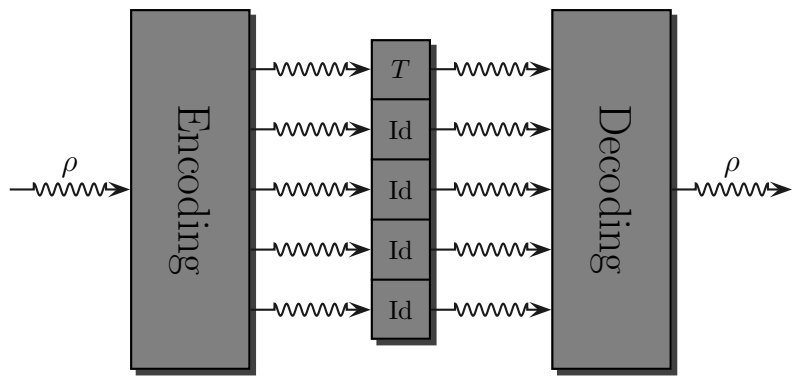
\begin{figure}[b]
  \begin{center}
    \begin{picture}(15,6)
      \psframe[fillcolor=meascolor,fillstyle=solid,shadow=true](3,0)(5,6)
      \psframe[fillcolor=meascolor,fillstyle=solid,shadow=true](7,0.5)(8,5.5)
      \psline{-}(7,1.5)(8,1.5)
      \psline{-}(7,2.5)(8,2.5)
      \psline{-}(7,3.5)(8,3.5)
      \psline{-}(7,4.5)(8,4.5)
      \psframe[fillcolor=meascolor,fillstyle=solid,shadow=true](10,0)(12,6)
      \pscoil[linecolor=qbitcolor,coilaspect=0,coilheight=1,coilwidth=.2]{->}(1,3)(3,3)
      \pscoil[linecolor=qbitcolor,coilaspect=0,coilheight=1,coilwidth=.2]{->}(12,3)(14,3)
      \pscoil[linecolor=qbitcolor,coilaspect=0,coilheight=1,coilwidth=.2]{->}(5,1)(7,1)
      \pscoil[linecolor=qbitcolor,coilaspect=0,coilheight=1,coilwidth=.2]{->}(5,2)(7,2)
      \pscoil[linecolor=qbitcolor,coilaspect=0,coilheight=1,coilwidth=.2]{->}(5,3)(7,3)
      \pscoil[linecolor=qbitcolor,coilaspect=0,coilheight=1,coilwidth=.2]{->}(5,4)(7,4)
      \pscoil[linecolor=qbitcolor,coilaspect=0,coilheight=1,coilwidth=.2]{->}(5,5)(7,5)
      \pscoil[linecolor=qbitcolor,coilaspect=0,coilheight=1,coilwidth=.2]{->}(8,1)(10,1)
      \pscoil[linecolor=qbitcolor,coilaspect=0,coilheight=1,coilwidth=.2]{->}(8,2)(10,2)
      \pscoil[linecolor=qbitcolor,coilaspect=0,coilheight=1,coilwidth=.2]{->}(8,3)(10,3)
      \pscoil[linecolor=qbitcolor,coilaspect=0,coilheight=1,coilwidth=.2]{->}(8,4)(10,4)
      \pscoil[linecolor=qbitcolor,coilaspect=0,coilheight=1,coilwidth=.2]{->}(8,5)(10,5)
      \rput(4,3){\LARGE \rotateright{Encoding}}
      \rput(11,3){\LARGE \rotateright{Decoding}}
      \rput(7.5,1){$\Id$}
      \rput(7.5,2){$\Id$}
      \rput(7.5,3){$\Id$}
      \rput(7.5,4){$\Id$}
      \rput(7.5,5){$T$}
      \rput[b](2,3.2){\large $\rho$}
      \rput[b](13,3.2){\large $\rho$}
    \end{picture}
    \caption{Five bit quantum code: Encoding one qubit into five and correcting one error.}
    \label{fig:five-bit-code}
  \end{center}
\end{figure}

Nevertheless quantum error correction is possible although we have to do it in a more subtle way than
just copying; this was observed for the first time independently in \Cite{CSQECC} and \Cite{StQECC}. Let
us consider first the general scheme and assume that $T :\scr{B}(\scr{K}) \to \scr{B}(\scr{K})$ is a noisy
quantum channel. To send quantum systems of type $\scr{B}(\scr{H})$ undisturbed through $T$ we need an
\emph{encoding channel} $E: \scr{B}(\scr{K}) \to \scr{B}(\scr{H})$ and a decoding channel $D:
\scr{B}(\scr{H}) \to \scr{B}(\scr{K})$ such that $ETD = \Id$ holds, respectively $D^*T^*E^* = \Id$
in the Schr{\"o}dinger picture; cf. Figure \ref{fig:five-bit-code}.

A powerful error correction scheme should not be restricted to one particular type of error, i.e. one
particular noisy channel $T$. Assume instead that $\goth{E} \subset \scr{B}(\scr{K})$ is a linear subspace of
``error operators'' and $T$ is any channel given by 
\begin{equation} \label{eq:58}
  T_*(\rho) = \sum_j F_j \rho F_j^*,\quad F_j \in \goth{E}.
\end{equation}
An isometry $V: \scr{H} \to \scr{K}$ is called an \emph{error correcting code} for $\goth{E}$ if for each
$T$ of the form (\ref{eq:58}) there is a decoding channel $D: \scr{B}(\scr{H}) \to \scr{B}(\scr{K})$ with
$D_*\bigl(T(V\rho V^*)\bigr) = \rho$ for all $\rho \in \scr{S}(\scr{H})$. By the theory of Knill and Laflamme
\Cite{KnLafQECC} this is equivalent to the factorization condition
\begin{equation} \label{eq:25}
  \langle V\psi, F_j^*F_kV\phi\rangle = \omega(F_j^*F_k)\langle\psi,\phi\rangle
\end{equation}
where $\omega(F_j^*F_k)$ is a factor which does not depend on the arbitrary vectors $\psi,\phi \in \scr{H}$.

The most relevant examples of error correcting codes are those which generalize the classical idea of
sending multiple copies in a certain sense. This means we encode a small number $N$ of $d$-level systems 
into a big number $M \gg N$ of systems of the same type, which are then transmitted and decoded back into
$N$ systems afterwards. During the transmission $K < M$ arbitrary errors are allowed. Hence we have
$\scr{H} = \scr{H}_1^{\otimes N}$, $\scr{K} = \scr{H}_1^{\otimes M}$ with $\scr{H}_1 = \Bbb{C}^d$ and $T$ is an
arbitrary tensor product of $K$ noisy channels $S_j$, $j=1,\ldots,K$ and $M-K$ ideal channels $\Id$. The
most well known code for this type of error is the ``five-bit code'' where one qubit is encoded into
five and one error is corrected \Cite{BDiVSW} (cf. Figure \ref{fig:five-bit-code} for $N=1, M=5$ and
$K=1$). To define the corresponding error space $\goth{E}$ consider the finite sets $X = \{1,\ldots,N\}$ and $Y
= \{1+N,\ldots,M+N\}$ and define first for each subset $Z \subset Y$:
\begin{multline} 
  \goth{E}(Z) = \SP \{ A_1 \otimes \cdots \otimes A_M \in \scr{B}(\scr{K}) \,|\\ A_j \in \scr{B}(\scr{H}_1)\ \text{arbitrary
    for $j+N \in Z$},\ A_j = \Bbb{1}\ \text{otherwise}\}.
\end{multline}
$\goth{E}$ is now the span of all $\goth{E}(Z)$ with $|Z| \leq K$ (i.e. the length of $Z$ is less or equal
to $K$). We say that an error correcting code for this particular $\goth{E}$ \emph{corrects $K$ errors}.

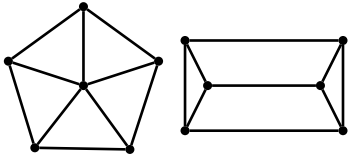
\begin{wrapfigure}{l}{5.5cm}
  \psset{unit=3mm}
    \begin{pspicture}(16,8)
      \rput(4,4){
        \SpecialCoor
        \cnode[fillstyle=solid,fillcolor=black](0,0){0.2}{A}
        \cnode[fillstyle=solid,fillcolor=black](3.5;90){0.2}{B}
        \cnode[fillstyle=solid,fillcolor=black](3.5;18){0.2}{C}
        \cnode[fillstyle=solid,fillcolor=black](3.5;-54){0.2}{D}
        \cnode[fillstyle=solid,fillcolor=black](3.5;162){0.2}{E}
        \cnode[fillstyle=solid,fillcolor=black](3.5;232){0.2}{F}
        \psset{linewidth=1pt}
        \ncline{-}{A}{B}
        \ncline{-}{A}{C}
        \ncline{-}{A}{D}
        \ncline{-}{A}{E}
        \ncline{-}{A}{F}
        \ncline{-}{B}{C}
        \ncline{-}{C}{D}
        \ncline{-}{B}{E}
        \ncline{-}{E}{F}
        \ncline{-}{F}{D}
        \NormalCoor
        }
      \rput(12,4){
        \cnode[fillstyle=solid,fillcolor=black](-3.5,-2){0.2}{A}
        \cnode[fillstyle=solid,fillcolor=black](-3.5,2){0.2}{B}
        \cnode[fillstyle=solid,fillcolor=black](3.5,-2){0.2}{C}
        \cnode[fillstyle=solid,fillcolor=black](3.5,2){0.2}{D}
        \cnode[fillstyle=solid,fillcolor=black](-2.5,0){0.2}{E}
        \cnode[fillstyle=solid,fillcolor=black](2.5,0){0.2}{F}
        \psset{linewidth=1pt}
        \ncline{-}{A}{B}
        \ncline{-}{B}{D}
        \ncline{-}{D}{C}
        \ncline{-}{C}{A}
        \ncline{-}{A}{E}
        \ncline{-}{B}{E}
        \ncline{-}{C}{F}
        \ncline{-}{D}{F}
        \ncline{-}{E}{F}
     }
    \end{pspicture}
  \caption{Two graphs belonging to (equivalent) five bit codes. The input node can be chosen in both
    cases arbitrarily.}
  \label{fig:graphs}
\end{wrapfigure}
There are several ways to construct error correcting codes (see
e.g. \Cite{GotQECC,CRSSQECC,AsKnQECC}). Most of these methods are somewhat involved however and require
knowledge from classical error correction which we want to skip. Therefore we will only present
the scheme proposed in \Cite{Schlingel}, which is quite easy to describe and admits a simple way to check
the error correction condition. Let us sketch first the general scheme. We start with an undirected Graph
$\Gamma$ with two  kinds of vertices: A set of input vertices, labeled by $X$ and a set of output vertices
labeled by $Y$. The links of the graph are given by the adjacency matrix, i.e. a $N+M \times N+M$ matrix $\Gamma$
with $\Gamma_{jk} = 1$ if node $k$ and $j$ are linked and $\Gamma_{jk} = 0$ otherwise. With respect to $\Gamma$ we can
define now an isometry $V_\Gamma: \scr{H}_1^{\otimes N}  \to  \scr{H}_1^{\otimes M}$ by   
\begin{equation} \label{eq:27}
  \langle j_{N+1} \ldots j_{N+M} | V_\Gamma | j_1 \ldots j_N\rangle = \exp\left(\frac{i\pi}{d} \vec{j} \cdot \Gamma \vec{j}\right), 
\end{equation}
with $\vec{j} = (j_1,\ldots,j_{N+M}) \in \Bbb{Z}_d^{N+M}$ (where $\Bbb{Z}_d$ denotes the cyclic group with $d$
elements). There is an easy condition under which $V_\Gamma$ is an error correcting code. To write it down we
need the following additional terminology: We say that an error correcting code $V: \scr{H}_1^{\otimes N}  \to
\scr{H}_1^{\otimes M}$ \emph{detects} the \emph{error configuration} $Z 
\subset Y$ if
\begin{equation}
  \langle V\psi, F V\phi\rangle = \omega(F)\langle\psi,\phi\rangle\quad \forall F \in \goth{E}(Z)
\end{equation}
holds. With Equation (\ref{eq:25}) it is easy to see that $V$ corrects $K$ errors iff it detects all
error configurations of length $2K$ or less. Now we have the following theorem:

\begin{thm}
  The quantum code $V_\Gamma$ defined in Equation (\ref{eq:27}) detects the error configuration $Z \subset Y$ if the
  system of equations
  \begin{equation}
    \sum_{l \in X \cup Z} \Gamma_{kl} g_l = 0,\quad k \in Y \setminus E,\ g_l \in \Bbb{Z}_d
  \end{equation}
  implies that
  \begin{equation}
    g_l = 0,\ l \in X\ \text{and}\ \sum_{l \in Z} \Gamma_{kl} g_l = 0,\ k \in X
  \end{equation}
  holds.
\end{thm}

We omit the proof, see \Cite{Schlingel} instead. Two particular examples (which are equivalent!) are
given in Figure \ref{fig:graphs}. In both cases we have $N=1$, $M=5$ and $K=1$ i.e. one input node, which
can be chosen arbitrarily, five output nodes and the corresponding codes correct one error. For a more
detailed survey on quantum error correction, in particular for more examples we refer to \Cite{BethDPG}.

\section{Quantum computing}
\label{sec:quantum-computing}

Quantum computing is without a doubt the most prominent and most far reaching application of quantum
information theory, since it promises on the one hand, ``exponential speedup'' for some problems which
are ``hard to solve'' with a classical computer, and gives completely new insights into classical
computing and complexity theory on the other. Unfortunately, an exhaustive discussion would require its
own review article. Hence we we are only able to give a short overview (see Part II of \Cite{NC} for a
more complete presentation and for further references). 

\subsection{The network model of classical computing}
\label{sec:classical-computing}

\begin{figure}[b]
  \begin{center}    
    \begin{minipage}{4cm}
      \begin{center}

        \includegraphics[scale=0.5]{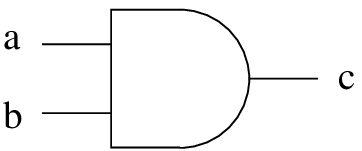}\\[1em]
    \end{center}
  \end{minipage}
  \begin{minipage}{4cm}
    \begin{center}
      
      \includegraphics[scale=0.5]{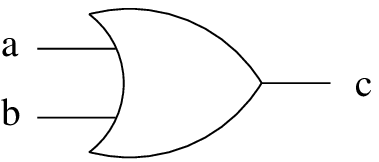}\\[1em]
  \end{center}
  \end{minipage}
  \begin{minipage}{4cm}
    \begin{center}
      \includegraphics[scale=0.5]{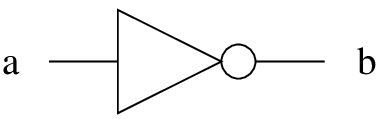}\\[1em]
    \end{center}
  \end{minipage}\\[1em]
  \begin{minipage}{4cm}
    \begin{center}
    \begin{tabular}{ll|l}
        a & b & c\\ \hline
        0 & 0 & 0\\
        1 & 0 & 0\\
        0 & 1 & 0\\
        1 & 1 & 1\\
      \end{tabular}
    \end{center}
  \end{minipage}
  \begin{minipage}{4cm}
    \begin{center}
      \begin{tabular}{ll|l}
        a & b & c\\ \hline
        0 & 0 & 0\\
        1 & 0 & 1\\
        0 & 1 & 1\\
        1 & 1 & 1\\
      \end{tabular}
    \end{center}
    \end{minipage}
    \begin{minipage}{4cm}
      \begin{center}
        \begin{tabular}{l|l}
          a & b\\ \hline
          0 & 1\\
          1 & 0\\
        \end{tabular}
      \end{center}
    \end{minipage}\\[1em]
    \begin{minipage}{4cm}
      \begin{center}
        {\small $c = ab$}\\[1ex]
        AND, $\land$
      \end{center}
    \end{minipage}
    \begin{minipage}{4cm}
      \begin{center}
        {\small $c = a + b - ab$}\\[1ex]
        OR, $\lor$
      \end{center}
    \end{minipage}
     \begin{minipage}{4cm}
       \begin{center}
         {\small $b = 1 - a$}\\[1ex]
         NOT, $\lnot$
       \end{center}
    \end{minipage}\\[1em]
    \caption{Symbols and definition for the three elementary gates AND, OR and  NOT.}
    \label{fig:gates}
  \end{center}
\end{figure}

\begin{figure}[h]
  \begin{center}
    \includegraphics[scale=0.8]{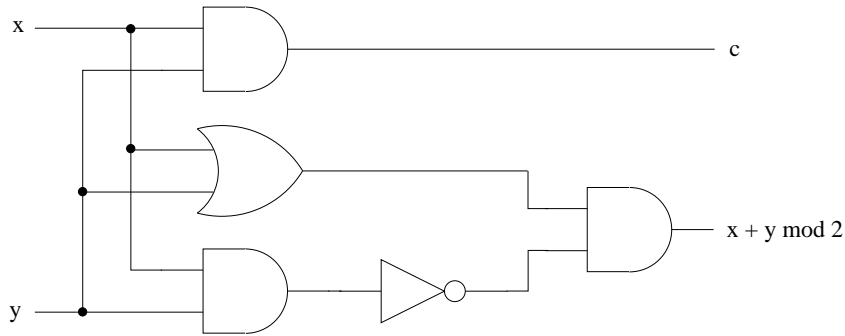}
    \caption{Half-adder circuit as an example for a boolean network.} 
    \label{fig:half-adder}
  \end{center}
\end{figure}

Let us start with a brief (and very informal) introduction to classical computing (for a more complete
review and hints for further reading see Chapter 3 of \Cite{NC}). What we need first is a
\emph{mathematical model} for computation. There are in fact several different choices and the Turing
machine \Cite{Turing36} is the most prominent one. More appropriate for our purposes is, however, the so
called \emph{network model}, since it allows an easier generalization to the quantum case.
The basic idea is to interpret a classical (deterministic) computation as the evaluation of a map $f:
\Bbb{B}^N \to \Bbb{B}^M$ (where $\Bbb{B} = \{0,1\}$ denotes the field with two elements) which maps $N$ input
bits to $M$ output bits. If $M=1$ holds $f$ is called a boolean function and it is for many purposes
sufficient to consider this special case -- each general $f$ is in fact a Cartesian product of boolean
functions. Particular examples are the three elementary gates AND, OR and NOT defined in Figure
\ref{fig:gates} and arbitrary algebraic expressions constructed from them: e.g. the XOR gate 
$(x,y) \mapsto x + y \mod 2$ which can be written as $(x \lor y) \land \lnot (x \land y)$. It is now a standard result of
boolean algebra that each boolean function can be represented in this way and there are in general many
possibilities to do this. A special case is the \emph{disjunctive normal form} of $f$; cf
\Cite{Wegener87}.  
To write such an expression down in form of equations is, however, somewhat confusing. $f$ is therefore
expressed most conveniently in graphical form as a \emph{circuit} or \emph{network}, i.e. a graph $C$
with nodes representing elementary gates and edges (``wires'') which determine how the gates should
be composed; cf. Figure \ref{fig:half-adder} for an example. A \emph{classical computation} can now be
defined as a circuit applied to a specified string of input bits.

Variants of this model arise if we replace AND, OR and NOT by another (finite) set $G$ of elementary
gates. We only have to guarantee that each function $f$ can be expressed as a composition of elements
from $G$. A typical example for $G$ is the set which contains only the NAND gate $(x,y) \mapsto x \uparrow y = \lnot (x \land
y)$. Since AND,  OR and NOT can be rewritten in terms of NAND (e.g. $\lnot x = x \uparrow x$) we can calculate each
boolean function by a circuit of NAND gates. 

\subsection{Computational complexity} 
\label{sec:comp-compl}

One of the most relevant questions within classical computing, and the central subject of computational
complexity, is whether a given  problem is easy to solve or not, where ``easy'' is defined in terms of
the scaling behavior of the resources needed in dependence of the size of the input data. We will give
in the following a rough survey over the most basic aspects of this field, while we refer the reader to
\Cite{Pap94} for a detailed presentation.


To start with, let us specify the basic question in greater detail. First of all the problems we want 
to analyze are \emph{decision problems} which only give the two possible values ``yes'' and ``no''. They
are mathematically described by boolean functions acting on bit strings of arbitrary size. A well known
example is the factoring problem given by the function $\fac$ with $\fac(m,l) = 1$ if $m$ (more precisely
the natural number represented by $m$) has a divisor less then $l$ and $\fac(m,l) = 0$ otherwise. Note
that many tasks of classical computation can be reformulated this way, so that we do not get a severe loss
of generality. The second crucial point we have to clarify is the question what exactly are the resources
we have mentioned above and how we have to quantify them. A natural physical quantity which come into 
mind immediately is the time needed to perform the computation (space is another candidate, which we do
not discuss here, however). Hence the question we have to discuss is how the computation time $t$ depends
on the size $L$ of the input data $x$ (i.e. the length $L$ of the smallest register needed to represent
$x$ as a bit string).  

However a precise definition of ``computation time'' is still model dependent. For a Turing machine we can
take simply the number of head movements needed to solve the problem, and in the network model we choose
the number of steps needed to execute the whole circuit, if gates which operate on different bits are
allowed to work simultaneously\footnote{Note that we have glanced over a lot of technical problems at
  this point. The crucial difficulty is that each circuit $C_N$ allows only the computation of a boolean
  function $f_N: \Bbb{B}^N \to \Bbb{B}$ which acts on input data of length $N$. Since we are interested in 
  answers for arbitrary finite length inputs a sequence $C_N$, $N \in \Bbb{N}$ of circuits with appropriate
  uniformity properties is needed; cf. \Cite{Pap94} for details.}. Even with a fixed type of model the
functional behavior of $t$ depends on the set of elementary operations we choose, e.g. the set of elementary
gates in the network model. It is therefore useful to divide computational problems into \emph{complexity
  classes} whose definitions do not suffer under model dependent aspects. The most fundamental one is
the class $\CCP$ which contains all problems which can be computed in ``polynomial time'', i.e. $t$ is, as
a function of $L$, bounded from above by a polynomial. The model independence of this class is basically
the content of the strong Church Turing hypotheses which states, roughly speaking, that each model of
computation can be simulated in polynomial time on a probabilistic Turing machine.  

Problems of class $\CCP$ are considered ``easy'', everything else is ``hard''. However even if a
(decision) problem is hard the situation is not hopeless. E.g. consider the factoring problem $\fac$
described above. It is generally believed (although not proved) that this problem is is not in class
$\CCP$. But if somebody gives us a divisor $p < l$ of $m$ it is easy to check whether $p$ is really a
factor, and if the answer is true we have computed $\fac(m,l)$. This example motivates the following
definition: A decision problem $f$ is in class $\CCNP$ (``nondeterministic polynomial time'') if there is
a boolean function $f'$ in class $\CCP$ such that $f'(x,y) = 1$ for some $y$ implies $f(x)$. In our
example $\fac'$ is obviously defined by $\fac'(m,l,p) = 1$ $\Leftrightarrow$ $p < l$ and $p$ is a devisor of $m$. It is
obvious that $\CCP$ is a subset of $\CCNP$ the other inclusion however is rather nontrivial. The
conjecture is that $\CCP \not= \CCNP$ holds and great parts of complexity theory are based on it. Its
proof (or disproof) however represents one of the biggest open questions of theoretical informatics.  

To introduce a third complexity class we have to generalize our point of view slightly. Instead of
a function $f: \Bbb{B}^N \to \Bbb{B}^M$ we can look at a noisy classical $T$ which sends the input
value $x \in \Bbb{B}^N$ to a probability distribution $T_{xy}$, $y \in \Bbb{B}^M$ on $\Bbb{B}^M$
(i.e. $T_{xy}$ is the transition matrix of the classical channel $T$; cf. Subsection
\ref{sec:classical-channels}). Roughly speaking, we can interpret such a channel as a \emph{probabilistic
  computation} which can be realized as a circuit consisting of ``probabilistic gates''. This means there
are several different ways to proceed at each step and we use a classical random number generator to decide
which of them we have to choose. If we run our device several times on the same input data $x$ we get
different results $y$ with probability $T_{xy}$. The crucial point is now that we can allow some of the
outcomes to be wrong as long as there is an easy way (i.e. a class $\CCP$ algorithm) to check the
validity of the results. Hence we define $\CCBPP$ (``bounded error probabilistic polynomial time'') as
the class of all decision problems which admit a polynomial time probabilistic algorithm with error
probability less than $1/2 -\epsilon$ (for fixed $\epsilon$). It is obvious that $\CCP \subset \CCBPP$ holds but the relation
between $\CCBPP$ and $\CCNP$ is not known.  

\subsection{Reversible computing}
\label{sec:reversible-computing}

In the last subsection we have discussed the time needed to perform a certain computation. Other
physical quantities which seem to be important are space and energy. Space can be treated in a similar
way as time and there are in fact space-related complexity classes (e.g $\mathbf{PSPACE}$ which stands
for ``polynomial space''). Energy, however, is different, because it turns surprisingly out that it is
possible to do any calculation \emph{without expending any energy}! One source of energy consumption in a
usual computer is the intrinsic irreversibility of the basic operations. E.g. a basic gate like AND
maps two input bits to one output bit, which implies obviously that the input can not be reconstructed
from the output. In other words: one bit of information is erased during the operation of the AND gate,
hence a small amount of energy is dissipated to the environment. A thermodynamic analysis, known as
Landauer's principle, shows that this energy loss is at least $k_B T \ln 2$, where $T$ is the temperature
of the environment \Cite{Lan61}. 

If we want to avoid this kind of energy dissipation we are restricted to reversible processes, i.e. it
should be possible to reconstruct the input data from the output data. This is called \emph{reversible
  computation} and it is performed in terms of \emph{reversible gates}, which in turn can be described by
invertible functions $f: \Bbb{B}^N \to \Bbb{B}^N$. This does not restrict the class of problems which can
be solved however: We can repackage a non-invertible function $f: \Bbb{B}^N \to \Bbb{B}^M$ into an
invertible one $f': \Bbb{B}^{N + M} \to \Bbb{B}^{N+M}$ simply by $f'(x,0) = (x,f(x))$ and an appropriate
extension to the rest of $\Bbb{B}^{N+M}$. It can be even shown that a reversible computer performs as
good as a usual one, i.e. an ``irreversible'' network can be simulated in polynomial time by a reversible
one.  This will be of particular importance for quantum computing, because a reversible computer is, as
we will see soon, a special case of a quantum computer.

\subsection{The network model of a quantum computer}
\label{sec:netw-model-quant}

Now we are ready to introduce a mathematical model for quantum computation. To this end we will
generalize the network model discussed in Subsection \ref{sec:classical-computing} to the network model
of quantum computation. 

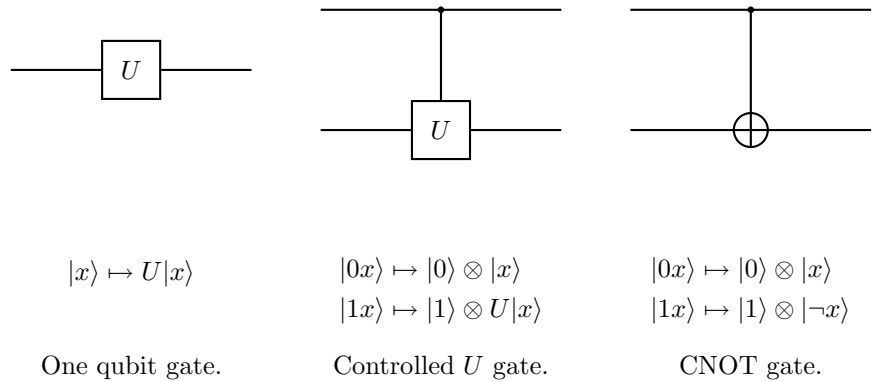
\begin{figure}[h]
  \begin{center}
    \begin{minipage}{4cm}
      \begin{center}
      \begin{picture}(4,4)
        \psline{-}(0,2)(1.5,2)
        \psframe(1.5,1.5)(2.5,2.5)
        \psline{-}(2.5,2)(4,2)
        \rput(2,2){$U$}
      \end{picture}
    \end{center}
    \end{minipage}
    \begin{minipage}{4cm}
      \begin{center}
      \begin{picture}(4,4)
        \psline{-}(0,3)(4,3)
        \psline{-}(0,1)(1.5,1)
        \psframe(1.5,0.5)(2.5,1.5)
        \psline{-}(2.5,1)(4,1)
        \psline{*-}(2,3)(2,1.5)
        \rput(2,1){$U$}
      \end{picture}
      \end{center}
    \end{minipage}
    \begin{minipage}{4cm}
      \begin{center}
      \begin{picture}(4,4)
        \psline{-}(0,3)(4,3)
        \psline{-}(0,1)(4,1)
        \pscircle(2,1){0.3}
        \psline{*-}(2,3)(2,0.7)
      \end{picture}
      \end{center}
    \end{minipage}\\[1em]
    \begin{minipage}{4cm}
      \begin{center}
        $\ket{x} \mapsto U \ket{x}$
      \end{center}
    \end{minipage}
    \begin{minipage}{4cm}
      \begin{center}
        \begin{align*}
        \ket{0x} &\mapsto \ket{0} \otimes \ket{x}\\
        \ket{1x} &\mapsto \ket{1} \otimes U \ket{x}
      \end{align*}
      \end{center} 
    \end{minipage}
    \begin{minipage}{4cm}
      \begin{center}
        \begin{align*}
        \ket{0x} &\mapsto \ket{0} \otimes \ket{x}\\
        \ket{1x} &\mapsto \ket{1} \otimes \ket{\lnot x}
      \end{align*}
      \end{center}
    \end{minipage}\\[1em]
    \begin{minipage}{4cm}
      \begin{center}
        One qubit gate.
      \end{center}
    \end{minipage}
    \begin{minipage}{4cm}
      \begin{center}
        Controlled $U$ gate.
      \end{center}
    \end{minipage}
    \begin{minipage}{4cm}
      \begin{center}
       CNOT gate.
      \end{center}
    \end{minipage}
    \caption{Universal sets of quantum gates.}
    \label{fig:q-gates}
  \end{center}
\end{figure}

A classical computer operates by a network of gates on a finite number of classical bits. A quantum
computer operates on a finite number of qubits in terms of a network of \emph{quantum gates} -- this is
the rough idea. To be more precise consider the Hilbert space $\scr{H}^{\otimes N}$ with $\scr{H} = \Bbb{C}^2$
which describes a \emph{quantum register} consisting of $N$ qubits. In $\scr{H}$ there is a preferred 
set $\ket{0}, \ket{1}$ of orthogonal states, describing the two values a classical bit can have. Hence we
can describe each possible value $x$ of a classical register of length $N$ in terms of the
\emph{computational basis} $\ket{x} = \ket{x_1} \otimes \cdots \otimes \ket{x_N}$, $x \in \Bbb{B}^N$. A quantum gate is now
nothing else but a unitary operator acting on a small number of qubits (preferably 1 or 2) and a quantum
network is a graph representing the composition of elementary gates taken from a small set $G$ of
unitaries. A \emph{quantum computation} can now be defined as the application of such a network to an
input state $\psi$ of the quantum register (cf. Figure \ref{fig:q-circuit} for an example). Similar to the
classical case the set $G$ should be universal; i.e. each unitary operator on a quantum register of
arbitrary length can be represented as a composition of elements from $G$. Since the group of unitaries
on a Hilbert space is continuous, it is not possible to do this with a finite set $G$. However we can find
at least suitably small sets which have the chance to be realizable technically (e.g. in an ion-trap)
somehow in the future. Particular examples are on the one hand the controlled $U$ operations and the set
consisting of CNOT and all one-qubit gates on the other (cf. Figure \ref{fig:q-gates}; for a proof of
universality see Section 4.5 of \Cite{NC}). 

\begin{figure}[h]
  \begin{center}
    \begin{picture}(14,5)
      \psline{-}(0,4)(0.5,4)
      \psframe(0.5,3.5)(1.5,4.5)
      \rput(1,4){$H$}
      \psline(1.5,4)(14,4)
      \psline{-}(0,3)(2,3)
      \psframe(2,2.5)(3,3.5)
      \psframe(3.25,2.5)(4.25,3.5)
      \psline(3,3)(3.25,3)
      \rput(2.5,3){$U_1$}
      \rput(3.75,3){$H$}
      \psline{*-}(2.5,4)(2.5,3.5)
      \psline(4.25,3)(14,3)
      \psline(0,2)(4.75,2)
      \psframe(4.75,1.5)(5.75,2.5)
      \psframe(6,1.5)(7,2.5)
      \psframe(7.25,1.5)(8.25,2.5)
      \psline(5.75,2)(6,2)
      \psline(7,2)(7.25,2)
      \rput(5.25,2){$U_2$}
      \rput(6.5,2){$U_1$}
      \rput(7.75,2){$H$}
      \psline{*-}(5.25,4)(5.25,2.5)
      \psline{*-}(6.5,3)(6.5,2.5)
      \psline{-}(8.25,2)(14,2)
      \psline{-}(0,1)(8.75,1)
      \psframe(8.75,0.5)(9.75,1.5)
      \psframe(10,0.5)(11,1.5)
      \psframe(11.25,0.5)(12.25,1.5)
      \psframe(12.5,0.5)(13.5,1.5)
      \psline(9.75,1)(10,1)
      \psline(11,1)(11.25,1)
      \psline(12.25,1)(12.5,1)
      \rput(9.25,1){$U_3$}
      \rput(10.5,1){$U_2$}
      \rput(11.75,1){$U_1$}
      \rput(13,1){$H$}
      \psline{*-}(9.25,4)(9.25,1.5)
      \psline{*-}(10.5,3)(10.5,1.5)
      \psline{*-}(11.75,2)(11.75,1.5)
      \psline(13.5,1)(14,1)
    \end{picture}
    \[ H = \frac{1}{\sqrt{2}}\left[\begin{array}{cc} 1 & 1\\ 1 & -1 \end{array}\right] \quad 
       U_k =\left[\begin{array}{cc} 1 & 0\\ 0 & e^{2^{-k} \pi} \end{array}\right] \]
    \caption{Quantum circuit for the discrete Fourier transform on a 4-qubit register.}
    \label{fig:q-circuit}
  \end{center}
\end{figure}

Basically we could have considered arbitrary quantum operations instead of only unitaries as gates. We
have seen however in Subsection \ref{sec:ideal-channels-noisy} that we can implement each operation
unitarily if we add an ancilla to the systems. Hence this kind of generalization is already covered by the
model. (As long as non-unitarily implemented operations are a desired feature. Decoherence effect due to
unavoidable interaction with the environment are a completely different story; we come back to this point
at the end of the Subsection.) The same holds for measurements at intermediate steps and subsequent
conditioned operations. In this case we get basically the same  result with a different network where all
measurements are postponed to the end. (Often it is however very useful to allow measurements at
intermediate steps as we will see in the next Subsection.)

Having a mathematical model of quantum computers in mind we are now ready to discuss how it would work in
principle.
\begin{enumerate}
\item 
  The first step is in most cases preprocessing of the input data on a classical computer. E.g. the
  Shor algorithm for the factoring problem does not work if the input number $m$ is a pure prime
  power. However in this case there is an efficient classical algorithm. Hence we have to check first
  whether $m$ is of this particular form and use this classical algorithm where appropriate. 
\item \label{item:1}
  Based on these preprocessed data we have to prepare the quantum register in the next step. This means
  in the most simple case to write classical data, i.e. to prepare the state $\ket{x} \in \scr{H}^{\otimes N}$ if
  the (classical) input is $x \in \Bbb{B}^N$. In many cases however it might be more intelligent to use a
  superposition of several $\ket{x}$, e.g. the state
  \begin{equation} \label{eq:28}
    \Psi = \frac{1}{\sqrt{2^N}} \sum_{x \in \Bbb{B}^N} \ket{x},
  \end{equation}
  which represents actually the superposition of all numbers the registers can represent -- this is indeed
  the crucial point of quantum computing and we come back to it below.
\item \label{item:2}
  Now we can apply the quantum circuit $C$ to the input state $\psi$ and after the calculation we get the
  output state $U\psi$, where $U$ is the unitary represented by $C$.
\item 
  To read out the data after the calculation we perform a von Neumann measurement in the computational
  basis, i.e. we measure the observable given by the one dimensional projectors $\kb{x}$, $x \in
  \Bbb{B}^N$. Hence we get $x \in \Bbb{B}^N$ with probability $P_N = |\langle\psi| x\rangle|^2$. 
\item 
  Finally we have to postprocess the measured value $x$ on a classical computer to end up with the final
  result $x'$. If, however, the output state $U\Psi$ is a proper superposition of basis vectors $\ket{x}$
  (and not just one $\ket{x}$) the probability $p_x$ to get this particular $x'$ is less than $1$. In
  other words we have performed a probabilistic calculation as described in the last paragraph of
  Subsection   \ref{sec:comp-compl}. Hence we have to check the validity of the results (with a class
  $\CCP$ algorithm  on a classical computer) and if they are wrong we have to go back to step
  \ref{item:1}.   
\end{enumerate}

So, why is quantum computing potentially useful? First of all, a quantum computer can perform at least as
good as a classical computer. This follows immediately from our discussion of reversible computing in
Subsection \ref{sec:reversible-computing} and the fact that any invertible function $f: \Bbb{B}^N \to
\Bbb{B}^N$ defines a unitary by $U_f : \ket{x} \mapsto \ket{f(x)}$ (the quantum CNOT gate in Figure
\ref{fig:q-gates} arises exactly in this way from the classical CNOT). But, there is on the other hand
strong evidence which indicates that a quantum computer can solve problems in polynomial time which a
classical computer can not. The most striking example for this fact is the Shor algorithm, which provides
a way to solve the factoring problem (which is most probably not in class $\CCP$) in polynomial
time. If we introduce the new complexity class $\CCBQP$ of decision problems which can be solved with high
probability and in polynomial time with a quantum computer, we can express this conjecture as $\CCBPP
\not= \CCBQP$.

 The mechanism which gives a quantum computer its potential power is the ability to
operate not just on one value $x \in \Bbb{B}^N$, but on whole superpositions of values, as already mentioned
in step \ref{item:1} above. E.g. consider a, not necessarily invertible, map $f: \Bbb{B}^N \to \Bbb{B}^M$
and the unitary operator $U_f$   
\begin{equation} \label{eq:29}
  \scr{H}^{\otimes N} \otimes \scr{H}^{\otimes M} \ni \ket{x} \otimes \ket{0} \mapsto U_f \ket{x} \otimes \ket{0} = \ket{x} \otimes \ket{f(x)} \in
  \scr{H}^{\otimes N} \otimes \scr{H}^{\otimes M}.
\end{equation}
If we let act $U_f$ on a register in the state $\Psi \otimes \ket{0}$ from Equation (\ref{eq:28}) we get the
result 
\begin{equation} \label{eq:30}
  U_f(\Psi \otimes \ket{0}) = \frac{1}{\sqrt{2^N}} \sum_{x \in \Bbb{B}^N} \ket{x} \otimes \ket{f(x)}.
\end{equation}
Hence a quantum computer can evaluate the function $f$ on all possible arguments $x \in \Bbb{B}^N$ at the
same time! To benefit from this feature -- usually called \emph{quantum parallelism} -- is, however, not
as easy as it looks like. If we perform a measurement on $U_f(\Psi \otimes \ket{0})$ in the computational basis we
get the value of $f$ for exactly one argument and the rest of the information originally contained in
$U_f(\Psi \otimes \ket{0})$ is destroyed. In other words it is not possible to read out all pairs $(x, f(x))$ from
$U_f(\Psi \otimes \ket{0})$ and to fill a (classical) lookup table with them. To take advantage from quantum
parallelism we have to use a clever algorithm within the quantum computation step (step \ref{item:2}
above). In the next section we will consider a particular example for this.

Before we come to this point, let us give some additional comments which link this section to other
parts of quantum information. The first point concerns entanglement. The state $U_f(\Psi \otimes \ket{0})$ is
highly entangled (although $\Psi$ is separable since $\Psi = \left[2^{-1/2}(\ket{0} + \ket{1})\right]^{\otimes N}$),
and this fact is essential for the ``exponential speedup'' of computations we could gain in a quantum
computer. In other words, to outperform a classical computer, entanglement is the most crucial resource
-- this will become more transparent in the next section. The second remark concerns error correction. Up
to now we have assumed implicitly all components of a quantum computer work perfectly without any
error. In reality however decoherence effects make it impossible to realize unitarily implemented
operations, and we have to deal with noisy channels. Fortunately it is possible within quantum
information to correct at least a certain amount of errors, as we have seen in Section
\ref{sec:quant-error-corr}). Hence unlike an analog computer\footnote{If an analog computer works
  reliably only with a certain accuracy, we can rewrite the algorithm into a digital one.} a quantum
computer can be designed fault tolerant, i.e. it can work with imperfectly manufactured components.

\subsection{Simons problem}
\label{sec:simons-problem}

We will consider now a particular problem (known as Simons problem; cf. \Cite{Simon94}) which shows
explicitly how a quantum computer can speed up a problem which is hard to solve with a classical
computer. It does not fit  however exactly into the general scheme sketched in the last subsection,
because a quantum ``oracle'' is involved, i.e. a black box which performs an (a priori unknown) unitary
transformation on an input state given to it. The term ``oracle'' indicates here that we are not
interested in the time the black box needs to perform the calculation but only in the number of times we
have to access it. Hence this example does not prove the conjecture $\CCBPP \not= \CCBQP$ stated 
above.  Other quantum algorithms which we have not the room here to discuss include: the Deutsch
\Cite{Deutsch85} and Deutsch-Josza problem \Cite{DJ92}, the Grover search algorithm
\Cite{Grover97,Grover97a} and of course Shor's factoring algorithm \Cite{Shor94,Shor97}. 

Hence let us assume that our black box calculates the unitary $U_f$ from Equation (\ref{eq:29}) with a
map $f: \Bbb{B}^N \to \Bbb{B}^N$ which is two to one and has period $a$, i.e. $f(x) = f(y)$ iff $y = x + a
\mod 2$. The task is to find $a$. Classically, this problem is hard, i.e. we have to query the oracle
exponentially often. To see this note first that we have to find a pair $(x,y)$ with $f(x) = f(y)$ and
the probability to get it with two random queries is $2^{-N}$ (since there is for each $x$ exactly one $y
\not= x$ with $f(x) = f(y)$). If we use the box $2^{N/4}$ times, we get  less than $2^{N/2}$ different
pairs. Hence the probability to get the correct solution is $2^{-N/2}$, i.e. arbitrarily small even with
exponentially many queries. 

Assume now that we let our box act on a quantum register $\scr{H}^{\otimes N} \otimes \scr{H}^{\otimes N}$ in the state $\Psi
\otimes \ket{0}$ with $\Psi$ from Equation (\ref{eq:28}) to get $U_f (\Psi \otimes \ket{0})$ from (\ref{eq:30}). Now we
measure the second register. The outcome is one of $2^{N-1}$ possible values (say $f(x_0)$), each of
which occurs equiprobable. Hence, after the measurement the first register is the state $2^{-1/2}(\ket{x}
+ \ket{x+a})$. Now we let a Hadamard gate $H$ (cf. Figure \ref{fig:q-circuit}) act on each qubit
of the first register and the result is (this follows with a short calculation) 
\begin{equation}
  \frac{1}{\sqrt{2}} H^{\otimes N} \bigl(\ket{x} + \ket{x+a}\bigr) = \frac{1}{\sqrt{2^{N - 1}}} \sum_{a \cdot y
    = 0} (-1)^{x \cdot y} \ket{y}
\end{equation}
where the dot denotes the ($\Bbb{B}$-valued) scalar product in the vector space $\Bbb{B}^N$. Now we
perform a measurement on the first register (in computational basis) and we get a $y \in \Bbb{B}^N$ with
the property $y \cdot a = 0$. If we repeat this procedure $N$ times and if we get $N$ linear independent values
$y_j$ we can determine $a$ as a solution of the system of equations $y_1 \cdot a = 0, \ldots, y_N \cdot a = 0$. The
probability to appear as an outcome of the second measurement is for each $y$ with $y \cdot a = 0$ given
by $2^{1-N}$. Therefore the success probability can be made arbitrarily big while the number of times
we have to access the box is linear in $N$. 

\section{Quantum cryptography}
\label{sec:quantum-cryptography}

Finally we want to have a short look on quantum cryptography -- another more practical application of
quantum information, which has the potential to emerge into technology in the not so distant future (see
e.g. \Cite{QCExp1,QCExp2,QCExp3} for some experimental realizations and \Cite{QCRev} for a more detailed
overview). Hence let us assume that Alice has a message $x \in \Bbb{B}^N$ which she wants to send secretly
to Bob over a public communication channels. One way to do this is the so called ``one-time pad'': Alice
generates randomly a second bit-string $y \in \Bbb{B}^N$ of the same length as $x$ sends $x+y$ instead of
$x$. Without knowledge of the key $y$ it is completely impossible to recover the message $x$ from
$x+y$. Hence this is a perfectly secure method to transmit secret data. Unfortunately it is completely
useless without a secure way to transmit the key $y$ to Bob, because Bob needs $y$ to decrypt the message
$x+y$ (simply by adding $y$ again). What makes the situation even worse is the fact that the key $y$ can
be used only once (therefore the name \emph{one-time} pad). If two messages $x_1$, $x_2$ are encrypted
with the same key we can use $x_1$ as a  key to decrypt $x_2$ and vice versa: $(x_1 + y) + (x_2 + y) =
x_1 + x_2$, hence both messages are partly compromised. 

Due to these problems completely different approaches, namely ``public key systems'' like DSA and RSA are
used today for cryptography. The idea is to use two keys instead of one: a \emph{private key} which
is used for decryption and only known to its owner and a \emph{public key} used for encryption, which is
publicly available (we do not discuss the algorithms needed for key generation, encryption and decryption
here, see \Cite{Singh99} and the references therein instead). To use this method, Bob generates a key
pair $(z,y)$, keeps his private key ($y$) at a secure place and sends the public one ($z$) to Alice
over a public channel. Alice encrypts her message with $z$ sends the result to Bob and he can decrypt it
with $y$. The security of this scheme relies on the assumption that the factoring problem is
computationally \emph{hard}, i.e. not in class $\CCP$, because to calculate $y$ from $z$ requires the
factorization of large integers. Since the latter is tractable on quantum computers via Shor's algorithm,
the security of public key systems breaks down if quantum computers become available in the
future. Another problem of more fundamental nature is the unproven status of the conjecture that
factorization is not solvable in polynomial time. Consequently, security of public key systems is not
proven either. 

The crucial point is now that quantum information provides a way to distribute a cryptographic key $y$ in
a secure way, such that $y$ can be used as a one-time pad afterwards. The basic idea is to use the no
cloning theorem to detect possible eavesdropping attempts. To make this more transparent, let us consider
a particular example here, namely the probably most prominent protocol proposed by Benett and Brassard in
1984 \Cite{BB84}.

\begin{enumerate}
\item 
  Assume that Alice wants to transmit bits from the (randomly generated) key $y \in \Bbb{B}^N$
  through an ideal quantum channel to Bob. Before they start they settle upon two orthonormal bases
  $e_0,e_1 \in \scr{H}$, respectively $f_0,f_1 \in \scr{H}$, which are mutually nonorthogonal,
  i.e. $|\langle e_j,f_k\rangle| \geq \epsilon > 0$ with $\epsilon$ big enough for each $j,k = 0,1$. If photons are used as
  information carrier a typical choice are linearly polarized photons with polarization direction rotated
  by 45$^\circ$ against each other.
\item 
  To send one bit $j \in \Bbb{B}$ Alice selects now at random one of the two bases, say $e_0, e_1$ and then
  she sends a qubit in the state $\kb{e_j}$ through the channel. Note that neither Bob nor a potential
  eavesdropper knows which bases she has chosen.
\item 
  When Bob receives the qubit he selects, as Alice before, at random a base and performs the
  corresponding von Neumann measurement to get one classical bit $k \in \Bbb{B}$, which he records together
  with the measurement method.
\item 
  Both repeat this procedure until the whole string $y \in \Bbb{B}^N$ is transmitted and then Bob tells
  Alice (through a classical, public communication channel) bit for bit which base he has used for the
  measurement (but not the result of the measurement). If he has used the same base as Alice both keep
  the corresponding bit otherwise they discard it. They end up with a   bit-string $y' \in \Bbb{B}^M$ of a
  reduced length $M$. If this is not sufficient they have to continue sending random bits until the key
  is long enough. For large $N$ the rate of successfully transmitted bits per bits sended is obviously
  $1/2$. Hence Alice has to send approximately twice as many bits as they need.  
\end{enumerate}

To see why this procedure is secure, assume now that the eavesdropper Eve can listen and modify the
information sent through the quantum channel and that she can listen on the classical channel but can not
modify it (we come back to this restriction in a minute). Hence Eve can intercept the qubits sent by
Alice and make two copies of it. One she forwards to Bob and the other she keeps for later analysis. Due
to the no cloning theorem however she has produced errors in both copies and the quality of her own
decreases if she tries to make the error in Bob's as small as possible. Even if Eve knows about the two
bases $e_0, e_1$ and $f_0, f_1$ she does not know which one Alice uses to send a particular
qubit\footnote{If Alice and Bob uses only one basis to send the data and Eve knows about it she can
  produce of course ideal copies of the qubits. This is actually the reason why two nonorthogonal bases
  are necessary.}. Hence Eve has to decide randomly which base to choose (as Bob). If $e_0, e_1$ and
$f_0, f_1$ are chosen optimal, i.e. $|\langle e_j,f_k\rangle|^2 = 0.5$ it is easy to see that the error rate Eve
necessarily produces if she randomly measures in one of the bases is $1/4$ for large $N$. To detect this
error Alice and Bob simply have to sacrify portions of the generated key and to compare randomly selected
bits using their classical channel. If the error rate they detect is too big they can decide to drop the
whole key and restart from the beginning.

So let us discuss finally a situation where Eve is able to intercept the quantum \emph{and} the classical
channel. This would imply that she can play Bob's part for Alice and Alice's for Bob. As a result she
shares a key with Alice and one with Bob. Hence she can decode all secret data Alice sends to Bob, read
it, and encode it finally again to forward it to Bob. To secure against such a ``woman in the middle
attack'', Alice and Bob can use classical authentication protocols which ensure that the correct person
is at the other end of the line. This implies that they need a small amount of initial secret material
which can be renewed however from the new key they have generated through quantum communication.

\chapter{Entanglement measures}
\label{cha:quant-theory-i}

We have seen in the last chapter that entanglement is an essential \emph{resource} for many tasks of
quantum information theory, like teleportation or quantum computation. This means that entangled states
are needed for the functioning of many processes and that they are consumed during operation. It is
therefore necessary to have \emph{measures} which tell us whether the entanglement contained in a number
of quantum systems is sufficient to perform a certain task. What makes this subject difficult, is the
fact that we can not restrict the discussion to systems in a maximally or at least highly entangled pure
state. Due to unavoidable decoherence effects realistic applications have to deal with imperfect systems
in mixed states, and exactly in this situation the question for the amount of available entanglement is
interesting.

\section{General properties and definitions}
\label{sec:gener-prop-defin}


The difficulties arising if we try to quantify entanglement can be divided, roughly speaking, into two
parts: First we have to find a reasonable quantity which describes exactly those properties which we are
interested in and second we have to calculate it for a given state. In this section we will discuss the
first problem and consider several different possibilities to define entanglement measures.

\subsection{Axiomatics}
\label{sec:axiomatics}

First of all, we will collect some general properties which a reasonable entanglement measure should
have (cf. also \Cite{BDiVSW,VPRK,VP98,Vidal00,H3LimEM}). To quantify entanglement, means nothing else but
to associate a positive real number to each state of (finite dimensional) two-partite systems. 

\begin{ax}{E0}{1} \label{ax:2}
  An entanglement measure is a function $E$ which assigns to each state $\rho$ of a finite dimensional
  bipartite system a positive real number $E(\rho) \in \Bbb{R}^+$.
\end{ax}

Note that we have glanced over some mathematical subtleties here, because $E$ is not just defined on the
state space of $\scr{B}(\scr{H} \otimes \scr{K})$ systems for particularly chosen Hilbert spaces $\scr{H}$ and
$\scr{K}$ -- $E$ is defined on any state space for arbitrary finite dimensional $\scr{H}$ and
$\scr{K}$. This is expressed mathematically most conveniently by a \emph{family of functions} which
behaves naturally under restrictions (i.e. the restriction to a subspace $\scr{H}' \otimes \scr{K}'$
coincides with the function belonging to $\scr{H}' \otimes \scr{K}'$). However we will see soon that we can
safely ignore this problem.  

The next point concerns the range of $E$. If $\rho$ is unentangled $E(\rho)$ should be zero of course and it
should be maximal on maximally entangled states. But what happens if we allow the dimensions of $\scr{H}$
and $\scr{K}$ to grow? To get an answer consider first a pair of qubits in a maximally entangled state
$\rho$. It should contain exactly one bit entanglement i.e. $E(\rho) = 1$ and $N$ pairs in the state $\rho^{\otimes N}$
should contain $N$ bits. If we interpret $\rho^{\otimes N}$ as a maximally entangled state of a $\scr{H} \otimes \scr{H}$
system with $\scr{H} = \Bbb{C}^{N}$ we get $E(\rho^{\otimes N}) = \log_2(\dim(\scr{H})) = N$, where we have to 
reshuffle in $\rho^{\otimes N}$ the tensor factors such that $(\Bbb{C}^2 \otimes \Bbb{C}^2)^{\otimes N}$ becomes 
$(\Bbb{C}^2)^{\otimes N} \otimes (\Bbb{C}^2)^{\otimes N}$ (i.e. ``all Alice particles to the left and all Bob particles to
the right'';  cf. Section \ref{sec:dist-entangl}.) This observation motivates the following.  

\begin{ax}{E1}{Normalization} \label{ax:3}
  $E$ vanishes on separable and takes its maximum on maximally entangled states. This means more precisely
  that $E(\sigma) \leq E(\rho) = \log_2(d)$ for $\rho, \sigma \in \scr{S}(\scr{H} \otimes \scr{H})$ and $\rho$ maximally entangled.
\end{ax}

One thing an entanglement measure should tell us, is how much quantum information can be \emph{maximally}
teleported with a certain amount of entanglement, where this maximum is taken over all possible
teleportation schemes and distillation protocols, hence it can not be increased further by additional
LOCC operations on the entangled systems in question. This consideration motivates the following Axiom. 

\begin{ax}{E2}{LOCC monotonicity} \label{ax:1} 
  $E$ can not increase under LOCC operation, i.e. $E[T(\rho)] \leq E(\rho)$ for all states $\rho$ and all LOCC 
  channels $T$.  
\end{ax} 

A special case of LOCC operations are of course local unitary operations $U \otimes V$. Axiom \ref{ax:1} implies
now that $E(U \otimes V \rho U^* \otimes V^*) \leq E(\rho)$ and on the other hand $E(U^* \otimes V^* \tilde{\rho} U \otimes V) \leq 
E(\tilde{\rho})$ hence with $\tilde{\rho} = U \otimes V \rho U^* \otimes V$ we get $E(\rho) \leq E(U \otimes V \rho V^* \otimes U^*)$ therefore 
$E(\rho) = E(U \otimes V \rho U^* \otimes V^*)$. We fix this property as a weakened version of Axiom \ref{ax:1}:  

\begin{ax}{E2a}{Local unitary invariance} 
  $E$ is invariant under local unitaries, i.e. $E(U \otimes V \rho U^* \otimes V^*) = E(\rho)$ for all states $\rho$ and all 
  unitaries $U$, $V$. 
\end{ax} 

This axiom shows why we do not have to bother about families of functions as mentioned above. If $E$ is 
defined on $\scr{S}(\scr{H} \otimes \scr{H})$ it is automatically defined on $\scr{S}(\scr{H}_1 \otimes \scr{H}_2)$ for 
all Hilbert spaces $\scr{H}_k$ with $\dim(\scr{H}_k) \leq \dim(\scr{H})$, because we can embed $\scr{H}_1 \otimes 
\scr{H}_2$ under this condition unitarily into $\scr{H} \otimes \scr{H}$.  

Consider now a convex linear combination $\lambda \rho + (1-\lambda)\sigma$ with $0 \leq \lambda \leq 1$. Entanglement can not be 
``generated''  by mixing two states, i.e. $E(\lambda \rho + (1-\lambda)\sigma) \leq \lambda E(\rho) + (1-\lambda)E(\sigma)$. 

\begin{ax}{E3}{Convexity} \label{ax:4}
  $E$ is a convex function, i.e. $E(\lambda \rho + (1-\lambda)\sigma) \leq \lambda E(\rho) + (1-\lambda)E(\sigma)$ for two states $\rho, \sigma$ and 
  $0 \leq \lambda \leq 1$. 
\end{ax}

The next property concerns the continuity of $E$, i.e. if we perturb $\rho$ slightly the change of $E(\rho)$ should
be small. This can be expressed most conveniently as continuity of $E$ in the trace norm. At this point
however it is not quite clear, how we have to handle the fact that $E$ is defined for arbitrary Hilbert
spaces. The following version is motivated basically by the fact that it is a crucial assumption in Theorem
\ref{thm:7} and \ref{thm:13}.

\begin{ax}{E4}{Continuity} \label{ax:5}
  Consider a sequence of Hilbert spaces $\scr{H}_N$, $N \in \Bbb{N}$ and two sequences of states $\rho_N, \sigma_N 
  \in \scr{S}(\scr{H}_N \otimes \scr{H}_N)$ with $\lim \|\rho_N - \sigma_N \|_1 = 0$. Then we have 
  \begin{equation}
    \lim_{N \to \infty} \frac{E(\rho_N) - E(\sigma_N)}{1 + \log_2 (\dim \scr{H}_N)} = 0.
  \end{equation}
\end{ax}

The last point we have to consider here are additivity properties: Since we are looking at entanglement 
as a resource, it is natural to assume that we can do with two pairs in the state $\rho$ twice as much as 
with one $\rho$, or more precisely $E(\rho \otimes \rho) = 2 E(\rho)$ (in $\rho \otimes \rho$ we have to reshuffle tensor factors 
again ;see above). 

\begin{ax}{E5}{Additivity} \label{ax:9}
  For any pair of two-partite states $\rho, \sigma \in \scr{S}(\scr{H} \otimes \scr{K})$ we have $E(\sigma \otimes \rho) = E(\sigma) +
  E(\rho)$. 
\end{ax}

Unfortunately this rather natural looking axiom seems to be too strong (it excludes reasonable 
candidates). It should be however always true that entanglement can not increase if we put two pairs 
together. 
 
\begin{ax}{E5a}{Subadditivity} \label{ax:7}
  For any pair of states $\rho, \sigma$ we have $E(\rho \otimes \sigma) \leq E(\rho)+ E(\sigma)$. 
\end{ax}

There are further modifications of additivity available in the literature. Most frequently used is the 
following, which restricts Axiom \ref{ax:9} to the case $\rho = \sigma$: 

\begin{ax}{E5b}{Weak additivity} \label{ax:6}
  For any state $\rho$ of a bipartite system we have $N^{-1} E(\rho^{\otimes N}) = E(\rho)$.
\end{ax}

Finally, the weakest version of additivity only deals with the behavior of $E$ for large tensor
products, i.e. $\rho^{\otimes N}$ for $N \to \infty$. 

\begin{ax}{E5c}{Existence of a regularization} \label{ax:8}
  For each state $\rho$ the limit
  \begin{equation}
    E^\infty(\rho) = \lim_{N \to \infty} \frac{E(\rho^{\otimes N})}{N}
  \end{equation}
  exists.
\end{ax}

\subsection{Pure states}
\label{sec:pure-states-1}

Let us consider now a pure state $\rho = \kb{\psi} \in \scr{S}(\scr{H} \otimes \scr{K})$. If it is entangled its
partial trace $\sigma = \tr_\scr{H} \kb{\psi} = \tr_\scr{K} \kb{\psi}$ is mixed and for a maximally entangled state
it is maximally mixed. This suggests to use the von Neumann entropy\footnote{We assume here and in the
  following that the reader is sufficiently familiar with entropies. If this is not the case we refer to
  \Cite{OP}.} of $\rho$, which measures how much a state is mixed, as an entanglement measure for mixed
states, i.e. we define \Cite{BBPS,BDiVSW} 
\begin{equation}
  \Evn(\rho) = - \tr\bigl[\tr_\scr{H} \rho \ln(\tr_\scr{H} \rho)\bigr].
\end{equation}
It is easy to deduce from the properties of the von Neumann entropy that $\Evn$ satisfies Axioms \ref{ax:2},
\ref{ax:3}, \ref{ax:4} and \ref{ax:6}. Somewhat more difficult is only Axiom \ref{ax:1} which follows
however from a nice theorem of Nielsen \Cite{Nielsen99} which relates LOCC operations (on pure states) to
the theory of majorization. To state it here we need first some terminology. Consider two probability
distributions $\lambda = (\lambda_1,\ldots,\lambda_M)$ and $\mu = (\mu_1, \ldots, \mu_N)$ both given in decreasing order (i.e. $\lambda_1 \geq \ldots \geq
\lambda_M$ and $\mu_1 \geq \ldots \geq \mu_N$). We say that $\lambda$ is \emph{majorized} by $\mu$, in symbols $\lambda \prec \mu$, if
\begin{equation}
  \sum_{j=1}^k \lambda_j \leq \sum_{j=1}^k \mu_j\quad \forall k = 1, \ldots, \min{M,N}
\end{equation}
holds. Now we have the following result (see \Cite{Nielsen99} for a proof).

\begin{thm}
  A pure state $\psi = \sum_j \lambda_j^{1/2} e_j \otimes e_j' \in \scr{H} \otimes \scr{K}$ can be transformed into another pure
  state $\phi = \sum_j \mu_j^{1/2} f_j \otimes f_j' \in \scr{H} \otimes \scr{K}$ via a LOCC operation, iff the Schmidt
  coefficients of $\psi$ are majorized by those of $\phi$, i.e. $\lambda \prec \mu$.  
\end{thm}

The von Neumann entropy of the restriction $\tr_\scr{H} \kb{\psi}$ can be immediately calculated from
the Schmidt coefficients $\lambda$ of $\psi$ by $\Evn(\kb{\psi}) = - \sum_j \lambda_j \ln(\lambda_j)$. Axiom \ref{ax:1} follows
therefore from the fact that the entropy $S(\lambda) = - \sum_j \lambda_j \ln(\lambda_j)$ of a probability distribution $\lambda$
is a \emph{Shur concave} function, i.e. $\lambda \prec \mu$ implies $S(\lambda) \geq S(\mu)$; see \Cite{Nielsen00a}.

Hence we have seen so far that $\Evn$ is one possible candidate for an entanglement measure on pure
states. In the following we will see that it is in fact the only candidate which is physically
reasonable. There are basically two reasons for this. The first one deals with distillation of
entanglement. It was shown by Bennett et. al. \Cite{BBPS} that each state $\psi \in \scr{H} \otimes \scr{K}$
of a bipartite system can be prepared out of (a possibly large number of) systems in an arbitrary entangled
state $\phi$ by LOCC operations. To be more precise, we can find a sequence of LOCC operations
\begin{equation}
  T_N: \scr{B}\bigl[(\scr{H} \otimes \scr{K})^{\otimes M(N)}\bigr] \to \scr{B}\bigl[(\scr{H} \otimes \scr{K})^{\otimes N}\bigr]  
\end{equation}
such that
\begin{equation}
  \lim_{N \to \infty} \|T_N^*(\kb{\phi}^{\otimes N}) - \kb{\psi}\|_1 = 0
\end{equation}
holds with a nonvanishing \emph{rate} $r = \lim_{N \to \infty} M(N)/N$. This is done either by distillation ($r
< 1$ if $\psi$ is higher entangled then $\phi$) or by ``diluting'' entanglement, i.e. creating many less
entangled states from few highly entangled ones ($r > 1$). All this can be performed in a \emph{reversible}
way: We can start with some maximally entangled qubits dilute them to get many less entangled states which can
be distilled afterwards to get the original states back (again only in an asymptotic sense). The crucial
point is that the asymptotic rate $r$ of these processes is given in terms of $\Evn$ by $r =
\Evn(\kb{\phi})/ \Evn(\kb{\psi})$. Hence we can say, roughly speaking that $\Evn(\kb{\psi})$ describes exactly the
amount of maximally entangled qubits which is contained in $\kb{\psi}$.

A second somewhat more formal reason is that $\Evn$ is the only entanglement measure on the set of pure
states which satisfies the axioms formulated above. In other words the following ``\emph{uniqueness
  theorem for entanglement measures}'' holds \Cite{PoRo97,Vidal00,DoHoRu}

\begin{thm} \label{thm:7}
  The reduced von Neumann entropy $\Evn$ is the only entanglement measure on pure states which satisfies
  Axioms \ref{ax:2} -- \ref{ax:9}.
\end{thm}

\subsection{Entanglement measures for mixed states}
\label{sec:entangl-meas-mixed}

To find reasonable entanglement measures for mixed states is much more difficult. There are in fact many
possibilities (e.g. the maximally entangled fraction introduced in Subsection \ref{sec:pure-states} can be
regarded as a simple measure) and we want to present therefore only four of the most reasonable
candidates. 
Among those measures which we do not discuss here are negativity quantities (\Cite{VidWe} and the
references therein) the ``best separable approximation'' \Cite{LeSa98}, the base norm associated with the
set of separable states \Cite{ViTa99,Rudolph00} and ppt-distillation rates \Cite{Rains00}.  

The first measure we want to present is oriented along the discussion of pure states: We define,
roughly speaking, the asymptotic rate with which maximally entangled qubits can be distilled at most out
of a state $\rho \in \scr{S}(\scr{H} \otimes \scr{K})$ as the \emph{Entanglement of Distillation} $\ED(\rho)$ of $\rho$;
cf \Cite{BBPSSW}. To be more precise consider all possible distillation protocols for $\rho$ (cf. Section
\ref{sec:dist-entangl}), i.e. all sequences of LOCC channels 
\begin{equation} \label{eq:54}
  T_N : \scr{B}(\Bbb{C}^{d_N} \otimes \Bbb{C}^{d_N}) \to \scr{B}(\scr{H}^{\otimes N} \otimes \scr{K}^{\otimes N})
\end{equation}
such that
\begin{equation} \label{eq:97}
  \lim_{N \to \infty} \| T_N^*(\rho^{\otimes N}) - \kb{\Omega_N}\, \|_1 = 0
\end{equation}
holds with a sequence of maximally entangled states $\Omega_N \in \Bbb{C}^{d_N}$. Now we can define 
\begin{equation}
  \ED(\rho) = \sup_{(T_N)_{N \in \Bbb{N}}} \limsup_{N \to \infty} \frac{\log_2 (d_N)}{N}, 
\end{equation}
where the supremum is taken over all possible distillation protocols $(T_N)_{N \in \Bbb{N}}$. It is not
very difficult to see that $\ED$ satisfies \ref{ax:2}, \ref{ax:3}, \ref{ax:1} and \ref{ax:6}. It is not
known whether continuity (\ref{ax:5}) and convexity (Axiom \ref{ax:4}) holds. It can be shown however
that $\ED$ is not convex (and not additive; Axiom \ref{ax:9}) if npt bound entangled states exist
(see \Cite{SST00}, cf. also Subsection \ref{sec:bound-entangl-stat-1}).

For pure states we have discussed beside distillation the ``dilution'' of entanglement and we can use,
similar to $\ED$, the asymptotic rate with which bipartite systems in a given state $\rho$ can be prepared
out of maximally entangled singlets \Cite{HHT}. Hence consider again a sequence of LOCC channels
\begin{equation} 
  T_N : \scr{B}(\scr{H}^{\otimes N} \otimes \scr{K}^{\otimes N}) \to \scr{B}(\Bbb{C}^{d_N} \otimes \Bbb{C}^{d_N})
\end{equation}
and a sequence of maximally entangled states $\Omega_N \in \Bbb{C}^{d_N}$, $N \in \Bbb{N}$, but now with the
property 
\begin{equation}
  \lim_{N \to \infty} \| \rho^{\otimes N} - T^*_N(\kb{\Omega_N})\, \|_1 = 0.
\end{equation}
Then we can define the \emph{entanglement cost} $\EC(\rho)$ of $\rho$ as
\begin{equation}
  \EC(\rho) = \inf_{(S_N)_{N \in \Bbb{N}}} \liminf_{N \to \infty} \frac{\log_2 (d_N)}{N},
\end{equation}
where the infimum is taken over all dilution protocols $S_N$, $N \in \Bbb{N}$. It is again easy to see that
$\EC$ satisfies \ref{ax:2}, \ref{ax:3}, \ref{ax:1} and \ref{ax:6}. In contrast to $\ED$ however it can be
shown that $\EC$ is convex (Axiom \ref{ax:4}), while it is not known, whether $\EC$ is continuous
(Axiom \ref{ax:5}); cf \Cite{HHT} for proofs. 

$\ED$ and $\EC$ are based directly on operational concepts. The remaining two measures we want to
discuss here are defined in a more abstract way. The first can be characterized as the minimal convex
extension of $\Evn$ to mixed states: We define the \emph{entanglement of formation} $\EOF$ of $\rho$ as
\Cite{BDiVSW} 
\begin{equation} \label{eq:75}
  \EOF(\rho) = \inf_{\rho = \sum_j p_j \kb{\psi_j}} \sum p_j \Evn(\kb{\psi_j}),
\end{equation}
where the infimum is taken over all decompositions of $\rho$ into a convex sum of pure states. $\EOF$
satisfies \ref{ax:2} - \ref{ax:5} and \ref{ax:7} (cf. \Cite{BDiVSW} for \ref{ax:1} and \Cite{Nielsen00}
for \ref{ax:5} the rest follows directly from the definition). Whether $\EOF$ is (weakly) additive (Axiom
\ref{ax:6}) is not known. Furthermore it is conjectured that $\EOF$ coincides with $\EC$. However proven
is only the identity $\EOF^\infty = \EC$, where the existence of the regularization $\EOF^\infty$ of $\EOF$ follows
directly from subadditivity. 

Another idea to quantify entanglement is to measure the ``distance'' of the (entangled) $\rho$ from the set
of separable states $\scr{D}$. It hat turned out \Cite{VPRK} that among all possible distance functions
the relative entropy is physically most
reasonable. Hence we define the \emph{relative entropy of entanglement} as
\begin{equation} \label{eq:79}
  \ER(\rho) = \inf_{\sigma \in \scr{D}} S(\rho|\sigma),\quad S(\rho|\sigma) = \bigl[ \tr \bigl( \rho \log_2 \rho - \rho \log_2 \sigma \bigr)\bigr],
\end{equation}
where the infimum is taken over all separable states. It can be shown that $\ER$ satisfies, as $\EOF$ the
Axioms \ref{ax:2} - \ref{ax:5} and \ref{ax:7}, where \ref{ax:3} and \ref{ax:1} are shown in \Cite{VPRK}
and \ref{ax:5} in \Cite{DH99}; the rest follows directly from the definition. It is shown in \Cite{VW1} that
$\ER$ does not satisfy \ref{ax:6}; cf. also Subsection \ref{sec:entangl-meas-under}. Hence the
regularization $\ER^\infty$ of $\ER$ differs from $\ER$.

Finally let us give now some comments on the relation between the measures just introduced. On pure states
all measures just discussed,
coincide with the reduced von Neumann entropy -- this follows
from Theorem \ref{thm:7} and the properties stated in the last Subsection. For mixed states the situation
is more difficult. It can be shown however that $\ED \leq \EC$ holds and that
all ``reasonable'' entanglement measures lie in between \Cite{H3LimEM}. 

\begin{thm} \label{thm:13}
  For each entanglement measure $E$ satisfying \ref{ax:2}, \ref{ax:3}, \ref{ax:1} and \ref{ax:6} and each
  state $\rho \in \scr{S}(\scr{H} \otimes \scr{K})$ we have $\ED(\rho) \leq E(\rho) \leq \EC(\rho)$. 
\end{thm}

Unfortunately no measure we have discussed in the last Subsection satisfies all the assumptions of the
theorem. It is possible however to get a similar statement for the regularization $E^\infty$ with weaker
assumptions on $E$ itself (in particular without assuming additivity); cf \Cite{DoHoRu}. 

\section{Two qubits}
\label{sec:two-qubits-1}

Even more difficult than finding reasonable entanglement measures are explicit calculations. All measures
we have discussed above involve optimization processes over spaces which grow exponentially with the
dimension of the Hilbert space. A direct numerical calculation for a general state $\rho$ is therefore
hopeless. There are however some attempts to get either some bounds on entanglement measures or to get
explicit calculations for special classes of states. 
We will concentrate this discussion to some relevant special cases. On the one hand we will concentrate
on $\EOF$ and $\ER$ and on the other we will look at two special classes of states where explicit
calculations are possible: Two qubit systems in this  section and states with symmetry properties in the
next one.  

\subsection{Pure states}
\label{sec:pure-states-2}

Assume for the rest of this section that $\scr{H} = \Bbb{C}^2$ holds and consider first a pure
state $\psi \in \scr{H} \otimes \scr{H}$. To calculate $\Evn(\psi)$ is of course not difficult and it is
straightforward to see that (cf. for all material of this and the following subsection \Cite{BDiVSW}): 
\begin{equation} \label{eq:59}
  \Evn(\psi) = H\left[ \frac{1}{2} \left( 1 + \sqrt{1-C(\psi)^2} \right)\right]
\end{equation}
holds, with
\begin{equation} \label{eq:65}
  H(x) = -x \log_2(x) - (1-x) \log_2(1-x)
\end{equation}
and the \emph{concurrence} $C(\psi)$ of $\psi$ which is defined by
\begin{equation} \label{eq:60}
  C(\psi) = \left| \sum_{j=0}^3 \alpha_j^2 \right|\ \text{with}\ \psi = \sum_{j=0}^3 \alpha_j \Phi_j,
\end{equation}
where $\Phi_j$, $j=0,\ldots,3$ denotes the Bell basis (\ref{eq:45}). Since $C$ becomes rather important in the
following let us reexpress it as $C(\psi) = |\langle\psi, \Xi \psi\rangle|$, where $\psi \mapsto \Xi\psi$ denotes complex conjugation in Bell
basis. Hence $\Xi$ is an antiunitary operator and it can be written as the tensor product $\Xi = \xi \otimes \xi$ of
the map $\scr{H} \ni \phi \mapsto \sigma_2 \bar{\phi}$, where $\bar{\phi}$ denotes complex conjugation in the canonical basis
and $\sigma_2$ is the second Pauli matrix. Hence local unitaries (i.e. those of the form $U_1 \otimes U_2$) commute
with $\Xi$ and it can be shown that this is not only a necessary but also a sufficient condition for a
unitary to be local \Cite{2Q}.

We see from Equations (\ref{eq:59}) and (\ref{eq:60}) that $C(\psi)$ ranges from $0$ to $1$ and that
$\Evn(\psi)$ is a monotone function in $C(\psi)$. The latter can be considered therefore as an entanglement
quantity in its own right. For a Bell state we get in particular $C(\Phi_j) = 1$ while a separable state
$\phi_1 \otimes \phi_2$ leads to $C(\phi_1 \otimes \phi_2) = 0$; this can be seen easily with the factorization $\Xi = \xi \otimes \xi$.  

Assume now that one of the $\alpha_j$ say $\alpha_0$ satisfies $|\alpha_0|^2 > 1/2$. This implies that $C(\psi)$ can not be
zero since 
\begin{equation}
  \left|\sum_{j=1}^3 \alpha_j^2\right| \leq 1 - |\alpha_0|^2
\end{equation}
must hold. Hence $C(\psi)$ is at least $1 - 2 |\alpha_0|^2$ and this implies for $\Evn$ \emph{and arbitrary} $\psi$ 
\begin{equation}
  \Evn(\psi) \geq h\bigl( |\langle\Phi_0,\psi\rangle|^2 \bigr)\ \text{with} \ h(x) = 
  \begin{cases} 
    H\left[\frac{1}{2} + \sqrt{x(1-x)}\right] & x \geq \frac{1}{2} \\
    0 & x < \frac{1}{2}
  \end{cases}.
\end{equation}
This inequality remains valid if we replace $\Phi_0$ by any other maximally entangled state $\Phi \in \scr{H} \otimes
\scr{H}$. To see this note that two maximally entangled states $\Phi, \Phi' \in \scr{H} \otimes \scr{H}$ are related
(up to a phase) by a local unitary transformation $U_1 \otimes U_2$ (this follows immediately from their Schmidt
decomposition; cf Subsection \ref{sec:pure-states}). Hence, if we replace the Bell basis in Equation
(\ref{eq:60}) by $\Phi_j' = U_1 \otimes U_2 \Phi_j$, $j=0,\ldots,3$ we get for the corresponding $C'$ the equation $C'(\psi)
= \langle U_1^* \otimes U_2^* \psi, \Xi U_1^* \otimes U_2^* \psi\rangle = C(\psi)$ since $\Xi$ commutes with local unitaries. We can even
replace $|\langle\Phi_0,\psi\rangle|^2$ with the supremum over all maximally entangled states and get therfore 
\begin{equation} \label{eq:61}
  \Evn(\psi)  \geq h\bigl[\scr{F}\bigl(\kb{\psi}\bigr)\bigr],
\end{equation}
where $\scr{F}\bigl(\kb{\psi}\bigr)$ is the maximally entangled fraction of $\kb{\psi}$ which we have introduced
in Subsection \ref{sec:pure-states}.

To see that even equality holds in Equation (\ref{eq:61}) note first that it is sufficient to consider
the case $\psi = a \ket{00} + b \ket{11}$ with $a,b \geq 0$, $a^2 + b^2 = 1$, since each pure state $\psi$ can be
brought into this form (this follows again from the Schmidt decomposition) by a local unitary
transformation which on the other hand does not change $\Evn$. The maximally entangled state which
maximizes $|\langle\psi,\Phi\rangle|^2$ is in this case $\Phi_0$ and we get $\scr{F}\bigl(\kb{\psi}\bigr) = (a+b)^2/2 = 1/2 +
ab$. Straightforward calculations show now that $h\bigl[\scr{F}\bigl(\kb{\psi}\bigr)\bigr]= h(1/2 + ab) =
\Evn(\psi)$ holds as stated.

\subsection{EOF for Bell diagonal states}
\label{sec:eof-bell-diagonal}

It is easy to extend the inequality (\ref{eq:61}) to mixed states if we use the convexity of $\EOF$ and
the fact that $\EOF$ coincides with $\Evn$ on pure states. Hence (\ref{eq:61}) becomes
\begin{equation} \label{eq:62}
  \EOF(\rho) \geq h\bigl[\scr{F}(\rho)\bigr].
\end{equation}
For general two qubit states this bound is not achieved however. This can be see with the example $\rho =
1/2 \bigl(\kb{\phi_1} + \kb{00}\bigr)$, which we have considered already in the last paragraph of Subsection
\ref{sec:pure-states}. It is easy to see that $\scr{F}(\rho) = 1/2$ holds hence $h\bigl[\scr{F}(\rho)\bigr] =
0$ but $\rho$ is entangled. Nevertheless we can show that equality holds in Equation (\ref{eq:62}) if we
restrict it to Bell diagonal states $\rho = \sum_{j=0}^3 \lambda j \kb{\Phi_j}$. To prove this statement we have to find
a convex decomposition $\rho = \sum_j \mu_j \kb{\Psi_j}$ of such a $\rho$ into pure states $\kb{\Psi_j}$ such that
$h\bigl[\scr{F}(\rho)\bigr] = \sum_j \mu_j \Evn(\kb{\Psi_j}$ holds. Since $\EOF(\rho)$ can not be smaller than
$h\bigl[\scr{F}(\rho)\bigr]$ due to inequality (\ref{eq:62}) this decomposition must be optimal and equality
is proven. 

To find such $\Psi_j$ assume first that the biggest eigenvalue of $\rho$ is greater than $1/2$, and let,
without loss of generality, be $\lambda_1$ this eigenvalue. A good choice for the $\Psi_j$ are then the eight pure
states
\begin{equation}
  \sqrt{\lambda_0} \Phi_0 + i \left( \sum_{j=1}^3 (\pm \sqrt{\lambda_j}) \Phi_j \right)
\end{equation}
The reduced von Neumann entropy of all these states equals $h(\lambda_1)$, hence $\sum_j \mu_j \Evn(\kb{\Psi_j}) =
h(\lambda_1)$ and therefore $\EOF(\rho) = h(\lambda_1)$. Since the maximally entangled fraction of $\rho$ is obviously
$\lambda_1$ we see that (\ref{eq:62}) holds with equality.

\begin{figure}[htbp]
  \begin{center}
    \begin{pspicture}(15,9)
    \rput(7.5,5){\includegraphics[scale=0.8]{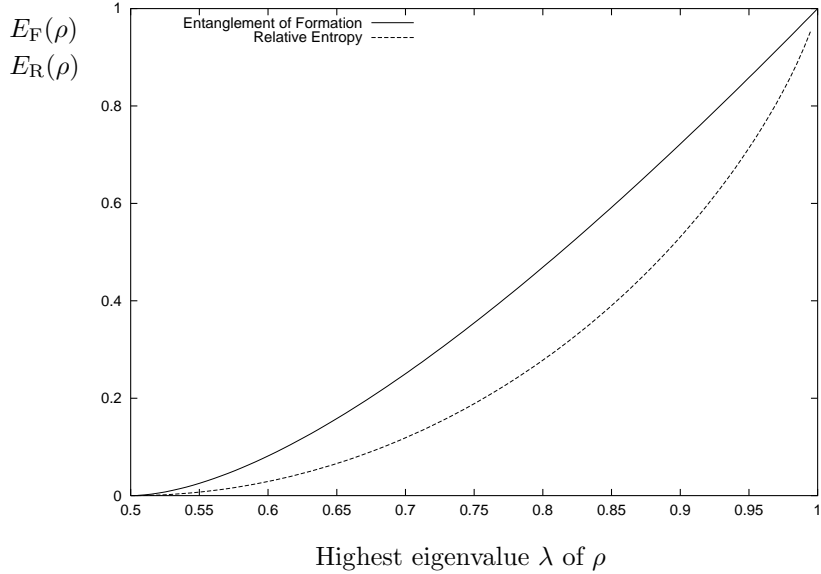}}
    \rput(7.5,0){Highest eigenvalue $\lambda$ of $\rho$}
    \rput[tl](0,9){$\EOF(\rho)$}
    \rput[tl](0,8.4){$\ER(\rho)$}
  \end{pspicture}
    \caption{Entanglement of Formation and Relative Entropy of Entanglement for Bell diagonal states,
      plotted as a function of the highest eigenvalue $\lambda$ of $\rho$} 
    \label{fig:em-bell-diag}
  \end{center}
\end{figure}

Assume now that the highest eigenvalue is less than $1/2$. Then we can find phase factors $\exp(i\phi_j)$
such that $\sum_{j=0}^3 \exp(i\phi_j) \lambda_j = 0$ holds and $\rho$ can be expressed as a convex linear combination of
the states
\begin{equation}
  e^{i\phi_0/2} \sqrt{\lambda_0} \Phi_0 + i \left( \sum_{j=1}^3 (\pm e^{i \phi_j/2} \sqrt{\lambda_j}) \Phi_j \right).
\end{equation}
The concurrence $C$ of all these states is $0$ hence their entanglement is $0$ by Equation (\ref{eq:59}),
which in turn implies $\EOF(\rho) = 0$. Again we see that equality is achieved in (\ref{eq:62}) since the
maximally entangled fraction of $\rho$ is less than $1/2$. Summarizing this discussion we have shown
(cf. Figure \ref{fig:em-bell-diag})

\begin{prop} \label{prop:5}
  A Bell diagonal state $\rho$ is entangled iff its highest eigenvalue $\lambda$ is greater than $1/2$. In this
  case the Entanglement of Formation of $\rho$ is given by
  \begin{equation}
      \EOF(\rho) = H\left[\frac{1}{2} + \sqrt{ \lambda (1 - \lambda )}\right].
  \end{equation}
\end{prop}

\subsection{Wootters formula}
\label{sec:wooters-formula}

If we have a general two qubit state $\rho$ there is a formula of Wootters \Cite{Wooters98} which
allows an easy calculation of $\EOF$. It is based on a generalization of the concurrence $C$ to mixed 
states. To motivate it rewrite $C^2(\psi) = |\langle \psi , \Xi \psi \rangle|$ as 
\begin{equation}
  C^2(\psi) = \tr\bigl(\kb{\psi} \kb{\Xi\psi}\bigr) = \tr\bigl(\rho \Xi  \rho \Xi\bigr) = \tr(R^2)
\end{equation}
with
\begin{equation}
  R = \sqrt{\sqrt{\rho} \Xi \rho \Xi \sqrt{\rho}}.
\end{equation}
Here we have set $\rho = \kb{\psi}$. The definition of the hermitian matrix $R$ however makes sense for
arbitrary $\rho$ as well. If we write $\lambda_j, j=1,\ldots,4$ for the eigenvalues of $R$ and $\lambda_1$ is without loss of
generality the biggest one we can define the \emph{concurrence} of an arbitrary two qubit state $\rho$ as
\Cite{Wooters98} 
\begin{equation} \label{eq:64}
  C(\rho) = \max\bigl(0,2 \lambda_1 - \tr(R)\bigr) = \max (0, \lambda_1 - \lambda_2 - \lambda_3 - \lambda_4).
\end{equation}
It is easy to see that $C(\kb{\psi})$ coincides with $C(\psi)$ from (\ref{eq:60}). The crucial point is now
that Equation (\ref{eq:59}) holds for $\EOF(\rho)$ if we insert $C(\rho)$ instead of $C(\psi)$:

\begin{thm}[Wootters Formula]
  The Entanglement of Formation of a two qubit system in a state $\rho$ is given by
  \begin{equation} \label{eq:66}
    \EOF(\rho) = H\left[ \frac{1}{2} \left( 1 + \sqrt{1-C(\rho)^2} \right)\right]
  \end{equation}
  where the concurrence of $\rho$ is given in Equation (\ref{eq:64}) and  $H$ denotes the binary entropy
  from (\ref{eq:65}). 
\end{thm}

To prove this theorem we have to find first a convex decomposition $\rho = \sum_j \mu_j \kb{\Psi_j}$ of $\rho$ into
pure states $\Psi_j$ such that the average reduced von Neumann entropy $\sum_j \mu_j \Evn(\Psi_j)$ coincides with the
right hand side of Equation (\ref{eq:66}). Second we have to show that we have really found the minimal
decomposition. Since this is much more involved than the simple case discussed in Subsection
\ref{sec:eof-bell-diagonal} we omit the proof and refer to \Cite{Wooters98} instead. Note however that
Equation (\ref{eq:66}) really coincides with the special cases we have derived for pure and Bell diagonal
states. Finally let us add the remark that there is no analogon of Wootters' formula for higher dimensional
Hilbert spaces. It can be shown \Cite{2Q} that the essential properties of the Bell basis $\Phi_j$,
$j=0,..,3$ which would be necessary for such a generalization are available only in $2 \times 2$ dimensions. 

\subsection{Relative entropy for Bell diagonal states}
\label{sec:relat-entr-bell}

To calculate the Relative Entropy of Entanglement $\ER$ for two qubit systems is more difficult. However
there is at least an easy formula for Bell diagonal states which we will give in the following;
\Cite{VPRK}.

\begin{prop}
  The Relative Entropy of Entanglement for a Bell diagonal state $\rho$ with highest eigenvalue $\lambda$ is given
  by (cf. Figure \ref{fig:em-bell-diag})
  \begin{equation}
    \ER(\rho) = 
    \begin{cases}
      1 - H(\lambda) & \lambda > \frac{1}{2} \\
      0 & \lambda \leq \frac{1}{2}
    \end{cases}
  \end{equation}
\end{prop}

\begin{proof}
  For a Bell diagonal state $\rho = \sum_{j=0}^3 \lambda_j \kb{\Phi_j}$ we have to calculate
  \begin{align}
    \ER(\rho) = & \inf_{\sigma \in \scr{D}} \bigl[ \tr \bigl( \rho \log_2 \rho - \rho \log_2 \sigma \bigr)\bigr]\\
    & = \tr (\rho \log_2 \rho) + \inf_{\sigma \in \scr{D}} \left[- \sum_{j=0}^3 \lambda_j \langle\Phi_j,\log_2(\sigma)\Phi_j\rangle\right]. \label{eq:67}
  \end{align}
  Since $\log$ is a concave function we have $- \log_2 \langle\Phi_j, \sigma \Phi_j\rangle \leq \langle \Phi_j, - \log_2(\sigma) \Phi_j\rangle$ and therefore
  \begin{equation}
    \ER(\rho) \geq \tr (\rho \log_2 \rho) + \inf_{\sigma \in \scr{D}} \left[ - \sum_{j=0}^3 \lambda_j  \log_2 \langle\Phi_j, \sigma \Phi_j\rangle \right].
  \end{equation}
  Hence only the diagonal elements of $\sigma$ in the Bell basis enter the minimization on the right hand
  side of this inequality and this implies that we can restrict the infimum to the set of separable Bell
  diagonal state. Since a Bell diagonal state is separable iff all its eigenvalues are less than $1/2$
  (Proposition \ref{prop:5}) we get
  \begin{equation}
    \ER(\rho) \geq \tr (\rho \log_2 \rho) + \inf_{p_j \in [0,1/2]} \left[ - \sum_{j=0}^3 \lambda_j \log_2 p_j\right], \
    \text{with}\  \sum_{j=0}^3 p_j = 1.  
  \end{equation}
  This is an optimization problem (with constraints) over only four real parameters and easy to solve. If
  the highest eigenvalue of $\rho$ is greater than $1/2$ we get $p_1 = 1/2$ and $p_j = \lambda_j / (2 - 2\lambda)$,
  where we have chosen without loss of generality $\lambda = \lambda_1$. We get a lower bound on $\ER(\rho)$ which
  is achieved if we insert the corresponding $\sigma$ in Equation (\ref{eq:67}). Hence we have proven the
  statement for $\lambda > 1/2$. which completes the proof, since we have seen already that $\lambda \leq 1/2$ implies
  that $\rho$ is separable (Proposition \ref{prop:5}).
\end{proof}

\section{Entanglement measures under symmetry}
\label{sec:entangl-meas-under}

The problems occuring if we try to calculate quantities like $\ER$ or $\EOF$ for general density matrices
arise from the fact that we have to solve optimization problems over very high dimensional spaces. One
possible strategy to get explicit results is therefore parameter reduction by symmetry arguments. This
can be done if the state in question admits some invariance properties like Werner, isotropic or
OO-invariant states; cf. Section \ref{sec:corr-entangl-finite}. We will give in the following some
particular examples for such calculations, while a detailed discussion of the general idea (together with
much more examples and further references) can be found in \Cite{VW1}.

\subsection{Entanglement of Formation}
\label{sec:entangl-form}
 
Consider a compact group of unitaries $G \subset \scr{B}(\scr{H} \otimes \scr{H})$ (where  $\scr{H}$ is again
arbitrary finite dimensional), the set of $G$-invariant states, i.e. all $\rho$ with $[V,\rho]=0$ for all
$V \in G$ and the corresponding twirl operation $P_G\sigma = \int_G V\sigma V^* dV$. Particular examples we are looking
at are: 1. Werner states where $G$ consists of all unitaries $U \otimes U$ 2. Isotropic states where each $V \in
G$ has the form $V = U \otimes \bar{U}$ and finally 3. OO-invariant states where $G$ consists of unitaries 
$U \otimes U$ with real matrix elements ($U = \bar{U}$) and the twirl is given in Equation (\ref{eq:70}).

One way to calculate $\EOF$ for a $G$-invariant state $\rho$ consists now of the following steps:
1. Determine the set $M_\rho$ of pure states $\Phi$ such that $P_G \kb{\Phi} = \rho$ holds. 2. Calculate the function
\begin{equation} \label{eq:76}
  P_G\scr{S} \ni \rho \mapsto \epsilon_G(\rho) = \inf \{ \Evn(\sigma) \, | \, \sigma \in M_\rho \} \in \Bbb{R},
\end{equation}
where we have denoted the set of $G$-invariant states with $P_G \scr{S}$. 3. Determine $\EOF(\rho)$ then in
terms of the \emph{convex hull} of $\epsilon$, i.e. 
\begin{multline} \label{eq:71}
  \EOF(\rho) = \inf \{ \mbox{$\sum_j$} \lambda_j \epsilon(\sigma_j) \, | \\
   \sigma_j \in P_G\scr{S},\ 0 \leq \lambda_j \leq 1,\ \ \rho=\mbox{$\sum_j$} \lambda_j \sigma_j,\ \mbox{$\sum_j$} \lambda_j = 1 \}.  
\end{multline}
The equality in the last Equation is of course a non-trivial statement which has to be proved. We skip
this point, however, and refer the reader to \Cite{VW1}. The advantage of this scheme 
relies on the fact that spaces of $G$ invariant states are in general very low dimensional (if $G$ is not
too small). Hence the optimization problem contained in step 3 has a much bigger chance to be tractable
than the one we have to solve for the original definition of $\EOF$. There is of course no guarantee
that any of this three steps can be carried out in a concrete situation. For the three examples mentioned
above, however, there are results available, which we will present in the following.

\subsection{Werner states}
\label{sec:werner-states}

Let us start with Werner states \Cite{VW1}. In this case $\rho$ is uniquely determined by its flip
expectation value $\tr(\rho F)$ (cf. Subsection \ref{sec:stat-under-symm}). To determine $\Phi \in \scr{H} \otimes
\scr{H}$ such that $P_\tUU \kb{\Phi} = \rho$ holds, we have to solve therefore the equation  
\begin{equation} \label{eq:72}
  \langle\Phi,F\Phi\rangle = \sum_{jk} \Phi_{jk} \overline{\Phi_{kj}} = \tr(F\rho),
\end{equation}
where $\Phi_{jk}$ denote components of $\Phi$ in the canonical basis. On the other hand the reduced density
matrix $\rho = \tr_1 \kb{\Phi}$ has the matrix elements $\rho_{jk} = \sum_l \Phi_{jl}\Phi_{kl}$. By exploiting $U \otimes U$
invariance we can assume without loss of generality that $\rho$ is diagonal. Hence to get the function
$\epsilon_\tUU$ we have to minimize
\begin{equation}
  \Evn\bigl(\kb{\Phi}\bigr) = \sum_j S \left[\sum_k |\Phi_{jk}|^2\right]
\end{equation}
under the constraint (\ref{eq:72}), where $S(x) = - x \log_2(x)$ denotes the von Neumann entropy. We skip
these calculations here (see \Cite{VW1} instead) and state the results only. For $\tr(F\rho) \geq 0$ we get
$\epsilon(\rho) = 0$ (as expected since $\rho$ is separable in this case) and with $H$ from (\ref{eq:65})
\begin{equation}
  \epsilon_\tUU(\rho) = H\left[ \frac{1}{2} \left( 1 - \sqrt{1 - \tr(F\rho)^2}\right)\right]
\end{equation}
for $\tr(F\rho) < 0$. The minima are taken for $\Phi$ where all $\Phi_{jk}$ except one diagonal element are zero
in the case $\tr(F\rho) \geq 0$ and for $\Phi$ with only two (non-diagonal) coefficients $\Phi_{jk}, \Phi_{kj}$, $j
\not= k$ nonzero if $\tr(\rho F) < 0$. The function $\epsilon$ is convex and coincides therefore with its convex
hull such that we get

\begin{figure}[t]
  \begin{center}
    \begin{pspicture}(15,9)
    \rput(7.5,5){\includegraphics[scale=0.8]{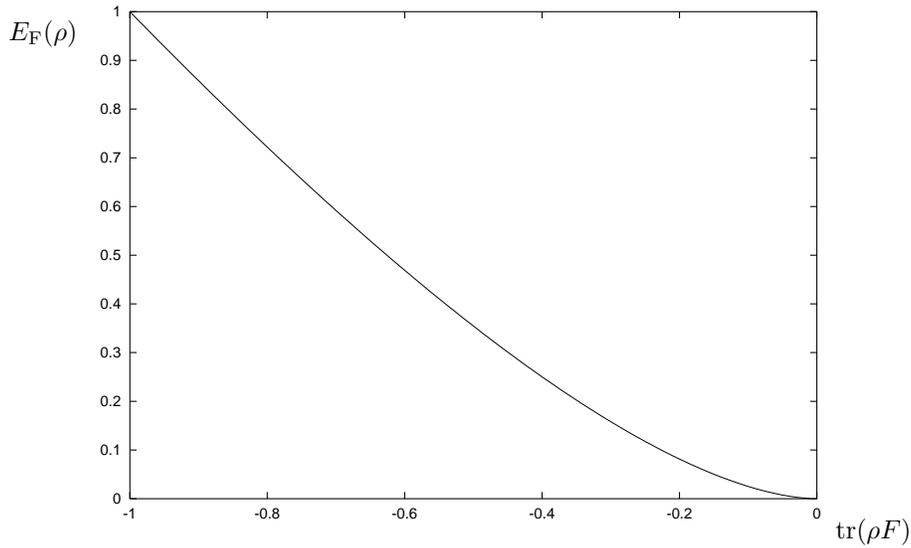}}
    \rput[r](15,0.5){$\tr(\rho F)$}
    \rput[tl](0,9){$\EOF(\rho)$}
  \end{pspicture}
    \caption{Entanglement of Formation for Werner states plotted as function of the flip expectation.} 
\label{fig:eof-werner}
  \end{center}
\end{figure}

\begin{prop} \label{prop:6}
  For any Werner state $\rho$ the Entanglement of Formation is given by (cf. Figure \ref{fig:eof-werner})
  \begin{equation} \label{eq:77}
    \EOF(\rho) = 
    \begin{cases}
      H\left[ \frac{1}{2} \left( 1 - \sqrt{1 - \tr(F\rho)^2}\right)\right] & \tr(F\rho) < 0 \\
      0 & \tr(F\rho) \geq 0.
    \end{cases}
  \end{equation}
\end{prop}

\subsection{Isotropic states}
\label{sec:isotropic-states-1}

\begin{figure}[t]
  \begin{center}
    \begin{pspicture}(15,9)
    \rput(7.5,5){\includegraphics[scale=0.8]{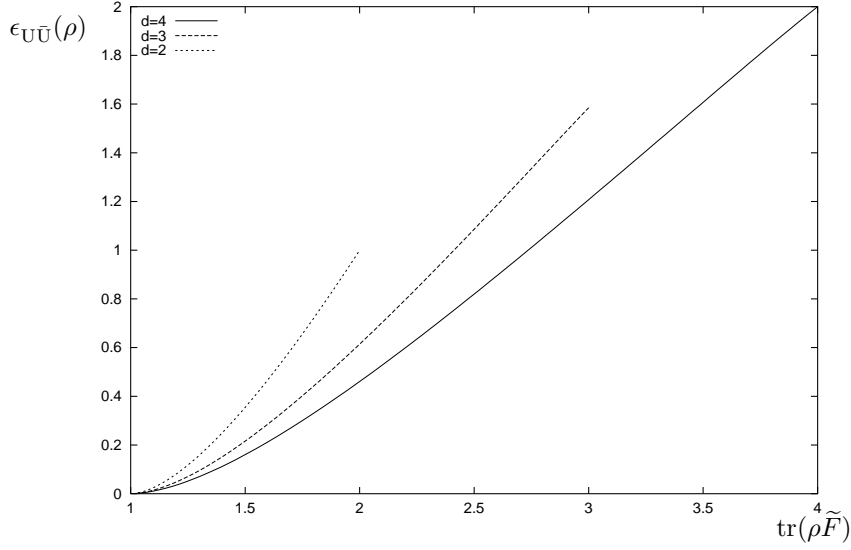}}
    \rput[r](14,0.5){$\tr(\rho\tilde{F})$}
    \rput[tl](0,9){$\epsilon_\tUUbar(\rho)$}
  \end{pspicture}
    \caption{$\epsilon$-function for isotopic states plotted as a function of the flip expectation. For $d > 2$
      it is not convex near the right endpoint.} 
    \label{fig:eof-iso}
  \end{center}
\end{figure}

Let us consider now isotropic, i.e. $U \otimes \bar{U}$ invariant states. They are determined by the
expectation value $\tr(\rho \tilde{F})$ with $\tilde{F}$ from Equation (\ref{eq:73}). Hence we have to look
first for pure states $\Phi$ with $\langle\Phi, \tilde{F} \Phi\rangle = \tr(\rho \tilde{F})$ (since this determines, as for Werner
states above, those $\Phi$ with $P_\tUUbar\bigl(\kb{\Phi}\bigr) = \rho$). To this end assume that $\Phi$ has the
Schmidt decomposition $\Phi = \sum_j \lambda_j f_j \otimes f_j' = U_1 \otimes U_2 \sum_j \lambda_j e_j \otimes e_j$ with appropriate unitary
matrices $U_1, U_2$ and the canonical basis $e_j$, $j=1,\ldots,d$. Exploiting the $U \otimes \bar{U}$ invariance of
$\rho$ we get   
\begin{align}
  \tr(\rho \tilde{F}) &=  \left\langle (\Bbb{1} \otimes V) \sum_j \lambda_j e_j \otimes e_j, \tilde{F} (\Bbb{1} \otimes V) \sum_k \lambda_k e_k \otimes e_k
  \right\rangle\\ 
  &= \sum_{j,k,l,m} \lambda_j \lambda_k \langle e_j \otimes V e_j, e_l \otimes e_l \rangle \langle e_m \otimes e_m, e_k \otimes V e_k\rangle\\
  &= \left| \sum_j \lambda_j \langle e_j, V e_j\rangle \right|^2 \label{eq:74}
\end{align}
with $V = U_1^T U_2$ and after inserting the definition of $\tilde{F}$. Following our general scheme, we
have to minimize $\Evn\left(\kb{\Phi}\right)$ under the constraint given in Equation (\ref{eq:74}). This is
explicitly done in \Cite{TV}. We will only state the result here, which leads to the function
\begin{equation} \label{eq:78}
  \epsilon_\tUUbar(\rho) =
  \begin{cases}
    H(\gamma) + (1 - \gamma) \log_2 (d-1) & \tr(\rho \tilde{F}) \geq \frac{1}{d} \\
    0 & \tr(\rho \tilde{F}) < 0
  \end{cases}
\end{equation}
with
\begin{equation}
  \gamma = \frac{1}{d^2} \left( \sqrt{\tr(\rho \tilde{F})} + \sqrt{[d-1][d- \tr(\rho \tilde{F})]} \right)^2.
\end{equation}
For $d \geq 3$ this function is not convex (cf. Figure \ref{fig:eof-iso}), hence we get 

\begin{prop} \label{prop:7}
  For any isotropic state the Entanglement of Formation is given as the convex hull
  \begin{equation}
    \EOF(\rho) = \inf \{ \mbox{$\sum_j$} \lambda_j \epsilon_\tUUbar(\sigma_j) \, | \, \rho = \mbox{$\sum_j$} \lambda_j \sigma_j,\ P_\tUUbar \sigma = \sigma \} 
  \end{equation}
  of the function $\epsilon_\tUUbar$ in Equation (\ref{eq:78}).
\end{prop}

\subsection{OO-invariant states}
\label{sec:oo-invariant-states-1}

The results derived for isotropic and Werner states can be extended now to a large part of the set of
OO-invariant states without solving new minimization problems. This is possible, because the definition
of
 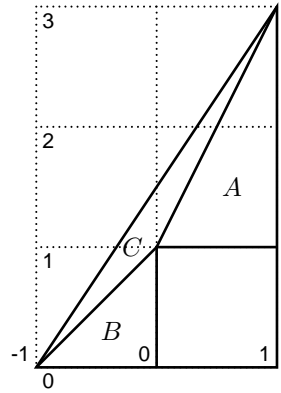
\begin{wrapfigure}{l}{4cm}
  \begin{center}
    \psset{unit=0.4cm}
    \begin{pspicture}(8,12)
      \rput[b](4,0){
        \psgrid[unit=1.6cm,subgriddiv=1,gridlabels=8pt,griddots=20](-1,0)(1,3)(-1,0)
        }
      \psset{linewidth=1pt}
      \pspolygon(0,0)(8,0)(8,12)
      \psline(4,0)(4,4)(8,4)
      \psline(0,0)(4,4)
      \psline(4,4)(8,12)
      \rput(2.5,1.2){$B$}
      \rput(6.5,6){$A$}
      \rput(3.2,4){$C$}
    \end{pspicture}
  \end{center}
  \caption{State space of OO-invariant states.}
  \label{fig:oo-triangle}
\end{wrapfigure}
$\EOF$ in Equation (\ref{eq:75}) allows under some conditions an easy extension to a suitable set of
non-symmetric states. If more precisely a nontrivial, minimizing decomposition $\rho = \sum_j p_j \kb{\psi_j}$ of
$\rho$ is known, all states $\rho'$ which are a convex linear combination of the same $\kb{\psi_j}$ but arbitrary
$p_j'$ have the same $\EOF$ as $\rho$ (see \Cite{VW1} for proof of the statement). For the general scheme we
have presented in Subsection \ref{sec:entangl-form} this implies the following: If we know the pure states
$\sigma \in M_\rho$ which solve the minimization problem for $\epsilon(\rho)$ in Equation (\ref{eq:76}) we get a minimizing
decomposition of $\rho$ in terms of $U \in G$ translated copies of $\sigma$. This follows from the fact that $\rho$ is by
definition of $M_\rho$ the twirl of $\sigma$. Hence any convex linear combination of pure states $U \sigma U^*$ with $U
\in G$ has the same $\EOF$ as $\rho$.

A detailed analysis of the corresponding optimization problems in the case of Werner and isotropic states
(which we have omitted here; see \Cite{VW1,TV} instead) leads therefore to the following results about
OO-invariant states: The space of OO-invariant states decomposes into four regions: The separable square
and three triangles $A,B,C$; cf. Figure \ref{fig:oo-triangle}. For all states $\rho$ in triangle $A$ we can
calculate $\EOF(\rho)$ as for Werner states in Proposition \ref{prop:6} and in triangle $B$ we have to apply
the result for isotropic states from Proposition \ref{prop:7}. This implies in particular that $\EOF$
depends in $A$ only on $\tr(\rho F)$ and in $B$ only on $\tr(\rho \tilde{F})$ and the dimension.

\subsection{Relative Entropy of Entanglement}
\label{sec:relat-entr-entagl}

To calculate $\ER(\rho)$ for a symmetric state $\rho$ is even easier as the treatment of $\EOF(\rho)$, because
we can restrict the minimization in the definition of $\ER(\rho)$ in Equation (\ref{eq:79}) to $G$-invariant
separable states, provided $G$ is a group of local unitaries. To see this assume that $\sigma \in \scr{D}$
minimizes $S(\rho|\sigma)$ for a $G$-invariant state $\rho$. Then we get $S(\rho|U\sigma U^*) = S(\rho|\sigma)$ for all $U \in G$ since
the relative entropy $S$ is invariant under unitary transformations of both arguments and due to its
convexity we even get $S(\rho|P_G\sigma) \leq S(\rho|\sigma)$. Hence $P_G\sigma$ minimizes $S(\rho|\,\cdot\,)$ as well, and since $P_G\sigma
\in \scr{D}$ holds for a group $G$ of local unitaries, we get $\ER(\sigma,\rho) = S(\rho|P_G\sigma)$ as stated. 

\begin{figure}[H]
  \begin{center}
    \begin{pspicture}(15,10)
    \rput(7.5,5){\includegraphics[scale=0.8]{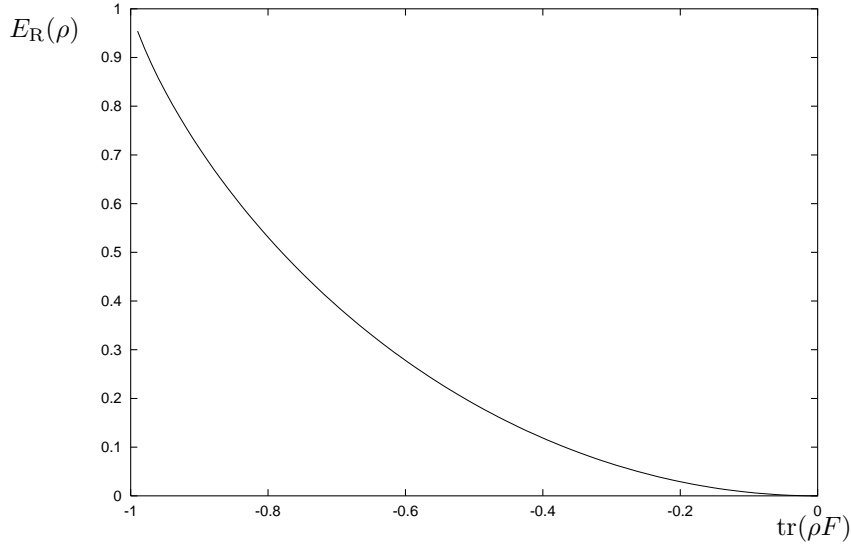}}
    \rput[r](14,0.5){$\tr(\rho F)$}
    \rput[tl](0,9){$\ER(\rho)$}
  \end{pspicture}
    \caption{Relative Entropy of Entanglement for Werner states, plotted as a function of the flip
      expectation.} 
    \label{fig:er-werner}
  \end{center}
\end{figure}

The sets of Werner and isotropic states are just intervals and the corresponding separable states form
subintervals over which we have to perform the optimization. Due to the convexity  of the relative
entropy in both arguments, however, it is clear that the minimum is attained exactly at the boundary
between entangled and separable states. For Werner states this is the state $\sigma_0$ with $\tr(F\sigma_0) = 0$,
i.e. it gives equal weight to both minimal projections. To get $\ER(\rho)$ for a Werner state $\rho$ we have to
calculate therefore only the relative entropy with respect to this state. Since all Werner states can be 
simultaneously diagonalized this is easily done and we get:
\begin{equation}
  \ER(\rho) = 1 - H\left(\frac{1+\tr(F\rho)}{2}\right)
\end{equation}
Similarly, the boundary point $\sigma_1$ for isotropic states is given by $\tr(\tilde{F}\sigma_1) = 1$ which leads
to 
\begin{equation}
  \ER(\rho) = \log_2d - \left(1 - \frac{\tr(\tilde{F}\rho)}{d}\right)\log_2(d-1) -
  S\left(\frac{\tr(\tilde{F}\rho)}{d}, \frac{1 - \tr(\tilde{F}\rho)}{d}\right)
\end{equation}
for each entangled isotropic state $\rho$, and $0$ if $\rho$ is separable. ($S(p_1,p_2)$ denotes here the
entropy of the probability vector $(p_1,p_2)$.)

\begin{figure}[H]
  \begin{center}
    \begin{pspicture}(15,10)
    \rput(7.5,5){\includegraphics[scale=0.8]{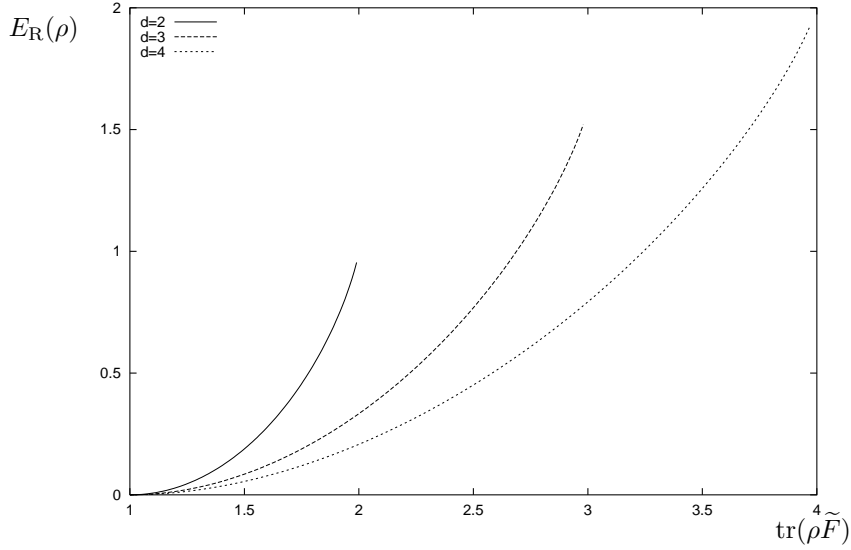}}
    \rput[r](14,0.5){$\tr(\rho\tilde{F})$}
    \rput[tl](0,9){$\ER(\rho)$}
  \end{pspicture}
    \caption{Relative Entropy of Entanglement for isotropic states and $d=2,3,4$, plotted as a function
      of $\tr(\rho\tilde{F})$.} 
\label{fig:er-sio}
  \end{center}
\end{figure}

Let us consider now OO-invariant states. As for EOF we divide the state space into the separable square
and the three triangles $A, B, C$; cf. Figure \ref{fig:oo-triangle}. The state at the coordinates $(1,d)$
is a maximally entangled state and all separable states on the line connecting $(0,1)$ with $(1,1)$
minimize the relative entropy for this state. Hence consider a particular state $\sigma$ on this line. The
convexity property of the relative entropy shows immediately that $\sigma$ is a minimizer for all states on
the line connecting $\sigma$ with the state at $(1,d)$. In this way it is easy to calculate $\ER(\rho)$ for all
$\rho$ in $A$. In a similar way we can treat the triangle $B$: We just have to draw a line from $\rho$ to the
state at $(-1,0)$ and find the minimizer for $\rho$ at the intersection with the separable border between
$(0,0)$ and $(0,1)$. For all states in the triangle $C$ the relative entropy is minimized by the
separable state at $(0,1)$.

An application of the scheme just reviewed is a proof that $\ER$ is not additive, i.e. it does not
satisfy Axiom \ref{ax:6}. To see this consider the state $\rho = \tr(P_-)^{-1} P_-$ where $P_-$ denotes
the projector on the antisymmetric subspace. It is a Werner state with flip expectation $-1$ (i.e. it
corresponds to the point $(-1,0)$ in Figure \ref{fig:oo-triangle}). According to our discussion above
$S(\rho|\,\cdot\,)$ is minimized in this case by the separable state $\sigma_0$ and we get $\ER(\rho) = 1$ independently of
the dimension $d$. The tensor product $\rho^{\otimes 2}$ can be regarded as a state in $\scr{S}(\scr{H}^{\otimes 2} \otimes
\scr{H}^{\otimes 2})$ with $U \otimes U \otimes V \otimes V$ symmetry, where $U, V$ are unitaries on $\scr{H}$.
 Note that the
corresponding state space of $UUVV$ invariant
states can be parameterized by the expectation of the three  operators $F \otimes \Bbb{1}$, $\Bbb{1} \otimes F$ and $F
\otimes F$ (cf. \Cite{VW1}) and we can apply the machinery just described to get the minimizer $\tilde{\sigma}$ of
$S(\rho|\,\cdot\,)$. If $d > 2$ holds it turns out that 
\begin{equation}
  \tilde{\sigma} = \frac{d+1}{2d \tr(P_+)^2} P_+ \otimes P_+ + \frac{d-1}{2d \tr(P_-)^2} P_- \otimes P_-
\end{equation}
holds (where $P_\pm$ denote the projections onto the symmetric and antisymmetric subspaces of $\scr{H} \otimes
\scr{H}$) and not $\tilde{\sigma} = \sigma_0 \otimes \sigma_0$ as one would expect. As a consequence we get the inequality
\begin{equation}
  \ER(\rho^{\otimes 2}) = 2 - \log_2\left(\frac{2d-1}{d}\right) < 2 = S(\rho^{\otimes2}|\sigma_0^{\otimes2}) = 2\ER(\rho).
\end{equation}
$d=2$ is a special case, where $\sigma_0^{\otimes 2}$ and $\tilde{\sigma}$ (and all their convex linear combination) give
the same value $2$. Hence for $d > 2$ the Relative Entropy of Entanglement is, as stated, not additive.



\chapter{Channel capacity}
\label{cha:quant-theory-ii}

In Section \ref{sec:quant-error-corr} we have seen that it is possible to send (quantum) information
undisturbed through a noisy quantum channel, if we encode one qubit into a (possibly long and highly
entangled) string of qubits. This process is wasteful, since we have to use many instances of the channel
to send just one qubit of quantum information. It is therefore natural to ask, which resources we need at
least if we are using the best possible error correction scheme. More precisely the question is: With
which maximal \emph{rate}, i.e. information sent per channel usage, we can transmit quantum information
undisturbed through a noisy channel? This question naturally leads to the concept of channel capacities
which we will review in this chapter.

\section{The general case}
\label{sec:general-definition}

We are mainly interested in classical and quantum capacities. The basic ideas behind both situations are 
however quite similar. In this section we will consider therefore a general definition of capacity which
applies to arbitrary channels and both kinds of information. (See also \Cite{Werner01} as a general
reference for this section.)

\subsection{The definition}
\label{sec:definition}

Hence consider two observable algebras $\scr{A}_1$, $\scr{A}_2$ and an arbitrary channel $T: \scr{A}_1 \to
\scr{A}_2$. To send systems described by a third observable algebra $\scr{B}$ undisturbed through
$T$ we need an \emph{encoding channel} $E: \scr{A}_2 \to \scr{B}$ and a \emph{decoding channel} $D: \scr{B}
\to \scr{A}_1$ such that $ETD$ equals the ideal channel $\scr{B} \to \scr{B}$, i.e. the identity on
$\scr{B}$. Note that the algebra $\scr{B}$ describing the systems to send, and the input respectively
output algebra of $T$ need not to be of the same type, e.g. $\scr{B}$ can be classical while $\scr{A}_1,
\scr{A}_2$ are quantum (or vice versa). 

In general (i.e. for arbitrary $T$ and $\scr{B}$) it is of course impossible to find such a pair $E$ and 
$D$. In this case we are interested at least in encodings and decodings which make the error produced
during the transmission as small as possible. To make this statement precise we need a measure for this 
error and there are in fact many good choices for such a quantity (all of them leading to equivalent
results, cf. Subsection \ref{sec:altern-defin}). We will use in the following the ``cb-norm difference''
$\| ETD - \Id\|_\cb$, where $\Id$ is the identity (i.e. ideal) channel on $\scr{B}$ and $\|\,\cdot\,\|_\cb$
denotes the norm of \emph{complete boundedness} (``cb-norm'' for short) 
\begin{equation}
  \|T\|_\cb = \sup_{n \in \Bbb{N}} \|T \otimes \Id_n\|,\quad \Id_n : \scr{B}(\Bbb{C}^n) \to \Bbb{B}(\Bbb{C}^n) 
\end{equation}
The cb-norm improves the sometimes annoying property of the usual operator norm that quantities like $\|T
\otimes \Id_{\scr{B}(\Bbb{C}^d)}\|$ may increase with the dimension $d$. On infinite dimensional observable algebras
$\|T\|_\cb$ can be infinite although each term in the supremum is finite. A particular example for a map
with such a behavior is the transposition on an infinite dimensional Hilbert space. A map with finite
cb-norm is therefore called completely bounded. In a finite dimensional setup each linear map is
completely bounded. For the transposition $\Theta$ on $\Bbb{C}^d$ we have in particular $\|\Theta\|_\cb = d$. The
cb-norm has some nice features which we will use frequently; this includes its multiplicativity $\|T_1 \otimes
T_2\|_\cb = \|T_1\|_\cb \|T_2\|_\cb$ and the fact that $\|T\|_\cb = 1$ holds for each (unital) channel. Another
useful relation is $\|T\|_\cb = \|T \otimes \Id_{\scr{B}(\scr{H})}\|$, which holds if $T$ is a map
$\scr{B}(\scr{H}) \to \scr{B}(\scr{H})$. For more properties of the cb-norm let us refer to \Cite{Paulsen}.

Now we can define the quantity 
\begin{equation} \label{eq:80}
  \Delta(T,\scr{B}) = \inf_{E,D}\| ETD - \Id_\scr{B}\|_\cb,
\end{equation}
where the infimum is taken over all channels $E: \scr{A}_2 \to \scr{B}$ and $D: \scr{B} \to \scr{A}_1$ and
$\Id_\scr{B}$ is again the ideal $\scr{B}$-channel. $\Delta$ describes, as indicated above, the smallest
possible error we have to take into account if we try to transmit \emph{one} $\scr{B}$ system through
\emph{one} copy of the channel $T$ using any encoding $E$ and decoding $D$. In Section
\ref{sec:quant-error-corr}, however, we have seen that we can reduce the error if we take $M$ copies of
the channel instead of just one. More generally we are interested in the transmission of ``codewords of
length'' $N$, i.e. $\scr{B}^{\otimes N}$ systems using $M$ copies of the channel $T$. Encodings and decodings are
in this case channels of the form $E: \scr{A}_2^{\otimes M}\to\scr{B}^{\otimes N}$ respectively $D: \scr{B}^{\otimes N} \to
\scr{A}_1^{\otimes M}$. If we increase the number $M$ of channels the error $\Delta(T^{\otimes M},\scr{B}^{\otimes N(M)})$
decreases provided the rate with which $N$ grows as a function of $M$ is not too large. A more precise
formulation of this idea leads to the following definition. 

\begin{defi} \label{def:2}
  Let $T$ be a channel and $\scr{B}$ an observable algebra. A number $c \geq 0$ is called \emph{achievable
    rate} for $T$ with respect to $\scr{B}$, if for any pair of sequences $M_j, N_j$, $j \in \Bbb{N}$ with
  $M_j \to \infty$ and $\limsup_{j \to \infty} N_j/M_j < c$ we have 
  \begin{equation}
    \lim_{j \to \infty} \Delta(T^{\otimes M_j}, \scr{B}^{\otimes N_j}) = 0.
  \end{equation}
  The supremum of all achievable rates is called the \emph{capacity} of $T$ with respect to $\scr{B}$
  and denoted by $C(T,\scr{B})$. 
\end{defi}

Note that by definition $c=0$ is an achievable rate hence $C(T,\scr{B}) \geq 0$. If on the other hand each
$c> 0$ is achievable we write $C(T,\scr{B}) = \infty$. At a first look it seems cumbersome to check all pairs
of sequences with given upper ratio when testing $c$. Due to some monotonicity properties of $\Delta$,
however, it can be shown that it is sufficient to check only one sequence provided the $M_j$ satisfy the
additional condition $M_j/(M_{j+1}) \to 1$.

\subsection{Simple calculations}
\label{sec:simple-calculations}

We see that there are in fact many different capacities of a given channel depending on the type of
information we want to transmit. However, there are only two different cases we are interested in:
$\scr{B}$ can be either classical or quantum. We will discuss both special cases in greater detail in the
next two sections. Before we do this, however, we will have a short look on some simple calculations which
can be done in the general case. To this end it is convenient to introduce the notations
\begin{equation}
  \scr{M}_d = \scr{B}(\Bbb{C}^d)\quad \text{and}\quad \scr{C}_d = \scr{C}(\{1,\ldots,d\})
\end{equation}
as shorthand notations for $\scr{B}(\Bbb{C}^d)$ and $\scr{C}(\{1,\ldots,d\})$ since some notations become
otherwise a little bit clumsy. First of all let us have a look on capacities of ideal channels. If
$\Id_{\scr{M}_f}$ and $\Id_{\scr{C}_f}$ denote the identity channels on the quantum algebra $\scr{M}_f$
respectively the classical algebra $\scr{C}_f$ we get
\begin{equation} \label{eq:82}
  C(\Id_{\scr{C}_f}, \scr{M}_d) = 0,\ C(\Id_{\scr{C}_f}, \scr{C}_d) =
  C(\Id_{\scr{M}_f},\scr{M}_d) = C(\Id_{M_f}, \scr{C}_d) = \frac{\log_2 f}{\log_2 d}.
\end{equation}
The first equation is the channel capacity version of the no-teleportation theorem: It is impossible to
transfer quantum information through a classical channel. The other equations follow simply by counting
dimensions.

For the next relation it is convenient to associate to a pair of channels $T$, $S$ the quantity
$C(T,S)$ which arises if we replace in Definition \ref{def:2} and Equation (\ref{eq:80}) the ideal
channel $\Id_\scr{B}$ by an arbitrary channel $S$. Hence $C(T,S)$ is a slight generalization of the
channel capacity which describes with which asymptotic rate the channel $S$ can be approximated by $T$
(and appropriate encodings and decodings). These generalized capacities satisfy the \emph{two step
  coding inequality}, i.e. for the three channels $T_1, T_2, T_3$ we have
\begin{equation} \label{eq:81}
  C(T_3,T_1) \geq C(T_2,T_1)C(T_3,T_2).
\end{equation}
 To prove it consider the relations
\begin{align}
  \| T_1^{\otimes N} &- E_1 E_2T_3^{\otimes K} D_2 D_1 \|_\cb \notag \\
  &= \| T_1^{\otimes N} - E_1 T_2^{\otimes M} D_1 + E_1 T_2^{\otimes M} D_1 -  E_1 E_2T_3^{\otimes K} D_2 D_1 \|_\cb \\
  & \leq \|  T_1^{\otimes N} - E_1 T_2^{\otimes M} D_1 \|_\cb + \|E_1\|_\cb \| T_2^{\otimes M} - E_2T_3^{\otimes K} D_2 \|_\cb \|D_1\|_\cb
  \\
  & \leq \|  T_1^{\otimes N} - E_1 T_2^{\otimes M} D_1 \|_\cb + \| T_2^{\otimes M} - E_2T_3^{\otimes K} D_2 \|_\cb
\end{align}
where we have used for the last inequality the fact that the cb-norm of a channel is one. If $c_1$ is an
achievable rate of $T_1$ with respect to $T_2$ such that $\limsup_{j\to\infty} M_j/N_j < c_1$ and $c_2$ is an
achievable rate of $T_2$ with respect to $T_3$ such that $\limsup_{j\to\infty} N_j/K_j < c_2$ we see that
\begin{equation}
  \limsup_{j\to\infty} \frac{M_j}{K_j} = \limsup_{j \to \infty} \frac{M_j}{N_j}\frac{N_j}{K_j} \leq \limsup_{j\to\infty}
  \frac{M_j}{N_j} \limsup_{k \to \infty} \frac{N_k}{K_k}.
\end{equation}
If we choose the sequences $M_j, N_j$ and $K_j$ clever enough (cf. the remark following Definition
\ref{def:2}) this implies that $c_1c_2$ is an achievable rate for $T_1$ with respect to $T_3$ and this
proves Equation (\ref{eq:81}).

As a first application of (\ref{eq:81}), we can relate all capacities $C(T,\scr{M}_d)$ (and
$C(T,\scr{C}_d)$) for different $d$ to one another. If we choose $T_3 = T$, $T_1 = \Id_{\scr{M}_d}$ 
and $T_2 = \Id_{\scr{M}_f}$ we get with (\ref{eq:82}) $C(T,\scr{M}_d) \leq \frac{\log_2 f}{\log_2
  d}C(T,\scr{M}_f)$, and exchanging $d$ with $f$ shows that even equality holds. A similar relation
can be shown for $C(T,\scr{C}_d)$. Hence the dimension of the observable algebra $\scr{B}$
describing the type of information to be transmitted, enters only via a multiplicative constant, i.e. it
is only a choice of units and we define the \emph{classical capacity} $C_c(T)$ and the \emph{quantum
  capacity} $C_q(T)$ of a channel $T$ as 
\begin{equation}
  C_c(T) = C(T,\scr{C}_2),\quad C_q(T) = C(T,\scr{M}_2).
\end{equation}

A second application of Equation (\ref{eq:81}) is a relation between the classical and the quantum
capacity of a channel. Setting $T_3 = T$, $T_1 = \Id_{\scr{C}_2}$ and $T_2 = \Id_{\scr{M}_2}$ we get again with
(\ref{eq:82})
\begin{equation}
  C_q(T) \leq C_c(T).
\end{equation}
Note that it is now not possible to interchange the roles of $\scr{C}_2$ and $\scr{M}_2$. Hence equality
does not hold here.

Another useful relation concerns concatenated channels: We transmit information of type $\scr{B}$ first
through a channel $T_1$ and then through a second channel $T_2$. It is reasonable to assume that the
capacity of the composition $T_2T_1$ can not be bigger than capacity of the channel with the smallest
bandwidth. This conjecture is indeed true and known as the ``\emph{Bottleneck inequality}'':
\begin{equation} \label{eq:83}
  C(T_2T_1,\scr{B}) \leq \min \{ C(T_1,\scr{B}), C(T_2,\scr{B})\}.
\end{equation}
To see this  consider an encoding and a decoding channel $E$ respectively $D$ for $(T_2T_1)^{\otimes M}$,
i.e. in the definition of $C(T_2T_1,\scr{B})$ we look at
\begin{equation}
  \| \Id_\scr{B}^{\otimes N} - E (T_2T_1)^{\otimes M} D \|_\cb = \| \Id_\scr{B}^{\otimes N} - (ET_2^{\otimes M})T_1^{\otimes M}D\|_\cb.
\end{equation}
This implies that $ET_2^{\otimes M}$ and $D$ are an encoding and a decoding channel for $T_1$. Something
similar holds for $D$ and $T_1^{\otimes M}D$ with respect to $T_2$. Hence each achievable rate for $T_2T_1$ is
also an achievable rate for $T_2$ and $T_1$, and this proves Equation (\ref{eq:83}). 

Finally we want to consider two channels $T_1$, $T_2$ in parallel, i.e. we consider the tensor product
$T_1 \otimes T_2$. If $E_j$, $D_j$, $j=1,2$ are encoding, respectively decoding channels for $T_1^{\otimes M}$ and
$T_2^{\otimes M}$ such that $\| \Id_\scr{B}^{\otimes N_j} - E_j T_j^{\otimes M} D_j \|_\cb \leq \epsilon$ holds, we get
\begin{align}
  \|\Id -& \Id \otimes (E_2 T^{\otimes M} D_2) + \Id \otimes ( E_2 T^{\otimes M}D_2) - E_1 \otimes E_2 (T_1 \otimes T_2)^{\otimes M} D_1 \otimes D_2\|_\cb \\
  &\leq \|\Id \otimes (\Id - E_2 T^{\otimes M} D_2 \|_\cb + \|(\Id - E_1 T_1^{\otimes M} D_1) \otimes E_2 T^{\otimes M}D_2\|_\cb \\
  &\leq \|\Id - E_2 T^{\otimes M} D_2\|_\cb + \|\Id - E_1 T_1^{\otimes M}D_1\|_\cb  \leq 2 \epsilon
\end{align}
Hence $c_1 + c_2$ is achievable for $T_1 \otimes T_2$ if $c_j$ is achievable for $T_j$. This implies the
inequality
\begin{equation}
  C(T_1 \otimes T_2,\scr{B}) \geq C(T_1,\scr{B}) + C(T_2,\scr{B}).
\end{equation}
When all channels are ideal, or when all systems involved are classical even equality holds, i.e. channel
capacities are \emph{additive} in this case. However, if quantum channels are considered, it is one of
the big open problems of the field, to decide under which conditions additivity holds.  

\section{The classical capacity}
\label{sec:classical-capacity}

In this section we will discuss the classical capacity $C_c(T)$ of a channel $T$. There are in fact three
different cases to consider: $T$ can be either classical or quantum and in the quantum case we can use
either ordinary encodings and decodings or a dense coding scheme (cf. Subsection \ref{sec:dense-coding}).

\subsection{Classical channels}
\label{sec:classical-channels-1}

Let us consider first a classical to classical channel $T: \scr{C}(Y) \to \scr{C}(X)$. This is basically
the situation of classical information theory and we will only have a short look here -- mainly to show
how this (well known) situation fits into the general scheme described in the last section\footnote{Please
  note that this implies in particular that we do not give a complete review of the foundations of
  classical information theory here; cf \Cite{Khinchin,Feinstein,CoTh} instead.}. 

First of all we have to calculate the error quantity $\Delta(T,\scr{C}_2)$ defined in Equation (\ref{eq:80}). As
stated in Subsection \ref{sec:classical-channels} $T$ is completely determined by its transition
probabilities $T_{xy}$, $(x,y) \in X \times Y$ describing the probability to receive $x \in X$ when $y \in Y$ was
sent. Since the cb-norm for a classical algebra coincides with the ordinary norm we get (we have set $X =
Y$ for this calculation): 
\begin{align}
  \|\Id - T\|_\cb &= \|\Id -T\| = \sup_{x,f} \left| \sum_y \left(\delta_{xy} - T_{xy}\right) f_y \right| \\
  &= 2 \sup_x \left( 1 - T_{xx} \right) \label{eq:84}
\end{align}
where the supremum in the first equation is taken over all $f \in \scr{C}(X)$ with $\|f\| = \sup_y |f_y| \leq 1$.
We see that the quantity in Equation (\ref{eq:84}) is exactly twice the \emph{maximal error probability},
i.e. the maximal probability of sending $x$ and getting anything different. Inserting this quantity for
$\Delta$ in Definition \ref{def:2} applied to a classical channel $T$ and the ``bit-algebra'' $\scr{B} =
\scr{C}_2$, we get exactly Shannons classical definition of the capacity of a discrete memoryless
channel \Cite{Shannon48}.

Hence we can apply Shannons \emph{noisy channel coding theorem} to calculate $C_c(T)$ for a classical
channel. To state it we have to introduce first some terminology. Consider therefore a state $p \in
\scr{C}^*(X)$ of the classical input algebra $\scr{C}(X)$ and its image $q = T^*(p) \in \scr{C}^*(Y)$ under
the channel. $p$ and $q$ are probability distributions on $X$ respectively $Y$ and $p_x$ can be
interpreted as the probability that the ``letter'' $x \in X$ was send. Similarly $q_y = \sum_x T_{xy} p_x$ is
the probability that $y \in Y$ was received and $P_{xy} = T_{xy} p_x$ is the probability that $x \in X$ was
sent and $y \in Y$ was received. The family of all $P_{xy}$ can be interpreted as a probability
distribution $P$ on $X \times Y$ and the $T_{xy}$ can be regarded as conditional probability of $P$
under the condition $x$. Now we can introduce the \emph{mutual information}
\begin{equation} \label{eq:85}
  I(p,T) = S(p) + S(q) - S(P) = \sum_{(x,y) \in X \times Y} P_{xy} \log_2\left( \frac{ P_{xy}}{p_x q_y}
    \right),
\end{equation}
where $S(p)$, $S(q)$ and $S(P)$ denote the entropies of $p, q$ and $P$. The mutual information describes,
roughly speaking, the information that $p$ and $q$ contain about each other. E.g. if $p$ and $q$ are
completely uncorrelated (i.e. $P_{xy} = p_x q_y$) we get $I(p,T) = 0$. If $T$ is on the other hand an
ideal bit-channel and $p$  equally distributed we have $I(p,T) = 1$. Now we can state Shannons Theorem
which expresses the classical capacity of $T$ in terms of mutual informations 
\Cite{Shannon48}: 

\begin{thm}[Shannon] \label{thm:8}
  The classical capacity of $C_c(T)$ of a classical communication channel $T: \scr{C}(Y) \to \scr{C}(X)$ is
  given by
  \begin{equation}
    C_c(T) = \sup_p I(p,T),
  \end{equation}
  where the supremum is taken over all states $p \in \scr{C}^*(X)$.
\end{thm}

\subsection{Quantum channels}
\label{sec:quantum-channels}

If we transmit classical data through a quantum channel $T: \scr{B}(\scr{H}) \to \scr{B}(\scr{H})$ the
encoding $E: \scr{B}(\scr{H}) \to \scr{C}_2$ is a parameter dependent preparation and the decoding $D: \scr{C}_2 \to
\scr{B}(\scr{H})$ is an observable. Hence the composition $ETD$ is a channel $\scr{C}_2 \to \scr{C}_2$,
i.e. a purely classical channel and we can calculate its capacity in terms of Shannons Theorem (Theorem
\ref{thm:8}). This observation leads to the definition of the ``\emph{one-shot}'' classical capacity of
$T$:
\begin{equation}
  C_{c,1}(T) = \sup_{E,D} C_c(ETD),
\end{equation}
where the supremum is taken over all encodings and decodings of classical bits. The term ``one-shot'' in
this definition arises from the fact that we need apparently only one invocation of the channel
$T$. However many uses of the channel are hidden in the definition of the classical capacity on the right
hand side. Hence $C_{c,1}(T)$ can be defined alternatively in the same way as $C_c(T)$ except that no
entanglement is allowed during encoding and decoding, or more precisely in Definition \ref{def:2} we
consider only encodings $E: \scr{B}(\scr{K})^{\otimes M} \to \scr{C}_2^{\otimes N}$ which prepare separable states and
only decodings $D: \scr{C}_2^{\otimes N} \to \scr{B}(\scr{H})^{\otimes M}$ which lead to separable observables. It is
not yet known, whether entangled codings can help to increase the transmission rate. Therefore we only
know that
\begin{equation}
  C_{c,1}(T) \leq C_c(T) = \sup_{M \in \Bbb{N}} \frac{1}{M} C_{c,1}(T^{\otimes M})
\end{equation}
holds. One reason why $C_{c,1}(T)$ is an interesting quantity relies on the fact that we have, due to the
following theorem by Holevo \Cite{Hol1Shot} a computable expression for it. 

\begin{thm} \label{thm:9}
  The one-shot classical capacity $C_{c,1}(T)$ of a quantum channel $T: \scr{B}(\scr{H}) \to
  \scr{B}(\scr{H})$ is given by
  \begin{equation} \label{eq:92}
    C_{c,1}(T) = \sup_{p_j, \rho_j} \left[ S\left(\sum_j p_j T^*[\rho_j]\right) - \sum_j p_j S\bigl(T^*[\rho_j]\bigr) \right],
  \end{equation}
  where the supremum is taken over all probability distributions $p_j$ and collections of density
  operators $\rho_j$.
\end{thm}

\subsection{Entanglement assisted capacity}
\label{sec:entangl-assist-capac}

Another classical capacity of a quantum channel arises, if we use dense coding schemes instead of simple
encodings and decodings to transmit the data through the channel $T$. In other words we can define the
\emph{entanglement enhanced classical capacity} $C_e(T)$ in the same way as $C_c(T)$ but by replacing the
encoding and decoding channels in Definition \ref{def:2} and Equation (\ref{eq:80}) by dense coding
protocols. Note that this implies that the sender Alice and the receiver Bob share an (arbitrary) amount
of (maximally) entangled states prior to the transmission. 

For this quantity a coding theorem was proven recently by Bennett and others \Cite{BSST} which we want
to state in the following. To this end assume that we are transmitting systems in the state $\rho \in
\scr{B}^*(\scr{H})$ through the channel and that $\rho$ has the purification $\Psi \in \scr{H} \otimes \scr{H}$,
i.e. $\rho = \tr_1\kb{\Psi} = \tr_2\kb{\Psi}$. Then we can define the \emph{entropy exchange}
\begin{equation} \label{eq:93}
  S(\rho,T) = S\Bigl[\bigl(T \otimes \Id\bigr)\bigl(\kb{\Psi}\bigr)\Bigr].
\end{equation}
The density operator $\bigl(T \otimes \Id\bigr)\bigl(\kb{\Psi}\bigr)$ has the output state $T^*(\rho)$  and the
input state $\rho$ as its partial traces. It can be regarded therefore as the quantum analog of the
input/output probability distribution $T_{xy}$ defined in Subsection \ref{sec:classical-channels-1}.
Another way to look at $S(\rho,T)$ is in terms of an ancilla representation of $T$: If $T^*(\rho) = \tr_\scr{K}
\left(U \rho \otimes \rho_\scr{K} U^*\right)$ with a unitary $U: \scr{H} \otimes \scr{K}$ and a pure environment state
$\rho_\scr{K}$ it can be shown \Cite{BaNiSch} that $S(\rho,T) = S\left[T_\scr{K}^*\rho\right]$ where $T_\scr{K}$
is the channel describing the information transfer into the environment, i.e. $T^*_\scr{K}(\rho) =
\tr_\scr{H} \left(U \rho \otimes   \rho_\scr{K} U^*\right)$, in other words $S(\rho,T)$ is the final entropy of the
environment. Now we can define  
\begin{equation}
  I(\rho,T) = S(\rho) + S(T^*\rho) - S(\rho,T)
\end{equation}
which is the quantum analog of the mutual information given in Equation (\ref{eq:85}). It has a number of
nice properties, in particular positivity, concavity with respect to the input state and additivity
\Cite{AdCe97} and its maximum with respect to $\rho$ coincides actually with $C_e(T)$ \Cite{BSST}.

\begin{thm} \label{thm:10}
  The entanglement assisted capacity $C_e(T)$ of a quantum channel $T: \scr{B}(\scr{H}) \to
  \scr{B}(\scr{H})$ is given by
  \begin{equation} \label{eq:90}
    C_e(T) = \sup_\rho I(\rho,T),
  \end{equation}
  where the supremum is taken over all input states $\rho \in \scr{B}^*(\scr{H})$.
\end{thm}

Due to the nice additivity properties of the quantum mutual information $I(\rho,T)$ the capacity $C_e(T)$ is
known to be additive as well. This implies that it coincides with the corresponding ``one-shot''
capacity, and this is an essential simplification compared to the classical capacity $C_c(T)$. 

\subsection{Examples} 
\label{sec:calc-gauss-chann}

\begin{figure}[h]
  \begin{center}
    \begin{pspicture}(15,9)
    \rput(7.5,5){\includegraphics[scale=0.8]{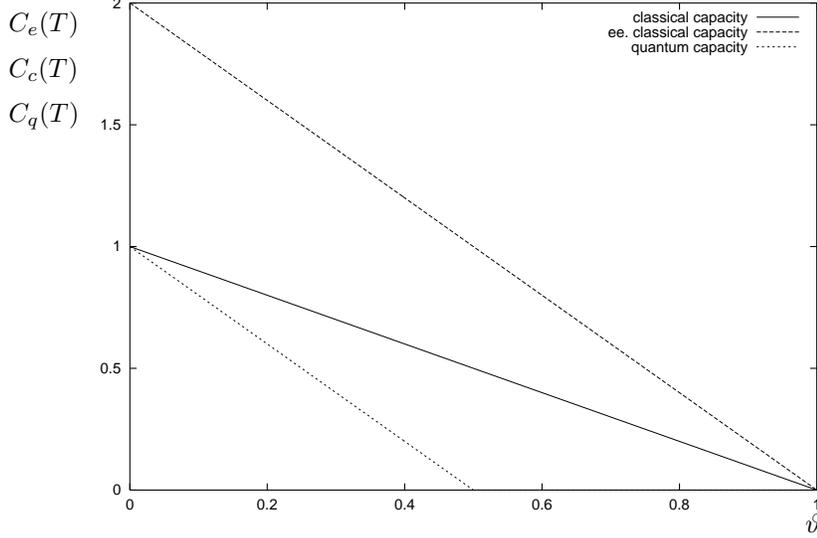}}
    \rput[r](13.5,0.5){$\vartheta$}
    \rput[tl](0,9){$C_{e}(T)$}
    \rput[tl](0,8.25){$C_{c}(T)$}
    \rput[tl](0,7.5){$C_{q}(T)$}
  \end{pspicture}
    \caption{Capacities of the quantum erasure channel plotted as a function of the error probability.}
    \label{fig:erasure}
  \end{center}
\end{figure}

Although the expressions in Theorem \ref{thm:9} and \ref{thm:10} are much easier then the original
definitions they involve still some optimization problems over possibly large parameter spaces.
Nevertheless there are special cases which allow explicit calculations. As a first example we will
consider the ``quantum erasure channel'' which transmits with probability $1-\vartheta$ the $d$-dimensional input
state intact while it is replaced with probability $\vartheta$ by an ``erasure symbol'', i.e. a $(d+1)^{\rm
  th}$ pure state $\psi_e$ which is orthogonal to all others \Cite{GrBePe}. In the Schr{\"o}dinger picture this
is 
\begin{equation} \label{eq:121}
  \scr{B}^*(\Bbb{C}^d) \ni \rho \mapsto T^*(\rho) = (1-\vartheta) \rho + \vartheta \tr(\rho) \kb{\psi_e} \in \scr{B}^*(\Bbb{C}^{d+1}).
\end{equation}
This example is very unusal, because all capacities discussed up to now (including the quantum capacity
as we will see in Subsection \ref{sec:upper-bounds-achi}) can be calculated explicitly: We get
$C_{c,1}(T) = C_c(T) = (1-\vartheta) \log_2(d)$ for the classical and $C_e(T) = 2C_c(T)$ for the entanglement
enhanced classical capacity \Cite{BeDiVSm97,BSST99}. Hence the gain by entanglement assistance is exactly
a factor two; cf. Figure \ref{fig:erasure}.

\begin{figure}[H]
  \begin{center}
    \begin{pspicture}(15,9)
    \rput(7.5,5){\includegraphics[scale=0.8]{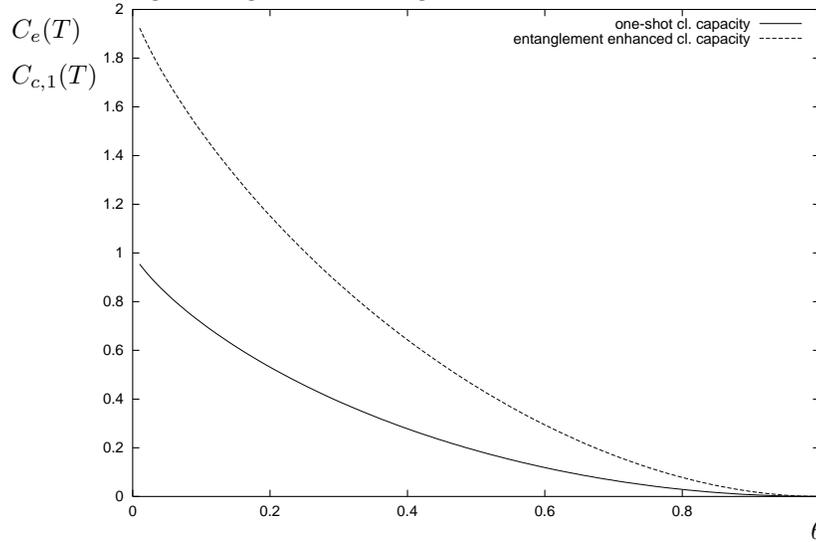}}
    \rput[r](13.5,0.5){$\theta$}
    \rput[tl](0,9){$C_{e}(T)$}
    \rput[tl](0,8.25){$C_{c,1}(T)$}
  \end{pspicture}
    \caption{Entanglement enhanced and one-shot classical capacity of a depolarizing qubit channel.}
    \label{fig:ccdepol}
  \end{center}
\end{figure}

\begin{figure}[h]
  \begin{center}
    \begin{pspicture}(15,9)
    \rput(7.5,5){\includegraphics[scale=0.8]{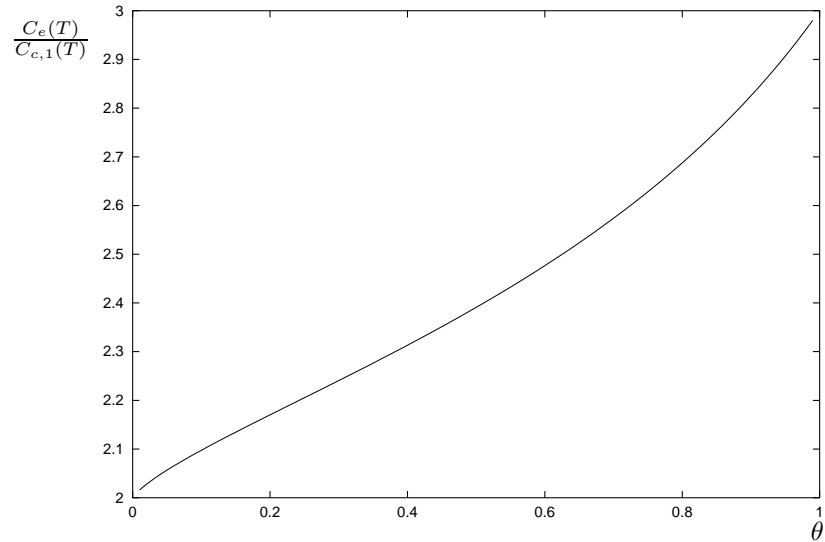}}
    \rput[r](13.5,0.5){$\theta$}
    \rput[tl](0,9){$\frac{C_{e}(T)}{C_{c,1}(T)}$}
  \end{pspicture}
    \caption{Gain of using entanglement assisted versus unassisted classical capacity for a depolarizing
      qubit channel.}
    \label{fig:depolgain}
  \end{center}
\end{figure}

Our next example is the depolarizing channel
\begin{equation} \label{eq:122}
  \scr{B}^*(\Bbb{C}^d) \ni \rho \mapsto T^*(\rho) = (1-\vartheta) \rho + \vartheta \tr(\rho) \frac{\Bbb{1}}{d} \in \scr{B}^*(\Bbb{C}^d),
\end{equation}
already discussed in Section \ref{sec:examples-channels}. It is more interesting and more difficult to
study. It is in particular not known whether $C_c$ and $C_{c,1}$ coincide in this case (i.e. the value of
$C_c$ is not known. Therefore we can compare $C_e(T)$ only with with $C_{c,1}$. Using the unitary
covariance of $T$ (cf. Subsection \ref{sec:chann-under-symm}) we see first that $I(U\rho U^*,T) = I(\rho,T)$
holds for all unitaries $U$ (to calculate $S(U\rho U^*,T)$ note that $U \otimes U \Psi$ is a purification of $U \rho U^*$
if $\Psi$ is a purification of $\rho$). Due to the concavity of $I(\rho,T)$ in the first argument we can average
over all unitaries and see that the maximum in Equation (\ref{eq:90}) is achieved on the maximally mixed
state. Straightforward calculation therefore shows that 
\begin{equation}
  C_e(T) = \log_2(d^2) + \left( 1 - \vartheta \frac{d^2 - 1}{d^2} \right) \log_2 \left( 1 - \vartheta \frac{d^2 - 1}{d^2}
  \right) + \vartheta \frac{d^2 - 1}{d^2} \log_2 \frac{\vartheta}{d^2}
\end{equation}
holds, while we have
\begin{equation}
  C_{c,1}(T) = \log_2(d) + \left( 1 - \vartheta \frac{d - 1}{d} \right) \log_2 \left( 1 - \vartheta \frac{d - 1}{d}
  \right) + \vartheta \frac{d - 1}{d} \log_2 \frac{\vartheta}{d},
\end{equation}
where the maximum in Equation (\ref{eq:92}) is achieved for an ensemble of equiprobable pure states taken
from an orthonormal basis in $\scr{H}$ \Cite{HolEECC}. This is plausible since the first term under the
$\sup$ in Equation (\ref{eq:92}) becomes maximal and the second becomes minimal: $\sum_j p_j T^*\rho_j$ is
maximally mixed in this case and its entropy is therefore maximal. The entropies of the $T^*\rho_j$ are on
the other hand minimal if the $\rho_j$ are pure. In Figure \ref{fig:ccdepol} we have plotted both
capacities as a function of the noise parameter $\vartheta$ and in Figure \ref{fig:depolgain} we have plotted the
quotient $C_e(T)/C_{c,1}(T)$ which gives an upper bound on the gain we get from entanglement assistance. 

As a third example we want to consider Gaussian channels defined in Subsection
\ref{sec:gaussian-channels}. Hence consider the Hilbert space $\scr{H} = \Lz(\Bbb{R})$ describing a
one-dimensional harmonic oscillator (or one mode of the electromagnetic field) and the
amplification/attenuation channel $T$ defined in Equation (\ref{eq:89}). The results we want to state
concern a slight modification of the original definitions of $C_{c,1}(T)$ and $C_e(T)$: We will consider
capacities for channels with \emph{constraint input}. This means that only a restricted class of states
$\rho$ on the input Hilbert space of the channel are allowed for encoding. In our case this means that we
will consider the constraint $\tr(\rho aa^*) \leq N$ for a positive real number $N >0$ and with the usual
creation and annihilation operators $a^*, a$. This can be rewritten as an energy constraint for a
quadratic Hamiltonian; hence this is a physically realistic restriction.  

\begin{figure}[h]
  \begin{center}
    \begin{pspicture}(15,9.5)
    \rput(7.5,5){\includegraphics[scale=0.8]{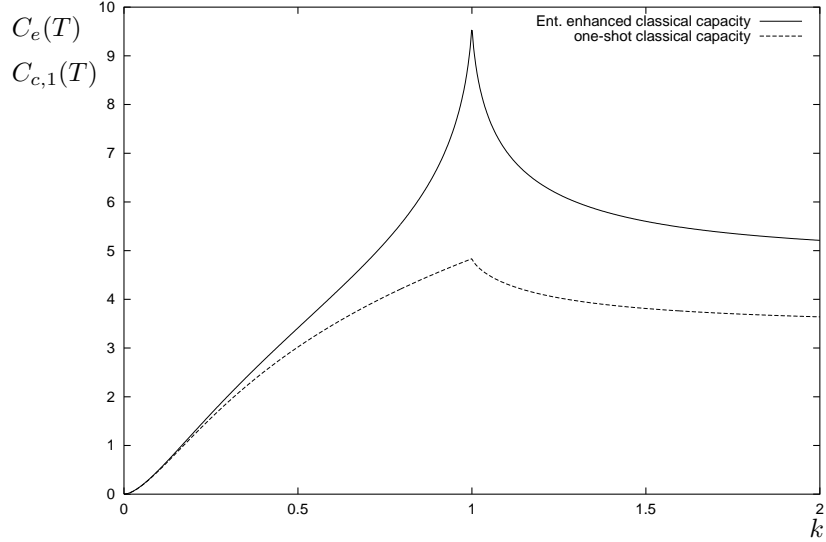}}
    \rput[r](13.5,0.5){$k$}
    \rput[tl](0,9){$C_{e}(T)$}
    \rput[tl](0,8.25){$C_{c,1}(T)$}
  \end{pspicture}
    \caption{One-shot and entanglement enhanced classical capacity of a Gaussian
      amplification/attenuation channel with $N_c=0$ and input noise $N=10$}
    \label{fig:ccgauss}
  \end{center}
\end{figure}

For the entanglement enhanced capacity it can be shown now that the maximum in Equation (\ref{eq:90}) is
taken on Gaussian states. To get $C_e(T)$ it is sufficient therefore to calculate the quantum mutual
information $I(T,\rho)$ for the Gaussian state $\rho_N$ from Equation (\ref{eq:88}). The details can be found
in \Cite{GaussCap} and \Cite{BSST}, we will only state the results here. With the abbreviation 
\begin{equation}
  g(x) = (x+1)\log_2(x+1) - x \log_2 x
\end{equation}
we get $S(\rho_N) = g(N)$ and $S\bigl(T[\rho_N]\bigr) = g(N')$ with $N' = k^2 N + \max \{ 0, k^2 -1 \} + N_c$
(cf. Equation (\ref{eq:91})) for the entropies of input and output states and
\begin{equation}
  S(\rho,T) = g\left(\frac{D+N'-N-1}{2}\right) + g\left(\frac{D-N'+N-1}{2}\right)
\end{equation}
with
\begin{equation}
  D = \sqrt{(N+N'+1)^2 -4 k^2 N(N+1)}
\end{equation}
for the entropy exchange. The sum of all three terms gives $C_e(T)$ which we have plotted in Figure
\ref{fig:ccgauss} as a function of $k$.

\begin{figure}[H]
  \begin{center}
    \begin{pspicture}(15,9)
    \rput(7.5,5){\includegraphics[scale=0.8]{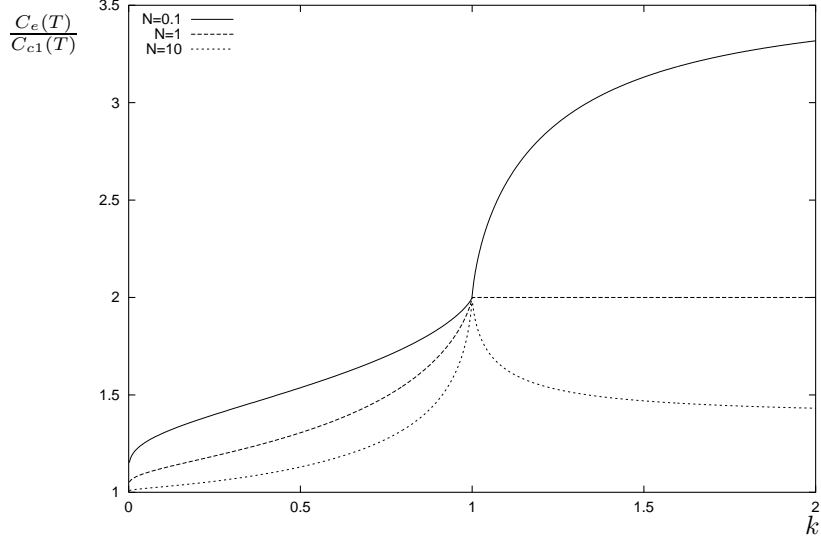}}
    \rput[r](13.5,0.5){$k$}
    \rput[tl](0,9){$\frac{C_e(T)}{C_{c1}(T)}$}
  \end{pspicture}
    \caption{Gain of using entanglement assisted versus unassisted classical capacity for a Gaussian 
      amplification/attenuation channel with $N_c=0$ and input noise $N=0.1, 1, 10$}
    \label{fig:gaussgain}
  \end{center}
\end{figure}

To calculate the one-shot capacity $C_{c,1}(T)$ the optimization in Equation (\ref{eq:92}) has to be
calculated over probability distributions $p_j$ and collections of density operators $\rho_j$ such that $\sum_j
p_j \tr(aa^*\rho_j) \leq N$ holds. It is conjectured but not yet proven \Cite{GaussCap} that the maximum is
achieved on coherent states with Gaussian probability distribution $p(x) = (\pi N)^{-1} \exp(-|x|^2/N)$. If
this is true we get
\begin{equation}
  C_{c,1}(T) = g(N') - g(N_0')\ \text{with}\ N_0' = \max \{ 0, k^2 -1 \} + N_c.
\end{equation}
The result is plotted as a function of $k$ in Figure \ref{fig:ccgauss} and the ratio $G = C_e/C_1$ in
Figure \ref{fig:gaussgain}. $G$ gives an upper bound on the \emph{gain} of using entanglement assisted
versus unassisted classical capacity. 

\section{The quantum capacity}
\label{sec:quantum-capacity}

The quantum capacity of a quantum channel $T: \scr{B}(\scr{H}) \to \scr{B}(\scr{H})$ is more difficult to
treat than the classical capacities discussed in the last section. There is in particular no coding
theorem available which would allow explicit calculations. Nevertheless there are partial results
available, which we will review in the following.

\subsection{Alternative definitions}
\label{sec:altern-defin}

Let us start with two alternative definitions of $C_q(T)$. The first one proposed by Bennett \Cite{BDiVSW}
differs only in the error quantity which should go to zero. Instead of the cb-norm the \emph{minimal
  fidelity} is used. For a channel $T: \scr{B}(\scr{H}) \to \scr{B}(\scr{H})$ and a subspace $\scr{H}' \subset
\scr{H}$ it is defined as  
\begin{equation}
  \scr{F}_p(\scr{H}',T) = \inf_{\psi \in \scr{H}'} \bigl\langle\psi, T\bigl[\kb{\psi}\bigr]\psi\bigr\rangle
\end{equation}
and if $\scr{H}' = \scr{H}$ holds we simply write $\scr{F}_p(T)$. Hence a number $c$ is an achievable
rate if  
\begin{equation} \label{eq:87}
  \lim_{j \to \infty} \scr{F}_p(E_j T^{\otimes M_j} D_j) = 1
\end{equation}
holds for sequences 
\begin{equation} \label{eq:86}
  E_j: \scr{B}(\scr{H})^{\otimes M_j} \to \scr{M}_2^{\otimes N_j},\ \scr{D}_j: \scr{M}_2^{\otimes N_j} \to \scr{B}(\scr{H})^{\otimes
    M_j},\ j \in \Bbb{N}
\end{equation}
of encodings and decodings and sequences of integers $M_j, N_j$, $j \in \Bbb{N}$ satisfying the same
constraints as in Definition \ref{def:2} (in particular $\lim_{j \to \infty} N_j/M_j < c$). The equivalence to
our version of $C_q(T)$ follows now from the estimates \Cite{Werner01}
\begin{gather}
  \| T - \Id\| \leq \| T - \Id\|_\cb \leq 4 \sqrt{\|T - \Id\|}\\ \| T - \Id\| \leq 4 \sqrt{1
    -\scr{F}_p(T)} \leq 4 \sqrt{\|T - \Id\|}.  
\end{gather}

A second version of $C_q(T)$ is given in \Cite{BaNiSch}. To state it let us define first a \emph{quantum
  source} as a sequence $\rho_N; N \in \Bbb{N}$ of density operators $\rho_N \in \scr{B}^*(\scr{K}^{\otimes N})$ (with an
appropriate Hilbert space $\scr{K}$) and the \emph{entropy rate} of this source as $\limsup_{N \to \infty}
S(\rho_N)/N$. In addition we need the \emph{entanglement fidelity} of a state $\rho$ (with respect to a channel
$T$)
\begin{equation}
  \scr{F}_e(\rho,T) = \bigl\langle \Psi, \bigl(T \otimes \Id\bigr)\bigl[\kb{\Psi}\bigr] \Psi \rangle,
\end{equation}
where $\Psi$ is the purification of $\rho$. Now we define $c \geq 0$ to be achievable if there is a quantum source
$\rho_N$, $N \in \Bbb{N}$ with entropy rate $c$ such that
\begin{equation}
  \lim_{n \to \infty} \scr{F}_e(\rho_N,E_N' T^{\otimes N} D_N') = 1
\end{equation}
holds with encodings and decodings
\begin{equation}
  E_N': \scr{B}(\scr{H})^{\otimes N} \to \scr{B}(\scr{K}^{\otimes N}),\ \scr{D}_N': \scr{B}(\scr{K}^{\otimes N}) \to
  \scr{B}(\scr{H})^{\otimes N},\ j \in \Bbb{N}. 
\end{equation}
Note that these $E_N'$, $D_N'$ play a slightly different role then the $E_j$, $D_j$ in Equation
(\ref{eq:86}) (and in Definition \ref{def:2}), because the number of tensor factors of the input and the
output algebra is always identical, while in Equation (\ref{eq:86}) the quotients of these numbers lead
to the achievable rate. To relate both definitions we have to derive an appropriately chosen family of
subspaces $\scr{H}'_N \subset \scr{K}^{\otimes N}$ from the $\rho_N$ such that the minimal fidelities
$\scr{F}_p(\scr{H}_N', E_N'T^{\otimes N} D_N')$ of \emph{these subspaces} go to $1$ as $N \to \infty$. If we
identify the $\scr{H}_N'$ with tensor products of $\Bbb{C}^2$ and the $E_j$, $D_j$ of Equation (\ref{eq:86})
with restrictions of $E_N'$, $D_N'$ to these tensor products we recover Equation (\ref{eq:87}). A precise
implementation of this rough idea can be found in \Cite{BaKniNi} and it shows 
that both definitions just discussed are indeed equivalent.

\subsection{Upper bounds and achievable rates}
\label{sec:upper-bounds-achi}

Although there is no coding theorem for the quantum capacity $C_q(T)$, there is a fairly good candidate
which is related to the \emph{coherent information}
\begin{equation}
  J(\rho,T) = S(T^*\rho) - S(\rho,T).
\end{equation}
Here $S(T^*\rho)$ is the entropy of the output state and $S(\rho,T)$ is the entropy exchange defined in
Equation (\ref{eq:93}). It is argued \Cite{BaNiSch} that $J(\rho,T)$ plays a role in quantum information
theory which is analogous to that of the (classical) mutual information (\ref{eq:85}) in classical
information theory. $J(\rho,T)$ has some nasty properties, however: it can be negative \Cite{CeAd97} and it 
is known to be not additive \Cite{DiVSS}. To relate it to $C_q(T)$ it is therefore not sufficient to
consider a one-shot capacity as in Shannons Theorem (Thm \ref{thm:8}). Instead we have to define 
\begin{equation}
  C_s(T) = \sup_N \frac{1}{N} C_{s,1}(T^{\otimes N})\ \text{with}\ C_{s,1}(T) = \sup_\rho J(\rho,T).
\end{equation}
In \Cite{BaNiSch} and \Cite{BaSmTe98} it is shown that $C_s(T)$ is an upper bound on $C_q(T)$. Equality, 
however, is conjectured but not yet proven, although there are good heuristic arguments
\Cite{LLoyd97},\Cite{H3qc}. 

A second interesting quantity which provides an upper bound on the quantum capacity uses the
transposition operation $\Theta$ on the output systems. More precisely it is shown in \Cite{GaussCap} that
\begin{equation}
  C_q(T) \leq C_\theta(T) = \log_2 \| T\Theta \|_\cb
\end{equation}
holds for any channel. In contrast to many other calculations in this field it is particular easy to
derive this relation from properties of the cb-norm. Hence we are able to give a proof here. We start
with the fact that $\|\Theta\|_\cb = d$ if $d$ is the dimension of the Hilbert space on which $\Theta$ operates. Assume
that $N_j/M_j \to c \leq C_q(T)$ and $j$ large enough such that $\|\Id_2^{N_j} - E_j T^{\otimes M_j} D_j\| \leq \epsilon$
with appropriate encodings and decodings $E_j, D_j$. We get
\begin{align}
  2^{N_j} &= \| \Id_2^{N_j} \Theta\|_\cb \leq  \|\Theta(\Id_2^{N_j} - E_j T^{\otimes M_j} D_j)\|_\cb + \|\Theta E_j T^{\otimes M_j}
  D_j\|_\cb\\
  & \leq 2^{N_j}\|\Id_2^{N_j} - E_j T^{\otimes M_j} D_j\|_\cb + \|\Theta E_j \Theta (\Theta T)^{\otimes M_j} D_j\|_\cb \\
  & \leq 2^{N_j} \epsilon + \|\Theta T\|_\cb^{M_j},
\end{align}
where we have used for the last equation the fact that $D_j$ and $\Theta E_j\Theta$ are channels and that the
cb-norm is multiplicative. Taking on both sides the logarithm we get
\begin{equation}
  \frac{N_j}{M_j} + \frac{\log_2(1 - \epsilon)}{M_j} \leq \log_2\|\Theta T\|_\cb.
\end{equation}
In the limit $j \to \infty$ this implies $c \leq \log_2\|\Theta T\|$ and therefore $C_q(T) \leq \log_2\|\Theta T\|_\cb = C_\theta(T)$ as
stated.

Since $C_\theta(T)$ is an upper bound on $C_q(T)$ it is particularly useful to check whether the quantum
capacity for a particular channel is zero. If, e.g. $T$ is classical we have $\Theta T = T$ since the
transposition coincides on a classical algebra $\scr{C}_d$ with the identity (elements of $\scr{C}_d$ are
just diagonal matrices). This implies $C_\theta(T) = \log_2\|\Theta T\|_\cb = \log_2 \|T\|_\cb = 0$, because the cb-norm
of a channel is $1$. We see therefore that the quantum capacity of a classical channel is $0$ -- this is
just another proof of the no-teleportation theorem. A slightly more general result concerns channels
$T=RS$ which are the composition of a preparation $R : \scr{M}_d \to \scr{C}_f$ and a subsequent
measurement $S: \scr{C}_f \to \scr{M}_d$. It is easy to see that $\Theta T = \Theta RS$ is a channel, because $\Theta R\Theta$ is
a channel and $\Theta$ is the identity on $\scr{C}_f$, hence $\Theta R\Theta = \Theta R$ and $\Theta R\Theta S = \Theta RS = \Theta T$. Again we
get $C_\theta(T) = 0$.

Let us consider now some examples. The most simple case is again the quantum erasure channel from
Equation (\ref{eq:121}). As for the classical capacities its quantum capacity can be explicitly
calculated \Cite{BeDiVSm97} and we have $C_q(T) = \max(0,(1-2\vartheta)\log_2(d))$; cf. Figure
\ref{fig:erasure}.  

\begin{figure}[b]
  \begin{center}
    \begin{pspicture}(15,9)
    \rput(7.5,5){\includegraphics[scale=0.8]{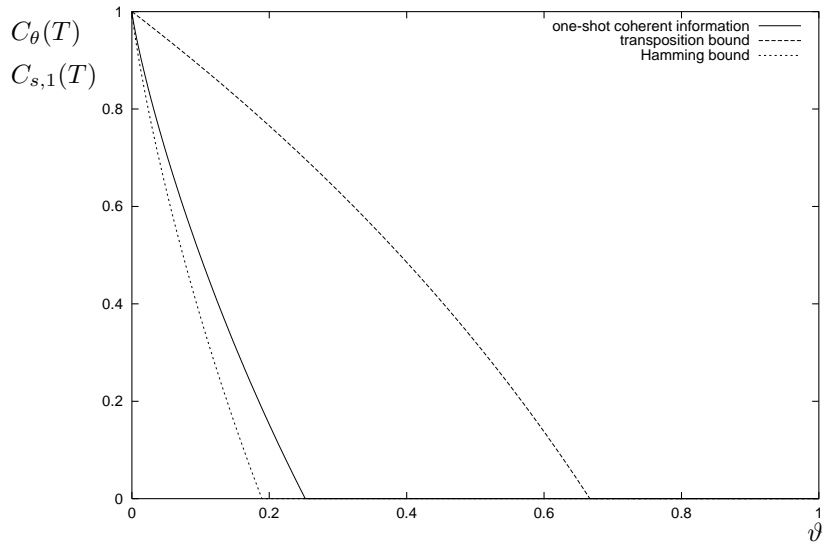}}
    \rput[r](13.5,0.5){$\vartheta$}
    \rput[tl](0,9){$C_\theta(T)$}
    \rput[tl](0,8.3){$C_{s,1}(T)$}
  \end{pspicture}
    \caption{$C_\theta(T)$, $C_s(T)$ and the Hamming bound of a depolarizing qubit channel plotted as function
      of the noise parameter $\vartheta$.}
    \label{fig:qcapdepol}
  \end{center}
\end{figure}

For the depolarizing channel (\ref{eq:122}) precise calculations of $C_q(T)$ are not availailable. Hence
let us consider first the coherent information. $J(T,\rho)$ inherits from $T$ its unitary covariance,
i.e. we have $J(U\rho U^*,T) = J(\rho,T)$. In contrast to the mutual information, however, it does not have
nice cocavity properties, which makes the optimization over all input states more difficult to
solve. Nevertheless, the calculation of $J(\rho,T)$ is straightforward and we get in the qubit case (if $\vartheta$
is the noise parameter of $T$ and $\lambda$ is the highest eigenvalue of $\rho$): 
\begin{multline} 
  J(\rho,T) = S\left(\lambda(1-\vartheta) + \frac{\vartheta}{2}\right) - S\left(\frac{1 -\vartheta/2 + A}{2}\right) - S\left(\frac{1 - \vartheta/2
      - A}{2}\right) \\- S\left(\frac{\lambda\vartheta}{2}\right) - S\left(\frac{(1-\lambda)\vartheta}{2}\right)
\end{multline}
where $S(x) = -x \log_2(x)$ denotes again the entropy function and
\begin{equation}
  A = \sqrt{(2\lambda -1)^2(1-\vartheta/2)^2  + 4 \lambda(1-\lambda)(1-\vartheta)^2}.
\end{equation}
Optimization over $\lambda$ can be performed at least numerically (the maximum is attained at
the left boundary ($\lambda = 1/2$) if $J$ is positive there, and the right boundary otherwise). The result is
plotted together with $C_\theta(T)$  in Figure \ref{fig:qcapdepol} as a function of $\theta$. The quantity $C_\theta(T)$
is much easier to compute and we get 
\begin{equation} \label{eq:98}
  C_\theta(T) = \max \{0, \log_2 \left(2 - \frac{3}{2} \theta\right) \}. 
\end{equation}

To get a lower bound on $C_q(T)$ we have to show that a certain rate $r \leq C_q(T)$ can be achieved with an
appropriate sequence 
\begin{equation}
  E_M: \scr{M}_d^{\otimes M} \to \scr{M}_2^{\otimes N(M)},\quad M, N(M) \in \Bbb{N}
\end{equation}
of error correcting codes and corresponding decodings $D_M$.  I.e. we need 
\begin{equation}
  \lim_{j \to \infty}  N(M)/M = r\ \text{and}\ \lim_{j \to \infty} \| E_M T^{\otimes M} D_M - \Id\|_\cb = 0.
\end{equation}
To find such a sequence note first that we can look at the depolarizing channel as a device
which produces an error with probability $\vartheta$ and leaves the quantum information intact otherwise. If more
and more copies of $T$ are used in parallel, i.e. if $M$ goes to infinity, the number of errors 
approaches therefore $\vartheta M$. In other words the probability to have more than $\vartheta M$ errors vanishes
asymptotically. To see this consider 
\begin{equation} \label{eq:123}
  T^{\otimes M} = \left( (\vartheta-1) \Id + \vartheta d^{-1} \tr(\,\cdot\,) \Bbb{1}\right)^{\otimes M} = \sum_{K = 1}^M (1-\vartheta)^K \vartheta^{N -K} T^{(M)}_K
\end{equation}
where $T^{(M)}_K$ denotes the sum of all $M$-fold tensor products with $d^{-1} \tr(\,\cdot\,) \Bbb{1}$ on $N$
places and $\Id$ on the $N-K$ remaining -- i.e. $T^{(M)}_K$ is a channel which produces exactly $K$
errors on $M$ transmitted systems. Now we have
\begin{align}
  \Bigl\|T^{\otimes M} - \sum_{K \leq \vartheta M} (1-\vartheta)^K & \vartheta^{N -K} T^{(M)}_K\Bigr\|_\cb \\ &= \left\| \sum_{K > \vartheta M} (1-\vartheta)^K \vartheta^{N -K}
    T^{(M)}_K\right\|_\cb \\ 
  & \leq \sum_{K > \vartheta M}^M (1-\vartheta)^K \vartheta^{N -K} \| T^{(M)}_K \|_\cb \\
  & \leq \sum_{K > \vartheta M}^M { M \choose K} (1-\vartheta)^K \vartheta^{N -K} = R.
\end{align}
The quantity $R$ is the tail a of Binomial series and vanishes therefore in the limit $M \to \infty$
(cf. e.g. Appendix B of \Cite{Purser95}). This shows that for $M \to \infty$ only terms $T^{(M)}_K$ with $K \leq
\vartheta M$ are relevant in Equation (\ref{eq:123}) -- in other words at most $\vartheta M$ errors occur asymptotically,
as stated. This implies that we need a sequence of codes $E_M$ which encode $N(M)$ qubits and correct
$\vartheta M$ errors on $M$ places. One way to get such a sequence is ``random coding'' -- the classical version
of this method is well known from the proof of Shannons theorem. The idea is, basically, to generate error
correcting codes of a certain type randomly. E.g. we can generate a sequence of random graphs with $N(M)$
input and $M$ output vertices (cf. Section \ref{sec:quant-error-corr}). If we can show that the
corresponding codes correct (asymptotically) $\vartheta M$ errors, the corresponding rate $r = \lim_{M\to\infty} N(M)/M$
is achievable. For the depolarizing channel\footnote{With a more thorough discussion similar results can be
  obtained for a much more general class of channels, e.g. all $T$ in a neighbourhood of the identity
  channel; cf. \Cite{MaUy01}.} such an analysis, using randomly generated stabilizer codes shows
\Cite{BDiVSW,GottD}
\begin{equation}
  C_q(T) \leq 1 - H(\vartheta) - \vartheta \log_23,
\end{equation}
where $H$ is the binary entropy from Equation (\ref{eq:65}). This bound can be further improved using a
more clever coding strategy; cf. \Cite{DiVSS}. 

\begin{figure}[h]
  \begin{center}
    \begin{pspicture}(15,9.5)
    \rput(7.5,5){\includegraphics[scale=0.8]{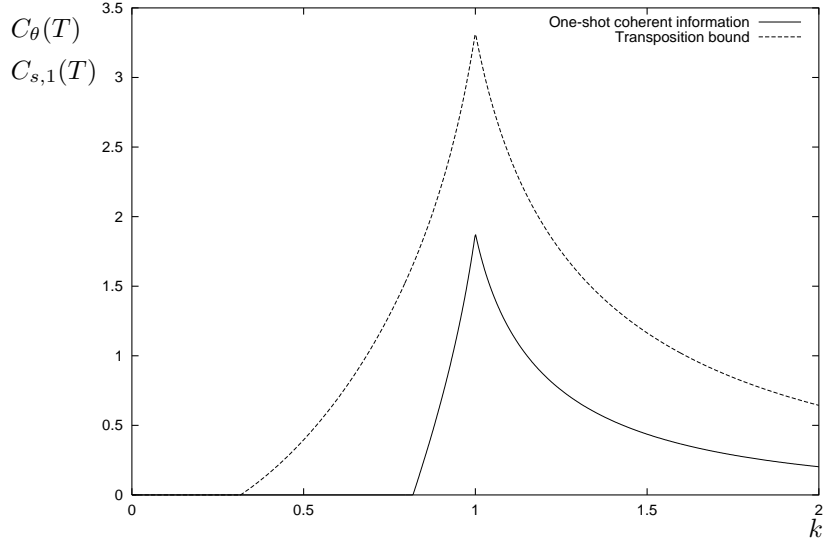}}
    \rput[r](13.5,0.5){$k$}
    \rput[tl](0,9){$C_\theta(T)$}
    \rput[tl](0,8.3){$C_{s,1}(T)$}
  \end{pspicture}
    \caption{$C_\theta(T)$ and $C_s(T)$ of a Gaussian amplification/attenuation channel as a function of
      amplification parameter $k$.}
    \label{fig:qcapgauss}
  \end{center}
\end{figure}

\begin{figure}[t]
  \begin{center}
    \begin{pspicture}(15,9)
    \rput(7.5,5){\includegraphics[scale=0.8]{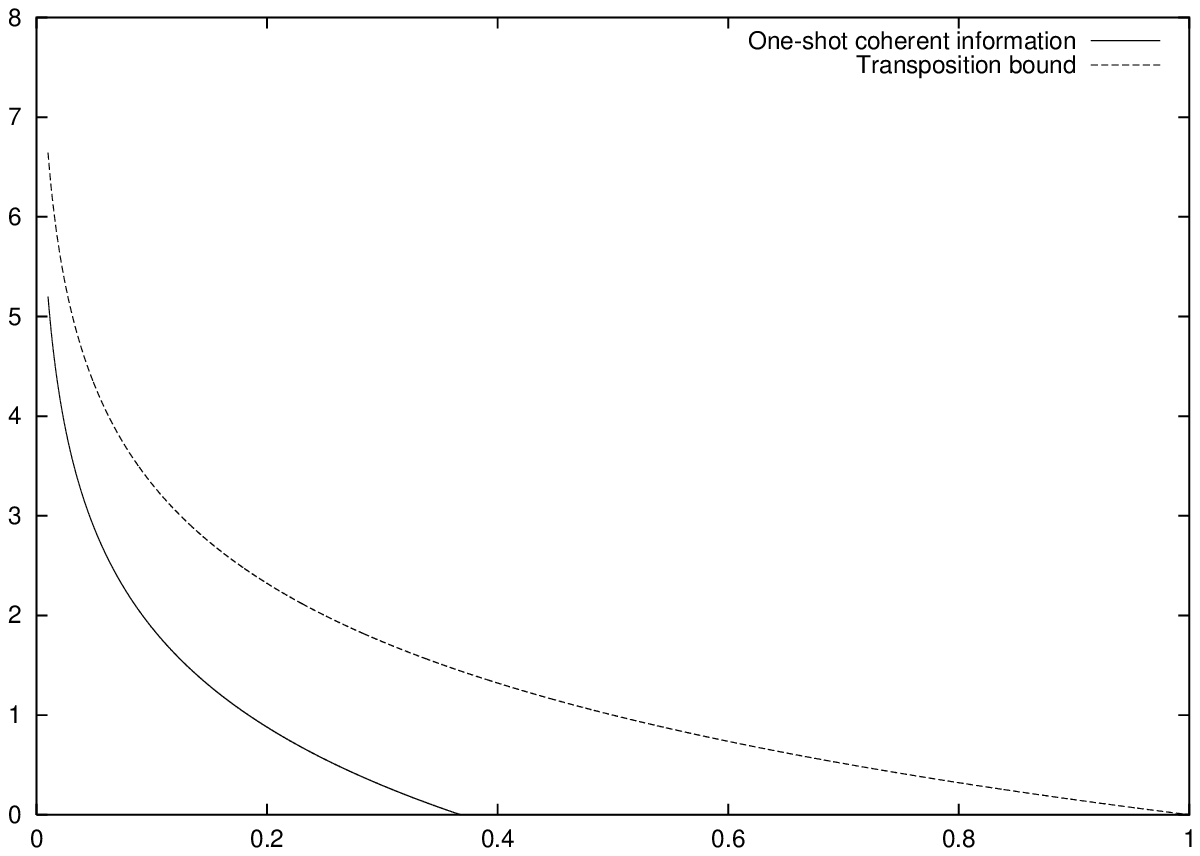}}
    \rput[r](13.5,0.5){$k$}
    \rput[tl](0,9){$C_\theta(T)$}
    \rput[tl](0,8.3){$C_{s,1}(T)$}
  \end{pspicture}
    \caption{$C_\theta(T)$ and $C_s(T)$ of a Gaussian amplification/attenuation channel as a function of the
      noise parameter $N_c$ (and with $k=1$).}
    \label{fig:qcapgauss2}
  \end{center}
\end{figure}

As a third example let us consider again the Gaussian channel studied already in Subsection 
\ref{sec:calc-gauss-chann}. For $C_\theta(T)$ we have (the corresponding calculation is not trivial
and uses properties of Gaussian channels which we have not discussed; cf. \Cite{GaussCap}.) 
\begin{equation}
  C_\theta(T) = \max \{0, \log_2 (k^2+1) -\log_2(|k^2-1|+2N_c)\}, 
\end{equation}
and we see that $C_\theta(T)$ and therefore $C_q(T)$ become zero if $N_c$ is large enough (i.e. $N_c \geq
\max\{1,k^2\}$). The coherent information for the Gaussian state $\rho_N$ from Equation (\ref{eq:88}) has the
form 
\begin{equation} \label{eq:124}
  J(\rho_N,T) = g(N') - g\left(\frac{D+N'-N-1}{2}\right) - g\left(\frac{D-N'+N-1}{2}\right)
\end{equation}
with $N', D$ and $g$ as in Subsection \ref{sec:calc-gauss-chann}. It increases with $N$ and we can
calculate therefore the maximum over all Gaussian states (which might differ from $C_S(T)$) as
\begin{equation}
  C_G(T) = \lim_{N \to \infty} J(\rho_N,T) = \log_2 k^2 - \log_2 |k^2 -1| - g \left(\frac{N_c}{k^2 - 1}\right).
\end{equation}
We have plotted both quantities in Figure \ref{fig:qcapgauss} as a function of $k$. 

Finally let us have a short look on the special case $k=1$, i.e. $T$ describes in this case only the
influence of classical Gaussian noise on the transmitted qubits. If we set $k=1$ in Equation (\ref{eq:124})
and take the limit $N \to \infty$ we get $C_G(T) = -\log_2(N_c e)$ and $C_\theta(T)$ becomes $C_\theta(T) = \max \{0,
-\log_2(N_c)\}$; both quantities are plotted in Figure \ref{fig:qcapgauss2}. This special case is
interesting because the one-shot coherent information $C_G(T)$ is achievable, provided the noise
parameter $N_c$ satisfies certain conditions\footnote{It is only shown that $\log_2(\lfloor1/(N_ce)\rfloor)$ can be
  achieved, where $\lfloor x\rfloor$ denotes the biggest integer less than $x$. It is very likely  however that
  this is only a restriction of the methods used in the proof and not of the result.} \Cite{HaPre01}.  
Hence there is strong evidence that the quantum capacity lies between the two lines in Figure
\ref{fig:qcapgauss2}. 

\subsection{Relations to entanglement measures}
\label{sec:relat-entangl-meas}

The duality lemma proved in Subsection \ref{sec:duality-lemma} provides an interesting way to derive
bounds on channel capacities and capacity like quantities from entanglement measures (and vice versa)
\Cite{BDiVSW,H3qc}: To derive a state of a bipartite system from a channel $T$ we can take a maximally 
entangled state $\Psi \in \scr{H} \otimes \scr{H}$, send one particle through $T$ and get a less entangled pair in
the state $\rho_T = (\Id \otimes T^*)\kb{\Psi}$. If on the other hand an entangled state $\rho \in \scr{S}(\scr{H} \otimes
\scr{H})$ is given, we can use it as a recource for teleportation and get a channel $T_\rho$. The two maps
$\rho \mapsto T_\rho$ and $T \mapsto \rho_T$ are, however, not inverse to one another. This can be seen easily from the
duality lemma (Theorem \ref{thm:6}): For each state $\rho \in \scr{S}(\scr{H} \otimes \scr{H})$ there is a channel
$T$ and a pure state $\Phi \in \scr{H} \otimes \scr{H}$ such that $\rho = (\Id \otimes T^*) \kb{\Phi}$ holds; but $\Phi$ is in
general not maximally entangled (and uniquely determined by $\rho$). Nevertheless, there are special
cases in which the state derived from $T_\rho$ coincides with $\rho$: A particular class of examples is given
by teleportation channels derived from a Bell-diagonal state.

On $\rho_T$ we can evaluate an entanglement measure $E(\rho_T)$ and get in this way a quantity which is related
to the capacity of $T$. A particularly interesting candidate for $E$ is the ``one-way LOCC'' distillation
rate $E_{D,\to}$. It is defined in the same way as the entanglement of distillation $E_D$, except that only
one-way LOCC operation are allowed in Equation (\ref{eq:97}). According to \Cite{BDiVSW} $E_{D,\to}$ is
related to $C_q$ by the inequalities $E_{D,\to}(\rho) \geq C_q(T_\rho)$ and $E_{D,\to}(T_\rho) \leq C_q(T)$. Hence if
$\rho_{T_\rho} = \rho$ we can calculate $E_{D,\to}(\rho)$ in terms of $C_q(T_\rho)$ and vice versa.

A second interesting example is the transposition bound $C_\theta(T)$ introduced in the last subsection. It is
related to the \emph{logarithmic negativity} \Cite{VidWe} 
\begin{equation}
  E_\theta(\rho_T) = \log_2 \| (\Id \otimes \Theta)\rho_T \|_1,
\end{equation}
which measures the degree with which the partial transpose of $\rho$ fails to be positive. $E_\theta$ can be
regarded as entanglement measure although it has some drawbacks: it is not LOCC monotone (Axiom
\ref{ax:1}), it is not convex (Axiom \ref{ax:4}) and most severe: It does not coincides with the reduced
von Neumann entropy on pure states, which we have considered as ``\emph{the}'' entanglement measure for
pure states. On the other hand it is easy to calculate and it gives bounds on distillation rates and
teleportation capacities \Cite{VidWe}. In addition $E_\theta$ can be used together with the relation between
depolarizing channels and isotropic states to derive Equation (\ref{eq:98}) in a very simple way.

\chapter{Multiple inputs}
\label{cha:quant-theory-iii}

We have seen in Chapter \ref{cha:basic-tasks} that many tasks of quantum information which are impossible
with one-shot operations can be approximated by channels which operate on a large number of equally prepared
inputs. Typical examples are approximate cloning, undoing noise and distillation of entanglement. There
are basically two questions which are interesting for a quantitative analysis: First we can search for
the optimal solutions for a fixed number $N$ of input systems and second we can ask for the asymptotic
behavior in the limit $N \to \infty$. In the latter case the asymptotic rate, i.e. the number of outputs (of a
certain quality) per input system is of particular interest. 

\section{The general scheme}
\label{sec:general-scheme}

Both types of questions just mentioned can be treated (up to certain degree) independently from the
(impossible) task we are dealing with and we will study in the following the corresponding general
scheme. Hence consider a channel $T: \scr{B}(\scr{H}^{\otimes M}) \to \scr{B}(\scr{H}^{\otimes N})$ which operates on
$N$ input systems and produces $M$ outputs of the same type. Our aim is to optimize a ``\emph{figure of
  merit}'' $\scr{F}(T)$ which measures the deviation of $T^*(\rho^{\otimes N})$ from the target functional we want to
approximate. The particular type of device we are considering is mainly fixed by the choice of
$\scr{F}(T)$ and we will discuss in the following the most relevant examples. (Note that we have
considered them already on a qualitative level in Chapter \ref{cha:basic-tasks}; cf. in particular
Section \ref{sec:estimating-copying} and \ref{sec:dist-entangl}).

\subsection{Figures of merit}
\label{sec:figures-merit}

Let us start with pure state cloning \Cite{GiMa97,BDiVEFMS,BEM98,BuHi98,Klo,Klo2}, i.e. for each
(unknown) pure input state $\sigma = \kb{\psi}$, $\psi \in \scr{H}$ the $M$ clones $T^*(\sigma^{\otimes N})$ produced by the
channel $T$ should approximate $M$ copies of the input in the common state $\sigma^{\otimes   M}$ as good as
possible. There are in fact two different possibilities to measure the distance of $T^*(\sigma^{\otimes N})$ to
$\sigma^{\otimes M}$. We can either check the quality of each clone separately or we can test in addition the
correlations between output systems. With the notation 
\begin{equation} \label{eq:21}
  \sigma^{(j)} = \Bbb{1}^{\otimes (j-1)} \otimes \sigma \otimes \Bbb{1}^{\otimes (M - j)} \in \scr{B}(\scr{H}^{\otimes M})
\end{equation}
a figure of merit for the first case is given by
\begin{equation} \label{eq:99}
  \scr{F}_{c,1}(T) = \inf_{j=1,\ldots,N} \inf_{\sigma\, \pure} \tr \bigl( \sigma^{(j)} T^*(\sigma^{\otimes N}) \bigr).
\end{equation} 
It measures the worst one particle fidelity of the output state $T^*(\sigma^{\otimes N})$. If we are interested in
correlations too, we have to choose  
\begin{equation} \label{eq:100}
  \scr{F}_{c,\all}(T) = \inf_{\sigma\, \pure} \tr \bigl( \sigma^{\otimes M} T^*(\sigma^{\otimes N}) \bigr)
\end{equation}
which is again a ``worst case'' fidelity, but now of the full output with respect to $M$ uncorrelated
copies of the input $\sigma$. 

Instead of fidelities we can consider other error quantities like trace-norm distances or
relative entropies. In general, however, we do not get significantly different results from such
alternative choices; hence we can safely ignore them. Real variants arise if we consider instead
of the infima over all pure states quantities which prefer a (possibly discrete or even finite) class of
states. Such a choice leads to ``state dependent cloning'', because the corresponding optimal devices
perform better as ``universal'' ones (i.e. those described by the figures of merit above) on \emph{some
  states} but much worse on the rest. We ignore state dependent cloning in this work, because the
universal case is physically more relevant and technically more challenging. Other cases which we do not
discuss either include ``asymmetric cloning'', which arises if we trade in Equation (\ref{eq:99}) the
quality of one particular output system against the rest (see \Cite{CerfAC}), and cloning of mixed
states. The latter is much more difficult then the pure state case and even for classical systems, where
it is related to the so called ``bootstrap'' technique \Cite{Efron93}, nontrivial.

Closely related to cloning is purification, i.e. undoing noise. This means we are considering $N$ systems
originally prepared in the same (unknown) pure state $\sigma$ but which have passed a depolarizing channel
\begin{equation} \label{eq:20}
  R^*\sigma = \vartheta \sigma + (1-\vartheta) \Bbb{1}/d 
\end{equation}
 afterwards. The task is now to find a device $T$ acting on $N$ of the
decohered systems such that $T^*(R^*\sigma)$ is as close as possible to the original pure state. We have the
same basic choices for a figure of merit as in the cloning problem. Hence we define
\begin{equation} \label{eq:101}
  \scr{F}_{R,1}(T) = \inf_{j=1,\ldots,N} \inf_{\sigma\, \pure} \tr \Bigl( \sigma^{(j)} T^*\bigl[(R^*\sigma)^{\otimes N}\bigr] \Bigr)
\end{equation}
and
\begin{equation} \label{eq:102}
  \scr{F}_{R,\all}(T) = \inf_{\sigma\, \pure} \tr \Bigl( \sigma^{\otimes M} T^*\bigl[(R^*\sigma)^{\otimes N}\bigr] \Bigr).
\end{equation}
These quantities can be regarded as generalizations of $\scr{F}_{c,1}$ and $\scr{F}_{c,\all}$ which we
recover if $R^*$ is the identity. 

Another task we can consider is the approximation of a map $\Theta$ which is positive but not completely
positive, like the transposition. Positivity and normalization imply that $\Theta^*$ maps states to states but
$\Theta$ can not be realized by a physical device. An explicit example is the universal not gate (UNOT) which
maps each pure qubit state $\sigma$ to its orthocomplement $\sigma^\bot$ \Cite{UNOT}. It is given the the anti-unitary
operator  
\begin{equation}
  \psi = \alpha \ket{0} + \beta \ket{1} \mapsto \Theta\psi = \bar{\alpha} \ket{0} - \bar{\beta} \ket{1}.
\end{equation}
Since $\Theta\sigma$ is a state if $\sigma$ is, we can ask again for a channel $T$ such that $T^*(\sigma^{\otimes N})$ approximates
$(\Theta\sigma)^{\otimes M}$. As in the two previous examples we have the choice to allow arbitrary correlations in the
output or not and we get the following figures of merit:
\begin{equation} \label{eq:103}
  \scr{F}_{\theta,1}(T) = \inf_{j=1,\ldots,N} \inf_{\sigma\, \pure} \tr \bigl( (\Theta\sigma)^{(j)} T^*(\sigma^{\otimes N}) \bigr)
\end{equation}
and
\begin{equation} \label{eq:104}
  \scr{F}_{\theta,\all}(T) = \inf_{\sigma\, \pure} \tr \bigl( (\Theta\sigma)^{\otimes M} T^*(\sigma^{\otimes N}) \bigr).
\end{equation}
Note that we can plug in for $\Theta$ basically any functional which maps states to states. In addition we can
combine Equation (\ref{eq:101}) and (\ref{eq:102}) on the one hand with (\ref{eq:103}) and (\ref{eq:104})
on the other. As result we would get a measure for devices which undo an operation $R$ and approximate
an impossible machine $\Theta$ at the same time. 

\subsection{Covariant operations}
\label{sec:covariant-operations}

All the functionals just defined give rise to optimization problems which we will study in greater
detail in the next Sections. This means we are interested in two things: First of all the maximal value of
$\scr{F}_{\#,\natural}$ (with $\# = c,R,\theta$ and $\natural=1,\all$) given by
\begin{equation}
  \scr{F}_{\#,\natural}(N,M) = \inf_T \scr{F}_{\#,\natural}(T), 
\end{equation}
where the supremum is taken over all channels $T: \scr{B}(\scr{H}^{\otimes M}) \to \scr{B}(\scr{H}^{\otimes N})$, and
second the particular channel $\hat{T}$ where the optimum is attained. At a first look a complete solution
of these problems seems to be impossible, due to the large dimension of the space of all $T$, which
scales exponentially in $M$ and $N$. Fortunately all $\scr{F}_{\#,\natural}(T)$ admit a large symmetry group which
allows in many cases the explicit calculation of the optimal values $\scr{F}_{\#,\natural}(N,M)$ and the
determination of optimizers $\hat{T}$ with a certain covariance behavior. Note that this is an immediate
consequence of our decision to restrict the discussion to ``universal'' procedures, which do not prefer
any particular input state.

Let us consider permutations of the input systems first: If $p \in \Sym_N$ is a permutation on $N$ places
and $V_p$ the corresponding unitary on $\scr{H}^{\otimes N}$ (cf. Equation (\ref{eq:105})) we get obviously
$T^*(V_p \rho^{\otimes N} V_p^*)= T^*(\rho^{\otimes N})$, hence 
\begin{equation} \label{eq:107}
  \scr{F}_{\#,\natural}\bigl[\alpha_p(T)\bigr] = \scr{F}_{\#,\natural}(T)\ \forall p \in \Sym_N\ \text{with}\ \bigl[\alpha_p(T)\bigr](A) =
  V_p^*T(A)V_p.
\end{equation}
In other words: $\scr{F}_{\#,\natural}(T)$ is invariant under permutations of the input systems. Similarly we can
show that $\scr{F}_{\#,\natural}(T)$ is invariant under permutations of the output systems:
\begin{equation} \label{eq:108}
  \scr{F}_{\#,\natural}\bigl[\beta_p(T)\bigr] = \scr{F}(T)\ \forall p \in \Sym_M\ \text{with}\ \bigl[\beta_p(T)\bigr](A) = T(V_p^*AV_p).
\end{equation}
To see this consider e.g. for $\#=c$ and $\natural=\all$
\begin{equation}
  \tr\bigl[\sigma^{\otimes M} V_p T^*(\rho^{\otimes N}) V_p^*\bigr] = \tr\bigl[V_p \sigma^{\otimes M} V_p^* T^*(\rho^{\otimes N})\bigr] =
  \tr\bigl[\sigma^{\otimes M} T^*(\rho^{\otimes N})\bigr].
\end{equation}
For the other cases similar calculations apply.

Finally, none of the $\scr{F}_{\#,\natural}(T)$ singles out a preferred direction in the one particle Hilbert
space $\scr{H}$. This implies that we can rotate $T$ by local unitaries of the form $U^{\otimes N}$
respectively $U^{\otimes M}$ without changing $\scr{F}_{\#,\natural}(T)$. More precisely we have 
\begin{equation} \label{eq:106}
  \scr{F}_{\#,\natural}\bigl[\gamma_U(T)\bigr] = \scr{F}_{\#,\natural}(T)\ \forall U \in \U(d)
\end{equation}
with
\begin{equation}
  \bigl[\gamma_U(T)\bigr](A) = U^{*\otimes N}T(U^{\otimes M}  A U^{* \otimes M}) U^{\otimes N}.
\end{equation}
 The validity of Equation (\ref{eq:106}) can be proven in the same way as (\ref{eq:107}) and
 (\ref{eq:108}). The details are therefore left to the reader.  

Now we can average over the groups $S_N, S_M$ and $\U(d)$. Instead of the operation $T$ we consider
\begin{equation}
  \bar{T} = \frac{1}{N!M!}\sum_{p \in S_N} \sum_{q \in S_M} \int_G \alpha_p \beta_q \gamma_U(T) dU,
\end{equation}
where $dU$ denotes the normalized, left invariant Haar measure on $\U(d)$. We see immediately that
$\bar{T}$ has the following symmetry properties  
\begin{equation}
  \alpha_p(\bar{T}) = \bar{T},\ \beta_q(\bar{T}) = \bar{T},\ \gamma_U(\bar{T}) = \bar{T},\ \forall p \in S_N,\ \forall q \in S_M,\ \forall U
  \in \U(d) 
\end{equation}
and we will call each operation $T$ \emph{fully symmetric}, if it satisfies this equation. The concavity
of $\scr{F}_{\#,\natural}$ implies immediately that it can not decrease if we replace $T$ by $\bar{T}$:
\begin{align}
  \scr{F}_{\#,\natural}(T) &= \scr{F}_{\#,\natural}\left(\frac{1}{N!M!}\sum_{p \in S_N} \sum_{q \in S_M} \int_G \alpha_p \beta_q \gamma_U(T) dU\right) \\
  &\geq \frac{1}{N!M!}\sum_{p \in S_N} \sum_{q \in S_M} \int_G \scr{F}_{\#,\natural}\bigl[\alpha_p \beta_q \gamma_U(T)\bigr] dU = \scr{F}_{\#,\natural}(T).
\end{align}
To calculate the optimal value $\scr{F}_{\#,\natural}(N,M)$ it is therefore completely sufficient to search a
maximizer for $\scr{F}_{\#,\natural}(T)$ only among fully symmetric $T$ and to evaluate $\scr{F}_{\#,\natural}(T)$ for
this particular operation. This simplifies the problem significantly because the size of the parameter
space is extremely reduced. Of course we do not know from this argument whether the optimum is attained
on non-symmetric operations, however this information is in general less important (and for some
problems like optimal cloning a uniqueness result is available).

\subsection{Group representations}
\label{sec:group-repr}

To get an idea how this parameter reduction can be exploited practically, let us reconsider Theorem
\ref{thm:1}: The two representations $U \mapsto U^{\otimes N}$ and $p \mapsto V_p$ of $\U(d)$ respectively $\Sym_N$ on
$\scr{H}^{\otimes N}$ are ``commutants'' of each other, i.e., any operator on $\scr{H}^{\otimes N}$ commuting with
all $U^{\otimes N}$ is a linear combination of the $V_p$, and conversely. This knowledge can be used to
decompose the representation $U^{\otimes N}$ (and $V_p$ as well) into irreducible components. To reduce the
group theoretic overhead, we will discuss this procedure first for qubits only and come back to
the general case afterwards.  

Hence assume that $\scr{H} = \Bbb{C}^2$ holds. Then $\scr{H}^{\otimes N}$ is the Hilbert space of $N$
(distinguishable) spin-1/2 particles and it can be decomposed in terms of eigenspaces of total angular
momentum. More precisely consider
\begin{equation}
  L_k = \frac{1}{2} \sum_j \sigma_k^{(j)},\ k=1,2,3  
\end{equation}
the $k$-component of total angular momentum (i.e. $\sigma_k$ is the $k^{\rm th}$ Pauli matrix and $\sigma^{(j)} \in
\scr{B}(\scr{H}^{\otimes N})$ is defined according to Equation (\ref{eq:21})) and $\vec{L}^2 = \sum_k L_k^2$. The
eigenvalue expansion of $\vec{L}^2$ is well known to be
\begin{equation}
  \vec{L} = \sum_j s(s+1) P_s,\ \text{with}\ s = 
  \begin{cases} 0,1,\ldots,N/2 & N\ \text{even}\\ 1/2,3/2,\ldots,N/2 & N\ \text{odd} \end{cases},
\end{equation}
where the $P_s$ denote the projections to the eigenspaces of $\vec{L}^2$. It is easy to see that both
representations $U \mapsto U^{\otimes N}$ and $p \mapsto V_p$ commute with $\vec{L}$. Hence the eigenspaces $P_s \scr{H}^{\otimes
  N}$ of $\vec{L}^2$ are invariant subspaces of $U^{\otimes N}$ and $V_p$ and this implies that the restriction
of $U^{\otimes N}$ and $V_p$ to them are representations of $\SU(2)$ respectively $\Sym_N$. Since $\vec{L}^2$ is
constant on $P_s \scr{H}^{\otimes N}$ the $\SU(2)$ representation we get in this way must be (naturally
isomorphic to) a multiple of the irreducible spin-$s$ representation $\pi_s$. It is defined by
\begin{equation}
  \pi_s\left[\exp\left(\frac{i}{2} \sigma_k\right)\right] = \exp\left(i L_k^{(s)}\right)\ \text{with}\ L_k^{(s)} =
\frac{1}{2} \sum_{j=1}^{2s} \sigma^{(j)}_k, 
\end{equation}
on the representation space
\begin{equation} \label{eq:35}
  \scr{H}_s = \scr{H}^{\otimes 2s}_+
\end{equation}
(the Bose-subspace of $\scr{H}^{\otimes 2s}$). Hence we get
\begin{equation}
  P_s \scr{H}^{\otimes N} \cong \scr{H}_s \otimes \scr{K}_{N,s}, \quad U^{\otimes N} \psi = (\pi_s(U) \otimes \Bbb{1}) \psi\ \forall \psi \in P_s \scr{H}^{\otimes N}.
\end{equation}
Since $V_p$ and $U^{\otimes N}$ commute the Hilbert space $\scr{K}_{N,s}$ carries a representation
$\widehat{\pi}_{N,s}(p)$ of $\Sym_N$ which is irreducible as well. Note that
$\scr{K}_{N,s}$ depends in contrast to $\scr{H}_s$ on the number $N$ of tensor factors and its dimension
is (see \Cite{pur} or \Cite{Simon96} for general $d$)
 \begin{equation} \label{eq:33}
  \dim\scr{K}_{N,s}=\frac{2s+1}{N/2+s+1}{N\choose N/2-s}. 
\end{equation}
Summarizing the discussion we get
\begin{equation} \label{eq:22}
  \scr{H}^{\otimes N} \cong \bigoplus_s \scr{H}_s \otimes \scr{K}_{N,s},\ U^{\otimes N} \cong \bigoplus_s \pi_s(U) \otimes \Bbb{1},\ V_p \cong \bigoplus_s \Bbb{1} \otimes
  \widehat{\pi}(p).
\end{equation}

Let us consider now a fully symmetric operation $T$. Permutation invariance ($\alpha_p(T) = T$ and $\beta_p(T) =
T$) implies together with Equation (\ref{eq:22}) that 
\begin{equation} \label{eq:3}
  T(A_j \otimes B_j) = \bigoplus_s \left[\frac{\tr(B_j)}{\dim \scr{K}_{N,j}} T_{sj}(A_j) \otimes \Bbb{1}\right]\ \text{with}\
  T_{sj} : \scr{B}(\scr{H}_j) \to \scr{B}(\scr{H}_s),
\end{equation}
holds if $A_j \otimes B_j \in \scr{B}(\scr{H}_j \otimes \scr{K}_{N,j})$. The operations $T_{sj}$ are unital and have,
according to $\gamma_U(T)=T$ the following covariance properties 
\begin{equation}
  \pi_s(U) T(A_j) \pi_s(U^*) = T\bigl[\pi_j(U)A_j \pi_j(U^*)\bigr]\ \forall U \in \SU(2).
\end{equation}
The classification of all fully symmetric channels $T$ is reduced therefore to the study of all these
$T_{sj}$.  

We can apply now the covariant version of Stinespring's theorem (Theorem \ref{thm:3}) to find that  
\begin{equation}
  T_{sj}(A_j) = V^* (A_j \otimes \Bbb{1}) V,\ V: \scr{H}_s \to \scr{H}_j \otimes \tilde{\scr{H}},\ V\pi_s(U) = \pi_j(U) \otimes
  \tilde{\pi}(U) V,
\end{equation}
where $\tilde{\pi}$ is a representation of $\SU(2)$ on $\tilde{\scr{H}}$. If $\tilde{\pi}$ is irreducible with
total angular momentum $l$ the ``intertwining operator'' $V$ is well known: Its components in a
particularly chosen basis concide with certain Clebsh-Gordon coefficients. Hence the corresponding
operation is uniquely determined (up to unitary equivalence) and we write
\begin{equation} 
  T_{sjl}(A_j) = \bigl[V_l (A_j \otimes \Bbb{1}) V_l\bigr],\quad V_l\pi_s(U) = \pi_j(U) \otimes\pi_l(U) V_l
\end{equation}
where $l$ can range from $|j-s|$ to $j + s$. Since a general representation $\tilde{\pi}$ can be decomposed
into irreducible components we see that each covariant $T_{sj}$ is a convex linear combination of the
$T_{sjl}$ and we get with Equation (\ref{eq:3})  
\begin{equation} \label{eq:23}
  T(A_j \otimes B_j) = \bigoplus_s \left[ \sum_l c_{jl} \bigl[ T_{sjl}(A_j) \otimes (\tr(B_j) \Bbb{1}) \bigr]\right]
\end{equation}
where the $c_{jl}$ are constrained by $c_{jl} > 0$ and $\sum_j c_{jl} = (\dim \scr{K}_{N,j})^{-1}$. In this
way we have parameterized the set of fully symmetric operations completely in terms of group theoretical
data and we can rewrite $\scr{F}_{\#,\natural}(T)$ accordingly. This leads to an optimization problem for a
quantity depending only on $s,j$ and $l$, which is at least in some cases solvable.

To generalize the scheme just presented to the case $\scr{H} = \Bbb{C}^d$ with arbitrary $d$ we only have
to find a replacement for the decomposition in Equation (\ref{eq:22}). This, however, is well known from
group theory:
\begin{equation} \label{eq:109}
  \scr{H}^{\otimes N} \cong \bigoplus_Y \scr{H}_Y \otimes \scr{K}_Y,\ U^{\otimes N} \cong \bigoplus_Y \pi_Y(U)\otimes \Bbb{1},\ V_p \cong \bigoplus_Y \Bbb{1} \otimes
  \widehat{\pi}_Y(p),  
\end{equation}
where $\pi_Y:\U(d) \to\scr{B}(\scr{H}_Y)$ and $\widehat{\pi}_Y:\Sym_N\to\scr{B}(\scr{K}_Y)$ are irreducible
representations. The summation index $Y$ runs over all Young frames with $d$ rows and $N$ boxes, i.e., by
the arrangements of $N$ boxes into $d$ rows of lengths $Y_1\geq Y_2\geq\cdots\geq Y_d\geq0$ with $\sum_k Y_k=N$. The relation
to total angular momentum $s$ used as the parameter for $d=2$ is given by $Y_1 - Y_2 = 2s$, which
determines $Y$ together with $Y_1 + Y_2 = N$ completely. The rest of the arguments applies without
significant changes, this is in particular the case for Equation (\ref{eq:23}) which holds for general
$d$ if we replace $s$, $j$ and $l$ by Young frames. However, the representation theory of $\U(d)$ becomes
much more difficult. The generalization of results available for qubits ($d=2$) to $d > 2$ is therefore by
no means straightforward. 

Finally let us give a short comment on Gaussian states here. Obviously the methods just described do not
apply in this case. However, we can consider instead of $U^{\otimes N}$-covariance, covariance with respect to
phase-space translations. Following this idea some results concerning optimal cloning of Gaussian states
are obtained (see \Cite{CIvA01} and the refences therein), but the corresponding general theory is not as
far developed as in the finite dimensional case. 

\subsection{Distillation of entanglement}
\label{sec:dist-entangl-1}

Finally let us have another look at distillation of entanglement. The basic idea is quite the same as for
optimal cloning: Use multiple inputs to approximate a task which is impossible with one-shot
operations. From a  more technical point of view, however, it does not fit into the general
scheme proposed up to now. Nevertheless, some of the arguments can be adopted in an easy way. First of
all we have to replace the ``one-particle'' Hilbert space $\scr{H}$ with a two-fold tensor product
$\scr{H}_A \otimes \scr{H}_B$ and the channels we have to look at are LOCC operations
\begin{equation} \label{eq:112}
  T: \scr{B}(\scr{H}_A^{\otimes M} \otimes \scr{H}_B^{\otimes M}) \to \scr{B}(\scr{H}_A^{\otimes N} \otimes \scr{H}_B^{\otimes N});
\end{equation}
cf. Section \ref{sec:dist-entangl}.
Our aim is to determine $T$ such that $T^*(\rho^{\otimes N})$ is for each distillable (mixed) state $\rho \in
\scr{B}^*(\scr{H}_A \otimes \scr{H}_B)$, close to the $M$-fold tensor product $\kb{\Psi}^{\otimes M}$ of a maximally
entangled state $\Psi \in \scr{H}_A \otimes \scr{H}_B$. A figure of merit with a similar structure as the
$\scr{F}_{\#,\all}$ studied above can be derived directly from the definition of the entanglement measure
$E_D$ in  Section \ref{sec:entangl-meas-mixed}: We define (replacing the trace-norm distance with a
fidelity)  
\begin{equation} \label{eq:110}
  \scr{F}_D(T) = \inf_\rho \inf_\Psi \langle\Psi^{\otimes M}, T^*(\rho^{\otimes N}) \Psi^{\otimes M}\rangle
\end{equation}
where the infima are taken over all maximally entangled states $\Psi$ and all distillable states
$\rho$. Alternatively we can look at state dependent measures, which seem to be particularly important if
we try to calculate $E_D(\rho)$ for some state $\rho$. In this case we simply get
\begin{equation} \label{eq:111}
  \scr{F}_{D,\rho}(T) = \inf_\Psi \langle\Psi^{\otimes M}, T^*(\rho^{\otimes N}) \Psi^{\otimes M}\rangle.
\end{equation}
To translate the group theoretical analysis of the last two subsections is somewhat more
difficult. As in the case of $\scr{F}_{\#,\natural}$ we can restrict the search for optimizers to permutation
invariant operations, i.e. $\alpha_p(T) = T$ and $\beta_p(T) = T$ in the terminology of Subsection
\ref{sec:covariant-operations}. Unitary covariance
\begin{equation}
  U^{\otimes N} T(A) U^{* \otimes N} = T(U^{\otimes M} A U^{* \otimes M})
\end{equation}
however, can not be assumed \emph{for all} unitaries $U$ of $\scr{H}_A \otimes \scr{H}_B$, but only for local
ones ($U = U_A \otimes U_B$) in the case of $\scr{F}_D$ or only for local $U$ which leave $\rho$ invariant for
$\scr{F}_{D,\rho}$. This makes the analogon of the decomposition scheme from Subsection \ref{sec:group-repr}
more difficult and 
such a study is (up to my knowledge) not yet done. A related subproblem arises if we consider $\scr{F}_{D,\rho}$
from Equation  (\ref{eq:111}) for a state $\rho$ with special symmetry properties; e.g. an OO-invariant
state.  The corresponding optimization might be simpler and a solution would be relevant for the
calculation of $E_D$.

\section{Optimal devices}
\label{sec:optimal-devices}

Now we can consider the optimization problems associated to the figures of merit discussed in the last
section. This means that we are searching for those devices which approximate the impossible tasks in
question in the best possible way. As pointed out at the beginning of this Chapter this can be done for
finite $N$ and in the limit $N \to \infty$. The latter is postponed to the next section.

\subsection{Optimal cloning}
\label{sec:optimal-cloning}

The quality of an optimal, pure state cloner is
defined by the figures of merit $\scr{F}_{c,\#}$ in Equations (\ref{eq:99}) and (\ref{eq:100}) and the
group theoretic ideas sketched in Subsection \ref{sec:group-repr} allow the complete solution of this
problem. We will demonstrate some of the basic ideas in the qubit case first and state the final result
afterwards in full generality.

The solvability of this problem relies in part on the special structure of the figures of merit
$\scr{F}_{c,\#}$, which allows further simplifications of the general scheme sketched in
Subsection \ref{sec:group-repr}. If we consider e.g. $\scr{F}_{c,1}(T)$ (the other case works similarly)
we get: 
\begin{align}
  \scr{F}_{c,1}(T) &= \inf_{j=1,\ldots,N} \inf_{\sigma\, \pure} \tr \bigl( \sigma^{(j)} T^*(\sigma^{\otimes N}) \bigr)\\
  &= \inf_{j=1,\ldots,N} \inf_{\sigma\, \pure} \tr \bigl(T(\sigma^{(j)}) \sigma^{\otimes N}) \bigr)\\
  &= \inf_{j=1,\ldots,N} \inf_{\psi} \langle \psi^{\otimes N}, T(\sigma^{(j)}) \psi^{\otimes N}\rangle.
\end{align}
Hence $\scr{F}_{c,\#}$ only depends on the $\scr{B}(\scr{H}^{\otimes N}_+)$ component (where $\scr{H}^{\otimes N}_+$
denotes again the Bose-subspace of $\scr{H}^{\otimes N}$) of $T$ and we can assume without loss
of generality that $T$ is of the form
\begin{equation} \label{eq:114}
  T: \scr{B}(\scr{H}^{\otimes M}) \to \scr{B}(\scr{H}^{\otimes N}_+).
\end{equation}
The restriction of $U^{\otimes N}$ to $\scr{H}^{\otimes N}_+$ is an irreducible representation (for any $d$) and in
the qubit case ($d=2$) we have $U^{\otimes N} \psi = \pi_s(U) \psi$ with $s=N/2$ for all $\psi \in \scr{H}^{\otimes N}_+$. The
decomposition of $T$ from Equation (\ref{eq:3}) contains therefore only those summands with $s = N/2$.
This simplifies the optimization problem significantly, since the number of variables needed to
parametrize all relevant cloning maps according to Equation (\ref{eq:23}) is reduced 
from 3 to 2. A more detailed (and non-trivial) analysis shows that the maximum for $\scr{F}_{c,1}$ and
$\scr{F}_{c,\all}$ is attained if all terms in (\ref{eq:23}) except the one with $s=N/2$, $j=N/2$ and
$l=(M-N)/2$ vanish. The precise result is stated in the following theorem (\Cite{GiMa97,BDiVEFMS,BEM98}
for qubits and \Cite{Klo,Klo2} for general $d$).        

\begin{thm} \label{thm:11}
  For each $\scr{H} = \Bbb{C}^d$ both figures of merit $\scr{F}_{c,1}$ and $\scr{F}_{c,\all}$ are
  maximized by the cloner 
  \begin{equation} \label{eq:36}
   \hat{T}^*(\rho) = \frac{d[N]}{d[M]} S_M (\rho \otimes \Bbb{1}) S_M
  \end{equation}
  where $d[N]$, $d[M]$ denote the dimensions of the symmetric tensor products $\scr{H}^{\otimes N}_+$
  respectively $\scr{H}^{\otimes M}_+$ and $S_M$ is the projection from $\scr{H}^{\otimes M}$ to $\scr{H}^{\otimes
    M}_+$. This implies for the optimal fidelities 
  \begin{equation}
    \scr{F}_{c,1}(N,M) = \frac{d-1}{d}\frac{N}{N+d}\frac{M+d}{M}
  \end{equation}
  and 
  \begin{equation}
    \scr{F}_{c,\all}(N,M) = \frac{d[N]}{d[M]}.
  \end{equation}
  $\hat{T}$ is the \emph{unique} solution for both optimization problems, i.e. there is no other
  operation $T$ of the form (\ref{eq:114}) which maximizes $\scr{F}_{c,1}$ or $\scr{F}_{c,\all}$.  
\end{thm}

There are two aspects of this result which deserve special attention. One is the relation to state
estimation which is postponed to Subsection \ref{sec:estim-pure-stat}. The second concerns the role of
correlations: It does not matter whether we are looking for the quality of each single clone
($\scr{F}_{c,1}$) only, or whether correlations are taken into account ($\scr{F}_{c,\all}$). In both cases
we get the same optimal solution. This is a special feature of pure states, however. Although there are
no concrete results for quantum systems, it can be checked quite easily in the classical case that
considering correlations changes the optimal cloner for arbitrary mixed states drastically.  

\subsection{Purification}
\label{sec:purification}

\begin{figure}[b]
  \begin{center}
    \begin{pspicture}(15,9)
    \rput(7.5,5){\includegraphics[scale=0.8]{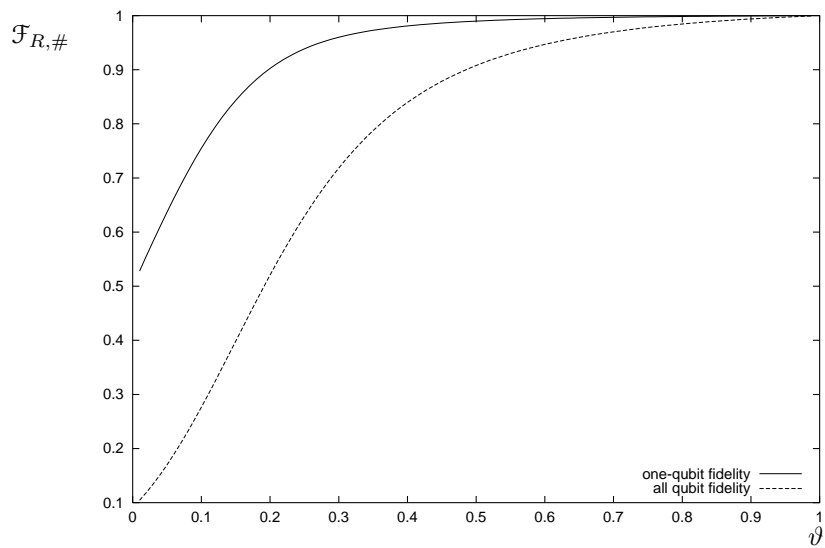}}
    \rput[r](13.5,0.5){$\vartheta$}
    \rput[tl](0,9){$\scr{F}_{R,\#}$}
  \end{pspicture}
    \caption{One- and all-qubit fidelities of the optimal purifier for $N=100$ and $M=10$. Plotted as a
      function of the noise parameter $\vartheta$.}
    \label{fig:purfid1}
  \end{center}
\end{figure}

To find an optimal purification device, i.e. maximizing $\scr{F}_{R,\#}$, is more difficult then the
cloning problem, because the simplification from Equation (\ref{eq:114}) does not apply. Hence we have to
consider all the summands in the direct sum decomposition of $T$ from Equation (\ref{eq:23}) and
solutions are available only for qubits. Therefore we will assume for the rest of this subsection that
$\scr{H} = \Bbb{C}^2$ holds. The $\SU(2)$ symmetry of the problem allows us to assume
without loss of generality that the pure initial  state $\psi$ coincides with one of the basis
vectors. Hence we get for the (noisy) input states of the purifier 
\begin{align}
  \rho(\beta) &=\frac{1}{2\cosh(\beta)} \exp\left(2\beta \frac{\sigma_3}{2}\right) = \frac{1}{e^\beta+e^{-\beta}}
\left(\begin{array}{ll}e^\beta&0\\0&e^{-\beta} \end{array}\right)  \\
  &=\tanh(\beta)\kb{\psi} +(1-\tanh(\beta)){1\over2}\idty, \quad \psi = \ket{0}
\end{align} 
The parameterization of $\rho$ in terms of the ``pseudo-temperature'' $\beta$ is chosen here, because it
simplifies some calculations significantly (as we will see soon). The relation to the form of $\rho = R^*\sigma$
initially given in Equation (\ref{eq:20}) is obviously $\vartheta = \tanh(\beta)$.

To state the main result of this subsection we have to decompose the product state $\rho(\beta)^{\otimes N}$ into
spin-$s$ components. This can be done in terms of Equation (\ref{eq:22}). $\rho(\beta)$ is not unitary 
of course. However we can apply (\ref{eq:22}) by analytic continuation, i.e. we treat $\rho(\beta)$ in the same
way as we would $\exp\left(i\beta \sigma_3\right)$. It is then straightforward to get 
\begin{equation} \label{eq:119}
  \rho(\beta)^{\otimes N}  = \bigoplus_s w_N(s) \rho_s(\beta) \otimes \frac{\Bbb{1}}{\dim \scr{K}_{N,s}},
\end{equation} with
\begin{equation} \label{eq:120}
   w_N(s) = \frac{\sinh\bigl((2s+1)\beta\bigr)}{\sinh(\beta)(2\cosh(\beta))^N} \dim \scr{K}_{N,s},
\end{equation}
 and
\begin{displaymath}
   \rho_s(\beta) = \frac{\sinh(\beta)}{\sinh\bigl((2s+1)\beta\bigr)} \exp(2\beta L_3^{(s)});
\end{displaymath}
where $L_3^{(s)}$ is the 3-component of angular momentum in the spin-$s$ representation and the dimension
of $\scr{K}_{N,s}$ is given in Equation (\ref{eq:33}). By (\ref{eq:35}) the representation space of $\pi_s$
coincides with the symmetric tensor product $\scr{H}^{2s}_+$. Hence we can interpret $\rho_s(\beta)$ as a state
of $2s$ (indistinguishable) particles. In other words the decomposition of $\rho(\beta)^{\otimes   N}$ leads in a
natural way to a family of operations 
\begin{equation}
  Q_s: \scr{B}(\scr{H}^{\otimes 2s}_+) \to \scr{B}(\scr{H}^{\otimes N}),\ \text{with}\ Q^*_s\bigl[\rho(\beta)^{\otimes
    N}\bigr] = \rho_s(\beta).
\end{equation}
We can think of the family $Q_s$, of operations as an instrument $Q$ which measures the number
of output systems and transforms $\rho(\beta)^{\otimes N}$ to the appropriate $\rho_s(\beta)$. The crucial point is now that
the purity of $\rho_s(\beta)$, measured in terms of fidelities with respect to $\psi$ increases provided $ s > 1/2$
holds.
Hence we can think of $Q$ as a purifier which arises naturally by reduction to irreducible spin
components \Cite{CEM99}. Unfortunately $Q$ does not produce a fixed number of output systems. The most
obvious way to construct a device which produces always the same number $M$ of outputs is to run the
optimal $2s \to M$ cloner $\hat{T}_{2s \to M}$ if $2s < M$ or to drop $2s - M$ particles if $M \leq 2s$
holds. More precisely we can define $\hat{Q}: \scr{B}(\scr{H}^{\otimes M}) \to \scr{B}(\scr{H}^{\otimes N})$ by
\begin{equation} \label{eq:43}
  \hat{Q}^*\bigl[\rho(\beta)^{\otimes N}\bigr] = \sum_s w_N(s) \hat{T}^*_{2s \to M}\bigr[\rho_s(\beta)\bigl],
\end{equation}
with 
\begin{equation}
  \hat{T}^*_{2s \to M}(\rho) = 
  \begin{cases}
    \frac{d[2s]}{d[M]} S_M (\rho \otimes \Bbb{1}) S_M & \text{for}\ M > 2s\\
    \tr_{2s-M} \rho & \text{for}\ M \leq 2s.
  \end{cases}
\end{equation}
$\tr_{2s-M}$ denotes here the partial trace over the $2s-M$ first tensor factors. Applying the general
scheme of Subsection \ref{sec:group-repr} shows that this is the best way to get exactly $M$ purified
qubits \Cite{pur}:

\begin{figure}[t]
  \begin{center}
    \begin{pspicture}(15,9)
    \rput(7.5,5){\includegraphics[scale=0.8]{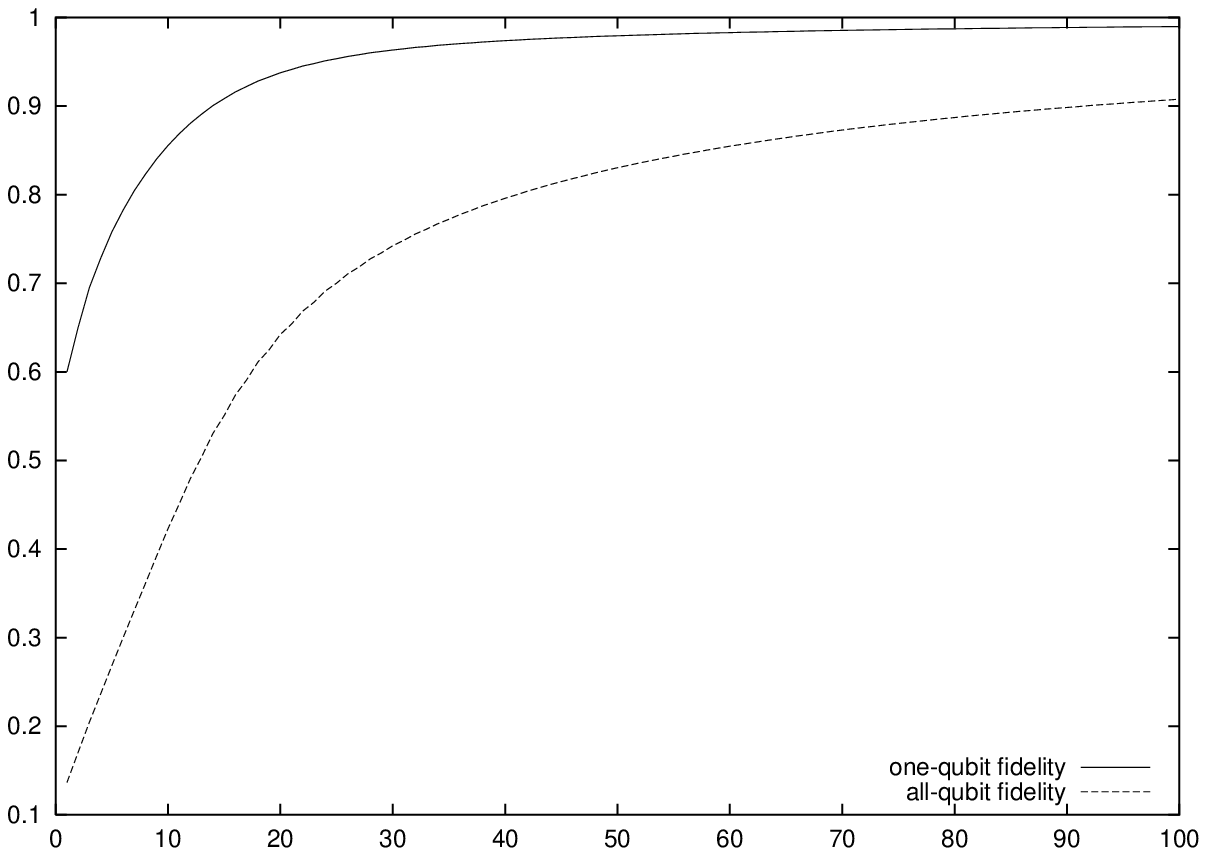}}
    \rput[r](13.5,0.5){$N$}
    \rput[tl](0,9){$\scr{F}_{R,\#}$}
  \end{pspicture}
    \caption{One- and all-qubit fidelities of the optimal purifier for $\vartheta=0.5$ and $M=10$. Plotted as a
      function of $N$.}
    \label{fig:purfid2}
  \end{center}
\end{figure}

\begin{thm}
  The operation $\hat{Q}$ defined in Equation (\ref{eq:43}) maximizes $\scr{F}_{R,1}$ and
  $\scr{F}_{R,\all}$. It is called therefore the \emph{optimal purifier}. The maximal values for
  $\scr{F}_{R,1}$ and $\scr{F}_{R,\all}$ are given by
  \begin{equation}
    \scr{F}_{R,1}(N,M) = \sum_s w_N(s) f_1(M,\beta,s),\ \scr{F}_{R,\all}(N,M) = \sum_s w_N(s) f_\all(M,\beta,s)
  \end{equation}
  with
  \begin{multline}  
    2 f_1(M,\beta,s) - 1 = \\
    =\begin{cases}
      \displaystyle {2s+1\over 2s}\coth\bigl((2s+1)\beta\bigr)-{1\over
        2s}\coth \beta & \mbox{for $2s > M$} \\[10pt]
      \displaystyle \frac{1}{2s+2} \frac{M+2}{M}
      \Bigl((2s+1)\coth\bigl((2s+1)\beta\bigr) - \coth\beta\Bigr) &
      \mbox{for $2s \leq M$.}
    \end{cases}
  \end{multline} 
  and
  \begin{multline} 
    f_{\rm all}(M,\beta,s) =
    \begin{cases}
      \displaystyle \frac{2s+1}{M+1}\
      \frac{1-e^{-2\beta}}{1-e^{-(4s+2)\beta}}
      & \mbox{$M \leq 2s$} \\[10pt]
      \displaystyle \frac{1-e^{-2\beta}}{1-e^{-(4s+2)\beta}} {2s \choose
        M}^{-1}\sum_K{K\choose M} e^{2\beta(K-s)} & \mbox{$M > 2s$.}
    \end{cases}
  \end{multline}
\end{thm}

\begin{figure}[t]
  \begin{center}
    \begin{pspicture}(15,9)
    \rput(7.5,5){\includegraphics[scale=0.8]{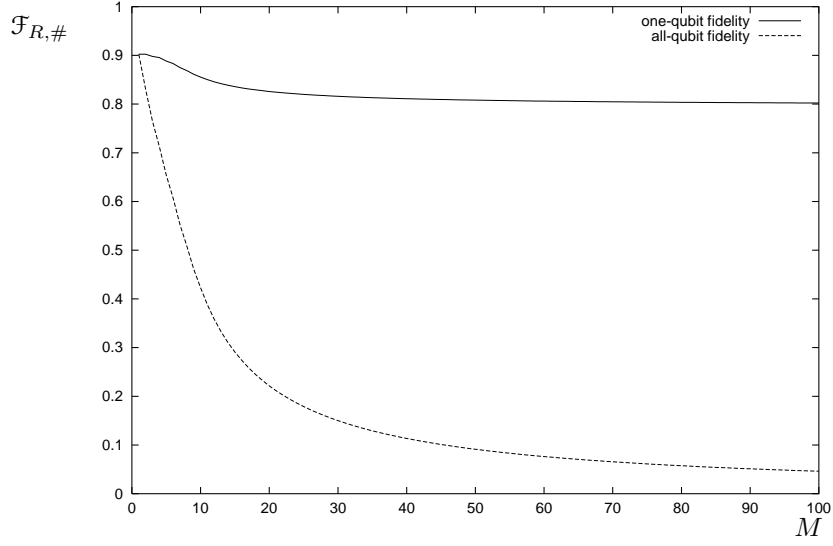}}
    \rput[r](13.5,0.5){$M$}
    \rput[tl](0,9){$\scr{F}_{R,\#}$}
  \end{pspicture}
    \caption{One- and all-qubit fidelities of the optimal purifier for $\vartheta=0.5$ and $N=10$. Plotted as a
      function of $M$.}
    \label{fig:purfid3}
  \end{center}
\end{figure}

The expression for the optimal fidelities given here look rather complicated and are not very
illuminating. We have plotted there both quantities as a function of $\vartheta$ (Figure \ref{fig:purfid1}) of
$N$ (Figure \ref{fig:purfid2}) and $M$ (Figure \ref{fig:purfid3}). While the first two plots looks quite
similar the functional behavior in dependence of $M$ seems to be very different. The study of the
asymptotic behavior in the next Section will give a precise analysis of this observation. 

\subsection{Estimating pure states}
\label{sec:estim-pure-stat}

We have already seen in Section \ref{sec:estimating-copying} that the cloning problem and state
estimation are closely related, because we can construct an approximate cloner $T$ from an estimator $E$
simply by running $E$ on the $N$ input states, and preparing $M$ systems according to the attained
classical information. In this section we want to go the other way round and show that the optimal cloner
derived in Theorem \ref{thm:11} leads immediately to an optimal pure state estimator; cf. \Cite{Bruss99}.

To this end let us assume that $E$ has the form (cf. Section \ref{sec:estimating-copying})
\begin{equation}
  \scr{C}(X) \ni f \mapsto E(f) = \sum_{\sigma \in X} f(\sigma) E_\sigma \in\scr{B}(\scr{H}^{\otimes N})
\end{equation}
where $X \subset \scr{B}^*(\scr{H})$ is a finite set\footnote{The generalization of the following
  considerations to continuous sets and a measure theoretic setup is straightforward and does not lead to
  a different result; i.e. we can not improve the estimation quality with continuous observables.} of
pure states. The quality of $E$ can be measured in analogy to Subsection \ref{sec:figures-merit} by a
fidelity-like quantity
\begin{equation}
  \scr{F}_s(E) = \inf_{\psi \in \scr{H}} \langle\psi, \rho_\psi \psi\rangle = \inf_{\psi \in \scr{H}} \sum_{\sigma \in X} \langle\psi^{\otimes N}, E_\sigma\psi^{\otimes N}\rangle \langle\psi, \sigma\psi\rangle
\end{equation}
where $\rho_\psi = \sum_\sigma \langle\psi^{\otimes N}, E_\sigma\psi^{\otimes n}\rangle \sigma$ is the (density matrix valued) expectation value of $E$ and the
infimum is taken over all pure states $\psi$. Hence $\scr{F}_s(E)$ measures the worst fidelity of $\rho_\psi$ with
respect to the input state $\psi$. If we construct now a cloner $T_E$ from $E$ by 
\begin{equation} \label{eq:113}
  T_E^*(\kb{\psi}^{\otimes N}) = \sum_\sigma \langle\psi^{\otimes N}, E_\sigma\psi^{\otimes n}\rangle \sigma^{\otimes M}
\end{equation}
 its one-particle fidelity $\scr{F}_{c,1}(T_E)$ coincides obviously with $\scr{F}_s(E)$. Since we can
 produce in this way arbitrary many clones of the same quality we see that $\scr{F}_s(E)$ is smaller than
 $\scr{F}_{c,1}(N,M)$ for all $M$ and therefore 
\begin{equation}
  \scr{F}_s(E) \leq \scr{F}_{c,1}(N,\infty) = \lim_{M \to \infty} \scr{F}_{c,1}(N,M) = \frac{d-1}{d}\frac{N}{N+d}
\end{equation}
where we can look at $\scr{F}_{c,1}(N,\infty)$ as the optimal quality of a cloner which produces arbitrary
many outputs from $N$ input systems.

To see that this bound can be saturated consider an asymptotically exact family
  \begin{equation} \label{eq:115}
  \scr{C}(X_M) \ni f \mapsto E^M(f) = \sum_{\sigma \in X} f(\sigma) E_\sigma^M \in\scr{B}(\scr{H}^{\otimes M}),\ X_M \subset \scr{S}(\scr{H})
\end{equation}
of estimators, i.e. the error probabilities (\ref{eq:50}) vanish in the limit $N \to \infty$. If the $E^M_\sigma \in
\scr{B}(\scr{H}^{\otimes M})$ are pure tensor products (i.e. the $E^M$ are realized by a ``quorum'' of
observables as described in Subsection \ref{sec:quant-state-estim}) they can not distinguish between the
output state $\hat{T}^*(\rho^{\otimes N})$ (which is highly correlated) and the pure product state
$\tilde{\rho}^{\otimes M}$ where $\tilde{\rho} \in \scr{B}^*(\scr{H})$ denotes the partial trace over $M-1$ tensor
factors (due to permutation invariance it does not matter which factors we trace away here). Hence if we
apply $E^M$ to the output of the optimal $N$ to $M$ cloner $\hat{T}_{N \to M}$ we get an estimate for
$\tilde{\rho}$ and in the limit $M \to \infty$ this estimate is exact. The fidelity $\langle\psi,\tilde{\rho}\psi\rangle$ of $\tilde{\rho}$
with respect to the pure input state $\psi$ of $\hat{T}_{N \to M}$ coincides however with
$\scr{F}_{c,1}(N,M)$. Hence the composition of $\hat{T}_{N \to M}$ with $E^M$ converges\footnote{Basically
  convergence must be shown here. It follows however easily from the corresponding property of the
  $E^M$.} to an estimator $E$ with $\scr{F}_e(E) =  \scr{F}_{c,1}(N,\infty)$. We can rephrase this result
roughly in the from: ``producing infinitely many optimal clones of a pure state $\psi$ is the same as
estimating $\psi$ optimally''. 

\subsection{The UNOT gate}
\label{sec:unot-gate}

The discussion of the last subsection shows that the optimal cloner $\hat{T}_{N \to M}$
produces better clones then any estimation based scheme (as in Equation (\ref{eq:113})), as long as we
are interested only in \emph{finitely many} copies. Loosely speaking we can say that the detour via
classical information is wasteful and destroys too much quantum information. The same is true for the
optimal purifier: We can first run an estimator on the mixed input state $\rho(\beta)^{\otimes N}$, apply the inverse
$(R^*)^{-1}$ of the channel map to the attained classical data and reprepare arbitrary many purified
qubits accordingly. The quality of output systems attained this way is, however worse, than those of the
optimal purifier from Equation (\ref{eq:43}) as long as the number $M$ of output systems is finite; this
can be seen easily from Figure \ref{fig:purfid3}. In this sense the UNOT gate is a harder task than
cloning and purification, because there is \emph{no quantum operation} which performs better than the
estimation based strategy. The following theorem can be proved again with the group theoretical scheme from
Subsection \ref{sec:group-repr} \Cite{UNOT}.

\begin{thm}
  Let $\scr{H} = \Bbb{C}^2$. Among all channels $T: \scr{B}(\scr{H}) \to \scr{B}(\scr{H}^{\otimes N}_+)$ the
  estimation based scheme just described attains the biggest possible value for the fidelity
  $\scr{F}_{\theta,\#}$, namely
  \begin{equation}
    \scr{F}_{\theta,1}(N,1) = \scr{F}_{\theta,\all}(N,1) = 1 - \frac{1}{N+2}.
  \end{equation}
\end{thm}

The dependence on the number $M$ of outputs is not interesting here,. because the optimal device produces
arbitrary many copies of the same quality. 

\section{Asymptotic behaviour}
\label{sec:asymptotic-behaviour}

If a device, such as the optimal cloner, is given which produces $M$ output system from $N$ inputs it is
interesting to ask for the maximal rate, i.e. the maximal ratio $M(N)/N$ in the limt $N \to \infty$ such that
the asymptotic fidelity $\lim_{N \to \infty} \scr{F}\bigl(N,M(N)\bigr)$ is above a certain threshold (preferably
equal to one). Note that this type of question was very important as well for distillation of
entanglement and channel capacities, but almost not computable in there. In the current context this type
of question is somewhat easier to answer. This relies on the one hand on the group theoretical structure
presented in the last section and on the other on the close relation to quantum state estimation. We
start this section therefore with a look on some aspects of the asymptotics of  mixed state estimation.

\subsection{Estimating mixed state}
\label{sec:mixed-state-cloning}

If we do not know a priori that the input systems are in a pure state much less is known about estimating
and cloning. It is in particular almost impossible to say anything about optimality for finitely many input
systems (only if $N$ is very small e.g. \Cite{Vidal99}). Nevertheless some strong results are available for
the behavior in the limit $N \to \infty$ and we will give here a short review of some of them. 

One quantity, interesting to be analyzed for a family of estimators $E^N$ in the limit $N \to \infty$ is the
variance of the $E^N$. To state some results in this context it is convenient to parameterize the state
space $\scr{S}(\scr{H})$ or parts of it in terms of $n$ real parameters $x = (x_1,\ldots,x_n) = \Sigma \subset \Bbb{R}^n$
and to write $\rho(x)$ as the corresponding state. If we want to cover all states, one particular
parameterization is e.g. the generalized Bloch ball from Subsection \ref{sec:quantum-mechanics}. An
estimator taking $N$ input systems is now a (discrete) observable $E_x^N \in \scr{B}(\scr{H}^{\otimes N})$, $ x \in
X_N$ with values in a (finite) subset $X_N$ of $\Sigma$. The expectation value of $E^N$ in the state $\rho(x)^{\otimes
  N}$ is therefore the vector $\langle E^N\rangle_x$ with components $\langle E^N\rangle_x,j$, $j=1,\ldots,n$ given by
\begin{equation}
  \langle E^N\rangle_{x,j} = \sum_{y \in X_N} y_j \tr\bigl( E^N_y \rho(x)^{\otimes N} \bigr)
\end{equation}
and the \emph{mean quadratic error} is described by the matrix
\begin{equation}
  V_{jk}^N(x) = \sum_{y \in X_N} \bigl(\langle E_N\rangle_{x,j} - y_j\bigr)\bigl(\langle E_N\rangle_{x,k} - y_k\bigr) \tr\bigl( E^N_y \rho(x)^{\otimes N} \bigr).
\end{equation}
For a good estimation strategy we expect that $V_{jk}(x)$ decreases as $1/N$, i.e.
\begin{equation}
  V_{jk}^N(x) \simeq \frac{W_{jk}(x)}{N},
\end{equation}
where the \emph{scaled} mean quadratic error matrix $W_{jk}(x)$ does not depend on $N$. The task is now
to find bounds on this matrix. We will state here two results taken from \Cite{GillMassar00}. To this
end we need the \emph{Hellstr{\"o}m quantum information matrix}
\begin{equation}
  H_{jk}(x) = \tr \bigl[\rho(x) \frac{\lambda j(x)\lambda_k(x) - \lambda_k(x)\lambda_j(x)}{2}\bigr]
\end{equation}
which is defined in terms of \emph{symmetric logarithmic derivatives} $\lambda_j$, which in turn are implicitly
given by
\begin{equation}
  \frac{\partial\rho(x)}{\partial x_j} = \frac{\lambda_j(x)\rho(x) + \rho(x) \lambda_j(x)}{2}.
\end{equation}
Now we have the following theorem \Cite{GillMassar00}:

\begin{thm}
  Consider a family of estimators $E^N$, $N \in \Bbb{N}$ as described above such that the following
  conditions hold:
  \begin{enumerate}
  \item 
    The scaled mean quadratic error matrix $N V^N_{jk}(x)$ converges uniformly in $x$ to $W_{jk}(x)$ as
    $N \to \infty$.
  \item 
    $W_{jk}(x)$ is continuous at a point $x_0 = x$. 
  \item 
    $H_{jk}(x)$ and its derivatives are bounded in a neighborhood of $x_0$.
  \end{enumerate}
  Then we have
  \begin{equation}
    \tr\bigl[ H^{-1}(x_0) W^{-1}(x_0)\bigr] \leq (d-1)
  \end{equation}
\end{thm}

For qubits this bound can be attained by a particular estimation strategy which measures on each qubit
separately. We refer to \Cite{GillMassar00} for details. 

A second quantity interesting to study in the limit $N \to \infty$ is the error probability defined in Section
\ref{sec:estimating-copying}; cf. Equation (\ref{eq:50}). For a good estimation strategy it should go to
zero of course, an additional question, however, concerns the rate with which this happens. We will
review here a result from \Cite{KWEst} which concerns the subproblem of \emph{estimating the
  spectrum}. Hence we are looking now at a family of observables $E^N: \scr{C}(X_N) \to \scr{B}(\scr{H}^{\otimes
  N})$, $N \in \Bbb{N}$ taking their values in a finite subset $X_N$ of the set
\begin{equation}
  \Sigma = \{ (x_1,\ldots,x_d) \in \Bbb{R}^d \, | \, x_1 \geq \cdots \geq x_d \geq 0,\, \mbox{$\sum_j$} x_j = 1 \}
\end{equation}
of ordered spectra of density operators on $\scr{H} = \Bbb{C}^d$. Our aim is to determine the behavior
of the error probabilities (cf. Equation (\ref{eq:50})
\begin{equation} \label{eq:116}
  K_N(\Delta) = \sum_{x \in \Delta \cap X_N} \tr(E^N_x \rho^{\otimes N})
\end{equation}
in the limit $N \to \infty$. Following the general arguments in Subsection \ref{sec:covariant-operations} we can
restrict our attention here to covariant observables, i.e. we can assume without loss of cloning quality
that the $E^N_x$ commute with all permutation unitaries $V_p$, $p \in \Sym_N$ and all local unitaries $U^{\otimes N}$, $U
\in \U(d)$. If we restrict our attention in addition to projection valued measures, which is suggestive for
ruling out unnecessary fuzziness, we see that each $E^N_x$ must coincide with a (sum of) projections
$P_Y$ from $\scr{H}^{\otimes N}$ onto the $\U(d)$ respectively  $V_p$ invariant subspace $\scr{H}_Y \otimes
\scr{K}_Y$, which is defined in Equation (\ref{eq:109}), where $Y = (Y_1,\ldots,Y_d)$ refers here to Young frames with
$d$ rows and $N$ boxes. The only remaining freedom for the $E^N$ is the assignment $x(Y) \in \Sigma$ of Young
frames (and therefore projections $E_N$) to points in $\Sigma$.  Since the Young frames themselves have up to 
normalization the same structure as the elements of $\Sigma$, one possibility for $s(Y)$ is just $s(Y) =
Y/N$. Written as quantum to classical channel this is
\begin{equation} \label{eq:117}
  \scr{C}(X_N) \ni f \mapsto \sum_Y f(Y/N) P_Y \in  \scr{B}(\scr{H}^{\otimes N}), 
\end{equation}
where $X_N \subset \Sigma$ is the set of normalized Young frames, i.e. all $Y/N$ if $Y$ has $d$ rows and $N$
boxes. It turns out, somewhat surprisingly that this choice leads indeed to an asymptotically exact
estimation strategy with exponentially decaying error probability (\ref{eq:116}). The following theorem
can be proven with methods from the theory of large deviations:

\begin{thm} \label{thm:12}
  The family of estimators $E^N$, $N \in \Bbb{N}$ given in Equation (\ref{eq:117}) is \emph{asymptotically
    exact}, i.e. the error probabilities $K_N(\Delta)$ vanish in the limit $N \to \infty$ if $\Delta$ is a complement of
  a ball around the spectrum $r \in \Sigma$ of $\rho$. If $\Delta$ is a set (possibly containing $r$) whose interior
  is   dense in its closure we have the asymptotic estimate for $K_N(\Delta)$:
  \begin{equation} \label{eq:118}
    \lim_{N\to\infty} \frac{1}{N} \ln\;K_N(\Delta) = \inf_{s \in \Delta} I(s),
  \end{equation}
  where the ``rate function'' $I : \Sigma \to \Bbb{R}$ is just the relative entropy between the two probability
  vectors $s$ and $r$
  \begin{equation}
  I(s) = \sum_j s_j\left(\ln s_j - \ln r_j\right).
\end{equation}
\end{thm}

To make this statement more transparent, note that we can rephrase (\ref{eq:118}) as
\begin{equation}
  K_N(\Delta) \approx \exp\left(- N \inf_{s \in \Delta} I(s)\right).
\end{equation}
Since the rate function $I$ vanishes only for $s = r$ we see that the probability measures $K_N$ converge
(weakly) to a point measure concentrated at $r \in \Sigma$. The rate of this convergence is exponential and
measured exactly by the function $I$. 

\subsection{Purification and cloning}
\label{sec:purification-cloning}

Let us come back now to the discussion of purification started in Subsection \ref{sec:purification}
(consequently we have $\scr{H} = \Bbb{C}^2$ again). Our aim is now to calculate the fidelities
$\scr{F}_{R,\#}\bigl(N,M(N)\bigl)$ in the limit $N \to \infty$ for a sequence $M(N)$, $N \in \Bbb{N}$ such that
$M(N)/N$ converges to a value $c \in \Bbb{R}$. The crucial step to do this is the application of
Theorem \ref{thm:12}. The density matrices $\rho_s(\beta)$ from Equation (\ref{eq:119}) can be defined
alternatively by 
\begin{equation}
  \rho_s(\beta) \otimes \frac{\Bbb{1}}{\dim \scr{K}_{N,s}} = w_N(s)^{-1} P_s \rho(\beta)^{\otimes N} P_s,\quad w_N(s) =
  \tr\bigl( \rho(\beta)^{\otimes N} P_s \bigr) 
\end{equation}
where $P_s$ is the projection from $\scr{H}^{\otimes N}$ to $\scr{H}_s \otimes \scr{K}_{N,s}$. In other words $P_s$
is equal to $P_Y$ from Equation (\ref{eq:117}) if we apply the reparametrization 
\begin{equation}
  (Y_1,Y_2) \mapsto (s,N) = \bigl((Y_1-Y_2)/2,Y_1+Y_2\bigr).  
\end{equation}
In a similar way we can rewrite the set of ordered spectra by $\Sigma \ni (x_1,x_2) \mapsto x_1 - x_2 \in [0,1]$ and
$K_N(\Delta)$ becomes a measure on $[0,1]$ (i.e. $\Delta \subset [0,1]$):
\begin{equation}
  K_N(\Delta) = \sum_{2s/N \in \Delta} \tr\bigl(\rho(\beta)^{\otimes N} P_s\bigr) = \sum_{2s/N \in \Delta} w_N(s)
\end{equation}
and the sum
\begin{equation}
  \scr{F}_{R,\#}\bigl(N,M(N)\bigr) = \sum_s w_N(s) f_\#\bigl(M(N),\beta,s\bigr)
\end{equation}
can be rephrased as the integral of a function $[0,1] \ni x \mapsto \tilde{f}_\#(N,\beta,x) \in \Bbb{R}$ with respect to
this measure, provided $\tilde{f}_\#$ is related to $f_\#$ by $\tilde{f}_\#(N,\beta,2s/N) =
f_\#\bigl(M(N),\beta,s\bigr)$. According to Theorem \ref{thm:12} the $K_N$ converge to a point measure
concentrated at the ordered spectrum of $\rho(\beta)$; but the latter corresponds, according to the
reparametrization above, to the noise parameter $\vartheta = \tanh \beta$. Hence if the sequence of functions
$\tilde{f}_\#(N,\beta,\,\cdot\,)$ converges for $N \to \infty$ uniformly (or at least uniformly on a neighborhood of $\vartheta$)
to $\tilde{f}_\#(\beta,\,\cdot\,)$ we get
\begin{equation}
  \lim_{N \to \infty} \scr{F}\bigl(N,M(N)\bigr) = \lim_{N \to \infty} \sum_s \tilde{f}_\#\bigl(N,\beta,s\bigr) = \tilde{f}_\#(\beta,\vartheta)
\end{equation}
for the limit of the fidelities. A precise formulation of this idea leads to the following theorem \Cite{pur}

\begin{figure}[t]
  \begin{center}
    \begin{pspicture}(15,9)
    \rput(7.5,5){\includegraphics[scale=0.8]{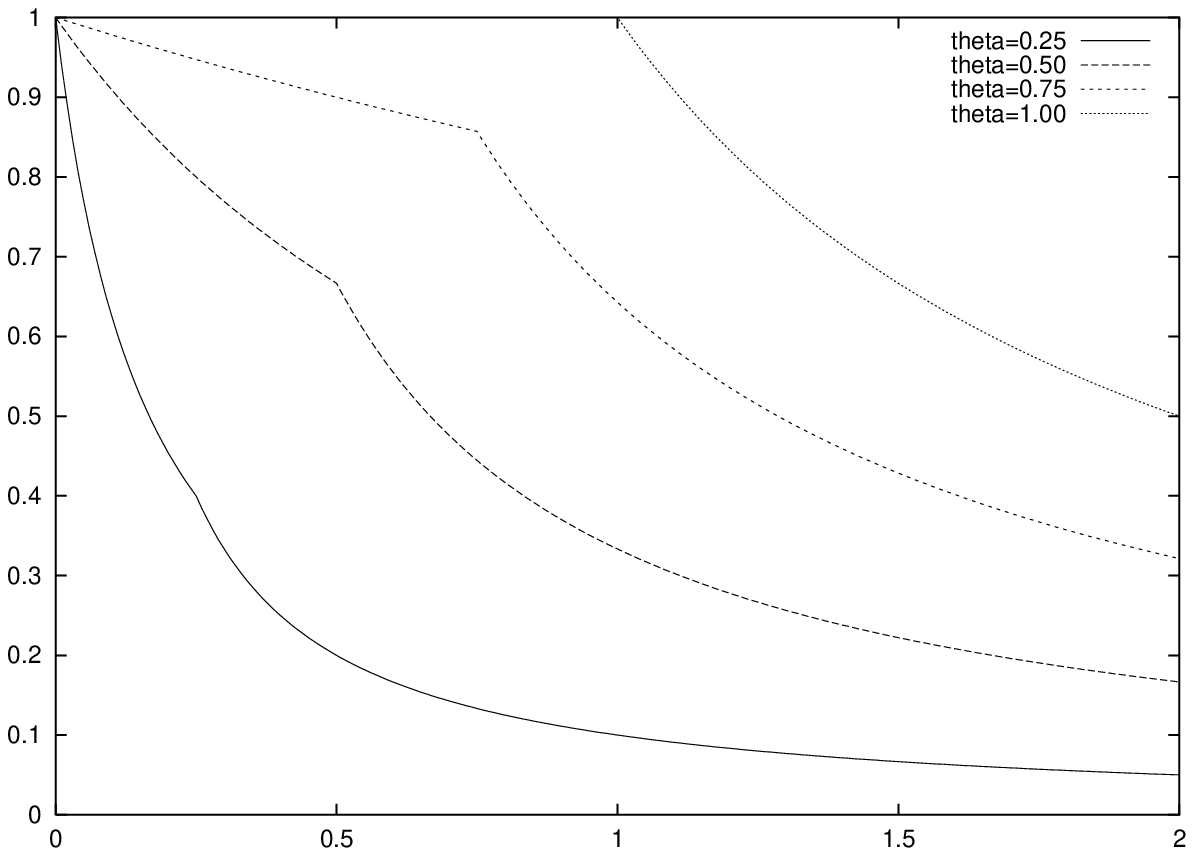}}
    \rput[r](13.5,0.5){$\mu$}
    \rput[tl](0,9){$\Phi(\mu)$}
  \end{pspicture}
    \caption{Asymptotic all-qubit fidelity $\Phi(\mu)$ plotted as function of the rate $\mu$.}
    \label{fig:purfid4}
  \end{center}
\end{figure}

\begin{thm}
  The two purification fidelities $\scr{F}_{R,\#}$ have the following limits
  \begin{equation}
    \lim_{N \to \infty} \lim_{M \to \infty} \scr{F}_{R,1}(N,M) = 1
  \end{equation}
  and
  \begin{equation}
    \Phi(\mu) = \lim_{N\to\infty\atop M/N\to\mu}\scr{F}_{R,\all}(N,M) = 
    \begin{cases}
      \displaystyle\frac{2\vartheta^2}{2\vartheta^2 + \mu(1-\vartheta)} &
      \mbox{if $\mu \leq \vartheta$}\\
      \displaystyle\frac{2\vartheta^2}{\mu(1+\vartheta)} &
      \mbox{if $\mu \geq \vartheta$.}\\
    \end{cases}
  \end{equation}
\end{thm}

If we are only interested in the quality of each qubit separately we can produce arbitrarily good
purified qubits at any rate. If on the other hand the correlations between the output systems should
vanish in the limit the rate is always zero. This can be seen from the function $\Phi$, which is the
asymptotic all-qubit fidelity which can be reached by a given rate $\mu$. We have plotted it in Figure
\ref{fig:purfid4}. Note finally that the results just stated contain the rates of optimal cloning
machines as a special case; we only have to set $\vartheta = 1$.



\pagestyle{special}
\bibliographystyle{mk}
\bibliography{qinf}

\end{document}